\documentclass[12pt,twoside,fleqn]{report}
\usepackage{epsfig}
\usepackage{axodraw}
\usepackage{thesis}
\usepackage{citesort}
\usepackage{ifthen}
\usepackage{fancyheadings}
\usepackage{palatino}
\renewcommand{\chaptermark}[1]{\markboth{#1}{#1}}

\def\lsim{\raise0.3ex\hbox{$\;<$\kern-0.75em\raise-1.1ex\hbox{$\sim\;$}}}
\def\gsim{\raise0.3ex\hbox{$\;>$\kern-0.75em\raise-1.1ex\hbox{$\sim\;$}}}
\newcommand{\slashed}[1]{\not\!#1}

\newcommand{\rb}[1]{\raisebox{1.5ex}[-1.5ex]{#1}}
\begin{document}
	\pagenumbering{arabic}
	\begin{titlepage}
\centering

%\vspace*{\fill}

\textbf{\huge CP Violation\\[3mm]
  in Production and Decay\\[4mm]
  of Supersymmetric Particles}

%\vspace{3cm}
\vspace{2cm}

Dissertation zur Erlangung des \\[.5em]
naturwissenschaftlichen Doktorgrades \\[.5em]
%der Bayerischen Julius-Maximilians-Universit\"at \\ [.5em]
%W\"urzburg \\[2.0cm]
der Bayerischen Julius-Maximilians-Universit\"at W\"urzburg \\[2.0cm]
vorgelegt von \\[.4cm]
{\large Olaf Kittel} \\[.2cm]
aus Erlangen \\[2.5cm]

%Betreuer der Dissertation\\
%\vspace{0.4cm}
%{\large Prof. Dr. rer. nat. Hans Fraas}\\
%\vspace{0.4cm}
%%%\setlength{\baselineskip}{0.7cm}
%am Institut f\"ur Theoretische Physik und Astrophysik\\[1cm]
%%%der Bayerischen Julius-Maximilians-Universit\"at\\
%%%W\"urzburg\\
\vspace*{\fill}

W\"urzburg \, 2004

\vspace{1cm}

\end{titlepage}

%\setcounter{page}{0}

%	\pagenumbering{arabic}
	\tableofcontents
	\markboth{Contents}{Contents}
\addcontentsline{toc}{chapter}{Prologue}
\markboth{Prologue}{Prologue}

{\phantom{Chapter 0}}

\vspace{1.8cm}

{ \huge\bf {\hfill} Prologue}\\
\vspace{3cm}

%{\Large\bf Supersymmetry and its discovery at colliders}\\
%{\Large\bf Supersymmetry and its search at colliders}\\
{\Large\bf Supersymmetry and the search for new particles}\\

\vspace{0.4cm}

Elementary particle physics has made enormous progress in the 
last decades. The electroweak and strong
interactions of the fundamental building blocks of matter,
the quarks and leptons, are now described by the so called
Standard Model (SM) of particle physics.

The development of the SM was possible due to
intensive  efforts and successful achievements
in experiment and theory.
On the one side, theoreticians have provided 
the physical models and the  mathematical techniques necessary 
to define and calculate observables. 
On the other side, experiments
with particle accelerators and detectors
%have not only allowed to test the models of the theoreticians, 
have not only allowed to find new fundamental particles,
but also high precision measurements 
%of particle interactions 
have made it possible to test the models.
%of the theoreticians.

However, 
%as successful as the SM is, 
%it is based on too many assumptions and 
%leaves many facts unexplained. 
there are general arguments which  point towards the
existence of a theory beyond the SM. One of the most
attractive candidates for such a more fundamental theory
is Supersymmetry (SUSY).
SUSY transformations change the spin of a particle field,
and thus bosonic and fermionic degrees of freedom get
related to each other. Therefore, new particles are 
predicted in SUSY models like in
the minimal supersymmetric extension of the SM.  
If SUSY is realized in nature, the supersymmetric partners 
of the SM particles  have to be discovered.

The next future colliders, like the 
Large Hadron Collider (LHC) at CERN or
a planned International Linear Collider (ILC),
%for electrons and positrons, 
are designed to find these particles.
Their properties will be measured with 
high precision in production and decay 
processes. The underlying physical model 
can then be determined by a comparison of 
experimental and phenomenological studies.

%\pagenumbering{arabic}

\chapter{Introduction}

\section{Motivation: Symmetries and models}

%\subsection{Models and symmetries}
Symmetries in physics have always played an important role in 
understanding the structure of the underlying theories. 
For instance, the existence of conservation laws can be 
explained by specific symmetry transformations under which a 
theory is invariant. Energy, momentum and angular momentum are 
conserved in field  theories with continuous 
spacetime symmetries. 

%High energy models of elementary particle interactions  have to be 
%formulated in a relativistic invariant way, called 
%Lorentz invariance. This invariance can be obtained by making the
%theory symmetric under transformations of the Poincar\'e group, 
%which are relativistic generalizations of the spacetime symmetries.
High energy models of elementary particle interactions  have to be 
invariant under the transformations of the Poincar\'e group, 
which are relativistic generalizations of the spacetime symmetries.
The interactions of the particles, the strong and 
electroweak forces, can be understood as a
consequence of the so called gauge symmetries.

But not only such continuous symmetries are crucial.
The discrete symmetries
\begin{itemize}
	\item Charge conjugation C: interchange of particles with antiparticles
\item Parity P: transformation of the space coordinates ${\bf x}\to{\bf -x}$ 
	\item Time reversal T: transformation of the parameter time $t\to{ -t}$
\end{itemize}
are also essential for the formulation of relativistic quantum field theories.
In particular, any Poincar\'e invariant local field theory has to be 
symmetric under the combined transformation of C, P and T, which is
called CPT invariance.

The discovery in the fifties that weak interactions violate
C and P maximally, was noted with substantial belief that a 
particle theory apparently still conserves the combined  
symmetry CP. However, a few years later, in 1964, 
CP violation was confirmed in the K-meson system. 

This lead to a powerful prediction in 1972.
The implementation of CP violation made it necessary to include a 
phase in the quark mixing matrix, 
the so called Kobayashi-Maskawa Matrix \cite{Kobayashi:1973fv}.
Such a phase is only CP violating if the matrix
is at least of dimension three, and thus a third generation of quarks and 
leptons was required. This prediction of a third family was made 
long before the final member of the second family, the charm quark, 
was found. 

Current experiments with B-mesons verify the existence
of one CP phase in the quark mixing matrix of the SM.
%the so called Kobayashi-Maskawa Matrix.
However, one phase alone cannot explain the observed baryon 
asymmetry of the universe, as shown in \cite{Csikor:1998eu}.
The fact that further sources of CP violation are needed
leads to the crucial prediction of CP violating phases in 
theories beyond the SM.
%Again it is the CP asymmetry, which 
%allows for crucial predictions. This time,
%it is the existence of further CP phases in theories 
%beyond the SM.

Apart from that, a further important symmetry of physics
was born in 1974: Supersymmetry (SUSY), which relates fermionic
and bosonic degrees of freedom. 
SUSY models, like the
the  Minimal Supersymmetric Standard Model (MSSM),
are one of the most attractive theories beyond the SM.
They give a natural solution to the hierarchy problem and
provide neutralinos as dark matter candidates. 
Furthermore, SUSY allows for grand unifications  
and for theories, which might include also gravity. 
%{\emph{maybe here some more words about SUSY and MSSM}}

%Therefore, the motivation of this wok is clear:
%The study of CP violation in the MSSM.

\section{CP violating phases and electric dipole moments
	\label{CP violating phases and electric dipole moments}}
The MSSM might 
%In the Minimal Supersymmetric Standard Model (MSSM) \cite{Haber-Kane} 
have several complex parameters, which cause 
CP violating effects. 
In the neutralino and chargino sector %of the MSSM,
these are the Higgsino mass parameter $\mu = |\mu|e^{i \varphi_{\mu}}$ 
and the $U(1)$ gaugino mass parameter 
$M_1 =|M_1|e^{i \varphi_{M_1}}$ \cite{Dugan:1984qf}. 
The $SU(2)$ gaugino mass parameter $M_2$  
can be made real by redefining the fields. 
In the sfermion sector of the MSSM, also the
trilinear scalar coupling parameter $A_f$ of the sfermion
$\tilde f$ can be complex, $A_f = |A_f| e^{i \varphi_{A_f}}$.
%Although SUSY particles have not yet been discovered,

\begin{figure}[h]
%	\fbox{
		\begin{minipage}{0.4\textwidth}
			\setlength{\unitlength}{0.035cm}
			\begin{picture}(110,80)(-60,-20)
				\ArrowLine(0,0)(25,0)
			\Vertex(25,0){2}
			\DashLine(25,0)(75,0){4}
			\Vertex(75,0){2}
			\Vertex(50,25){2}
			\ArrowLine(75,0)(100,0)
			\CArc(50,0)(25,0,180)
			\Photon(50,25)(50,50){4}{4}
			\Text(10,-12)[lb]{$e^-$}
			\Text(85,-12)[lb]{$e^-$}
			\Text(45,-15)[lb]{$\tilde\nu_e$}
			\Text(45,8)[lb]{$\tilde\chi^-_j$}
			\Text(35,35)[lb]{$\gamma$}
			\end{picture}
	\end{minipage}
	%	\fbox{
	\begin{minipage}{0.1\textwidth}
			\hspace*{2mm}
	\end{minipage}
%		\fbox{
		\begin{minipage}{0.4\textwidth}
		\setlength{\unitlength}{0.035cm}	
			\begin{picture}(110,80)(-40,-20)
			\ArrowLine(0,0)(25,0)
			\Vertex(25,0){2}
			\Line(25,0)(75,0)
			\Vertex(75,0){2}
			\Vertex(50,25){2}
			\ArrowLine(75,0)(100,0)
			\DashCArc(50,0)(25,0,180){5}
			\Photon(50,25)(50,50){4}{4}
			\Text(10,-12)[lb]{$e^-$}
			\Text(85,-12)[lb]{$e^-$}
			\Text(45,-18)[lb]{$\tilde\chi_j^0$}
			\Text(45,8)[lb]{$\tilde e_{1,2}^-$}
			\Text(35,35)[lb]{$\gamma$}
			\end{picture} 
	\end{minipage}
	\caption{SUSY contributions to the electron EDM.
			\label{EDM contributions}}
\end{figure}
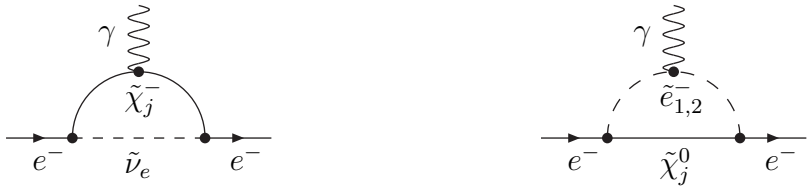

The CP violating phases are constrained by
electric dipole moments (EDMs)  \cite{Masiero:xj}
%The phases of complex parameters in the SM or in SUSY
%models are constrained or correlated by
%the experimental upper limits on the electric dipole
%moments (EDMs)  \cite{Masiero:xj}
%of electron $e$, neutron $n$,  $^{199}$Hg and $^{205}$Tl atoms,
%respectively:
%
%The limiting bounds of the  EDMs \cite{Masiero:xj}
of electron $e$, neutron $n$,  $^{199}$Hg and $^{205}$Tl atoms.
Their upper  bounds are, respectively:
\begin{eqnarray}
|d_e| &< &4.3\times10^{-27}~ e~{\rm cm}~ \cite{boundele},\\
|d_n| &< &6.3\times10^{-26}~ e~{\rm cm}~ \cite{boundneu},\\
|d_{Hg}| &<& 2.1\times10^{-28}~ e~{\rm cm}~ \cite{boundHg},\\ 
|d_{Tl}| &<& 1.3\times10^{-24}~ e~{\rm cm}~ \cite{boundTl}.
\end{eqnarray}
The CP phase in the quark sector of the SM give 
contributions to EDMs, which generally arise 
at two loop level, and respect    
%These SM contributions are too small to violate 
the  bounds of the  EDMs.
In SUSY models, however, neutralino and chargino 
contributions to the electron EDM can occur at one loop
level, see Fig.~\ref{EDM contributions}. 
For the neutron EDM in addition also  
gluino exchange contributions are present due to a
phase of the gluino mass parameter.
The phases of the SUSY parameters are thus constrained 
by the experimental upper limits of the EDMs. 
In the literature, three solutions are being proposed \cite{nath}:
%The usual consensus in the literature is, that one of 
%the following three conditions has to be realized \cite{nath}: 
\begin{itemize}
\item{
		The SUSY phases are severely 
		suppressed \cite{abeletal,smallphases}.} 
\item{
		SUSY particles of the first two generations are 
		rather heavy, with masses of the order of a 
		TeV \cite{Dimopoulos:1995mi}.} 
\item{
		There are strong cancellations between the different 
		SUSY contributions to the EDMs,
		allowing a SUSY particle spectrum of the order of a 
		few 100~GeV \cite{cancel,edms,BMPW}.}
\end{itemize}
%which can weaken the restriction on the phases. 
Due to such cancellations, for example, in the 
constrained MSSM \cite{edms},
the phase $\varphi_{M_1}$ is not restricted
%It is only correlated to 
but the phase of ${\mu}$ is still
constrained with $|\varphi_{\mu}|\lsim0.1\pi$ \cite{edms}.
If lepton flavor violating terms are 
included \cite{BMPW}, also the restriction on $\varphi_{\mu}$ 
may disappear.

The restrictions on the SUSY phases are thus very model dependent. 
Independent methods for their measurements are desirable,
in order to clarify the situation. In order to determine the 
phases unambiguously, measurements of CP sensitive observables 
are necessary. Such observables are non-zero only if CP is violated,
i.e. they are proportional to the sine of the phases.

%The goal of the thesis is to define such observables 
%and to calculate them in the MSSM for adequate
%supersymmetric processes. 
%Finally, we have to address the question, 
%whether the phases can be constrained in future electron-positron
%collisions at a linear collider (LC).
%Such a machine will operate at a center of mass energy of 
%500 GeV to 800 GeV. It will have high luminosity and an option 
%for longitudinally polarized beams. 
%An LC of this kind is thus an ideal tool for measuring the 
%properties of SUSY particles, like charginos and neutralinos, 
%with high precision.

\section{Methods for analyzing CP violating phases}

\subsection{Neutralino and chargino polarizations}

For measuring SUSY phases,
the study of neutralino and chargino  production
at an $e^+e^-$ linear collider with longitudinally polarized 
beams \cite{LC} will play an important role.
By measurements of the chargino masses and production cross sections,
a method has been developed \cite{choichargino,choigaiss,Kneur:1999nx}
to determine  $\cos(\varphi_{\mu})$,
in addition to the other parameters $M_2$, $|\mu|$ and $\tan\beta$
of the chargino sector. For  neutralino production 
%$e^+e^-\to\tilde\chi^0_i\tilde\chi^0_j$, 
analogous methods have been proposed
in~\cite{Kneur:1999nx,Choineutralinosystem,Choineutraiseofcross,Gounaris:2002pj} 
to determine also $\cos(\varphi_{M_1})$ and  $M_1$,
besides $\cos(\varphi_{\mu})$, $M_2$, $|\mu|$ and $\tan\beta$.

However, in order to determine also the sign of 
$\varphi_{\mu}$ and $\varphi_{M_1}$,
the transverse neutralino and chargino polarizations 
perpendicular to the production plane have to be taken into account 
\cite{choineutralino,choichargino,choigaiss}.
They are only present if there are CP violating phases
in the neutralino/chargino sector, 
%$\varphi_{\mu}$ and/or $\varphi_{M_1}$ 
and if a pair of different neutralinos/charginos is produced.
At tree level, their polarizations lead to 
triple-product asymmetries of the decay products
\cite{oshimoneut,oshimochargino}.
%which we define in the next section.
Energy distributions and polar angle distributions 
of the neutralinos and charginos or their decay products
are not CP sensitive at tree level, since they do not 
depend on the transverse neutralino or chargino polarizations,
see e.g. \cite{Bilenky:1985wu} for neutralino production.

In order to include %take into account
the particle polarizations in our calculations, 
we use the spin density matrix formalism of \cite{spinhaber}. 
For an introduction into this 
formalism and for our conventions and definitions used, 
see Appendices \ref{Neutralino production and decay matrices}
and \ref{Chargino production and decay matrices}.

\subsection{T-odd and CP-odd triple-product asymmetries}

The SUSY phases give rise to T-odd and CP-odd observables 
which involve triple products of momenta~\cite{tripleprods}.
%\cite{Choi:1999cc,Kizukuri:zp,Bartl:2002hi,Oshimo}.
They allow us to define various T and CP asymmetries which are
sensitive to the different SUSY phases.
On the one hand, these observables can be large because they
are present at tree level. On the other hand, they also allow a
determination of the sign of the phases, which is impossible
if only CP-even observables were studied.

We consider neutralino or chargino production
\begin{eqnarray}\label{intro:prod}
e^++e^- &\to& \tilde\chi_i+\tilde\chi_j
\end{eqnarray}
followed by the two-body decay of one 
neutralino or chargino into a SM 
particle $A$ (e.g. lepton or $W,Z$ boson)
and a SUSY particle $\tilde X$ (e.g. slepton or $\tilde\chi_1^0$):
 \begin{eqnarray}\label{intro:decay1}
\tilde\chi_i&\to& A+\tilde X.
\end{eqnarray}

The momenta of electron, chargino (or neutralino),
and particle $A$ define the triple product
\begin{eqnarray}\label{intro:triple1}
	{\mathcal T} &=& ({\bf p}_{e^-} \times {\bf p}_{\chi_i})
							\cdot {\bf p}_A,
\end{eqnarray}
%The triple product ${\mathcal T}$  
which is T-odd, i.e. changes sign under time reversal.
The T-odd asymmetry of the cross section $\sigma$ of 
production~(\ref{intro:prod}) and decay~(\ref{intro:decay1})
is then defined as 
\begin{eqnarray}\label{intro:AT1}
	 {\mathcal A}^{\rm T} &=& 
	 \frac{\sigma({\mathcal T}>0)-\sigma({\mathcal T}<0)}
	{\sigma({\mathcal T}>0)+\sigma({\mathcal T}<0)}.
\end{eqnarray}
%of the cross section $\sigma$ of production~(\ref{intro:prod}) 
%and decay~(\ref{intro:decay1}). 
The  asymmetry can be expressed by the angular 
distribution of particle $A$ 
\begin{eqnarray}\label{intro:ATN}
{\mathcal A}^{\rm T}
	 =\frac{\int^{0}_{1}   \frac{d\sigma}{d\cos\theta} d\cos\theta 
            -\int^{-1}_{0}\frac{d\sigma}{d\cos\theta} d\cos\theta}
          {\int^{0}_{1}   \frac{d\sigma}{d\cos\theta} d\cos\theta
				 +\int^{-1}_{0}\frac{d\sigma}{d\cos\theta} d\cos\theta}
		= \frac{N_+ - N_- }{N_+ + N_-},
\end{eqnarray}
where 
\begin{eqnarray}
\cos\theta&:=&\frac{{\bf p}_{e^-} \times {\bf p}_{\chi_i}}
{|{\bf p}_{e^-} \times {\bf p}_{\chi_i}|} 
\cdot \frac{{\bf p}_{A}}{|{\bf p}_{A}|},
\end{eqnarray}
and thus ${\mathcal A}^{\rm T}$ is the difference of the number of 
events with particle 
$A$ above $(N_+)$ and below $(N_-)$ the production plane, 
defined by ${\bf p}_{e^-} \times {\bf p}_{\chi_i}$,
normalized by the total number of events $N=N_++N_-$.

The T-odd asymmetry is not only sensitive to CP phases,
but also to absorptive contributions, which could enter 
via s-channel resonances or final state interactions at loop level.
%In order to eliminate the contributions from the absorptive parts
%Due to CPT invariance, the asymmetry 
%${\mathcal A}^{\rm T}$ is actually CP-odd, if absorptive phases
%are neglected. The absorptive contributions to ${\mathcal A}^{\rm T}$
%are due to s-channel resonances in the production process or
%final state interactions. 
%As they only arise at loop level, we will neglect them.  
Although the absorptive contributions are a higher order effect,
and thus expected to be small, they do not signal CP violation.  
However, they can be eliminated in the CP-odd asymmetry
\begin{eqnarray}\label{intro:ACP1}
{\mathcal A}^{\rm CP} &=& 
	\frac{1}{2}({\mathcal A}^{\rm T}-\bar{\mathcal A}^{\rm T}),
\end{eqnarray}
where $\bar{\mathcal A}^{\rm T}$ denotes the asymmetry for the
CP-conjugated process.
%However, in the following we will neglect absorptive contributions,
%as they only arise at loop level.

Note that the triple product ${\mathcal T}$~(\ref{intro:triple1})
%that these T or CP asymmetries 
requires the identification of the neutralino (or chargino) 
momentum ${\bf p}_{\chi_i}$, which could be reconstructed 
by measuring the decay of the other $\tilde\chi_j$.
%This can be done if the corresponding information from the decay
%of the other  $\tilde\chi^-_j$ on the opposite side is
%also available. This is the case if, for example, the 
%$\tilde\chi^-_j$ decays like 
%$\tilde\chi^-_j \to \tilde\chi_1^- Z^0$,
%$\tilde\chi^-_j \to \tilde\chi^0_1 W^-$ or
%$\tilde\chi^-_j \to \tilde\chi_1^- H_1^0$
%and $Z,W,H_1^0$ decay hadronically,
%$Z^0\to q\,\bar q $, $W^-\to q\,\bar q' $, $H_1^0\to b \,\bar b$.
Therefore, the masses of the neutralinos/charginos  
as well as the masses of their decay products have to be known.

To avoid the reconstruction of ${\bf p}_{\chi_i}$,
we can also define triple products
in which ${\bf p}_{\chi_i}$ is replaced by a momentum of
the decay products of particles A, if A is a $W$ or $Z$ boson, 
or $\tilde X$, if $\tilde X$ is a slepton.
In the first case, $A =W$ or $Z$ and $\tilde X =\tilde\chi_1^0$, 
the decay of the boson into two quarks
\begin{eqnarray}\label{intro:decay2}
A\to& q + q',
\end{eqnarray}
defines the triple product
\begin{eqnarray}
	{\mathcal T} &=& ({\bf p}_{e^-} \times {\bf p}_{q})
	\cdot {\bf p}_{q'}.
\end{eqnarray}
In the second case, $A =\ell$ and $\tilde X = \tilde\ell$, the decay 
\begin{eqnarray}\label{intro:decay3}
	\tilde X\to& \ell + \tilde\chi^0_1,
\end{eqnarray}
defines the triple product
\begin{eqnarray}
	{\mathcal T} &=& ({\bf p}_{e^-} \times {\bf p}_{A})
	\cdot {\bf p}_{\ell}.
\end{eqnarray}
These triple products define then corresponding 
T or CP asymmetries, which do not require the momentum reconstruction
of $\tilde\chi_i$. However, for these triple products
the leptons have to be distinguished by their energy 
distributions \cite{neut1}, and 
the quarks have to be distinguished by flavor 
tagging \cite{flavortaggingatLC,Aubert:2002rg,flavortagginginWdecays}. 

Triple-product asymmetries can also be defined and analyzed for
three-body decays of neutralinos 
\cite{oshimoneut,choineutralino,karldipl,karlneutralino} 
and charginos \cite{oshimochargino,karldipl,Wachter}.

\subsection{Statistical error and significance}

The T-odd and CP-odd asymmetries, as defined 
in~(\ref{intro:AT1}) and~(\ref{intro:ACP1}), 
could be measured in neutralino and chargino production
at future linear collider experiments,
and would allow us to determine the values of the SUSY phases.
In order to decide whether an asymmetry, and thus a CP phase 
can be measured, we have to calculate its statistical error.
Also, we have to consider the statistical significance of 
the asymmetry. 

The relative statistical error of
the asymmetry is given by 
\begin{equation}\label{errorofA}
	\delta {\mathcal A} := 
	\frac{\Delta {\mathcal A}}{|{\mathcal A}|} = 
	\frac{1}{|{\mathcal A}| \sqrt{N}},
\end{equation}
with the number of events 
$N={\mathcal L} \cdot\sigma$, where
${\mathcal L}$ is the integrated luminosity of the linear collider.
%and $\sigma$ the cross section. 
Formula~(\ref{errorofA}) follows from~(\ref{intro:ATN}),
%since the T-odd asymmetry is the difference of the number of events with
%particle $A$ above $(N_+)$ and below $(N_-)$ the production
%plane, normalized by the total number of events $N=N_++N_-$:
%\begin{eqnarray}
% {\mathcal A}^{\rm T}
%	&=& \frac{N_+ - N_- }{N_+ + N_-}.
% \end{eqnarray}
%The absolute statistical error of a number of events  $N$  
%is given by $\Delta N = \sqrt{N}$.
with the estimate 
$\Delta N_{\pm} = \sqrt{N_{\pm}}\approx\sqrt{N/2}$.

The statistical significance of the 
asymmetry~(\ref{intro:AT1}) is then defined as
\begin{equation}\label{significanceofAT}
	S = |{\mathcal A}^{\rm T}| \sqrt{{\mathcal L} \cdot\sigma}.
\end{equation}
For $S=1$, the asymmetry can be measured at the
$68\%$ confidence level (CL), for $S=1.96$ at the
$95\%$ CL, etc.
The significance for the CP-odd asymmetry~(\ref{intro:ACP1})
is given by
\begin{equation}\label{significanceofACP}
	S = |{\mathcal A}^{\rm CP}| \sqrt{2{\mathcal L} \cdot\sigma},
\end{equation}
since $\Delta {\mathcal A}^{\rm CP} =\Delta{\mathcal A}^{\rm
	T}/\sqrt{2}$, which follows 
from~(\ref{intro:ACP1}).

Also background and detector
simulations have to be taken into account to predict the 
expected accuracies for the asymmetries, see 
e.g. \cite{Aguilar-Saavedra}.
However, this would imply detailed Monte Carlo studies,
which is beyond the scope of the present work.

\section{Organization of the work}

The goal of the thesis is to 
%define CP sensitive observables 
analyze CP violating effects of MSSM phases 
in production and/or two-body decay processes of
neutralinos, charginos and sfermions. 
We will therefore define and calculate
T-odd  and CP-odd asymmetries
for the different supersymmetric processes.
%analyze the CP violating effects of the MSSM phases 

We study neutralino and chargino production in 
electron-positron collisions at a future 
linear collider (LC)
%Such a machine will operate at 
with a center of mass energy of 
$500$~GeV to $800$~GeV, high luminosity and
longitudinally polarized beams.
%. It will have high luminosity and an option 
%for longitudinally polarized beams. 
A LC of this kind is an ideal tool for measuring the 
properties of SUSY particles
%, like charginos and neutralinos, 
with high precision.

Finally, we address the question, whether the phases can 
be constrained at the LC. We thus calculate the statistical 
significances for measuring the asymmetries.
Our analyses will also have particular emphasis on the 
beam-polarization dependence of the asymmetries and 
cross sections.

In most of the numerical examples we  choose 
$\varphi_{M_1}=\pm \pi/2$, $\varphi_{\mu}=0$, 
which is allowed by the constraints from the electron and 
neutron EDMs. In order to show the full phase dependences 
of the asymmetries in some examples we study their $\varphi_{\mu}$ 
behavior in the whole $\varphi_{\mu}$ range, relaxing in this 
case the restrictions from the EDMs. 
This is justified e.g. in theories with lepton flavor 
violation \cite{BMPW}, where the constraints on $\varphi_{\mu}$ disappear.

\begin{itemize}
\item
Chapter~\ref{CP violation in production and decay of neutralinos} 
contains 
%the discussion of 
neutralino production $e^+ e^-\to\tilde\chi^0_i~\tilde\chi^0_j$ and decay: 
%in electron-positron collisions: 

\begin{itemize}
	\item
In Section~\ref{T odd asymmetries in neutralino production 
and decay into sleptons} 
we discuss neutralino decay into sleptons:
$\tilde\chi^0_i\to\ell~\tilde\ell$ for $\ell=e,\mu,\tau$.
\item
In Section~\ref{A CP asymmetry in neutralino production 
and decay into staus with tau polarization}
we discuss neutralino decay into a stau-tau pair:
$\tilde\chi^0_i\to\tau~\tilde\tau$,
including the $\tau$ polarization.
\item
In Section~\ref{CP observables in neutralino production and decay
	into the Z boson}
we discuss neutralino decay into a $Z$ boson:
$\tilde\chi^0_i\to\tilde\chi^0_n ~Z$.
\end{itemize}

\item
Chapter~\ref{CP violation in production and decay of charginos}  
deals with chargino production 
$e^+ e^-\to\tilde\chi^{\pm}_i~\tilde\chi^{\mp}_j$ and decay: 
%in electron-positron collisions: 

\begin{itemize}
	\item
	In Section~\ref{CP violation in chargino production and 
		decay into the sneutrino}  
	we discuss chargino decay into sneutrinos:
	$\tilde\chi^+_i\to \ell^+~\tilde\nu_{\ell},$ $\ell=e,\mu,\tau$.
	\item

	In Section~\ref{CP observables in chargino production 
		and decay into the W boson}
	we discuss chargino decay into a $W$ boson:
	$\tilde\chi^+_i\to \tilde\chi^0_n ~W^+$.
\end{itemize}

\item
In chapter~\ref{CP violation in sfermion decays} 
we analyze CP violation in the two-body decay chain of
a sfermion:
$ \tilde f\to f~\tilde\chi^0_j,~
 \tilde\chi^0_j \to  \tilde\chi^0_1~Z, ~
Z\to f~\bar f$.

\item
Chapter~\ref{Conclusions} contains a summary and conclusions.
\item
In the Appendices we give a short account on the MSSM,
with emphasis on its complex parameters. We discuss details of particle
kinematics and phase space and give the analytical formulae for
the production and decay amplitudes squared. Finally, 
we give useful spin-formulae for fermions and bosons 
and a formulary of our definitions and conventions,
%where we also define the SM input parameters 
used for our numerical calculations.

\end{itemize}

%\subsection{not a section:: loop-corr}
%
%from Z paper:
%
%Our numerical results presented below are obtained at tree level.
%One-loop corrections to $e^+e^-\to\tilde\chi^0_i \tilde\chi^0_j$
%have been given in \cite{HEPHY} for real MSSM parameters. They are
%of the order of a few percent and may reach values up to 10\%. As
%the bulk of the one-loop corrections are presumably CP-even, we
%expect that they will not significantly change our tree-level result
%for ${\mathcal A}_{f}$. For an appropriate analysis of the one-loop
%corrections to ${\mathcal A}_{f}$ it would be necessary to adopt
%the formulae of  \cite{HEPHY} to the case of complex MSSM
%parameters, which is beyond the scope of the present paper.

	\chapter{CP violation in production and decay of neutralinos
	\label{CP violation in production and decay of neutralinos}}

{\Large\bf Overview}\\
\vspace{0.3cm}

We study neutralino production with longitudinally 
polarized beams
$e^+\,e^- \to\tilde{\chi}^0_i \, \tilde{\chi}^0_j$
with the subsequent leptonic decay of one neutralino
$\tilde{\chi}^0_i \to \tilde{\ell} \, \ell$; 
$ \tilde{\ell} \to \tilde{\chi}^0_1 \, \ell$, for
$ \ell= e,\mu,\tau$, \cite{neut1}
or the decay into the $Z$ boson
$\tilde\chi^0_i \to \chi^0_n Z$; 
$Z \to \ell \bar\ell (q\bar q)$ \cite{neut2}.
These decay modes allow the definition of CP observables
which are sensitive to the phases $\varphi_{M_1}$
and $\varphi_{\mu}$.

For the leptonic decay of the neutralino into the tau
$\tilde{\chi}^0_i \to \tilde \tau^{\pm} \, \tau^{\mp}$,
we propose the transverse $\tau^{\mp}$ polarization as a CP
sensitive observable \cite{neut3}. 
This asymmetry is also sensitive to the phase $\varphi_{A_{\tau}}$.

We present numerical results for the asymmetries, cross sections
and branching ratios for a linear electron-positron collider 
in the 500 GeV - 800 GeV range.
The asymmetries can go up to 60\% and we  estimate the event rates 
which are necessary to observe the asymmetries. 
Polarized electron and positron beams 
can significantly enhance the asymmetries and cross sections.

\section{T-odd asymmetries in neutralino production and decay into sleptons
	\label{T odd asymmetries in neutralino production and decay into sleptons}}

%We study two triple-product asymmetries for
%neutralino production 
%$e^+\,e^- \to\tilde\chi^0_i \, \tilde\chi^0_j$
%and the subsequent leptonic decay of one neutralino
%$\tilde\chi^0_i \to \tilde\ell \, \ell$; 
%$ \tilde\ell \to \tilde\chi^0_1 \, \ell$ for
%$ \ell= e,\mu,\tau$.		
%We present numerical results for the asymmetries,
%which are sensitive to the phases $\varphi_{M_1}$
%and $\varphi_{\mu}$. 
%At a linear electron-positron collider in the 500 GeV range
%the asymmetries can go up to 25\%.
%Polarized electron and positron beams 
%can significantly enhance the asymmetries and cross sections.
%In addition, we show how the two
%decay leptons can be distinguished by making use of their energy
%distributions.

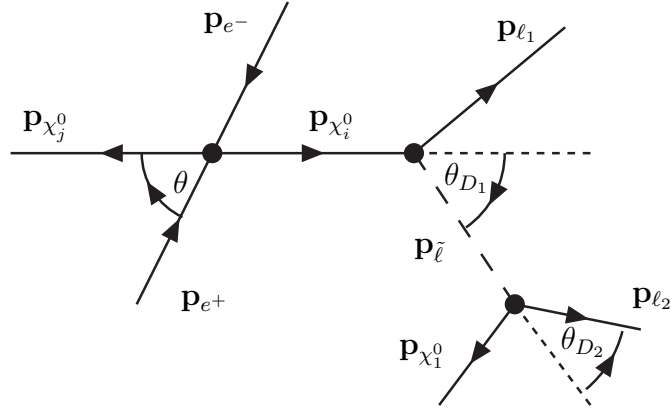
\begin{figure}[h]
%\fbox{
\begin{picture}(5,6.5)(-2,.5)
	\put(1,4.7){${\bf p}_{\chi_j^0}$}
   \put(3.4,6){${\bf p}_{e^- }$}
   \put(3,3.8){$\theta $}
   \put(3.1,2.3){${\bf p}_{e^+}$}
   \put(4.8,4.7){${\bf p}_{\chi_i^0}$}
   \put(7.3,5.9){${\bf p}_{\ell_1}$}
	\put(6.6,3.9){$ \theta_{D_1}$}
%	\put(7.2,4.6){$ \theta_{1}$}
   \put(6.2,3.){${\bf p}_{\tilde{ \ell} }$}
   \put(9.1,2.35){${\bf p}_{\ell_2}$}
   \put(6,1.65){${\bf p}_{\chi_1^0}$}
	\put(8.15,1.72){$\theta_{D_2} $}
%	\put(8.5,4.3){$ \phi_{D_1}$}
%	\put(9.,4.8){$ \phi_{1}$}
%	\put(9.8,1.3){$\phi_{2} $}
\end{picture}
\scalebox{1.9}{
\begin{picture}(0,0)(1.3,-0.25)
\ArrowLine(40,50)(0,50)
\Vertex(40,50){2}
\ArrowArcn(40,50)(14,245,180)
\ArrowLine(55,80)(40,50)
\ArrowLine(25,20)(40,50)
\ArrowLine(40,50)(80,50)
\ArrowLine(80,50)(110,75)
\DashLine(80,50)(115,50){1.5}
%\ArrowArc(115,50)(6, 65,255)
%\ArrowArc(115,50)(12,65,130)
%\ArrowArc(80,50)(22,0,40)
\ArrowArcn(80,50)(18,0,305)
\DashLine(80,50)(100,20){4}
\Vertex(80,50){2}
\ArrowLine(100,20)(125,15)
\ArrowArc(100,20)(22,310,346)
\ArrowLine(100,20)(85,0)
\DashLine(100,20)(115,0){1.5}
%\ArrowArc(115,0)(16,15,55)
\Vertex(100,20){2}
\end{picture}}
\caption{\label{picture neutralinos to sleptons}
Schematic picture of the neutralino production and decay process.}
\end{figure}

%\subsection{Introduction}

For neutralino production
\begin{eqnarray}
	e^+ + e^-&\to& \tilde{\chi}^0_i+\tilde{\chi}^0_j
  \label{production_neut} 
\end{eqnarray}
with longitudinally polarized beams and the subsequent leptonic 
two-body decay of one of the neutralinos
\begin{eqnarray}
   \tilde{\chi}^0_i&\to& \tilde{\ell} + \ell_1, 
  \label{decay_neut1} 
\end{eqnarray}
we introduce the triple-product 
 \begin{eqnarray}\label{AT1_neut}
	 {\mathcal T}_{I} &=& ({\bf p}_{e^-} \times {\bf p}_{\chi_i^0}) 
	 \cdot {\bf p}_{\ell_1},
 \end{eqnarray}
and define the T-odd asymmetry
 \begin{eqnarray}
	 {\mathcal A}_{I}^{\rm T} = \frac{\sigma_I({\mathcal T}_{I}>0)
						 -\sigma_I({\mathcal T}_{I}<0)}
							{\sigma_I({\mathcal T}_{I}>0)+
							\sigma_I({\mathcal T}_{I}<0)},
\label{TasymmetryI_neut}
\end{eqnarray}
where $\sigma_I$ is the cross section 
%(\ref{crossection_neut}) 
for reactions (\ref{production_neut}) and (\ref{decay_neut1}).

With the subsequent leptonic decay of the slepton
\begin{eqnarray}
  \tilde{\ell}&\to&\tilde{\chi}^0_1+\ell_2; \quad \ell= e,\mu,\tau,
  \label{decay_neut2}
\end{eqnarray}
%For neutralino production (\ref{production_neut}) and  the 
%two-body decay chain of the neutralino (\ref{decay_neut1}) and
%(\ref{decay_neut2}) 
%
we can construct a further asymmetry which does not require the 
identification of the neutralino momentum. We  replace the 
neutralino momentum ${\bf p}_{\chi_i^0}$ in~(\ref{AT1_neut}) by the 
momentum ${\bf p}_{\ell_2}$  of the lepton from the slepton decay 
%and define the triple product: 
 \begin{eqnarray}
{\mathcal T}_{II} &=& ({\bf p}_{e^-} \times {\bf p}_{\ell_2}) 
\cdot {\bf p}_{\ell_1}
	\label{AT2_neut}
\end{eqnarray}
and define the asymmetry
\begin{eqnarray}
	{\mathcal A}_{II}^{\rm T} = \frac{\sigma_{II}({\mathcal T}_{II}>0)
						 -\sigma_{II}({\mathcal T}_{II}<0)}
							{\sigma_{II}({\mathcal T}_{II}>0)+
							\sigma_{II}({\mathcal T}_{II}<0)},
\label{TasymmetryII_neut}
\end{eqnarray}
where $\sigma_{II}$ is  the cross 
section for reactions (\ref{production_neut}) -
(\ref{decay_neut2}).

%Due to CPT invariance these T-odd asymmetries are CP-odd 
%if the widths of the exchanged particles and final state interactions 
%are neglected, which is done in this work.

These T-odd observables in the production of 
neutralinos at  tree level are due to spin effects. 
Only if there are CP-violating phases $\varphi_{M_1}$
and $\varphi_{\mu}$ in the neutralino
sector and  if two different neutralinos are
%Only if $M_1$ and/or $\mu$ attain non-vanishing physical phases and 
produced, each of them has a polarization perpendicular to the production 
plane \cite{choineutralino,Choineutralinosystem,gudineutralino}. 
This polarization  leads to
asymmetries in the angular distributions of the decay products, 
as defined in~(\ref{TasymmetryI_neut}) and~(\ref{TasymmetryII_neut}).

\subsection{Cross sections
     \label{Cross sections}}

In order to calculate the production and decay amplitudes, 
%for the neutralino production and decay    cross sections
%$\sigma_{I}$ and $\sigma_{II}$, 
we use the spin density matrix formalism 
of~\cite{spinhaber,gudineutralino}, see 
Appendix~\ref{Neutralino production and decay matrices}.
For neutralino production (\ref{production_neut}) and 
decay (\ref{decay_neut1}), the amplitude squared can be written as
\begin{eqnarray}       \label{amplitude1_neut}
|T_I|^2 &=& |\Delta(\tilde{\chi}^{0}_i)|^2
	\sum_{\lambda_i \lambda_i'}~
	\rho_P(\tilde{\chi}^{0}_i)^{\lambda_i \lambda_i'}
	\rho_{D_1}(\tilde{\chi}^{0}_i)_{\lambda_i'\lambda_i},
\end{eqnarray}
with the neutralino propagator $\Delta(\tilde{\chi}^{0}_i)$, 
the spin-density matrix of neutralino 
production $\rho_P(\tilde{\chi}^{0}_i)$, 
the decay matrix $\rho_{D_1}(\tilde{\chi}^{0}_i)$, 
and the neutralino helicities $\lambda_i, \lambda_i'$.
Inserting the expansions of the density matrices 
$\rho_P(\tilde{\chi}^{0}_i)$, see~(\ref{neut:rhoP}),  
and $\rho_{D_1}(\tilde{\chi}^{0}_i)$, see~(\ref{neut:rhoD}),
into (\ref{amplitude1_neut}) gives
   \begin{eqnarray} \label{amplitude2}
		|T_I|^2 &=& 4~|\Delta(\tilde{\chi}^{0}_i)|^2~  
		( P D_1 + \vec \Sigma_P \vec\Sigma_{D_1} ).
\end{eqnarray}
Analogously, the amplitude squared for  the complete process
of neutralino production, followed by the two-body decays (\ref{decay_neut1})
and (\ref{decay_neut2}), can be written as
\begin{eqnarray}       \label{amplitude2_neut}
	|T_{II}|^2 &=& \phantom{4~}
	|\Delta(\tilde{\chi}^{0}_i)|^2~|\Delta(\tilde{\ell})|^2
	\sum_{\lambda_i \lambda_i'}~
	\rho_P(\tilde{\chi}^{0}_i)^{\lambda_i \lambda_i'}
	\rho_{D_1}(\tilde{\chi}^{0}_i)_{\lambda_i'\lambda_i}~D_2\\
	&=& 
4~|\Delta(\tilde{\chi}^{0}_i)|^2~  
		|\Delta(\tilde{\ell})|^2
	  ( P D_1 + \vec \Sigma_P \vec\Sigma_{D_1} )\;D_2,
\end{eqnarray}
where $D_2$ is the factor for the slepton decay, given in 
(\ref{sleptonD_2}).

The cross sections and distributions in the laboratory system are then 
obtained by integrating the squared amplitudes 
\begin{equation}\label{crossection_neut}
	d\sigma_{I,II}=\frac{1}{2 s}|T_{I,II}|^2 d{\rm Lips}_{I,II}
\end{equation}
over the Lorentz invariant phase space elements 
\begin{eqnarray}\label{lipsI,II}
	d{\rm Lips}_{I} &:=&
	d{\rm Lips}(s;p_{\chi_j^0},p_{\ell_1},p_{\tilde\ell}),\\
	d{\rm Lips}_{II} &:=&
	d{\rm Lips}(s;p_{\chi_j^0},p_{{\ell}_1},p_{{\ell}_2},p_{\chi_1^0}),
\end{eqnarray}
given in (\ref{Lipsleptonic1}) and (\ref{Lipsleptonic2}), respectively.

The contributions of the spin correlation terms 
$\vec \Sigma_P \vec\Sigma_{D_1} $  to the total cross section vanish.
Their contributions to the energy distributions of the leptons
$\ell_1$ and $\ell_2$ from decay (\ref{decay_neut1}) and (\ref{decay_neut2})
vanish due to the Majorana properties of the neutralinos 
\cite{pectovgudi} if CP is conserved. In our case of CP violation, 
they  vanish to leading order perturbation theory \cite{pectovgudi},
and thus the contributions can be neglected since they are
proportional to the widths of the exchanged particles.

\subsection{T-odd asymmetries
	\label{T-odd asymmetry}}

Inserting the cross sections (\ref{crossection_neut}) in the definitions of the
asymmetries (\ref{TasymmetryI_neut}) and (\ref{TasymmetryII_neut}) we obtain 
 \begin{eqnarray} \label{properties_neut}
	 {\mathcal A}_{I,II}^{\rm T}
	 = \frac{\int {\rm Sign}[{\mathcal T}_{I,II}]
		 |T|^2 d{\rm Lips}_{I,II}}
           {\int |T_{I,II}|^2 d{\rm Lips}_{I,II}}
	=  \frac{\int {\rm Sign}[{\mathcal T}_{I,II}]
	\Sigma_P^2 \Sigma_{D_1}^2 d{\rm Lips}_{I,II}}
          {\int  P D_1 d{\rm Lips}_{I,II}},
\end{eqnarray}
where we have used the narrow width approximation
for the propagators.
In the numerator only the spin-correlation term  
$\Sigma_P^2 \Sigma_{D_1}^2$ remains, since only
this term contains the triple products (\ref{AT1_neut}) or
(\ref{AT2_neut}). Thus, the contributions to 
${\mathcal A}_{I,II}^{\rm T}$ are directly proportional to 
the neutralino polarization $\Sigma_P^2$ perpendicular to the 
production plane.

In  case the neutralino decays into a scalar tau,
we take stau mixing into account and the asymmetries are reduced
due to their dependence on the 
$\tilde \chi^0_i$-$\tilde \tau_k$-$\tau$ couplings
\begin{eqnarray}\label{Amixing}
	 {\mathcal A}_{I,II}^{\rm T}\propto
	 \frac{|a^{\tilde \tau}_{ki}|^2-|b^{\tilde \tau}_{ki}|^2}
	 {|a^{\tilde \tau}_{ki}|^2+|b^{\tilde \tau}_{ki}|^2},
\end{eqnarray}
which can be seen from the expressions of $D_1$ and $\Sigma_{D_1}^2$,
given in Appendix~(\ref{Neutralino decay into sleptons}). Since the 
asymmetries are proportional to the absolute values of 
$a^{\tilde \tau}_{ki},b^{\tilde \tau}_{ki}$,
they are not sensitive to the phase $\varphi_{A_{\tau}}$ of $A_{\tau}$.
As an observable which is sensitive to $\varphi_{A_{\tau}}$, 
we will consider in 
Section~\ref{A CP asymmetry in neutralino production and decay into staus
		with tau polarization} an asymmetry which involves the transverse 
$\tau$ polarization.

\subsection{Numerical results
	\label{slep:Numerical results}}

We analyze 
%In the following numerical analysis we study 
%for $\sqrt{s} = 500$~GeV
%and longitudinally polarized beams with $P_-=0.8$ and $P_+=-0.6$, 
the dependence of the asymmetries
${\mathcal A}_I^{\rm T}$ and ${\mathcal A}_{II}^{\rm T}$,
the neutralino production cross sections 
$\sigma_P(e^+e^-\to\tilde{\chi}^0_i\tilde{\chi}^0_j)$ and 
the branching ratios ${\rm BR}(\tilde{\chi}^0_i\to \tilde{\ell} \ell)$ 
on the parameters
$\mu = |\mu| \, e^{ i\,\varphi_{\mu}}$, 
$M_1 = |M_1| \, e^{ i\,\varphi_{M_1}}$ and $M_2$
for $\tan \beta=10$.
%We choose a center of mass energy of $\sqrt{s} = 500$~GeV
%and longitudinal beam polarization $P_{e^-}=0.8$ and $P_{e^+}=-0.6$.
In order to reduce the number of parameters, we assume 
$|M_1|=5/3 M_2 \tan^2\theta_W $ and fix the universal scalar
mass parameter $m_0=100$ GeV. The renormalization group equations for  
the slepton masses are given in~(\ref{mselR}) and~(\ref{mselL}).
Since the pair production of equal  neutralinos is not CP sensitive,
we discuss the lightest pairs 
$\tilde{\chi}^0_1 \,\tilde{\chi}^0_2$,
$\tilde{\chi}^0_1 \,\tilde{\chi}^0_3$  and 
$\tilde{\chi}^0_2 \,\tilde{\chi}^0_3$,
for which we choose  a center of mass energy of $\sqrt{s} = 500$~GeV
and longitudinal beam polarization $P_{e^-}=0.8$ and $P_{e^+}=-0.6$.

%In most of our numerical examples below we have chosen 
%$\varphi_{M_1}=\pm \pi/2$, $\varphi_{\mu}=0$, 
%which agrees with the constraints from the electron and 
%neutron EDMs. In order to show the full phase dependences 
%of the asymmetries in some examples we study their $\varphi_{\mu}$ 
%behavior in the whole $\varphi_{\mu}$ range, relaxing in this 
%case the restrictions from the EDMs. 
%This is justified e.g. in theories with lepton flavor 
%violation \cite{BMPW}, where the constraints on $\varphi_{\mu}$ disappear.

\subsubsection{Production of $\tilde\chi^0_1 \, \tilde\chi^0_2$ }

\begin{figure}
 \begin{picture}(20,20)(0,-2)
	\put(2.5,16.5){\fbox{$\sigma_P(e^+e^- \to\tilde{\chi}^0_1 
			\tilde{\chi}^0_2)$ in fb}}
\put(0,9){\includegraphics{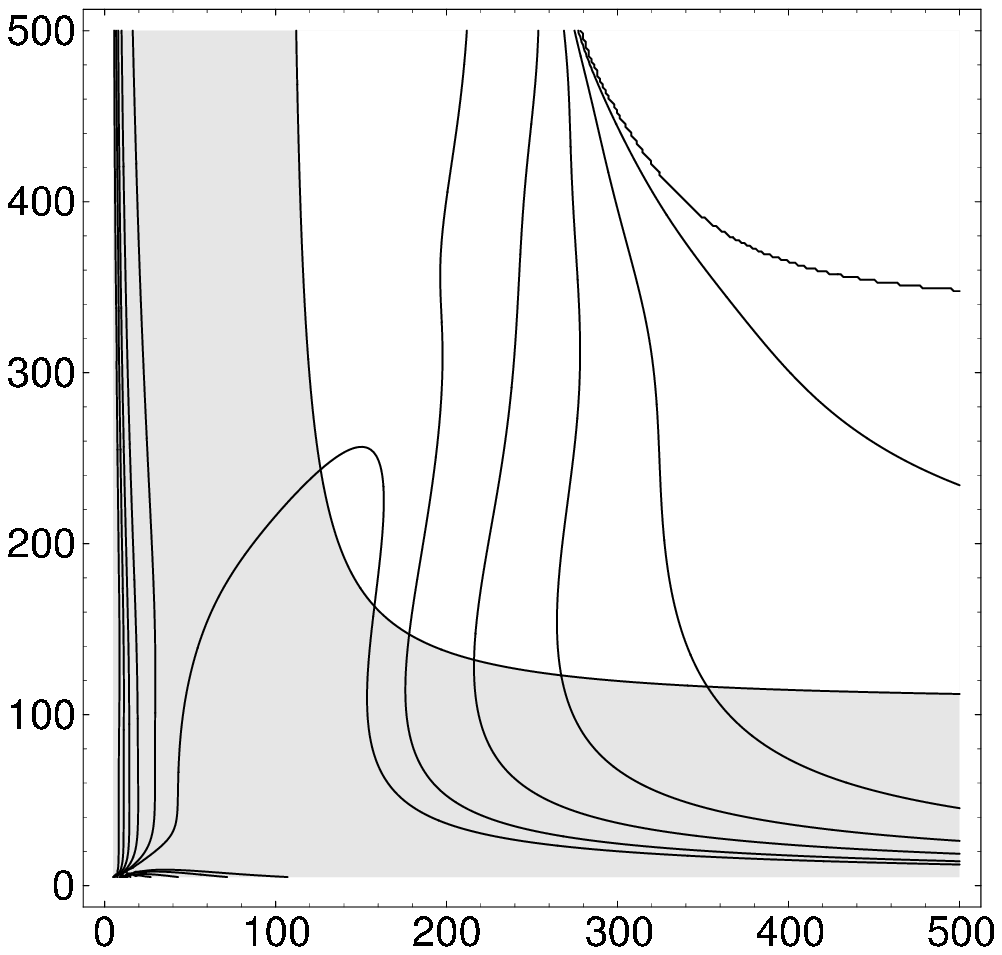}}
\put(5.5,8.8){$|\mu|~[{\rm GeV}]$}
   \put(0,16.3){$M_2~[{\rm GeV}]$ }
	\put(6,15){\begin{picture}(1,1)(0,0)
			\CArc(0,0)(6,0,380)
			\Text(0,0)[c]{{\scriptsize A}}
		\end{picture}}
\put(1.8,12.1){ \footnotesize$300$}
\put(2.6,13.5){\footnotesize$200$}
\put(3.2,13){\scriptsize$100$}
\put(3.7,12.5){\footnotesize$50$}
\put(4.3,12){\footnotesize$25$}
\put(5.9,12.5){\footnotesize$10$}
\put(0.5,8.8){Fig.~\ref{plots_12}a}
   \put(8,9){\includegraphics{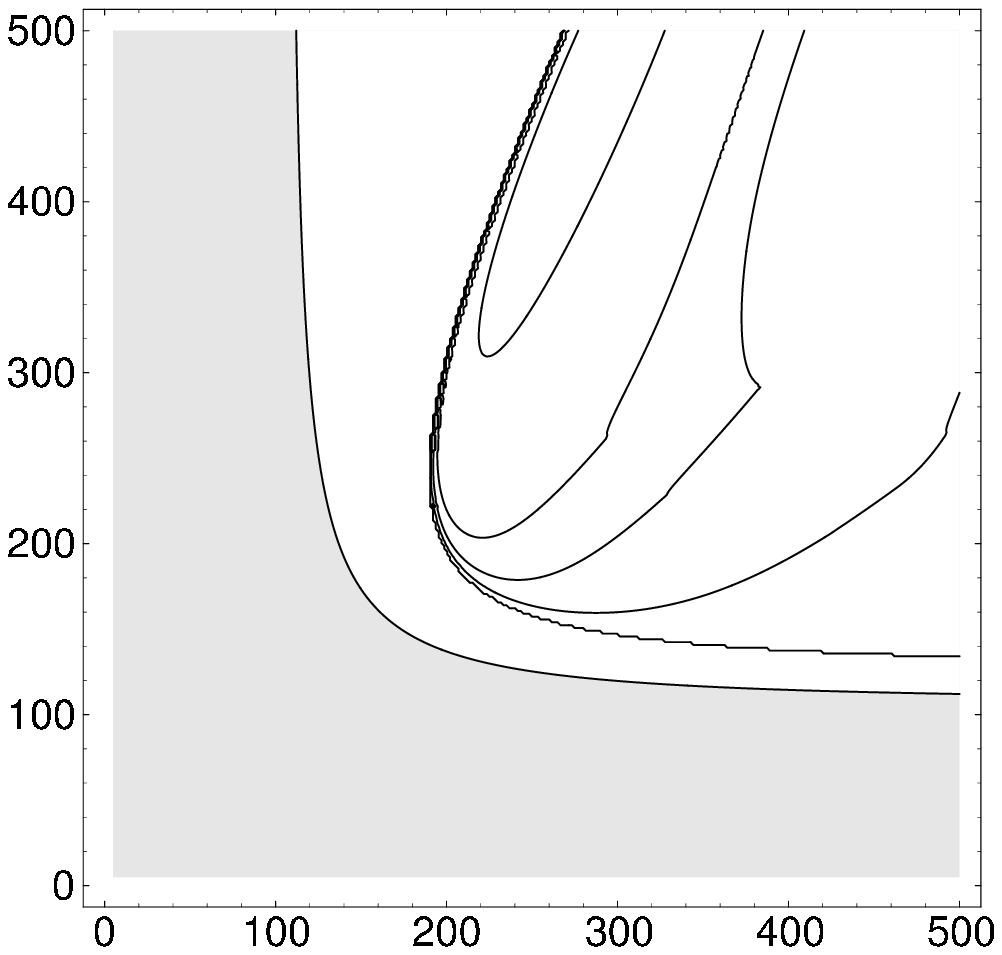}}
	\put(10.5,16.5){\fbox{${\rm BR}(\tilde{\chi}^0_2 \to \tilde{\ell}_R\ell_1)$ in \%}}
   \put(13.5,8.8){$|\mu|~[{\rm GeV}]$}
	\put(8,16.3){$M_2~[{\rm GeV}]$}
	\put(10.85,15){\begin{picture}(1,1)(0,0)
			\CArc(0,0)(6,0,380)
			\Text(0,0)[c]{{\scriptsize B}}
		\end{picture}}
\put(11.8,14.7){$64$}
\put(12.1,13.5){$40$}
\put(12.7,13){$20$}
\put(14,12.5){$4$}
\put(8.5,8.8){Fig.~\ref{plots_12}b}
	\put(0,0){\includegraphics{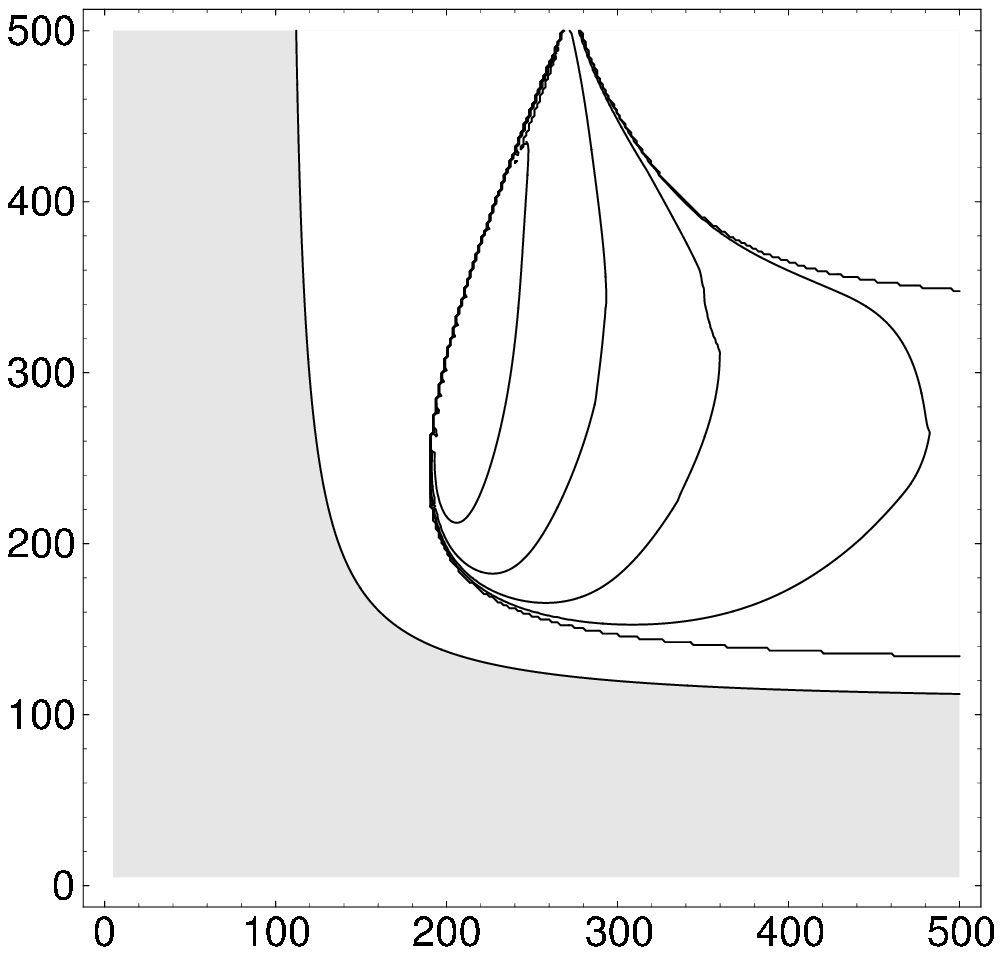}}
	\put(2.,7.5){\fbox{$\sigma(e^+e^- \to\tilde{\chi}^0_1
			\tilde{\chi}^0_1 \ell_1 \ell_2 )$ in fb}}
	\put(5.5,-0.3){$|\mu|~[{\rm GeV}]$}
	\put(0,7.3){$M_2~[{\rm GeV}]$}
		\put(6,6){\begin{picture}(1,1)(0,0)
			\CArc(0,0)(6,0,380)
			\Text(0,0)[c]{{\scriptsize A}}
	\end{picture}}
		\put(2.85,6){\begin{picture}(1,1)(0,0)
			\CArc(0,0)(6,0,380)
			\Text(0,0)[c]{{\scriptsize B}}
		\end{picture}}
	\put(3.2,4.2){\footnotesize $60$}
   \put(3.8,3.8){\footnotesize $20$}
   \put(4.6,3.5){\footnotesize $4 $}
	\put(5.8,3.4){\footnotesize $0.4 $}
	\put(0.5,-0.3){Fig.~\ref{plots_12}c}
   \put(8,0){\includegraphics{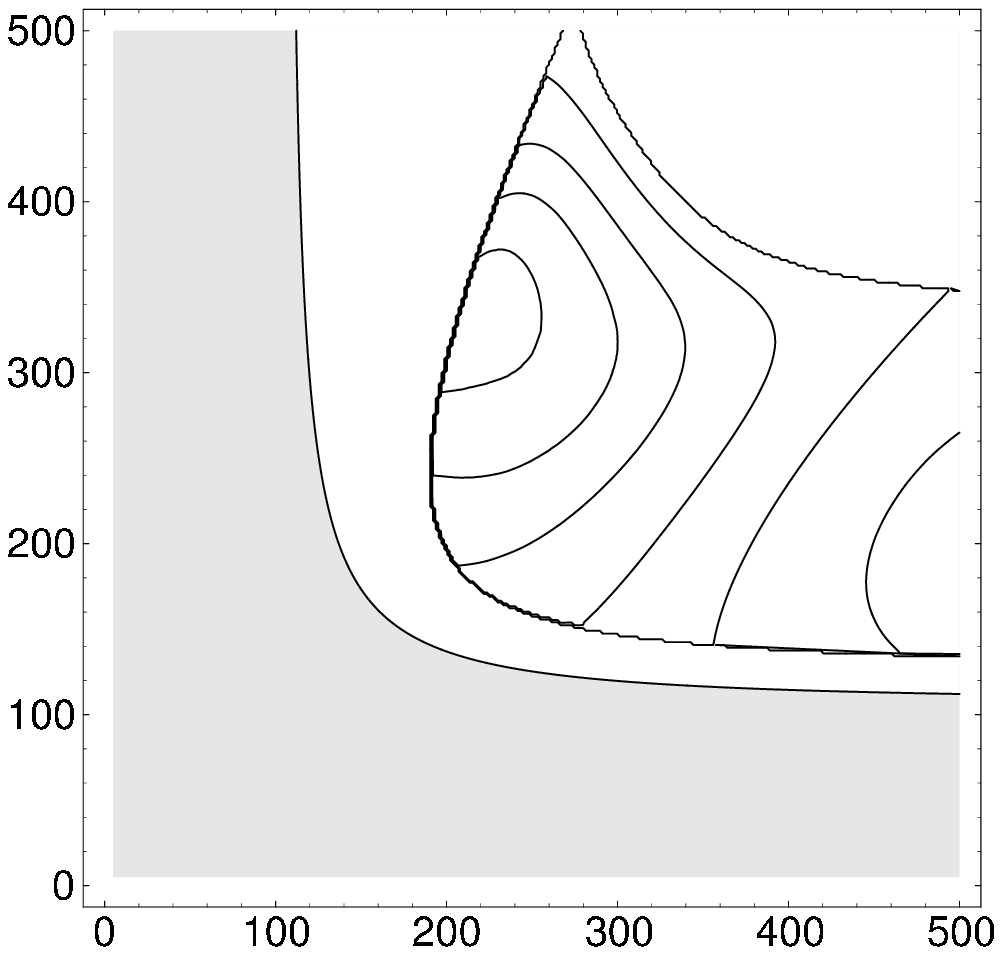}}
	\put(11.5,7.5){\fbox{${\mathcal A}_{II}^{\rm T}$ in \% }}
	\put(8,7.3){$M_2~[{\rm GeV}]$}
	\put(13.5,-0.3){$|\mu|~[{\rm GeV}]$}
		\put(14,6){\begin{picture}(1,1)(0,0)
			\CArc(0,0)(6,0,380)
			\Text(0,0)[c]{{\scriptsize A}}
	\end{picture}}
		\put(10.85,6){\begin{picture}(1,1)(0,0)
			\CArc(0,0)(6,0,380)
			\Text(0,0)[c]{{\scriptsize B}}
		\end{picture}}
	\put(11.3,4.5){$ 9.5 $}
	\put(11.7,3.9){$ 8 $}
	\put(12.,3.5){$6 $}
	\put(12.4,3.2){$3 $}
   \put(13,3){$0 $}
   \put(14.3,2.8){$-3 $}
   \put(8.5,-0.3){Fig.~\ref{plots_12}d}
 \end{picture}
\vspace*{-1.5cm}
\caption{
	Contour lines of    
	\ref{plots_12}a: $\sigma_P(e^+e^- \to\tilde{\chi}^0_1\tilde{\chi}^0_2)$, 
	\ref{plots_12}b: ${\rm BR}(\tilde{\chi}^0_2 \to \tilde{\ell}_R\ell_1)$,
	$ \ell= e,\mu$,
	\ref{plots_12}c: $\sigma_P(e^+e^-\to\tilde\chi^0_1\tilde\chi^0_2 ) \times
	{\rm BR}(\tilde \chi^0_2\to\tilde\ell_R\ell_1)\times
	{\rm BR}(\tilde\ell_R\to\tilde\chi^0_1\ell_2)$
	with ${\rm BR}(\tilde\ell_R \to\tilde\chi^0_1\ell_2)=1$,
	\ref{plots_12}d: the asymmetry ${\mathcal A}_{II}^{\rm T}$,
	in the $|\mu|$--$M_2$ plane for $\varphi_{M_1}=0.5\pi $, 
	$\varphi_{\mu}=0$,  $\tan \beta=10$, $m_0=100$ GeV,
	$A_{\tau}=-250$ GeV, $\sqrt{s}=500$ GeV and $(P_{e^-},P_{e^+})=(0.8,-0.6)$.
	The area A (B) is kinematically forbidden by
	$m_{\chi^0_1}+m_{\chi^0_2}>\sqrt{s}$
	$(m_{\tilde\ell_R}>m_{\chi^0_2})$.
	The gray  area is excluded by $m_{\chi_1^{\pm}}<104 $ GeV. 
	\label{plots_12}}
\end{figure}

In Fig.~\ref{plots_12}a we show the cross section for
$\tilde\chi^0_1 \tilde\chi^0_2$ production
for $\varphi_{\mu}=0$ and $\varphi_{M_1}=0.5~\pi$ in the $|\mu|$--$M_2$
plane. The cross section reaches values up to 300 fb.
For $|\mu|  \lsim 250 $ GeV the right selectron exchange dominates
so that our choice of polarizations $(P_{e^-},P_{e^+})=(0.8,-0.6)$
enhances the cross section by a factor up to 2.5 compared 
to the unpolarized case. For  $|\mu| \gsim 300$ GeV the left 
selectron exchange dominates because of the larger 
$\tilde\chi^0_2-\tilde e_L$ coupling.
In this region a sign reversal of both polarizations, i.e. 
$(P_{e^-},P_{e^+})=(-0.8,0.6)$, would enhance the
cross section up to a factor of 20.

The branching ratio 
${\rm BR} (\tilde\chi^0_2 \to\tilde\ell_R\ell_1 )$
for the neutralino two-body decay into right
selectrons and smuons, 
%${\rm BR} (\tilde\chi^0_2 \to\tilde\ell_R\ell_1 )$, 
summed over both signs of charge, is shown in Fig.~\ref{plots_12}b. 
It reaches values up to 64\% and decreases
with increasing $|\mu|$ when the two-body decays
into the lightest neutral Higgs boson $H_1^0$  and/or the
$Z$ boson are kinematically allowed. The decays 
$\tilde\chi^0_2\to\tilde\chi^{\pm}_1W^{\mp}$ are not allowed.
With our choice $m_0=100$~GeV, the decays into left selectrons 
and smuons can be neglected because these channels are either not open or 
the branching ratio is smaller than 1\%. 
As we assume that the squarks and the other
Higgs bosons are heavy, the decay into the stau is competing,
and  dominates for $M_2 \lsim 200$ GeV in our scenario,
see Fig.~\ref{plotsstau_12}a, which is discussed below. 
%In our scenario this decay mode dominates 
%for $M_2 \lsim 200$ GeV, see Fig.~\ref{plotsstau_12}a.
The resulting cross section 
%$\sigma_P(e^+e^-\to\tilde\chi^0_1\tilde\chi^0_2 ) \times
%{\rm BR}(\tilde \chi^0_2\to\tilde\ell_R\ell_1)\times
%{\rm BR}(\tilde\ell_R\to\tilde\chi^0_1\ell_2)$
%with BR($ \tilde\ell_R \to\tilde\chi^0_1\ell_2$) = 1
is shown in Fig.~\ref{plots_12}c.

Fig.~\ref{plots_12}d shows the $|\mu|$--$M_2$ dependence of the 
asymmetry ${\mathcal A}_{II}^{\rm T}$ for $\varphi_{M_1}=0.5~\pi$ and 
$\varphi_{\mu}=0$.
In the region $|\mu|  \lsim 250 $ GeV, where the right selectron
exchange dominates, the asymmetry reaches $9.5\%$ for our choice of beam 
polarization. This enhances the asymmetry up to a factor of 2 compared to the
case of unpolarized beams. With increasing  $|\mu|$ the asymmetry
decreases and finally changes sign. This is due to the increasing
contributions of the left selectron exchange which contributes to the 
asymmetry with opposite sign and dominates for $|\mu| \gsim 300$ GeV.
In this region the asymmetry could be enhanced up to a factor 2 
by reversing the sign of both beam polarizations.
%%%%%%%%%%%%%%%%%%%%%%%%%%%%%%%%%%%%%%%%%%%%%%%%%%%%%%%%%%%%%%%%%%
%            P L O T     1
%%%%%%%%%%%%%%%%%%%%%%%%%%%%%%%%%%%%%%%%%%%%%%%%%%%%%%%%%%%%%%%%%
%
%
%

%%%%%%%%%%%%%%%%%%%%%%%%%%%%%%%%%%%%%%%%%%%%%%%%%%%%%%%%%%%%%%%%%%
%            P L O T     2
%%%%%%%%%%%%%%%%%%%%%%%%%%%%%%%%%%%%%%%%%%%%%%%%%%%%%%%%%%%%%%%%%
\begin{figure}
 \begin{picture}(20,20)(0,-2)
	 \put(2.,16.5){\fbox{$\sigma(e^+\,e^- \to\tilde{\chi}^0_1 \,
			\tilde{\chi}^0_1 \ell_1 \,\ell_2 )$ in fb}}
	\put(0,9){\includegraphics{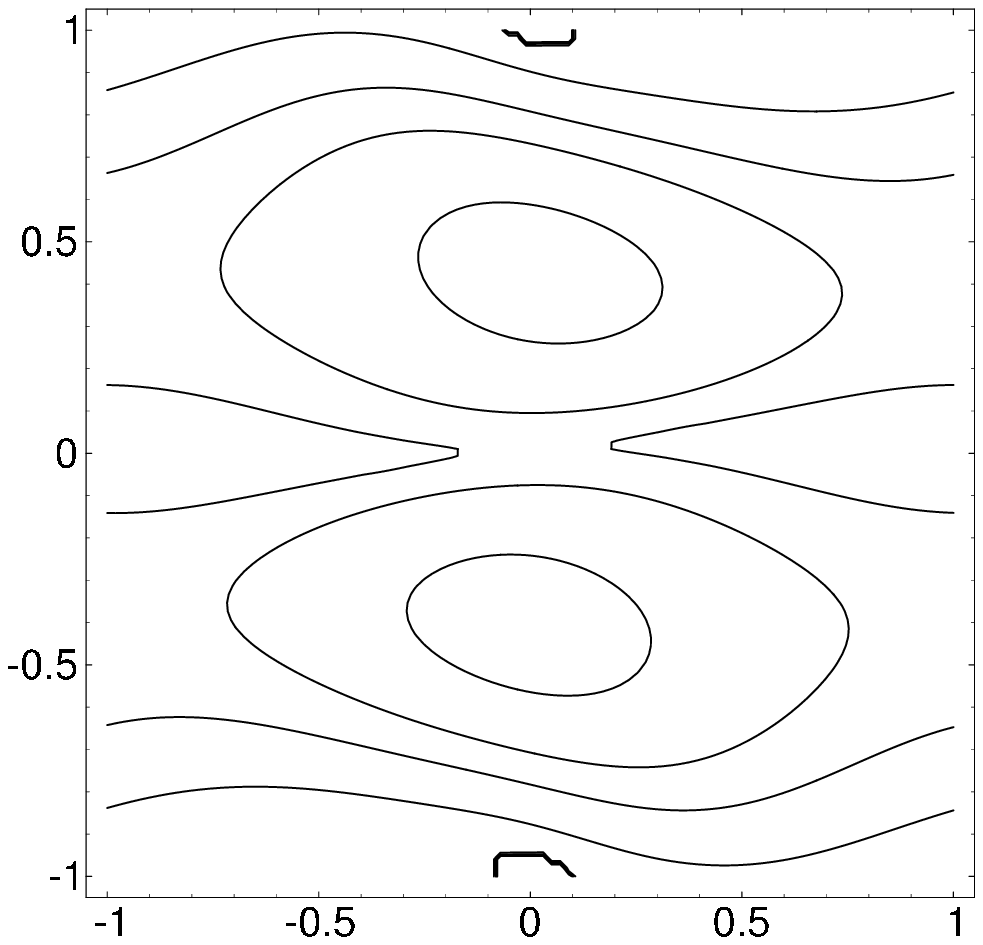}}
\put(6.5,8.8){$\varphi_{\mu}~[\pi]$}
\put(0,16.3){$ \varphi_{M_1}~[\pi]$ }
\put(4.,14.1){\footnotesize$64$}
\put(5.05,14.15){\footnotesize$56$}
\put(6.,14.4){\footnotesize$48$}
\put(6.6,14.95){\footnotesize$32$}
\put(6.3,12.9){\footnotesize$48$}
	\put(4.,11.2){\footnotesize$64$}
	\put(5.05,10.8){\footnotesize$56$}
	\put(5.8,10.55){\footnotesize$48$}
	\put(6.6,10.25){\footnotesize$32$}
	\put(1.3,12.45){\footnotesize$48$}
\put(0.5,8.8){Fig.~\ref{varphases_12}a}
\put(8,9){\includegraphics{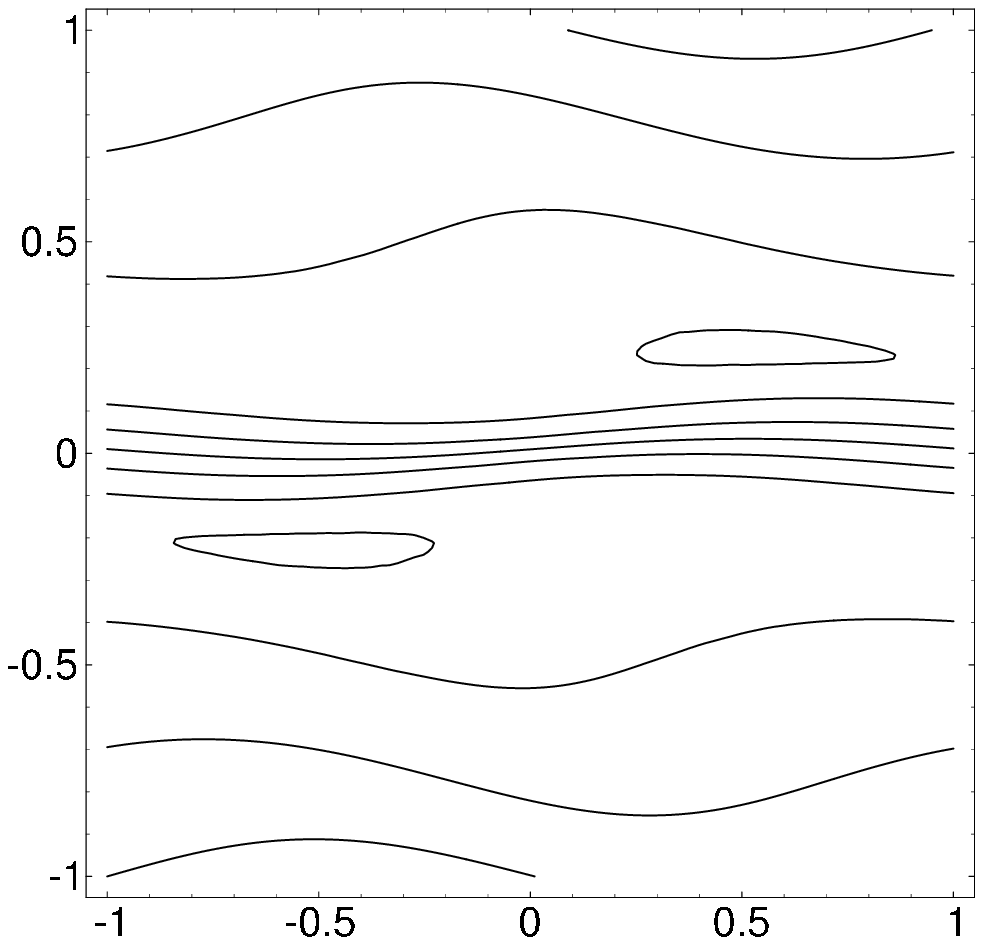}}
	\put(11.5,16.5){\fbox{${\mathcal A}_{II}^{\rm T}$ in \% }}
   \put(14.5,8.8){$\varphi_{\mu}~[\pi]$}
   \put(8,16.3){$ \varphi_{M_1}~[\pi]$ }
	\put(13.5,15.6){\footnotesize$0$}
   \put(11.,15.1){\footnotesize$5$}
   \put(12.,14.1){\footnotesize$8$}
	\put(13.3,13.4){\footnotesize$8.9$}
	\put(12.,13.1){\footnotesize$8$}
	\put(10.,12.){\footnotesize$-8.9$}
	\put(12.,12.3){\footnotesize$-8$}
	   \put(12.,11.3){\footnotesize$-8$}
		\put(13,10.4){\footnotesize$-5$}
		\put(10.2,9.7){\footnotesize$0$}
		\put(8.5,8.8){Fig.~\ref{varphases_12}b}
	\put(8,0){\includegraphics{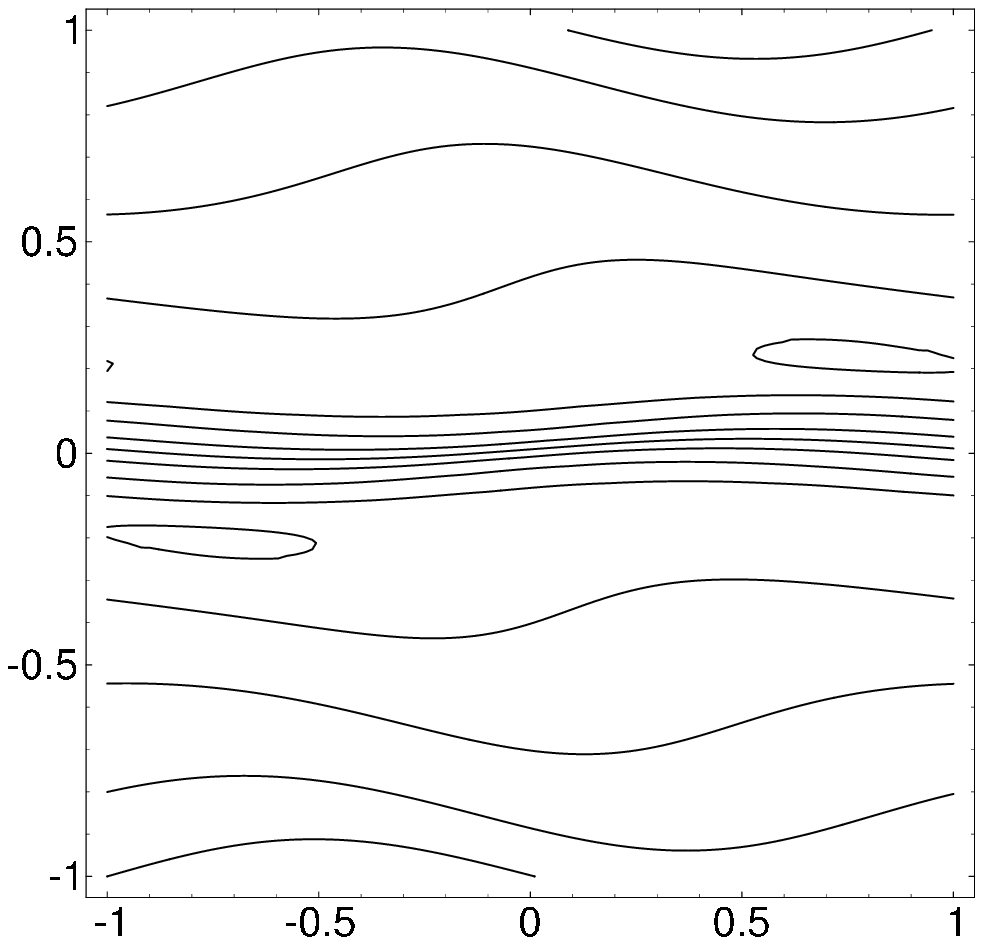}}
	\put(11.6,7.5){\fbox{${\mathcal A}_{I}^{\rm T}$ in \% }}
	\put(14.5,-.3){$\varphi_{\mu}~[\pi]$}
	\put(8,7.3){$ \varphi_{M_1}~[\pi]$}
	\put(11.2,6.25){\footnotesize$10$}
\put(11.5,5.6){\footnotesize$20$}
\put(12,4.8){\footnotesize$25$}
\put(14,4.35){\scriptsize$27$}
\put(12.6,4.2){\footnotesize$25$}
\put(13.5,6.6){\footnotesize$0$}
	\put(9.3,3.05){\scriptsize$-27$}
	\put(11.3,3.2){\footnotesize$-25$}
	\put(11.3,2.6){\footnotesize$-25$}
	\put(12.3,1.8){\footnotesize$-20$}
	\put(13.3,1.2){\footnotesize$-10$}
	\put(10.3,0.7){\footnotesize$0$}
	\put(0.5,-0.3){Fig.~\ref{varphases_12}c}
		\put(0,0){\includegraphics{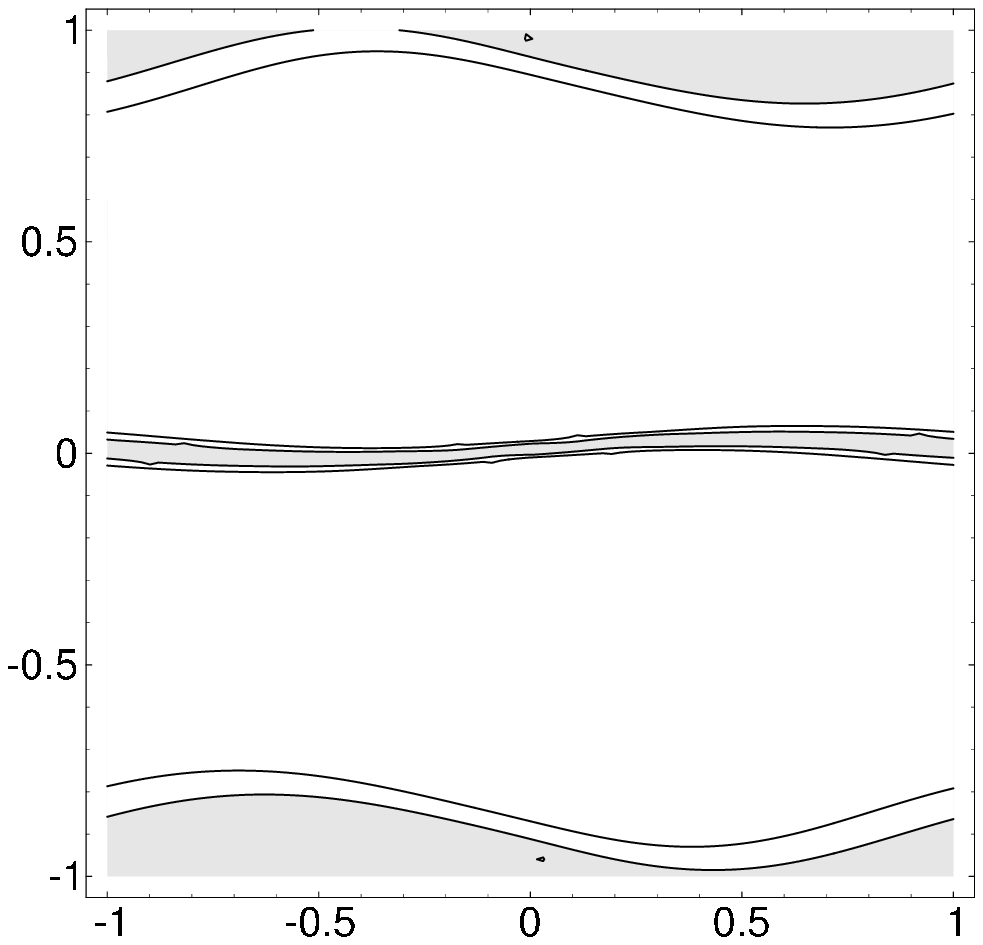}}
	\put(3.,7.5){\fbox{ $S={\mathcal A}_{II}^{\rm T}\sqrt{N}$}}
	\put(6.5,-.3){$\varphi_{\mu}~[\pi]$}
	\put(0,7.3){$ \varphi_{M_1}~[\pi]$}
	\put(5.6,6.3){\footnotesize$3$}
	\put(5,5.75){\footnotesize$5$}
\put(3.8,3.3){\footnotesize$5$}
\put(4.2,4.1){\footnotesize$5$}
	\put(2.8,1.55){\footnotesize$5$}
	\put(2.3,0.95){\footnotesize$3$}
	\put(8.5,-0.3){Fig.~\ref{varphases_12}d}
 \end{picture}
\vspace*{-1.5cm}
\caption{
	Contour lines of  
	\ref{varphases_12}a: $\sigma_P(e^+e^-\to\tilde\chi^0_1\tilde\chi^0_2)\times
	{\rm BR}(\tilde \chi^0_2\to\tilde\ell_R\ell_1)\times
	{\rm BR}(\tilde\ell_R\to\tilde\chi^0_1\ell_2)$
	with ${\rm BR}(\tilde\ell_R \to\tilde\chi^0_1\ell_2)=1$,
	\ref{varphases_12}b: the asymmetry ${\mathcal A}_{II}^{\rm T}$,
	\ref{varphases_12}c: the significance $S$,
	\ref{varphases_12}d: the asymmetry ${\mathcal A}_{I}^{\rm T}$,
	in the $\varphi_{\mu}$--$\varphi_{M_1}$ plane 
	for  $M_2=400$ GeV, $|\mu|=240$ GeV, $\tan \beta=10$, $m_0=100$ GeV,
	$A_{\tau}=-250$ GeV, $\sqrt{s}=500$ GeV and $(P_{e^-},P_{e^+})=(0.8,-0.6)$.
	In the gray shaded area of \ref{varphases_12}c we have $S<3$.
	For $\varphi_{M_1},\varphi_{\mu}=0$ we have
	$m_{\tilde \ell_R}=221$ GeV, $m_{\chi_1^0}=178$ GeV and
	$m_{\chi_2^0}=243$ GeV.
\label{varphases_12}}
\end{figure}

The sensitivity of the cross section $\sigma$ and the asymmetry 
${\mathcal A}_{II}^{\rm T}$ on the CP phases is shown by contour 
plots in the
$\varphi_{\mu}$--$\varphi_{M_1}$ plane  for $|\mu|=240$ GeV
and  $M_2=400$ GeV (Fig.~\ref{varphases_12}).
In our scenario the variation of the cross section, 
Fig.~\ref{varphases_12}a, is more than 100$\%$. 
In addition to the CP sensitive observables, the cross section may 
serve to constrain the phases. For unpolarized beams, 
the cross section would be reduced by a factor 0.4.
The asymmetry ${\mathcal A}_{II}^{\rm T}$ (Fig.~\ref{varphases_12}b)
varies between  -8.9$\%$ and 8.9$\%$. It is remarkable
that these maximal values are not necessarily obtained 
for maximal  CP phases. In our scenario the asymmetry is 
much more sensitive to variations of the phase $\varphi_{M_1}$ 
around  $0$. The reason is that ${\mathcal A}_{II}^{\rm T}$ is 
proportional to a product of
a CP odd ($\Sigma_P^2$) and a CP even factor ($\Sigma_{D_1}^2$),
see (\ref{properties_neut}). The  CP odd (CP even) factor has
as sine-like (cosine-like) dependence on the phases.
Thus the maximum of ${\mathcal A}_{II}^{\rm T}$ is shifted  
towards $\varphi_{M_1}=0$ in Fig.~\ref{varphases_12}b.
On the other hand, the asymmetry is rather 
insensitive to $\varphi_{\mu}$. For unpolarized beams this asymmetry 
would be reduced roughly by a factor 0.33.

%The relative statistical error of each asymmetry 
%${\mathcal A}$ can be calculated to $\delta {\mathcal A} = 
%\Delta {\mathcal A}/{\mathcal A} = S/({\mathcal A} \sqrt{N})$,
%with $S$ standard deviations,
%assuming a Gaussian distribution of the asymmetry ${\mathcal A}$.
%Here, $N={\mathcal L} \sigma$ is the number of events with 
%${\mathcal L}$ the total integrated luminosity and 
%$\sigma$ the total cross section. Assuming $\delta {\mathcal A}
%\approx1$, it follows $S \approx {\mathcal A} \sqrt{N}$.
%For example, in order to measure an asymmetry of 5\% with S=2
%(confidence level of 95\%), one would need at least $1.5\times 10^3$ events.
%This corresponds to a total cross section for reactions
%(\ref{production})-(\ref{decay_2}) of 3.1 fb with 
%${\mathcal L}=500~{\rm fb}^{-1}$. 

The statistical significance for measuring  each asymmetry
is given by $S= |{\mathcal A}^{\rm T}|\sqrt{N}$
(\ref{significanceofAT}), with $N={\mathcal L} \sigma$ 
is the number of events with ${\mathcal L}$ the total integrated 
luminosity.
%and $\sigma$ the total cross section.
We show the contour lines for  
$S=3$ and $5$ for ${\mathcal A}_{II}^{\rm T}$ in 
Fig.~\ref{varphases_12}c with  
${\mathcal L}= 500$ fb$^{-1}$. 

In Fig.~\ref{varphases_12}d we also show the asymmetry 
${\mathcal A}_{I}^{\rm T}$ which is a factor 2.9  larger than 
${\mathcal A}_{II}^{\rm T}$, because in ${\mathcal A}_{II}^{\rm T}$ 
the CP-violating effect from the production is partly washed out by the 
kinematics of the slepton decay. However, for a measurement of 
${\mathcal A}_{I}^{\rm T}$ the reconstruction of the $\tilde{\chi}^0_2$ 
momentum is necessary. The asymmetry ${\mathcal A}_{I}^{\rm T}$ 
shows a similar dependence on the phases as ${\mathcal A}_{II}^{\rm T}$ 
because both are due to the non vanishing neutralino 
polarization perpendicular to the production plane.
It is interesting to note that the asymmetries can be sizable 
for small values of $\varphi_{\mu}$, which is suggested by the 
EDM constraints, see 
Section~\ref{CP violating phases and electric dipole moments}.

Next we comment on the neutralino decay into the scalar tau and
discuss the main differences from the decay into the selectron and smuon.
In some regions of the parameter space, the decay
of the neutralino into the lightest stau $\tilde\tau_1$ may dominate over
that into the right selectron and smuon, and may even be the 
only decay channel. In Fig.~\ref{plotsstau_12}a we show contour lines 
of the branching ratio $BR (\tilde\chi^0_2 \to \tilde\tau_1\tau)$ in the
$|\mu|$--$M_2 $ plane for $A_{\tau} = -250$ GeV,
$\varphi_{M_1}=0.5 \, \pi $ and $\varphi_{\mu}=0$.
For $M_2<200$ GeV the branching ratio 
$BR (\tilde\chi^0_2 \to \tilde\tau_1\tau)$ is larger than 80\%. 
However, due to the mixing in the stau sector the asymmetry 
${\mathcal A}_{II}^{\rm T}$, Fig.~\ref{plotsstau_12}b, 
is reduced compared to that
in the selectron and smuon channels, see Fig.~\ref{plots_12}d. 
The reason is the suppression factor 
$(|a^{\tilde \tau}_{ki}|^2-|b^{\tilde \tau}_{ki}|^2)/(
  |a^{\tilde \tau}_{ki}|^2+|b^{\tilde \tau}_{ki}|^2)$ 
(\ref{Amixing}), which may be small or even be zero. 
%This may lead to a reduced or vanishing asymmetry, respectively,
%even in the case of non-zero CP phases. 

%%%%%%%%%%%%%%%%%%%%%%%%%%%%%%%%%%%%%%%%%%%%%%%%%%%%%%%%%%%%%%%%%%
%            P L O T     3
%%%%%%%%%%%%%%%%%%%%%%%%%%%%%%%%%%%%%%%%%%%%%%%%%%%%%%%%%%%%%%%%%
%

\begin{figure}[h]
	\begin{picture}(10,8)(0,0)
	\put(2.5,7.5){\fbox{${\rm BR}(\tilde{\chi}^0_2 \to\tilde{\tau}_1\tau)$ in \%}}
	\put(0,0){\includegraphics{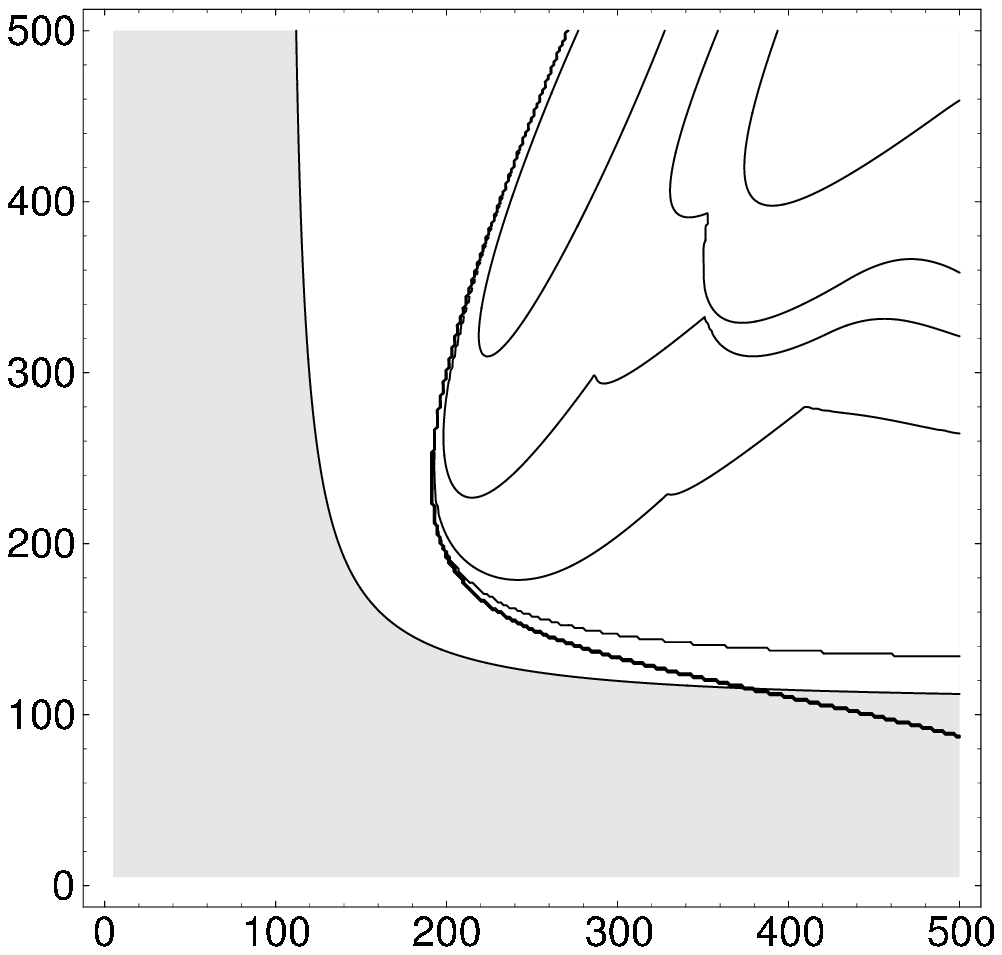}}
	\put(5.5,-0.3){$|\mu|~[{\rm GeV}]$}
	\put(0,7.3){$M_2~[{\rm GeV}]$}
\put(5.7,5.8){\footnotesize$15$}
\put(5.3,4.9){\footnotesize$40$}
\put(3.5,5.){\footnotesize$40$}
\put(4.5,4.1){\footnotesize$50$}
\put(5.5,3.6){\footnotesize$80$}
\put(6.,2.1){\footnotesize$100$}
		\put(2.85,6){\begin{picture}(1,1)(0,0)
			\CArc(0,0)(6,0,380)
			\Text(0,0)[c]{{\scriptsize B}}
			\end{picture}}
\put(0.5,-0.3){Fig.~\ref{plotsstau_12}a}
   \put(8,0){\includegraphics{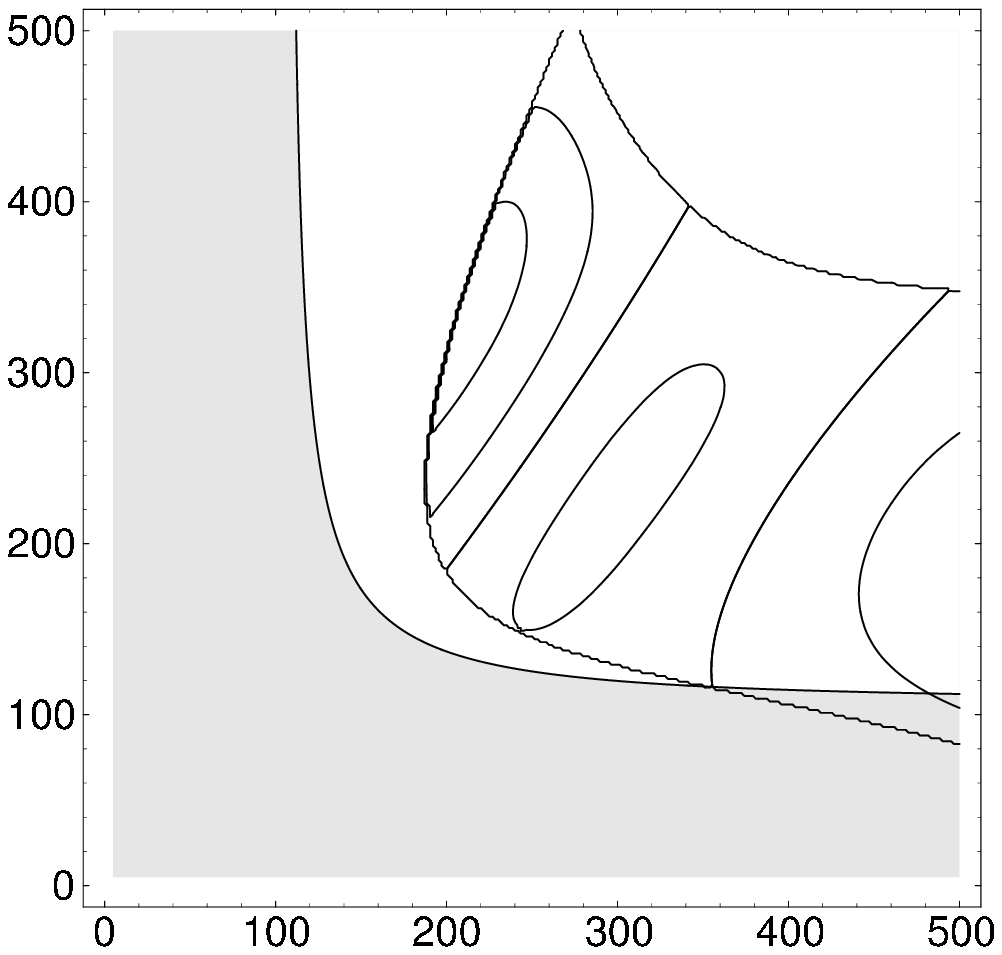}}
	\put(11.5,7.5){\fbox{${\mathcal A}_{II}^{\rm T}$ in \% }}
	\put(13.5,-0.3){$|\mu|~[{\rm GeV}]$}
	\put(8,7.3){$M_2~[{\rm GeV}]$}
	\put(11.45,5.2){\scriptsize$ -6 $}
	\put(12.2,5.5){\scriptsize$ -3 $}
	\put(12.1,4){\footnotesize$0 $}
	\put(12.1,3.3){\footnotesize$3 $}
   \put(13,3){\footnotesize$0 $}
	\put(14.2,2.8){\footnotesize$-3 $}
			\put(14,6){\begin{picture}(1,1)(0,0)
			\CArc(0,0)(6,0,380)
			\Text(0,0)[c]{{\scriptsize A}}
	\end{picture}}
		\put(10.85,6){\begin{picture}(1,1)(0,0)
			\CArc(0,0)(6,0,380)
			\Text(0,0)[c]{{\scriptsize B}}
			\end{picture}}
	\put(8.5,-0.3){Fig.~\ref{plotsstau_12}b}
 \end{picture}
 \vspace*{0.5cm}
\caption{
	Contour lines of  
	\ref{plotsstau_12}a: ${\rm BR}(\tilde{\chi}^0_2 \to\tilde{\tau}_1\tau)$ and
	\ref{plotsstau_12}b: the asymmetry  ${\mathcal A}_{II}^{\rm T}$,
	in the $|\mu|$--$M_2$ plane for $\varphi_{M_1}=0.5\pi $, 
	$\varphi_{\mu}=0$, $A_{\tau}=-250$ GeV, $\tan \beta=10$, 
	$m_0=100$ GeV, $\sqrt{s}=500$ GeV and $(P_{e^-},P_{e^+})=(0.8,-0.6)$.
	The area A (B) is kinematically forbidden by
	$m_{\chi^0_1}+m_{\chi^0_2}>\sqrt{s}$
	$(m_{\tilde\tau_1}>m_{\chi^0_2})$.
	The gray  area is excluded by $m_{\chi_1^{\pm}}<104 $ GeV.
\label{plotsstau_12}}
\end{figure}

\subsubsection{Production of $\tilde\chi^0_1 \tilde\chi^0_3$ }

\begin{figure}[h]
 \begin{picture}(10,8)(0,0)
	\put(0,0){\includegraphics{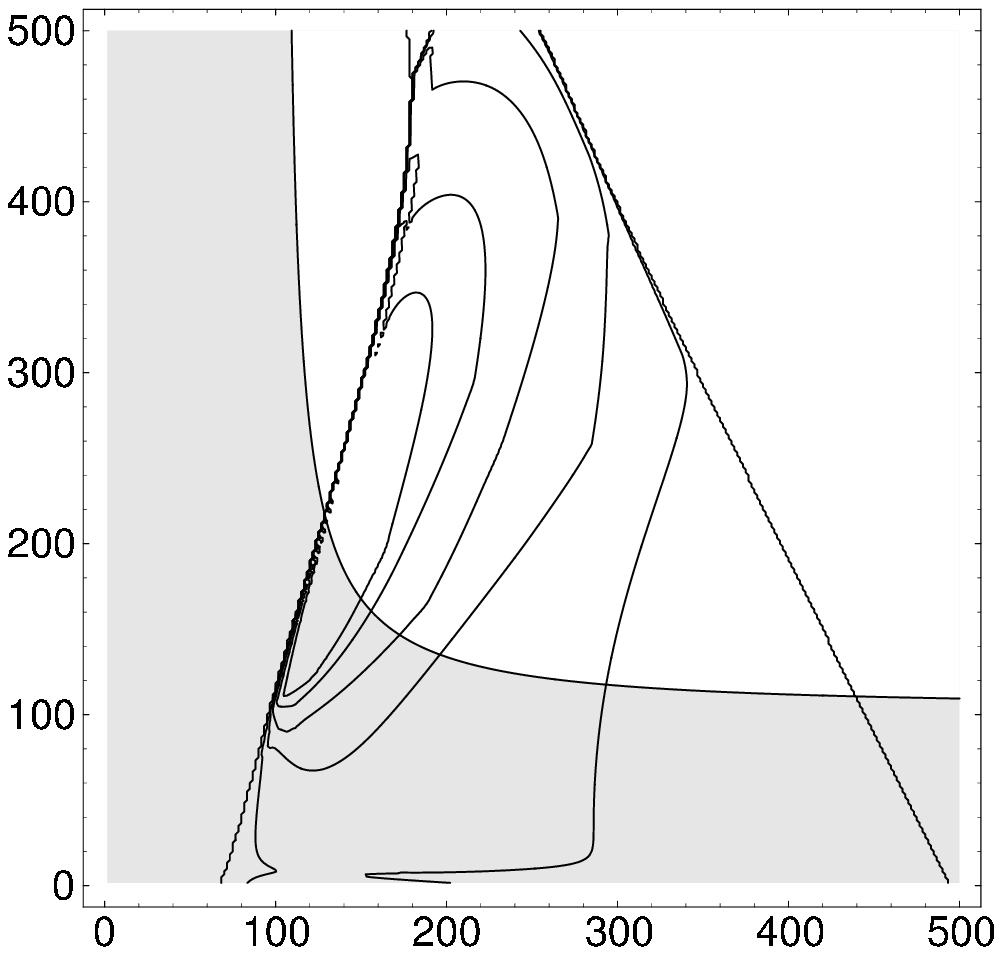}}
	\put(1.9,7.5){\fbox{$\sigma(e^+\,e^- \to\tilde{\chi}^0_1 \,
			\tilde{\chi}^0_1 \ell_1 \,\ell_2 )$ in fb}}
	\put(5.5,-0.3){$|\mu|~[{\rm GeV}]$}
	\put(0,7.3){$M_2~[{\rm GeV}]$}
	\put(2.4,3.6){\scriptsize$ 100 $}
	\put(3.,5.2){\footnotesize$60 $}
	\put(3.5,5.7){\footnotesize$20 $}
	\put(3.9,3.7){\footnotesize$4 $}
	\put(4.,3.){\footnotesize$0.4 $}
			\put(6,6){\begin{picture}(1,1)(0,0)
			\CArc(0,0)(6,0,380)
			\Text(0,0)[c]{{\scriptsize A}}
	\end{picture}}
		\put(2.5,6){\begin{picture}(1,1)(0,0)
			\CArc(0,0)(6,0,380)
			\Text(0,0)[c]{{\scriptsize B}}
			\end{picture}}
	\put(0.5,-0.3){Fig.~\ref{plots_13}a}
   \put(8,0){\includegraphics{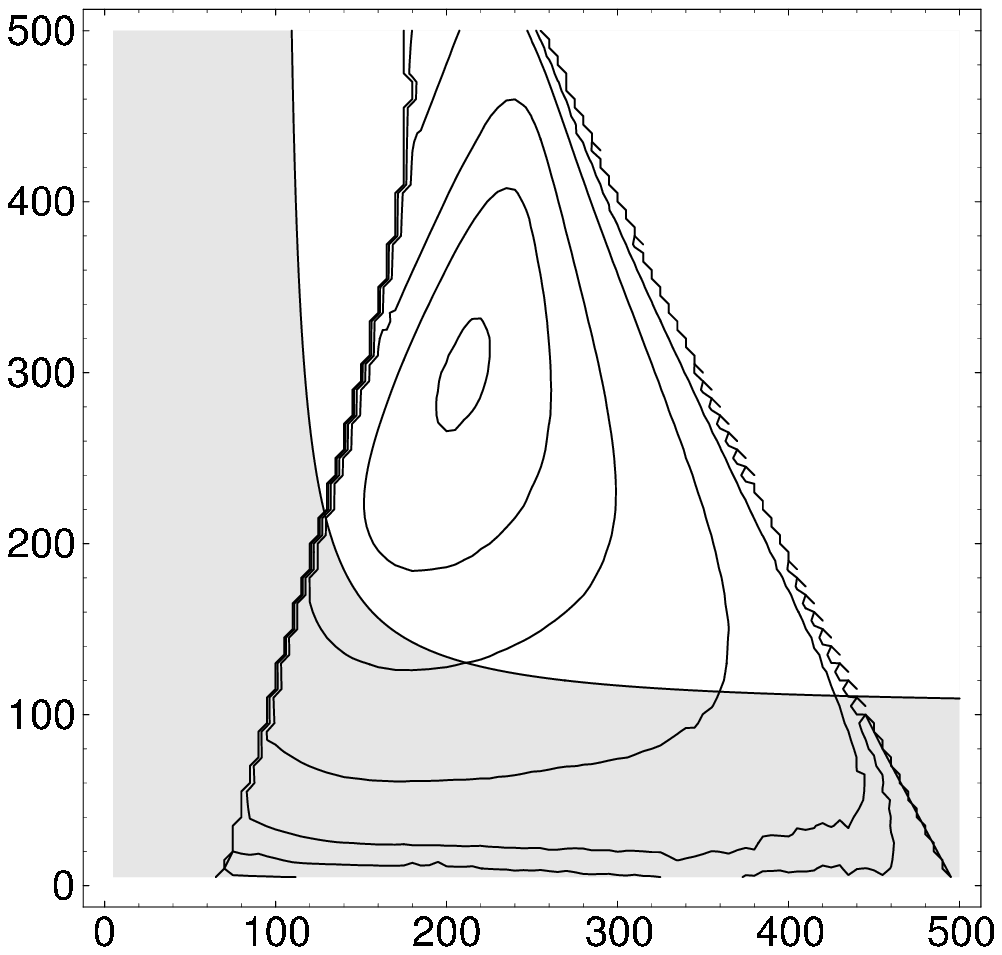}}
	\put(11.5,7.5){\fbox{${\mathcal A}_{II}^{\rm T}$ in \% }}
	\put(13.5,-0.3){$|\mu|~[{\rm GeV}]$}
	\put(8,7.3){$M_2~[{\rm GeV}]$}
	\put(11.,4.3){\scriptsize$ -9.5 $}
	\put(11.2,3.4){\footnotesize$-8 $}
	\put(11.7,2.9){\footnotesize$-6 $}
   \put(12.6,2.3){\footnotesize$-3 $}
	\put(13.4,1.3){\footnotesize$-1 $}
	\put(14.4,0.8){\scriptsize$0 $}
			\put(14,6){\begin{picture}(1,1)(0,0)
			\CArc(0,0)(6,0,380)
			\Text(0,0)[c]{{\scriptsize A}}
	\end{picture}}
		\put(10.5,6){\begin{picture}(1,1)(0,0)
			\CArc(0,0)(6,0,380)
			\Text(0,0)[c]{{\scriptsize B}}
			\end{picture}}
	\put(8.5,-0.3){Fig.~\ref{plots_13}b}
 \end{picture}
\vspace*{0.5cm}
\caption{
	Contour lines of  
	\ref{plots_13}a: $\sigma_P(e^+e^-\to\tilde\chi^0_1\tilde\chi^0_3) \times
	{\rm BR}(\tilde \chi^0_3\to\tilde\ell_R\ell_1)\times
	{\rm BR}(\tilde\ell_R\to\tilde\chi^0_1\ell_2)$
	with ${\rm BR}(\tilde\ell_R \to\tilde\chi^0_1\ell_2)=1$ and $\ell= e,\mu$,
	\ref{plots_13}b: the asymmetry ${\mathcal A}_{II}^{\rm T}$,
	in the $|\mu|$--$M_2$ plane for $\varphi_{M_1}=0.5\pi $, 
	$\varphi_{\mu}=0$, $\tan \beta=10$, $m_0=100$ GeV,
	$A_{\tau}=-250$ GeV, $\sqrt{s}=500$ GeV and $(P_{e^-},P_{e^+})=(0.8,-0.6)$.
	The area A (B) is kinematically forbidden by
	$m_{\chi^0_1}+m_{\chi^0_3}>\sqrt{s}$
	$(m_{\tilde\ell_R}>m_{\chi^0_3})$.
	The gray  area is excluded by $m_{\chi_1^{\pm}}<104 $ GeV. 
\label{plots_13}}
\end{figure}

We show in Fig.~\ref{plots_13}a and b contour lines of the cross section 
$\sigma_P(e^+e^-\to\tilde\chi^0_1\tilde\chi^0_3 ) \times
{\rm BR}(\tilde\chi^0_3\to\tilde\ell_R\ell_1)\times
{\rm BR}(\tilde\ell_R\to\tilde\chi^0_1\ell_2)$
with ${\rm BR}( \tilde\ell_R \to\tilde\chi^0_1\ell_2)=1$,
and of the asymmetry ${\mathcal A}_{II}^{\rm T}$, respectively.
The cross section with polarized beams reaches more than 100~fb,
which is up to a factor 2.5 larger than for unpolarized
beams. The asymmetry ${\mathcal A}_{II}^{\rm T}$, shown in 
Fig.~\ref{plots_13}b, 
reaches -9.5\%. For unpolarized beams this value would be reduced 
by a factor 0.75.  For our choice of parameters the cross section and
the asymmetry for $\tilde\chi^0_1  \tilde\chi^0_3$ 
production and decay show a similar dependence on $M_2$ and $|\mu|$ 
as for $\tilde\chi^0_1 \tilde\chi^0_2$ production, however, the
kinematically allowed regions are different.
We also studied the $\varphi_{\mu}$ dependence of ${\mathcal A}_{II}^{\rm T}$.
For $\varphi_{\mu}=0.5\pi (0.1\pi)$ and $\varphi_{M_1}=0$,
the maximal values of ${\mathcal A}_{II}^{\rm T}$ in the
$M_2$--$|\mu|$ plane are $|{\mathcal A}_{II}^{\rm T}|<3\%(1\%)$.
%%%%%%%%%%%%%%%%%%%%%%%%%%%%%%%%%%%%%%%%%%%%%%%%%%%%%%%%%%%%%%%%%%
%            P L O T     4
%%%%%%%%%%%%%%%%%%%%%%%%%%%%%%%%%%%%%%%%%%%%%%%%%%%%%%%%%%%%%%%%%
%

\subsubsection{Production of $\tilde \chi ^0_2 \tilde \chi^0_3$ }

\begin{figure}[h]
 \begin{picture}(20,20)(0,-2)
	\put(2.4,16.5){\fbox{$\sigma_P(e^+\,e^- \to\tilde\chi^0_2 \,
			\tilde\chi^0_3)$ in fb}}
	\put(0,9){\includegraphics{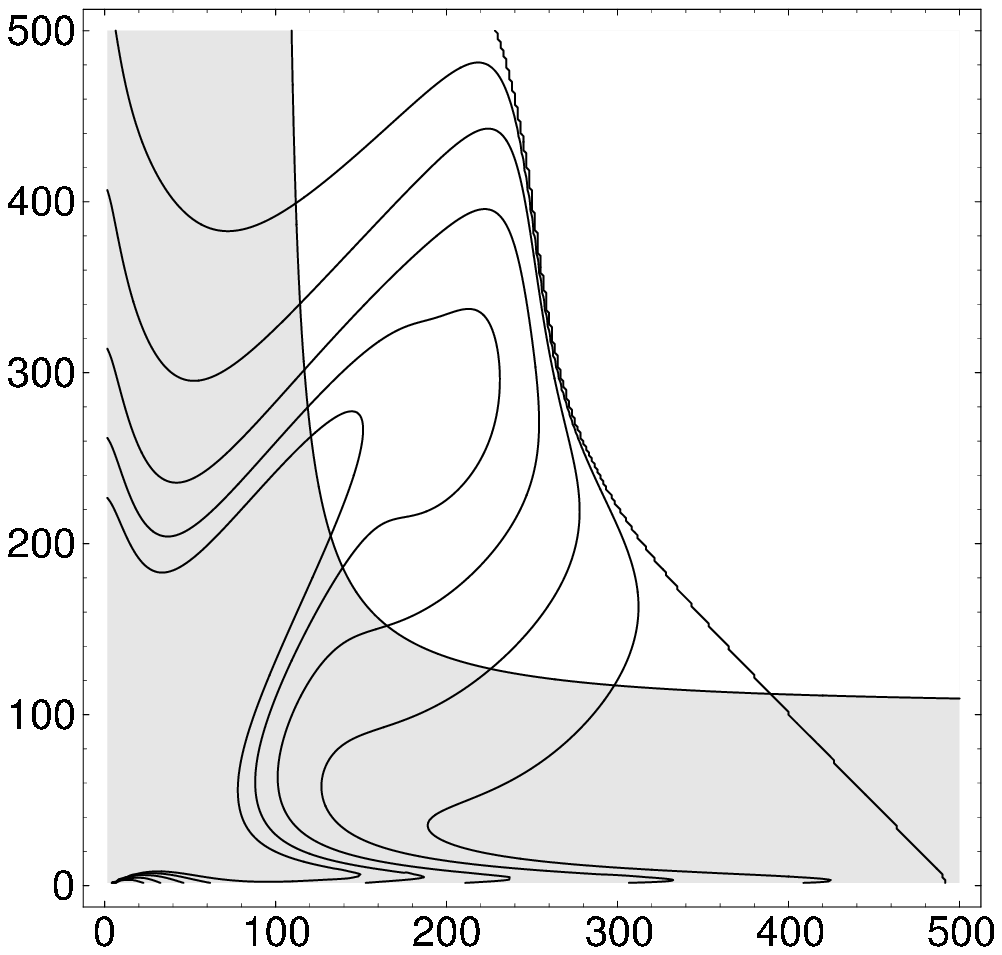}}
   \put(5.5,8.8){$|\mu|~[{\rm GeV}]$}
   \put(0,16.3){$M_2~[{\rm GeV}]$}
\put(1.8,12.3){\footnotesize$100$}
\put(2.9,12.65){\footnotesize$75$}
\put(3.1,12.1){\footnotesize$50$}
\put(3.5,11.7){\footnotesize$25$}
\put(4.1,11.3){\footnotesize$10$}
	\put(5.5,14){\begin{picture}(1,1)(0,0)
			\CArc(0,0)(6,0,380)
			\Text(0,0)[c]{{\scriptsize A}}
	\end{picture}}
\put(0.5,8.8){Fig.~\ref{plots_23}a}
   \put(8,9){\includegraphics{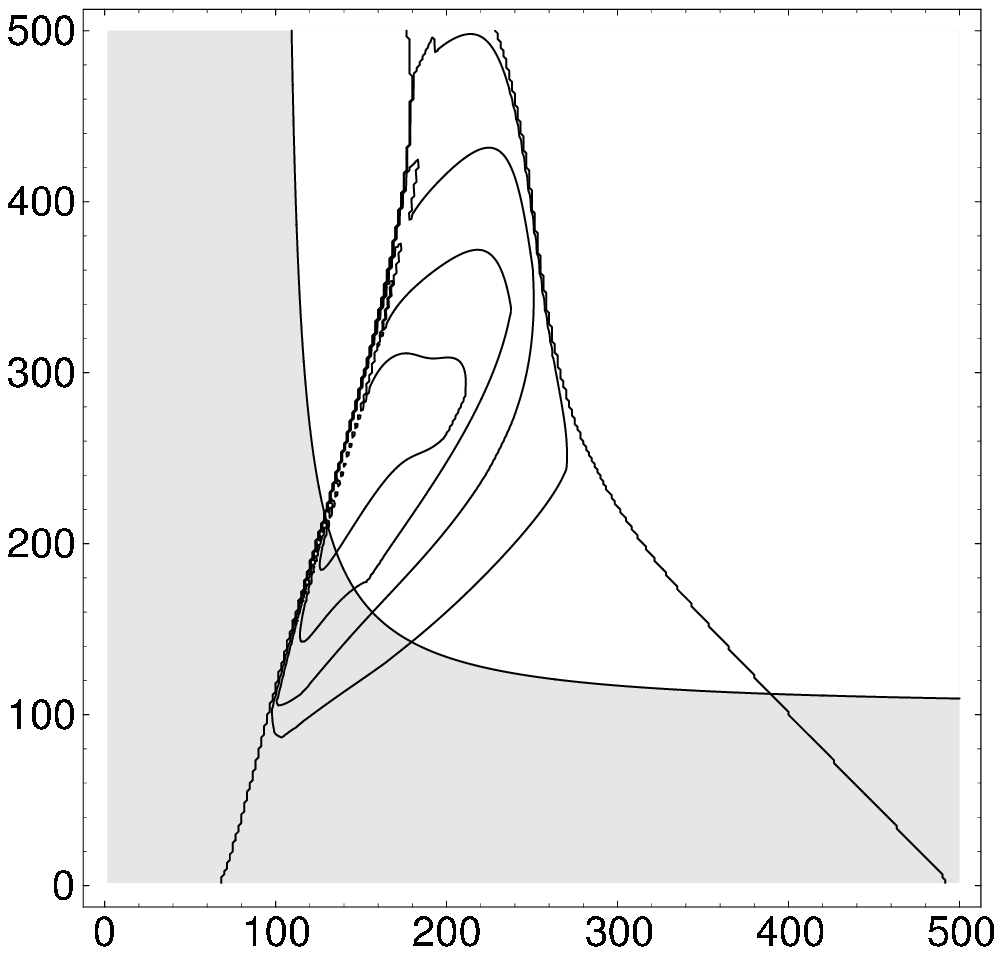}}
	\put(9.9,16.5)
{\fbox{$\sigma(e^+\,e^- \to\tilde\chi^0_2 \,
			\tilde\chi^0_1 \ell_1 \,\ell_2 )$ in fb}}
   \put(13.5,8.8){$|\mu|~[{\rm GeV}]$}
   \put(8,16.3){$M_2~[{\rm GeV}]$}
\put(11.3,15.4){\footnotesize$4$}
\put(11.3,14.5){\footnotesize$20$}
\put(11.1,13.8){\footnotesize$40$}
\put(10.8,13.1){\footnotesize$52$}
	\put(13.5,14){\begin{picture}(1,1)(0,0)
			\CArc(0,0)(6,0,380)
			\Text(0,0)[c]{{\scriptsize A}}
	\end{picture}}
		\put(10.5,15){\begin{picture}(1,1)(0,0)
			\CArc(0,0)(6,0,380)
			\Text(0,0)[c]{{\scriptsize B}}
			\end{picture}}
\put(8.5,8.8){Fig.~\ref{plots_23}b}
	\put(0,0){\includegraphics{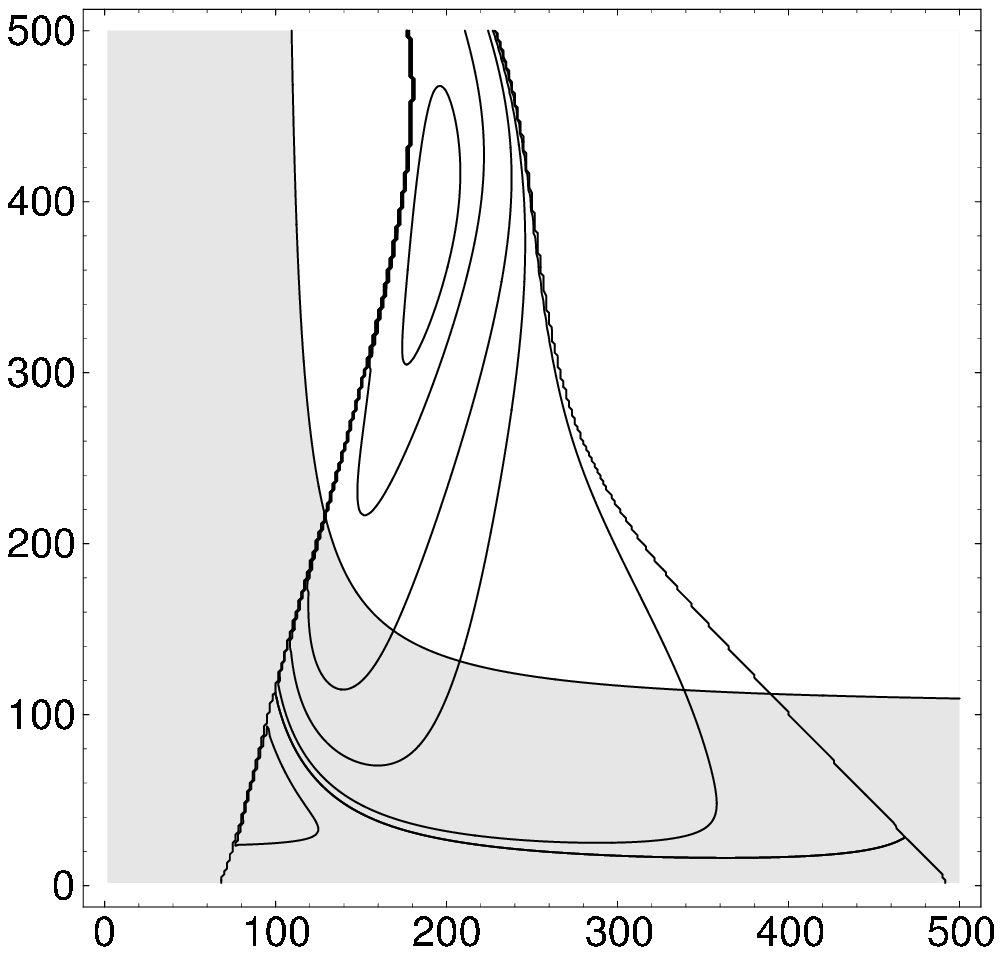}}
	\put(3.5,7.5){\fbox{${\mathcal A}_{I}^{\rm T}$ in \% }}
	\put(5.5,-0.3){$|\mu|~[{\rm GeV}]$}
	\put(0,7.3){$M_2~[{\rm GeV}]$}
	\put(2.9,5.3){\footnotesize$ 25 $}
   \put(2.6,4){\footnotesize$15 $}
	\put(2.6,3.){\footnotesize$10 $}
	\put(3.2,2.7){\footnotesize$5 $}
	\put(4.5,2.2){\footnotesize$2 $}
	\put(5.5,1.05){\footnotesize$0 $}
	\put(1.7,1.15){\footnotesize$-3 $}
				\put(5.5,5){\begin{picture}(1,1)(0,0)
			\CArc(0,0)(6,0,380)
			\Text(0,0)[c]{{\scriptsize A}}
	\end{picture}}
		\put(2.5,6){\begin{picture}(1,1)(0,0)
			\CArc(0,0)(6,0,380)
			\Text(0,0)[c]{{\scriptsize B}}
			\end{picture}}
	\put(0.5,-0.3){Fig.~\ref{plots_23}c}
   \put(8,0){\includegraphics{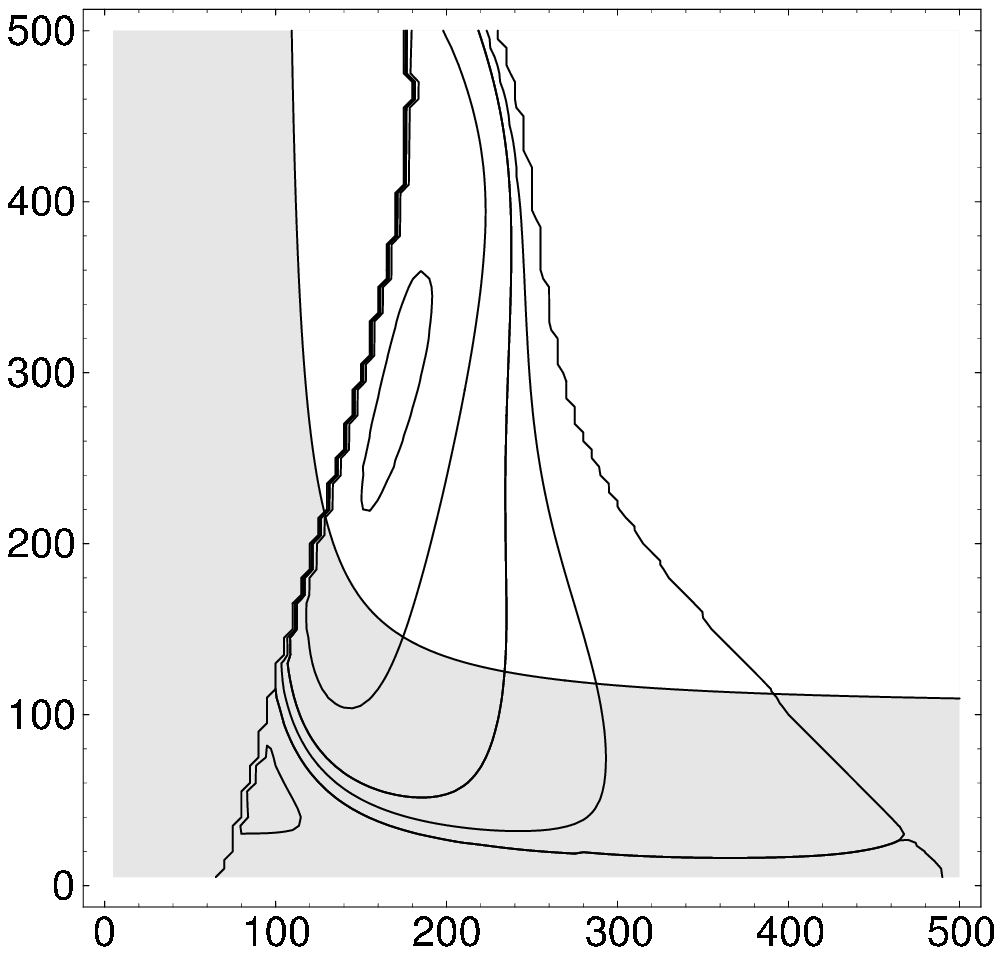}}
	\put(11.5,7.5){\fbox{${\mathcal A}_{II}^{\rm T}$ in \% }}
	\put(13.5,-0.3){$|\mu|~[{\rm GeV}]$}
	\put(8,7.3){$M_2~[{\rm GeV}]$}
	\put(10.6,4.1){\footnotesize$ 10 $}
	\put(10.8,3){\footnotesize$5$}
	\put(11.4,2.55){\footnotesize$2 $}
	\put(11.9,2.3){\footnotesize$1 $}
	\put(13.5,1.1){\footnotesize$0 $}
	\put(9.6,1.2){\footnotesize$-2 $}
				\put(13.5,5){\begin{picture}(1,1)(0,0)
			\CArc(0,0)(6,0,380)
			\Text(0,0)[c]{{\scriptsize A}}
	\end{picture}}
		\put(10.5,6){\begin{picture}(1,1)(0,0)
			\CArc(0,0)(6,0,380)
			\Text(0,0)[c]{{\scriptsize B}}
			\end{picture}}
\put(8.5,-0.3){Fig.~\ref{plots_23}d}
 \end{picture}
\vspace*{-1.5cm}
\caption{
	Contour lines of  
	\ref{plots_23}a: $\sigma_P(e^+e^- \to\tilde{\chi}^0_2\tilde{\chi}^0_3)$, 
	\ref{plots_23}b: $\sigma_P(e^+e^-\to\tilde\chi^0_2\tilde\chi^0_3 ) \times
	{\rm BR}(\tilde \chi^0_3\to\tilde\ell_R\ell_1)\times
	{\rm BR}(\tilde\ell_R\to\tilde\chi^0_1\ell_2)$ for $ \ell= e,\mu$,
	and ${\rm BR}(\tilde\ell_R\to\tilde\chi^0_1\ell_2)=1$,
	\ref{plots_23}c: the asymmetry ${\mathcal A}_{I}^{\rm T}$,
	\ref{plots_23}d: the asymmetry ${\mathcal A}_{II}^{\rm T}$,
	in the $|\mu|$--$M_2$ plane for $\varphi_{M_1}=0.5\pi $, 
	$\varphi_{\mu}=0$, $\tan \beta=10$, $m_0=100$ GeV,
	$A_{\tau}=-250$ GeV, $\sqrt{s}=500$ GeV and $(P_{e^-},P_{e^+})=(0.8,-0.6)$.
	The area A (B) is kinematically forbidden by
	$m_{\chi^0_2}+m_{\chi^0_3}>\sqrt{s}$
	$(m_{\tilde\ell_R}>m_{\chi^0_3})$.
	The gray  area is excluded by $m_{\chi_1^{\pm}}<104 $ GeV.
\label{plots_23}}
\end{figure}

The production of the neutralino pair 
$ e^+e^-\to\tilde\chi^0_2 \tilde\chi^0_3$
could make it easier to reconstruct the production plane
because both neutralinos decay. This
allows one to determine also asymmetry ${\mathcal A}_{I}^{\rm T}$, 
which is a factor 2-3 larger than ${\mathcal A}_{II}^{\rm T}$. 
We discuss the decay of the heavier neutralino 
$\tilde\chi^0_3$, which has a larger kinematically allowed region
in the $|\mu|$--$M_2$ plane than that of $\tilde\chi^0_2$. 
In Fig.~\ref{plots_23} we show the production cross section
$\sigma_P(e^+e^-\to\tilde\chi^0_2\tilde\chi^0_3 )$ which reaches 
100 fb. The cross section
$\sigma_P(e^+e^-\to\tilde\chi^0_2\tilde\chi^0_3 ) \times
{\rm BR}(\tilde\chi^0_3\to\tilde\ell_R\ell_1)\times
{\rm BR}(\tilde\ell_R\to\tilde\chi^0_1\ell_2)$
with ${\rm BR}(\tilde\ell_R \to\tilde\chi^0_1\ell_2)=1$,
is shown in Fig.~\ref{plots_23}b.
The asymmetry ${\mathcal A}_{II}^{\rm T}$ is shown in Fig.~\ref{plots_23}d.
As to the $\varphi_{\mu}$ dependence of ${\mathcal A}_{I}^{\rm T}$,
we found that for $\varphi_{\mu}=0.5\pi (0.1\pi)$ and $\varphi_{M_1}=0$,
$|{\mathcal A}_{I}^{\rm T}|$ can reach 25\% (2\%) in the 
$|\mu|$--$M_2$ plane.
%%%%%%%%%%%%%%%%%%%%%%%%%%%%%%%%%%%%%%%%%%%%%%%%%%%%%%%%%%%%%%%%%%
%            P L O T     5
%%%%%%%%%%%%%%%%%%%%%%%%%%%%%%%%%%%%%%%%%%%%%%%%%%%%%%%%%%%%%%%%%

\subsubsection{Energy distributions of the leptons
	  \label{Energy distributions of the leptons}}

In order to measure the asymmetries 
${\mathcal A}_{I}^{\rm T}$~(\ref{TasymmetryI_neut}) and 
${\mathcal A}_{II}^{\rm T}$~(\ref{TasymmetryII_neut}),
the two leptons $\ell_1$ and $\ell_2$ from  the neutralino 
(\ref{decay_neut1}) and slepton decay (\ref{decay_neut2})
have to be distinguished.
We therefore calculate the energy distributions of the leptons
from the first and second decay vertex in the laboratory system,
see Appendix \ref{Energy distributions of the decay leptons}.
One can distinguish between the two leptons
event by event, if their energy distributions do not overlap.
If their energy distributions do overlap, only those leptons can 
be distinguished, whose energies are not both 
in the overlapping region.

\begin{figure}[h]
 \begin{picture}(20,19)(0,0)
\put(-1,17){\includegraphics{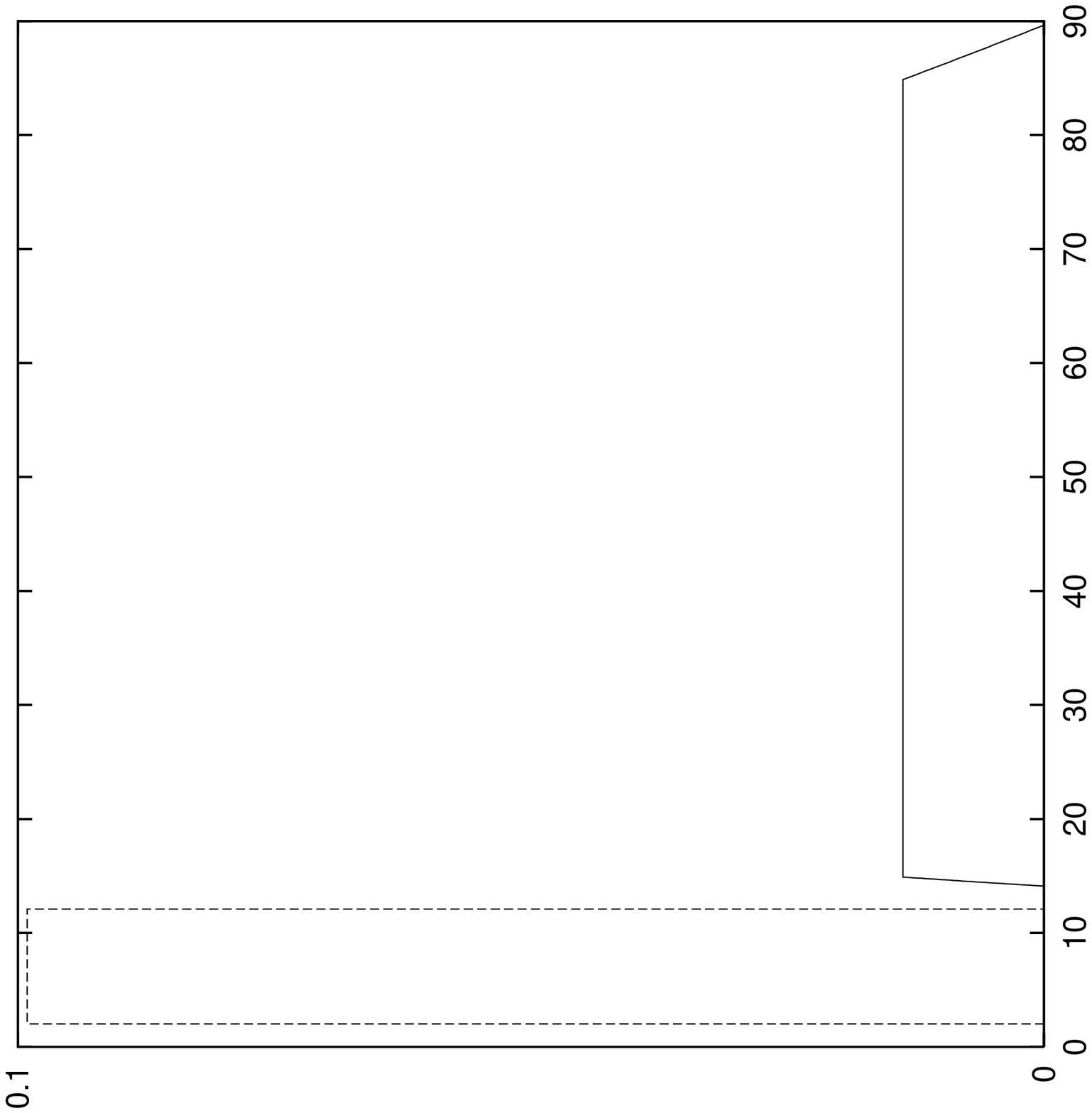}}
   \put(5.5,8.8){$E~[{\rm GeV}]$}
	\put(0,16.6){$ \frac{1}{\sigma}\frac{d\sigma}{dE}$}
\put(1.5,13.3){$\ell_1$}
\put(4.,10.7){$\ell_2$}
\put(0.5,8.8){Fig.~\ref{plotedist}a}
   \put(7,17){\includegraphics{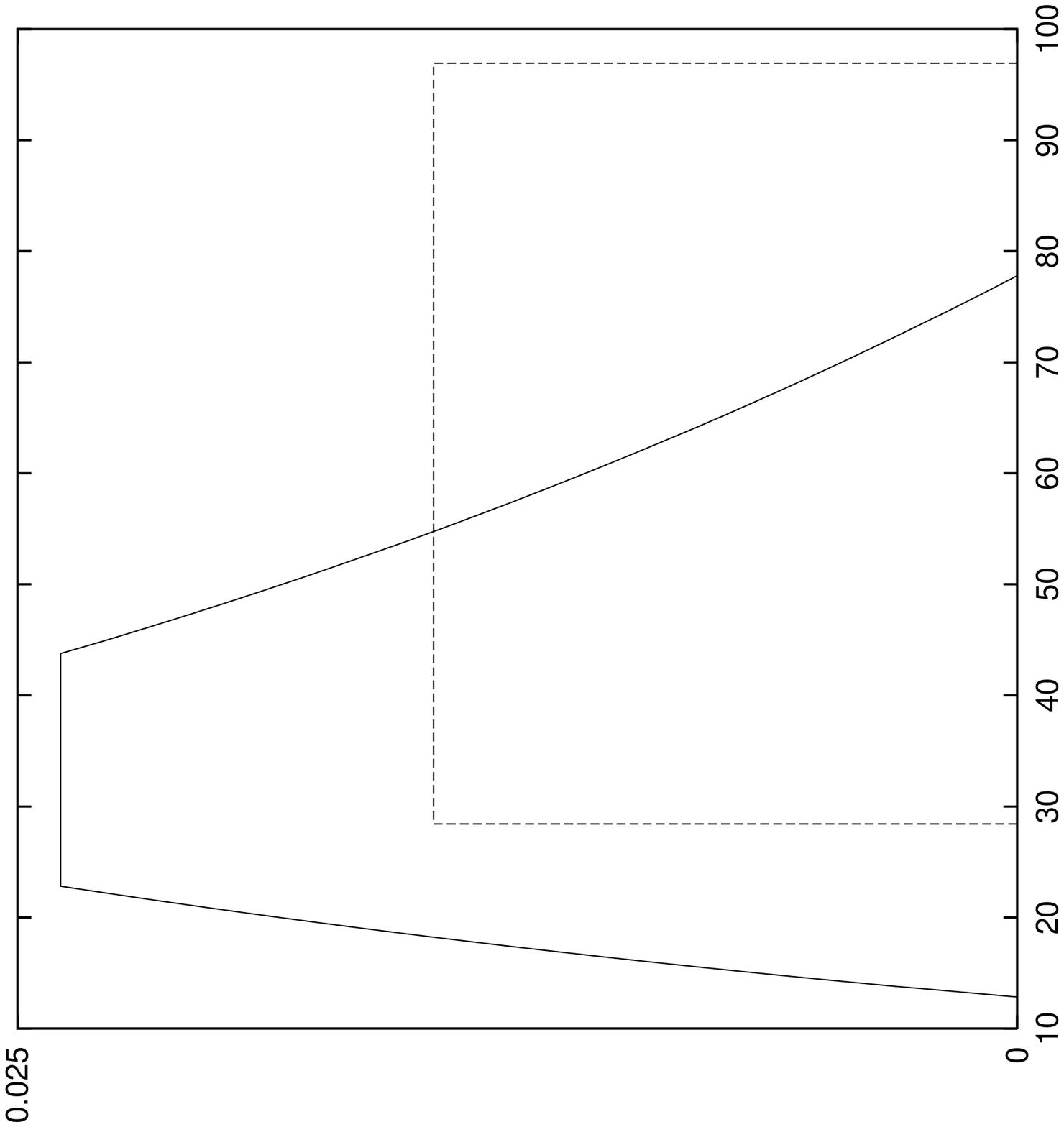}}
   \put(13.5,8.8){$E~[{\rm GeV}]$}
   \put(8,16.6){$ \frac{1}{\sigma}\frac{d\sigma}{dE} $ }
\put(10.3,15.3){$\ell_2$}
\put(12.5,13.6){$\ell_1$}
\put(8.5,8.8){Fig.~\ref{plotedist}b}
	\put(3,8){\includegraphics{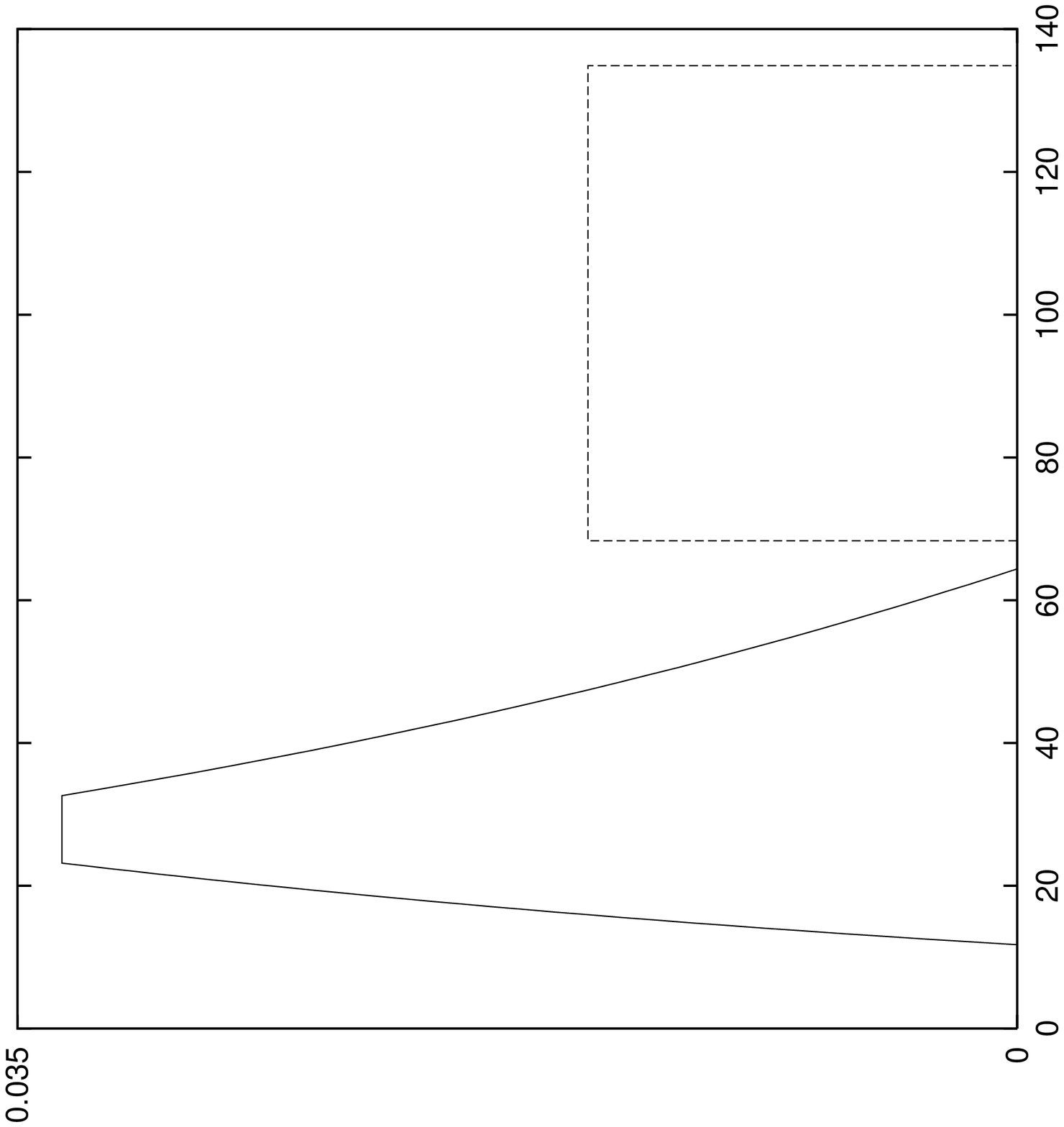}}
	\put(9.5,-0.3){$E~[{\rm GeV}]$}
	\put(4,7.6){$  \frac{1}{\sigma}\frac{d\sigma}{dE}$ }
	\put(5.8,6.3){$ \ell_2 $}
	\put(9.2,3.5){$\ell_1 $}
	\put(4.5,-0.3){Fig.~\ref{plotedist}c}

 \end{picture}
\vspace*{.5cm}
\caption{
Energy distributions in the laboratory system
for $\ell_1$ (dashed line) and $\ell_2$ (solid line)
for $e^+  e^-\to \tilde\chi^0_1\tilde\chi^0_2$
and the subsequent decays $\tilde\chi^0_2\to \tilde\ell_R \ell_1$ 
and $\tilde\ell_R\to\tilde\chi^0_1\ell_2$,
for $M_2=300$ GeV, $m_{\tilde\ell_R}=180$ GeV,
$\tan\beta=10$ and $\{|\mu|,m_{\chi_1^0},m_{\chi_2^0}\}/{\rm GeV}=
\{200,140,185\},\;
\{300,145,240\},\;
\{500,150,300\}$ in a, b, c respectively.
\label{plotedist}}
\end{figure}

We show in Figs.~\ref{plotedist}a - c different types of energy 
distributions for lepton $\ell_1$ (dashed line), and lepton $\ell_2$
(solid line), $\ell=e,\mu$, for
$e^+  e^-\to \tilde\chi^0_1\tilde\chi^0_2$
and the subsequent decays 
$\tilde\chi^0_2\to \tilde\ell \ell_1$ 
and $\tilde\ell\to\tilde\chi^0_1\ell_2$.
The parameters
$\tan \beta= 10$, $M_2=300$ GeV, $\varphi_{\mu}=0$, 
$\varphi_{M_1}=0.5 \pi$, and for $|\mu|=200$, $300$ and $500$~GeV,
are chosen such that the
slepton mass $m_{\tilde\ell_R}=180$~GeV is constant, the
LSP mass $m_{\chi_1^0}=140,145,150 $~GeV is almost constant
whereas the neutralino mass $m_{\chi_2^0}=185,240,300$~GeV is increasing.
The mass difference between $\tilde\ell_R$ and $\tilde\chi^0_1$  
decreases ($\Delta m=40,35,30$~GeV), whereas
the mass difference between $\tilde\chi^0_2$ and $\tilde\ell_R$
increases ($\Delta m=5,60,120$~GeV).
The endpoints of the energy distributions of the decay leptons 
depend on  these mass differences. Thus, in 
Fig.~\ref{plotedist}a, the second lepton is more energetic 
than the first lepton. The energy distributions 
do not overlap and thus the two leptons can be distinguished
by measuring their energies.
This also holds for Fig.~\ref{plotedist}c, where 
the first  lepton is more energetic than the second one.
In Fig.~\ref{plotedist}b the two distributions overlap 
because the mass differences between $\tilde\chi^0_1$, $\tilde\ell_R$
and $\tilde\chi^0_2$ are similar. One has to apply cuts  
to distinguish the leptons, which reduce the number of events. 

%A potentially large background may be due to slepton production
%$ e^+e^-\to\tilde\ell^+ \tilde\ell^-
%\to\ell^+\ell^-\tilde\chi^0_1\tilde\chi^0_1$.
%However, these reactions would lead generally to ''two-sided
%events``, whereas the events from
%$ e^+e^-\to\tilde\chi^0_1 \tilde\chi^0_i
%\to\ell^+\ell^-\tilde\chi^0_1\tilde\chi^0_1$
%are ''one-sided events``. Moreover, the background
%reaction is CP-even and will not give rise to a CP asymmetry, because
%the sleptons are scalars and their decay is a two-body one.

%%%%%%%%%%%%%%%%%%%%%%%%%%%%%%%%%%%%%%%%%%%%%%%%%%%%%%%%%%%%%%%%%%
%            P L O T     6
%%%%%%%%%%%%%%%%%%%%%%%%%%%%%%%%%%%%%%%%%%%%%%%%%%%%%%%%%%%%%%%%%

\subsection{Summary of Section \ref{T odd asymmetries in neutralino production and decay into sleptons}
	\label{Summary of section x}}

We have considered two triple-product asymmetries in 
neutralino production
$e^+e^- \to\tilde\chi^0_i \tilde\chi^0_j$
and the subsequent leptonic two-body decay chain of one  neutralino
$\tilde\chi^0_i \to \tilde\ell  \ell$,
$ \tilde\ell \to \tilde\chi^0_1  \ell$ for
$ \ell= e,\mu,\tau$. These asymmetries
are present  already at tree level and are due to spin effects in the 
production and decay process of two different neutralinos. 
The asymmetries are sensitive to  CP-violating phases of the 
gaugino and Higgsino mass parameters $M_1$ and/or $\mu$  
in the neutralino production process.

For the process 
$e^+e^- \to\tilde\chi^0_1 \tilde\chi^0_2$
and neutralino decay into a right slepton
$\tilde\chi^0_2 \to \tilde\ell_R \ell$,
we have shown that the asymmetries can be as large as 25\%.
They can be enhanced using polarized beams,
and can be sizable even for a small phases, 
$\varphi_{\mu}, \varphi_{M_1}\approx 0.1\pi$,
which is suggested by the experimental limits on EDMs. 
%The asymmetries are simlar for the processes 
%$e^+e^- \to\tilde\chi^0_1 \tilde\chi^0_3$ and
%$e^+e^- \to\tilde\chi^0_2 \tilde\chi^0_3$.
%Depending on the MSSM scenario, 
%our proposed asymmetries should be accessible in future  
%electron-positron linear collider experiments in the
%500 GeV range. Longitudinally polarized electron and positron
%beams can considerably enhance both asymmetries
%and production cross sections.

		\section{CP asymmetry in neutralino production and decay into
	polarized taus
	\label{A CP asymmetry in neutralino production and decay into staus
		with tau polarization}}

%We propose a CP-odd asymmetry in the  process
%$e^+\,e^- \to\tilde\chi^0_i \, \tilde\chi^0_j$
%and the subsequent decay of one neutralino into a stau tau pair
%$\tilde\chi^0_i \to \tilde\tau^{\pm} \, \tau^{\mp}$
% by means of the transverse $\tau^{\mp}$ polarization.
% In the Minimal Supersymmetric Standard Model with 
% complex parameters $\mu$, $M_1$ and $A_{\tau}$,
% we calculate the asymmetry and cross sections at a future
% 500~GeV collider with longitudinal polarized beams.
% The asymmetry can reach values up to 60\% and we  
% estimate the sensitivity for measuring the $\tau$
% polarization which is necessary to probe the CP asymmetry. 
%
%
%\subsection{Introduction}

For the two-body decays of neutralinos into sleptons, where the lepton
polarizations are summed, we have shown in the last section 
that the asymmetries have only CP-odd contributions from 
the neutralino production process.
For neutralino decay into a tau, the tau polarization allows to define
an asymmetry, which has also CP-odd contributions from the
neutralino decay process.
This is particularly interesting  since an asymmetry can be defined,
which is sensitive to the CP phase of the trilinear scalar coupling 
parameter $A_{\tau}$.

We consider neutralino production
\begin{equation} \label{stau:production}
	e^++e^-\to \tilde\chi^0_i + \tilde\chi^0_j; \quad i,j=1,\dots,4,
\end{equation}
and the subsequent two-body decay of one neutralino into a tau
\begin{equation} \label{stau:decay}
	\tilde\chi^0_i \to \tilde\tau_m^{\pm} +\tau^{\mp}; \quad m=1,2.
\end{equation}
The $\tau^-$ polarization is given by \cite{Renard}
\begin{equation} \label{stau:polvector}
	{\bf P} =\frac{  {\rm Tr}(\rho
			\mbox{\boldmath$ \sigma$ })}{{\rm Tr}(\rho)},
\end{equation}
with $\rho$ the hermitean spin density matrix
of the $\tau^-$ and $\sigma_i$ the Pauli matrices.
The component $P_3$ of the polarization vector 
${\bf P}=(P_1,P_2,P_3)$ is the longitudinal polarization,  
$P_1$ is the transverse polarization in the plane 
and $P_2$ is the transverse polarization perpendicular to the plane
defined by the momenta ${\bf p}_{\tau}$ and ${\bf p}_{e^-}$.
The transverse polarization $P_2$ is proportional to the triple product
%formed by 
%${\bf p}_{e^-}$ and ${\bf p}_{\tau}$.
%The component $P_2$ is the polarization perpendicular to
%${\bf p}_{\tau}$ and ${\bf p}_{e^-}$ and
%is proportional to the triple product
\begin{equation} \label{stau:triplepol}
	{\mathcal T}_{\tau}={\bf s}_{\tau}\cdot({\bf p}_{\tau}\times {\bf p}_{e^-}),
\end{equation}
where ${\bf s}_{\tau}$ is the $\tau^-$ spin 3-vector.
For its definition in the laboratory system, see
\ref{stau:polvec}.
%Since under time reversal the triple-product changes sign,
%the transverse $\tau^-$ polarization $P_2$ is a T-odd observable.
%Due to CPT invariance, $P_2$ is actually a CP-odd observable if 
%absorbtive phases from  final-state interactions are 
%neglected. 
In order to eliminate absorptive phases, we define the 
CP-odd asymmetry 
\begin{equation} \label{stau:asy}
	{\mathcal A}_{\rm CP}=\frac{1}{2}(P_2-\bar{P}_2),
\end{equation}
where 
%$\bf P$ is the $\tau^-$ polarization vector in the 
%decay $\tilde\chi^0_i\to \tilde\tau_m^+ \tau^-$ and 
$\bar{\bf P}$ denotes the 
$\tau^+$ polarization in the charge conjugated process
$\tilde\chi^0_i\to \tilde\tau_m^- \tau^+$.
The asymmetry ${\mathcal A}_{\rm CP}$ is sensitive to the phase 
$\varphi_{A_{\tau}}$
%of the trilinear scalar coupling parameter $A_{\tau}$ 
in the stau sector, as well as to the phases
$\varphi_{\mu}$ and $\varphi_{M_1}$ in the neutralino sector.

Note, that for the two-body decay~(\ref{stau:decay}),
the transverse $\tau$ polarization $P_2$ is the only observable
which is sensitive to $\varphi_{A_{\tau}}$.
The asymmetries defined in 
Section~\ref{T odd asymmetries in neutralino production and 
decay into sleptons}, where the $\tau$ polarization is
summed, are only sensitive to CP violation 
due to $\varphi_{\mu}$ and $\varphi_{M_1}$ in the production process. 
%We stress that without measuring the 
%transverse $\tau^{\mp}$ polarization no sensitivity
%to the phase $\varphi_{A_{\tau}}$ of $A_{\tau}$ can
%be obtained, because (\ref{stau:decay}) is a two-body decay.
%When summing over the $\tau^-$ polarization, we are sensitive
%only to CP violation in the production process, see 
%Section~\ref{T odd asymmetries in neutralino production and decay into sleptons}.

\subsection{Tau spin-density matrix and cross section 
	\label{Tau spin-density matrix}}

%The unnormalized spin-density matrix
%of the $\tau^-$ is defined by:
%\begin{equation} \label{stau:spindesity}
%	\rho^{\lambda_k\lambda'_k}\equiv
%	\int (|T|^2)^{\lambda_k\lambda'_k}
%	d{\rm Lips}(s;p_{\chi_j },p_{\tau},p_{\tilde\tau_m}),
%\end{equation}
%where $|T|^2$ is the amplitude squared
%and $d{\rm Lips}$ is the Lorentz invariant phase-space 
%element~(\ref{Lipsleptonic1}) 
%for neutralino production (\ref{stau:production}) and decay (\ref{stau:decay}). The $\tau^-$ helicities are denoted by $\lambda_k$ and $\lambda'_k$.

In the spin density matrix formalism, see 
Appendix~\ref{Neutralino production and decay matrices},
the unnormalized spin-density matrix of the $\tau^-$
can be written as
\begin{equation} \label{stau:matrixelement}
\rho_P(\tau^-)^{\lambda_k\lambda'_k}=
	|\Delta(\tilde\chi^0_i)|^2~\sum_{\lambda_i,\lambda'_i}~
	{\rho_P(\tilde\chi^0_i)}^{\lambda_i\lambda'_i}~
	{\rho_D(\tilde\chi^0_i)}_{\lambda'_i\lambda_i}^{\lambda_k\lambda'_k}.
\end{equation}
It is composed of the neutralino propagator $\Delta(\tilde\chi^0_i)$, 
the spin density matrices $\rho_P(\tilde\chi^0_i)$ for neutralino 
production~(\ref{stau:production}) and $\rho_D(\tilde\chi^0_i)$ 
for neutralino decay~(\ref{stau:decay}).
%the propagator  $ \Delta(\ti\chi^0_i ) = 
%1/[p^2_{\chi_i} -m^2_{\chi_i}
%+im_{\chi_i}\Gamma_{\chi_i}]$, with 
%$p_{\chi_i}, m_{\chi_i}, \Gamma_{\chi_i}$
%being the four-momenta, masses and widths of the decaying neutralino, 
%respectively.
The $\tilde\chi^0_i$ helicities are denoted by 
$\lambda_i$ and $\lambda'_i$, and the $\tau^-$ helicities 
are denoted by $\lambda_k$ and $\lambda'_k$.
The neutralino production matrix $\rho_P(\tilde\chi^0_i)$ is defined 
in~(\ref{neut:rhoP}) and the neutralino decay matrix
$\rho_D(\tilde\chi^0_i)$ in~(\ref{stau:rhoD}).
Inserting these density matrices
%~(\ref{neut:rhoP}) and~(\ref{stau:rhoD}) 
into~(\ref{stau:matrixelement}) gives
\begin{equation} \label{stau:matrixelement2}
	\rho_P(\tau^-)^{\lambda_k\lambda'_k}=
	4 |\Delta(\tilde\chi^0_i)|^2~
	\lbrack (P D + \Sigma^a_P \Sigma^a_{D})\delta_{\lambda_k\lambda'_k}+
	(P D^b+\Sigma^a_P\Sigma^{ab}_{D})
	\sigma^b_{\lambda_k\lambda'_k})\rbrack.
\end{equation}
The last term of the coefficient 
$\Sigma^{ab}_{D}$, see~(\ref{stau:sab1}), contains for $b=2$ the triple 
product~(\ref{stau:triplepol}). This term is proportional
to the product of the $\tilde \chi^0_i$-$\tilde \tau_k$-$\tau$ couplings
${\rm Im}({b^{\tilde \tau}_{mi}}^*a^{\tilde \tau}_{mi})$  
and is therefore sensitive to the phases $\varphi_{A_{\tau}},
\varphi_{\mu}$ and $\varphi_{M_1}$.

The amplitude squared is obtained by summing over the $\tau$ helicities
in~(\ref{stau:matrixelement2})
\begin{eqnarray} \label{stau:amplitude}
|T|^2&=& 4 |\Delta(\tilde\chi^0_i)|^2~
	\lbrack P~ (2D) + \Sigma^a_P~ (2\Sigma^a_{D})\rbrack,
\end{eqnarray}
where the CP sensitive term $\Sigma^{ab}_{D}$ drops out.
The cross section is then given by
\begin{equation}\label{stau:crossection}
	d\sigma=\frac{1}{2 s}|T|^2 d{\rm Lips}
	(s;p_{\chi_j^0},p_{\tau},p_{\tilde\tau}),
\end{equation}
with the phase space element $d{\rm Lips}$ as defined
in~(\ref{Lipsleptonic1}).

\subsection{Transverse tau polarization and CP asymmetry 
	\label{Transverse tau polarization and CP asymmetry}}

From~(\ref{stau:matrixelement2}) we obtain for the 
transverse $\tau^-$ polarization~(\ref{stau:polvector})
%\begin{equation} \label{stau:polasy1}
%P_2= \frac{\int|\Delta(\tilde\chi^0_i)|^2~\Sigma^a_P \Sigma^{a2}_{D}
%~d{\rm Lips}}
%{\int |\Delta(\tilde\chi^0_i)|^2 ~P D~d{\rm Lips}},
%\end{equation}
\begin{equation} \label{stau:polasy1}
P_2= \frac{\int \Sigma^a_P \Sigma^{a2}_{D}
~d{\rm Lips}}
{\int P D~d{\rm Lips}},
\end{equation}
which follows since 
we have used the narrow width approximations for the
propagators and in the numerator 
$\int|\Delta(\tilde\chi^0_i)|^2
~P D^2~d{\rm Lips}=0$ and in the denominator 
$\int|\Delta(\tilde\chi^0_i)|^2~\Sigma^a_P \Sigma^a_D
~d{\rm Lips}=0$.

As can be seen from~(\ref{stau:polasy1}), 
$P_2$ is proportional to the spin correlation term
$\Sigma^{a2}_{D}$~(\ref{stau:sab1}), which contains the 
CP-sensitive part ${\rm Im}({b^{\tilde \tau}_{mi}}^*a^{\tilde \tau}_{mi})
\epsilon_{\mu\nu\rho\sigma}\,
p_{\tau}^{\mu} \,p_{\tilde\chi_i^0}^{\nu}\,s^{a,\,\rho}_{\chi_i^0}\,
s^{b,\,\sigma}_{\tau}$.
In order to study the dependence of $P_2$ on the parameters, 
we expand for $\tilde\tau_1$
\begin{eqnarray}\label{stau:Im}
&&{\rm Im}({b^{\tilde \tau}_{1i}}^*a^{\tilde \tau}_{1i})= 
g^2\cos^2\theta_{\tilde \tau} Y_{\tau}{\rm Im}(f^{\tau}_{Li} N_{i3})+
g^2\sin^2\theta_{\tilde \tau} Y_{\tau}\sqrt{2}\tan\theta_W{\rm Im}(N_{i1}N_{i3})
\nonumber\\
%[3mm]
&&{}+g^2\sin^2\theta_{\tilde \tau}\cos^2\theta_{\tilde \tau}
	[Y^2_{\tau}{\rm Im}(N_{i3}N_{i3}e^{i\varphi_{\tilde \tau}})+
g^2\sqrt{2}\tan\theta_W
{\rm Im}(f^{\tau}_{Li} N_{i1}e^{-i\varphi_{\tilde \tau}})],
\end{eqnarray}
using the definitions of the couplings in the stau sector, 
see Appendix~\ref{Lagrangian}.

For $\varphi_{\mu},\varphi_{M_1}=0$,
%If only $\varphi_{A_{\tau}}\neq 0,\pi$,
we find from (\ref{stau:Im}) that 
$P_2\propto \sin 2\theta_{\tilde \tau} \sin\varphi_{\tilde \tau}$.
We note that the dependence of $\varphi_{\tilde \tau}$
on $\varphi_{A_{\tau}}$ is weak if 
$|A_{\tau}|\ll |\mu|\tan\beta$, see~(\ref{eq:phtau}). Thus, we expect
that $P_2$ increases with increasing $|A_{\tau}|$.

In order to measure $P_2$ and the CP asymmetry 
${\mathcal A}_{\rm CP}$~(\ref{stau:asy}),
the $\tau^-$ from the neutralino decay
$\tilde\chi^0_i \to \tilde\tau_m^+ \tau^-$
and the $\tau^+$ from the subsequent $\tilde\tau_m^+$
decay $\tilde\tau_m^+\to\tilde\chi_1^0\tau^+$ have to be distinguished.
This can be accomplished by 
%measuring the energies of the $\tau$'s and making use of 
their different energy distributions, 
see Appendix~\ref{Energy distributions of the decay leptons}.

%A potentially large background may be due to stau production
%$ e^+e^-\to\tilde\tau_l^+ \tilde\tau_m^-
%\to\tau^+\tau^-\tilde\chi^0_1\tilde\chi^0_1$.
%However, these reactions would generally lead to ''two-sided
%events``, whereas the events from
%$ e^+e^-\to\tilde\chi^0_1 \tilde\chi^0_i
%\to\tau^+\tau^-\tilde\chi^0_1\tilde\chi^0_1$
%are ''one-sided events``. Moreover, the background
%reaction is CP-even and will not give rise to a CP asymmetry, because
%the staus are scalars with a two-body decay.

\subsection{Numerical results\label{stau:numerics}}

%In the following we demonstrate the effect of CP
%violation by means of the CP asymmetry in Eq.~\rf{eq:asy}.
We present numerical results for
$e^+e^-\to \tilde\chi^0_1\tilde\chi^0_2$ and the subsequent 
decay of the neutralino into the lightest stau 
$\tilde\chi^0_2\to \tilde\tau_1\tau $ for a linear collider  with
$\sqrt{s}=500$ GeV and longitudinally polarized beams with
$(P_{e^-},P_{e^+})=(\pm0.8,\mp0.6)$.
%or $(P_{e^-},P_{e^+})=(-0.8,0.6)$.
This choice favors right or left selectron exchange in the neutralino
production process, respectively.

We study the dependence of the asymmetry ${\mathcal A}_{\rm CP}$ 
and the production cross sections 
$\sigma= \sigma_P(e^+e^-\to  \tilde\chi^0_1 \tilde\chi^0_2)\times 
{\rm BR}( \tilde\chi^0_2\to\tilde\tau_1^+\tau^-)$ on the parameters 
$\varphi_{\mu}$, $|\mu|$, $\varphi_{M_1}$, $|M_1|$,
$\varphi_{A_{\tau}}$, $|A_{\tau}|$ and $\tan\beta$.
We assume $|M_1|=5/3~M_2\tan^2\theta_W $, 
%with $M_2$ real,
and use $m_0=100$~GeV for the universal scalar mass parameter
in the renormalization group equations of the selectron masses, 
see~(\ref{eq:mll}) and~(\ref{eq:mrr}).
%\cite{hall}, $M_{\tilde L}^2 = m^2_0+0.79 M^2_2$ 
%and $M_{\tilde E}^2 = m^2_0+0.23 M^2_2$, taking $m_0=100$~GeV. 
We take into account the restrictions on  $|A_{\tau}|$ due to the tree-level
vacuum stability conditions \cite{casas}. 
%The restrictions on the masses of the
%SUSY particles are $m_{\chi^{\pm}_1}>104$~GeV, $m_{\T_1}>100$~GeV and 
%$m_{\tilde\tau_1}>m_{\chi^0_1}$. 

For the calculation of the branching ratio 
${\rm BR}(\tilde\chi^0_2\to\tilde\tau_1^+\tau^-)$
we include the two-body decays
%we concentrate on the parameter domain where two-body decays are allowed
%and neglect three-body decays. We consider the two-body decays
%
\begin{eqnarray}
\tilde\chi^0_2 &\to& \tilde\tau_m\tau,~
\tilde\ell_{R,L}\ell,~ 
\tilde\chi^0_1 Z,~
\tilde\chi^{\mp}_n W^{\pm},~
\tilde\chi^0_1 H_1^0,
\quad \ell=e,\mu, \quad m,n=1,2,
\end{eqnarray}
with $m_{A}=1$~TeV,
such that the neutralino decays into the charged Higgs bosons 
$\tilde\chi^0_2 \to \tilde\chi^{\pm}_n H^{\mp}$,
as well as decays into the heavy neutral Higgs
bosons $\tilde\chi^0_2 \to \tilde\chi^0_1~H_{2,3}^0$,
are excluded in our scenarios.
%with $H_1^0$ being the lightest neutral Higgs boson.
%The Higgs mass parameter is chosen as $m_{A}=1$~TeV,
%which means that explicit CP violation is not important
%for the lightest Higgs state \cite{ref3}.

\begin{figure}[t]
\setlength{\unitlength}{0.035cm}
%	\fbox{
			\begin{minipage}{0.47\textwidth}
 \begin{picture}(100,210)(0,0)
	\put(0,0){\includegraphics{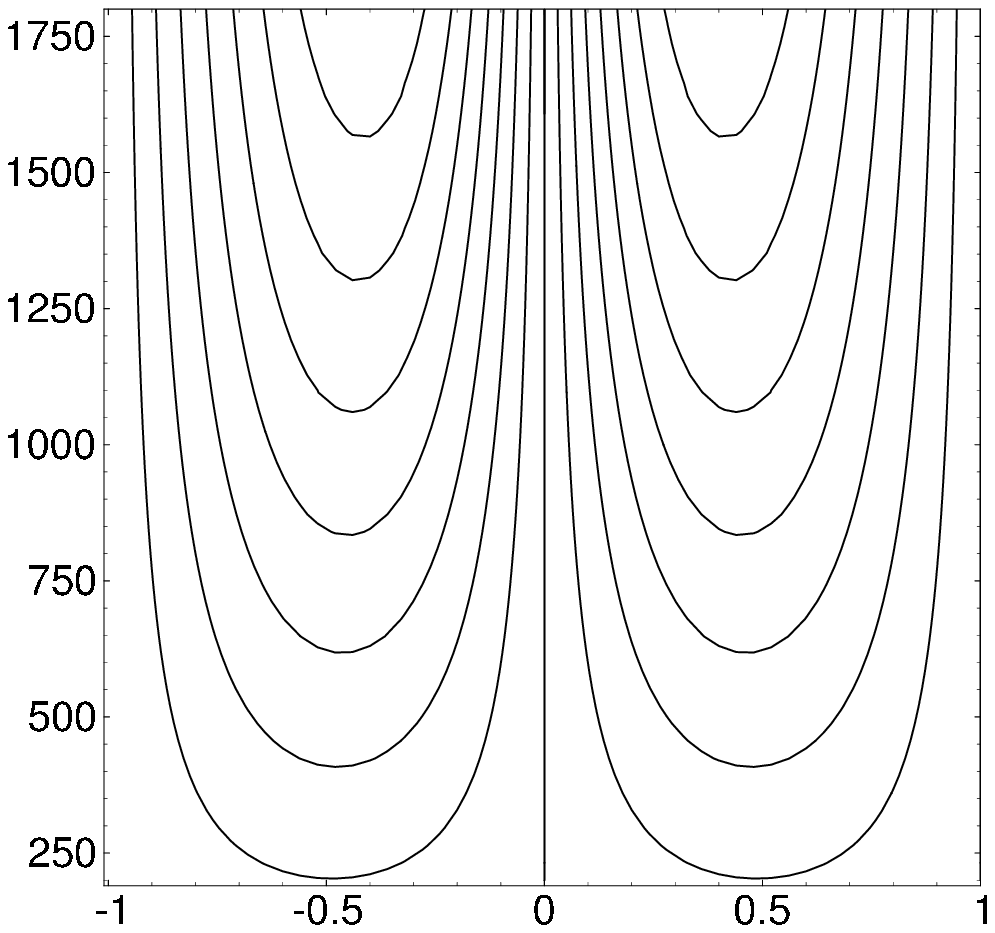}}
	\put(80,210){\fbox{${\mathcal A}_{\rm CP}$ in \% }}
	\put(170,-5){$\varphi_{A_{\tau}}~[\pi]$}
	\put(0,210){$ A_{\tau}~[{\rm GeV}]$}
	\put(60,23){\footnotesize{$-2$}}
	\put(60,50){\footnotesize{$-4$}}
	\put(61,75){\footnotesize{$-6$}}
	\put(61,100){\footnotesize{$-8$}}
	\put(60,125){\footnotesize{$-10$}}
	\put(60,150){\footnotesize{$-12$}}
	\put(61,185){\footnotesize{$-14$}}
	\put(150,23){\footnotesize{$2$}}
	\put(150,50){\footnotesize{$4$}}
	\put(150,75){\footnotesize{$6$}}
	\put(150,100){\footnotesize{$8$}}
	\put(146,125){\footnotesize{$10$}}
	\put(145,150){\footnotesize{$12$}}
	\put(144,185){\footnotesize{$14$}}
	\put(102,22){\footnotesize{$0$}}
%	\put(192,22){\footnotesize{$0$}}
 \end{picture}
\vspace*{.3cm}
\caption{Contour lines of  ${\mathcal A}_{\rm CP}$ 
for $M_2=200$~GeV, $|\mu|=250$~GeV, 
$\tan\beta=5$, $\varphi_{M_1}=\varphi_{\mu}=0$ and 
$(P_{e^-},P_{e^+})=(0.8,-0.6)$.
\label{fig1}}
\end{minipage}
%}
\hspace*{0.5cm}
\begin{minipage}{0.47\textwidth}
 \begin{picture}(100,210)(0,0)
	\put(0,0){\includegraphics{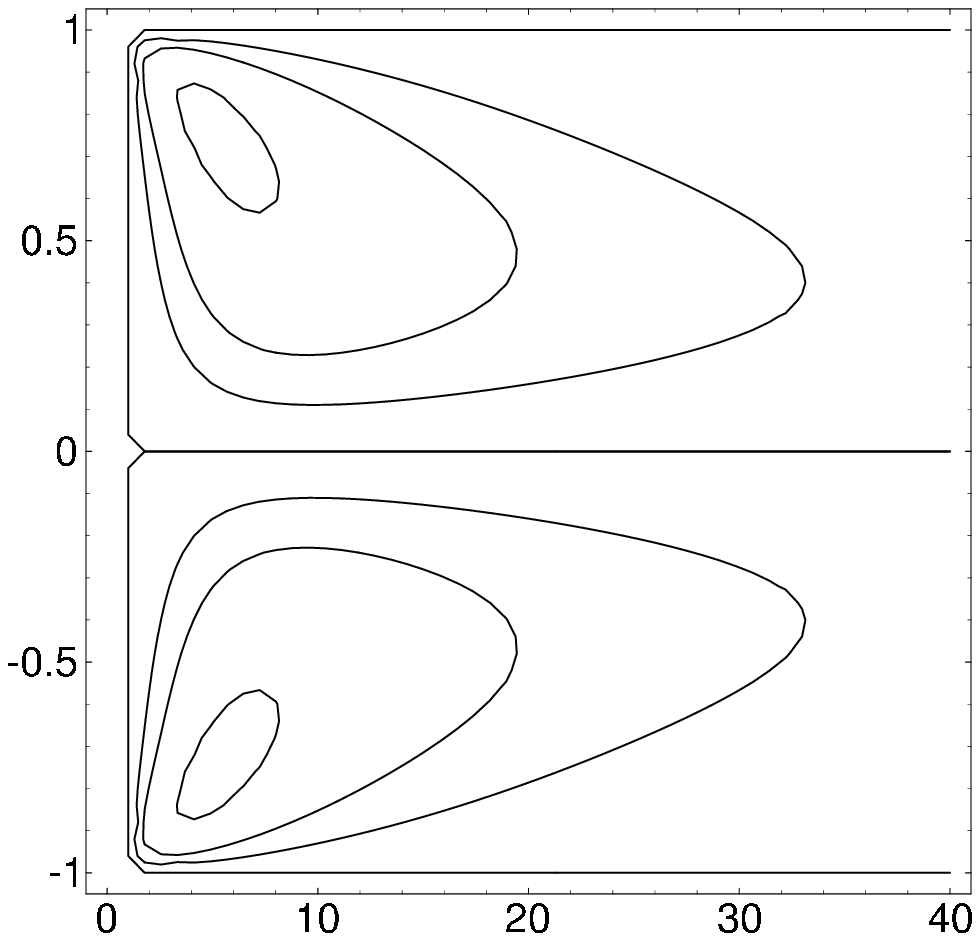}}
	\put(80,210){\fbox{${\mathcal A}_{\rm CP}$ in \% }}
	\put(170,-5){$\tan\beta$}
	\put(0,210){$ \varphi_{M_{1}}~[\pi]$ }
	\put(35,40){\footnotesize{$-20$}}
	\put(85,62){\footnotesize{$-10$}}
	\put(144,70){\footnotesize{$-5$}}
	\put(175,110){\footnotesize{$0$}}
		\put(45,165){\footnotesize{$20$}}
		\put(90,145){\footnotesize{$10$}}
		\put(154,140){\footnotesize{$5$}}
 \end{picture}
\vspace*{.3cm}
\caption{Contour lines of  ${\mathcal A}_{\rm CP}$ 
for $A_{\tau}=1$~TeV, $M_2=300$~GeV, $|\mu|=250$~GeV, 
$\varphi_{A_{\tau}}=\varphi_{\mu}=0$
and $(P_{e^-},P_{e^+})=(0.8,-0.6)$.
\label{fig2}}
\end{minipage}
\vspace*{.5cm}
\end{figure}

In Fig.~\ref{fig1} we show the contour lines of ${\mathcal A}_{\rm CP}$
in the $\varphi_{A_{\tau}}$--$|A_{\tau}|$ plane.
The asymmetry ${\mathcal A}_{\rm CP}$ is proportional to
$\sin 2\theta_{\tilde\tau} \sin\varphi_{\tilde\tau}$, 
and  increases with increasing
$|A_{\tau}|\gg |\mu|\tan\beta$, which is expected from~(\ref{stau:Im}).
Furthermore, in the parameter region shown 
the cross section $\sigma$ varies between 20~fb and 30~fb.

In Fig.~\ref{fig2} we show the dependence
of ${\mathcal A}_{\rm CP}$ on $\tan\beta$ and $\varphi_{M_1}$.
Large values up to $\pm 20\%$ are obtained for $\tan\beta\approx 5$. 
Note that these values are obtained for $\varphi_{M_1}\approx \pm0.8\pi$
rather than for maximal $\varphi_{M_1}\approx\pm0.5\pi$.
%The reason is that ${\mathcal A}_{II}^{\rm T}$ is 
%proportional to a product of
%a CP odd ($\Sigma_P^2$) and a CP even factor ($\Sigma_{D_1}^2$),
%see (\ref{properties_neut}). The  CP odd (CP even) factor has
%as sine-like (cosine-like) dependence on the phases.
%Thus the maximum of ${\mathcal A}_{II}^{\rm T}$ is shifted  
%towards $\varphi_{M_1}=0$ in Fig.~\ref{varphases_12}b.
This is due to the interplay of CP-even and CP-odd 
contributions to the spin correlation terms
in~(\ref{stau:polasy1}). In the region shown in Fig.~\ref{fig2},
the cross section $\sigma$ varies between 10~fb and 30~fb.

\begin{figure}
	\setlength{\unitlength}{0.035cm}
\begin{picture}(120,220)(0,0)
\put(0,0){\includegraphics{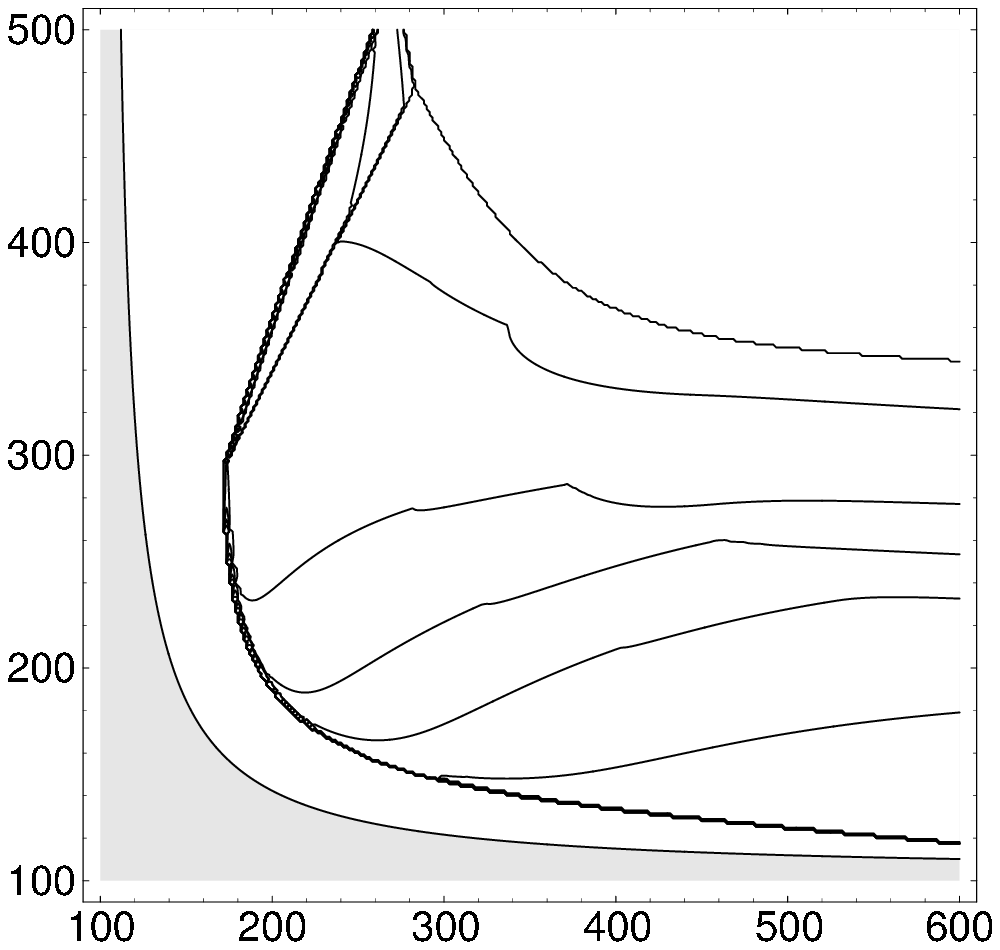}}
   \put(165,-8){$|\mu|~[{\rm GeV}]$}
	\put(-5,210){$M_2~[{\rm GeV}]$}
\put(50,210){\fbox{$\sigma(e^+e^-\to\tilde\chi^0_1\tilde\tau_1^+\tau^-)$ in fb}}
	\put(170,55){\footnotesize{100}}
	\put(135,72){\footnotesize{50}}
	\put(110,82){\footnotesize{25}}
	\put(85,98){\footnotesize{10}}
	\put(80,146){\footnotesize{1}}
		\put(20,-8){Fig.~\ref{fig3}a}
\put(225,0){\includegraphics{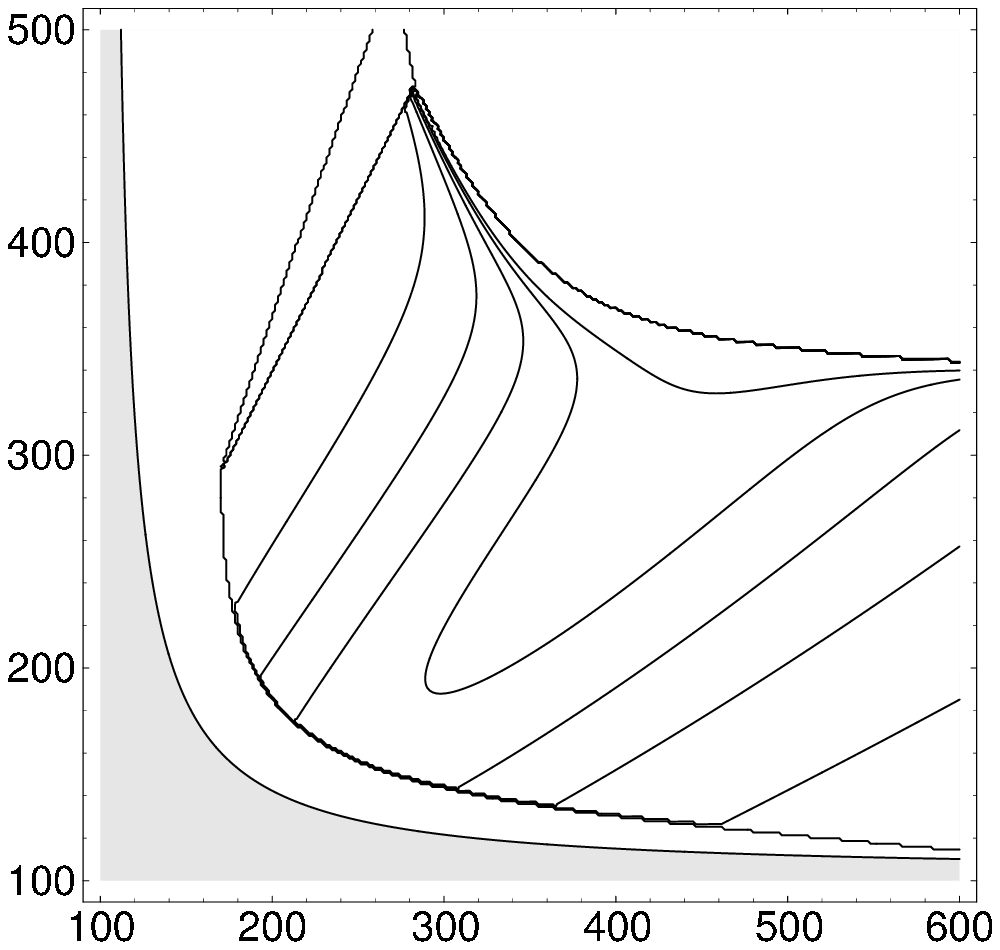}}
   \put(390,-8){$|\mu|~[{\rm GeV}]$}
	\put(225,210){$M_2~[{\rm GeV}]$}
	\put(300,210){\fbox{${\mathcal A}_{\rm CP}$ in \% }}
	\put(362,122){\scriptsize{-15}}
	\put(320,68){\scriptsize{-12}}
	\put(340,60){\scriptsize{-9}}
	\put(360,55){\scriptsize{-6}}
	\put(380,43){\scriptsize{-3}}
	\put(300,87){\scriptsize{-9}}
	\put(292,96){\scriptsize{-6}}
	\put(283,110){\scriptsize{-3}}
	\put(293,160){\scriptsize{0}}
	\put(245,-8){Fig.~\ref{fig3}b }
\end{picture}
\vspace*{.3cm}
 \caption{Contour lines of $\sigma$ and ${\mathcal A}_{\rm CP}$
in the $|\mu|$--$M_2$ plane for
 $\varphi_{A_{\tau}}=0.5\pi$, $\varphi_{M_1}=\varphi_{\mu}=0$,
$A_{\tau}=1$~TeV, $\tan \beta=5$ and $(P_{e^-},P_{e^+})=(-0.8,0.6)$.
The blank area outside the area of the contour lines is kinematically
forbidden since here either $\sqrt{s}<m_{\chi_1^0}+m_{\chi_2^0}$ or
$m_{\tilde\tau_1}+m_{\tau}>m_{\chi_2^0}$. The gray area is excluded by
$m_{\chi^{\pm}_1}<104$~GeV.
\label{fig3}}
\end{figure}

Figs.~\ref{fig3}a and \ref{fig3}b show, 
for $\varphi_{A_{\tau}}=0.5\pi$ and $\varphi_{M_1}=\varphi_{\mu}=0$,
the $|\mu|$--$M_2$ dependence of  
the cross section $\sigma$ and the asymmetry 
${\mathcal A}_{\rm CP}$, respectively.
%for $\tan \beta=5$, $A_{\tau}=1000$ GeV, $\varphi_{A_{\tau}}=0.5\pi$, 
%$\varphi_{M_1}=\varphi_{\mu}=0$ and $P_{e^-}=-0.8$, $P_{e^+}=0.6$. 
The asymmetry reaches values up to $-15 \%$ due to the 
large value of $|A_{\tau}|=1$ TeV and the choice of 
the beam polarization $(P_{e^-},P_{e^+})=(-0.8,0.6)$. 
This choice also enhances the cross section, which reaches 
values of more than $100$~fb. 
%The gray shaded area excludes 
%chargino masses $m_{\chi^{\pm}_1}<104$~GeV.
%In the blank area either the sum of the masses of the produced neutralinos
%exceeds $\sqrt{s}=500$ GeV or the two-body decay
%$\tilde\chi^0_2\to \tilde\tau_1^+\tau^- $ is not open.

\begin{figure}[t]
	\setlength{\unitlength}{0.035cm}
\begin{picture}(120,220)(0,0)
\put(0,0){\includegraphics{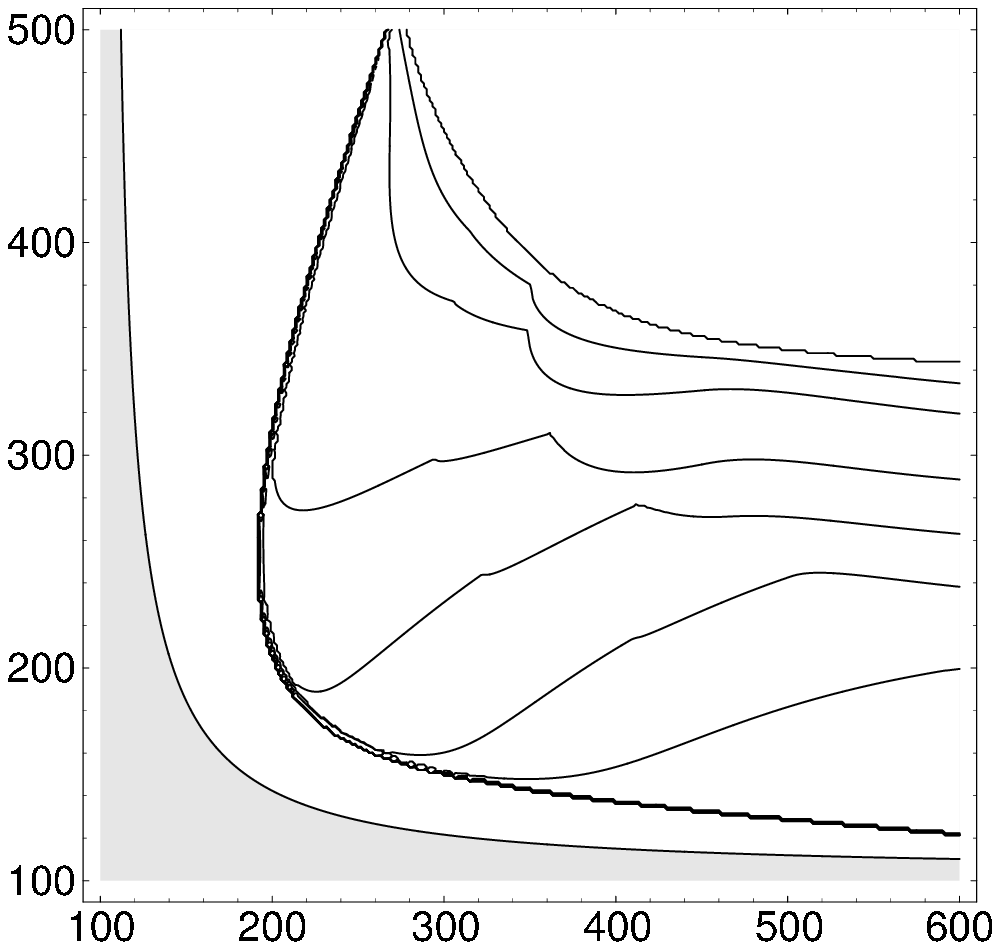}}
   \put(165,-8){$|\mu|~[{\rm GeV}]$}
	\put(-5,210){$M_2~[{\rm GeV}]$}
\put(50,210){\fbox{$\sigma(e^+e^-\to\tilde\chi^0_1\tilde\tau_1^+\tau^-)$ in fb}}
		\put(170,49){\footnotesize{150}}
	\put(135,64){\footnotesize{100}}
	\put(102,74){\footnotesize{50}}
	\put(82,94){\footnotesize{25}}
	\put(85,127){\footnotesize{10}}
	\put(86,152){\footnotesize{5}}
	\put(20,-8){Fig.~\ref{fig4}a}
\put(225,0){\includegraphics{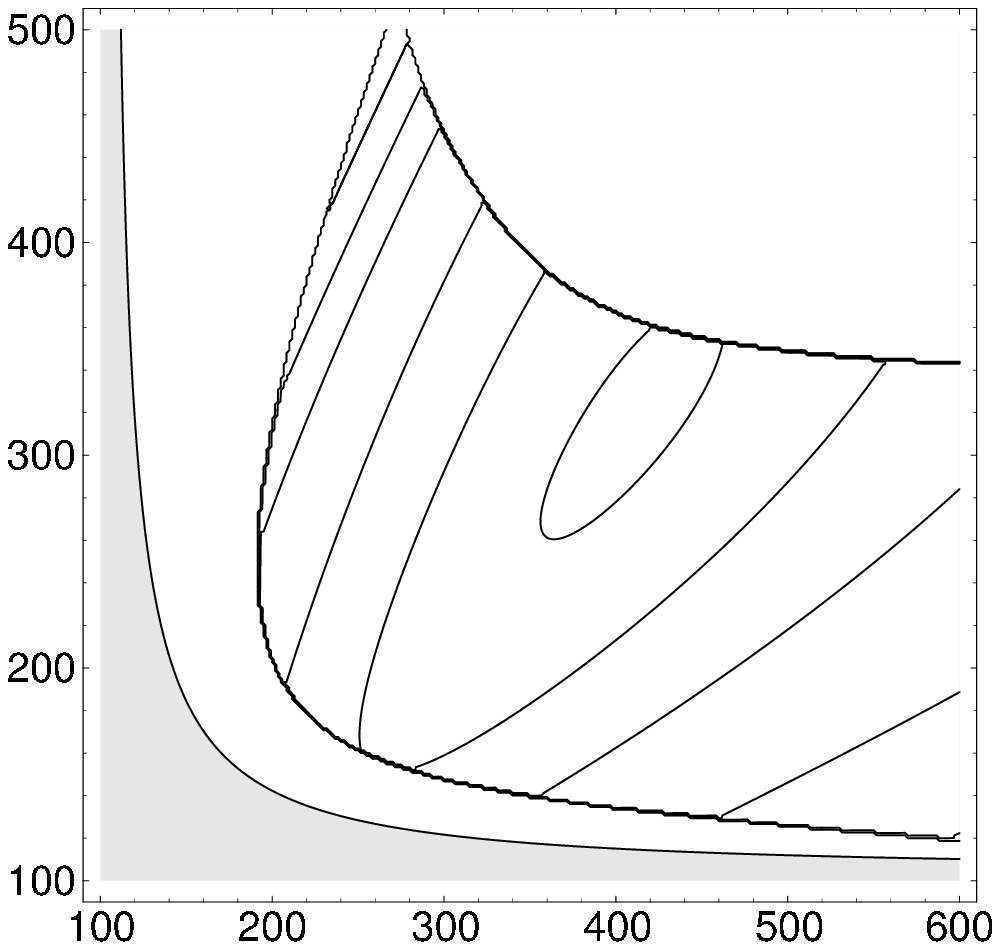}}
   \put(390,-8){$|\mu|~[{\rm GeV}]$}
	\put(225,210){$M_2~[{\rm GeV}]$}
	\put(300,210){\fbox{${\mathcal A}_{\rm CP}$ in \% }}
	\put(340,105){\footnotesize{-38}}
	\put(360,73){\footnotesize{-30}}
	\put(385, 65){\footnotesize{-20}}
	\put(400, 40){\footnotesize{-10}}
	\put(320,110){\footnotesize{-30}}
	\put(305,125){\footnotesize{-20}}
	\put(280,110){\footnotesize{-10}}
	\put(295,157){-5}
	\put(245,-8){Fig.~\ref{fig4}b }
\end{picture}
\vspace*{.25cm}
 \caption{Contour lines of $\sigma$  and ${\mathcal A}_{\rm CP}$
in the $|\mu|$--$M_2$ plane for
 $\varphi_{M_1}=0.5\pi$, $\varphi_{A_{\tau}}=\varphi_{\mu}=0$,
 $A_{\tau}=250$~GeV, $\tan \beta=5$ and $(P_{e^-},P_{e^+})=(-0.8,0.6)$.
The blank area outside the area of the contour lines is kinematically
forbidden since here either $\sqrt{s}<m_{\chi_1^0}+m_{\chi_2^0}$ or
$m_{\tilde\tau_1}+m_{\tau}>m_{\chi_2^0}$. The gray area is excluded by
$m_{\chi^{\pm}_1}<104$~GeV.
\label{fig4}}
\end{figure}

For $\varphi_{M_1}= 0.5\pi$ and
$\varphi_{\mu}=\varphi_{A_{\tau}}=0$ we show in Figs.~\ref{fig4}a,b 
the contour lines of $\sigma$  and ${\mathcal A}_{\rm CP}$,
respectively, in the $|\mu|$-$M_2$ plane.
It is remarkable that in a large region the asymmetry is larger than
$-10\%$ and reaches values up to $-40\%$. 
Unpolarized beams would reduce ${\mathcal A}_{\rm CP}$
only marginally, however the largest values of 
$\sigma$ would be reduced by a factor $3$.

\begin{figure}
	\setlength{\unitlength}{0.035cm}
\begin{picture}(120,220)(0,0)
\put(0,0){\includegraphics{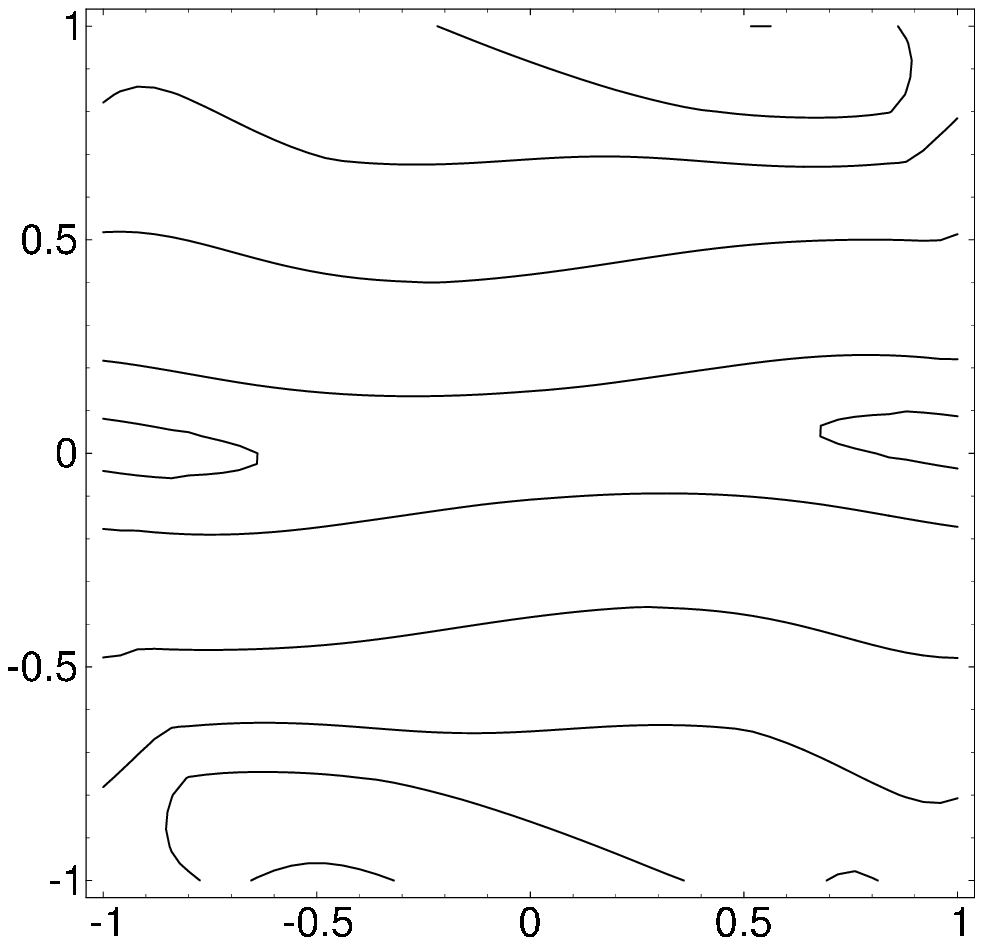}}
   \put(165,-8){$\varphi_{\mu}~[\pi]$}
	\put(-5,210){$\varphi_{M_1}~[\pi]$ }
	\put(42,210){\fbox{$\sigma(e^+e^-\to \tilde\chi^0_1\tilde\tau_1^+\tau^-)$ in fb }}
	\put(125,30){\footnotesize{15}}
	\put(130,55){\footnotesize{10}}
	\put(110,74){\footnotesize{5}}
	\put(100,96){\footnotesize{1}}
	\put(100,120){\footnotesize{1}}
	\put(90,143){\footnotesize{5}}
	\put(70,168){\footnotesize{10}}
	\put(140,180){\footnotesize{15}}
	\put(30,105){\footnotesize{0.2}}
	\put(185,105){\footnotesize{0.2}}
		\put(20,-8){Fig.~\ref{fig5}a}
\put(225,0){\includegraphics{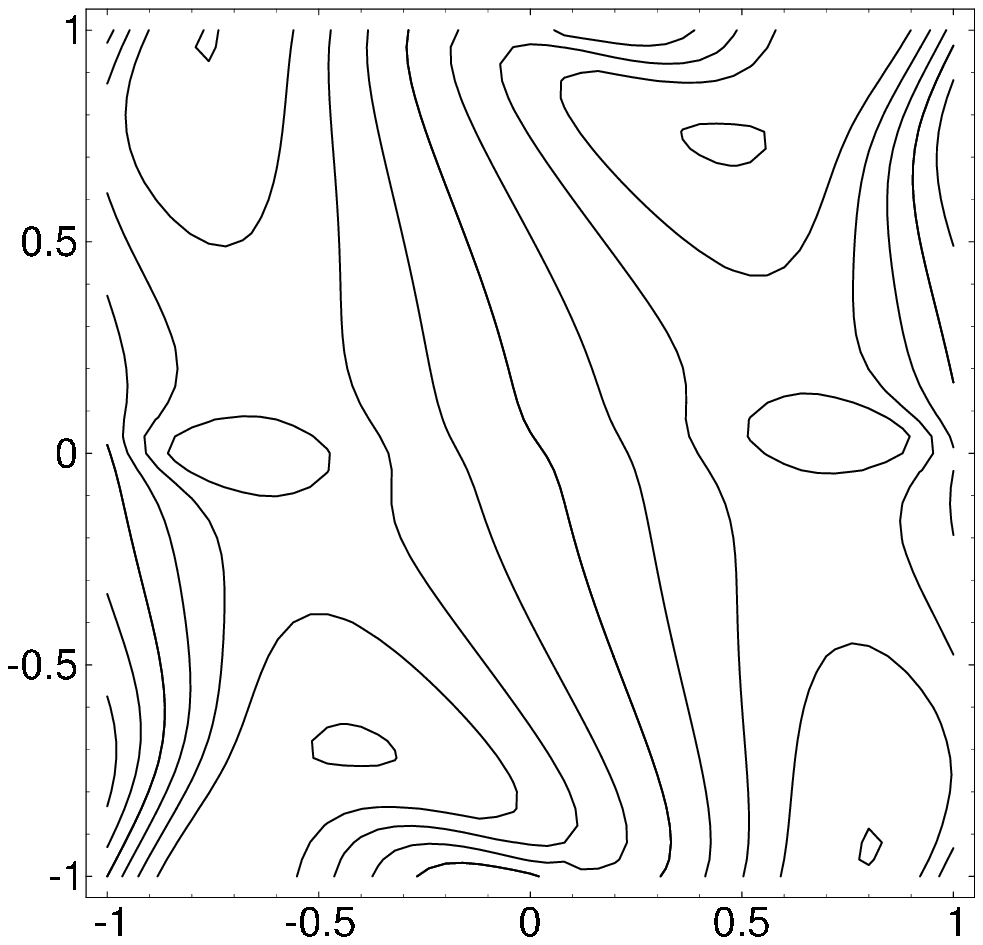}}
   \put(390,-8){$\varphi_{\mu}~[\pi]$}
	\put(225,210){$\varphi_{M_1}~[\pi]$ }
	\put(300,210){\fbox{${\mathcal A}_{\rm CP}$ in \% }}
	\put(296,45){\scriptsize{65}}
	\put(290,66){\scriptsize{45}}
	\put(277,80){\scriptsize{30}}
	\put(254,85){\scriptsize{0}}
	\put(275,103){\scriptsize{45}}
	\put(290,130){\scriptsize{30}}
	\put(251,135){\scriptsize{15}}
	\put(265,153){\scriptsize{45}}

	\put(323,80){\scriptsize{15}}
	\put(340,90){\scriptsize{0}}
	\put(350,100){\scriptsize{-15}}
	\put(403,30){\scriptsize{-65}}
	\put(395,58){\scriptsize{-45}}
	\put(375,83){\scriptsize{-30}}
	\put(393,107){\scriptsize{-45}}
	\put(391,125){\scriptsize{-30}}
	\put(405,135){\scriptsize{-15}}
	\put(420,125){\scriptsize{0}}
	\put(373,146){\scriptsize{-45}}
	\put(371,167){\scriptsize{-65}}
	\put(245,-8){Fig.~\ref{fig5}b }
\end{picture}
\vspace*{.25cm}
 \caption{Contour lines of $\sigma$  and ${\mathcal A}_{\rm CP}$
	 in the $\varphi_{\mu}$--$\varphi_{M_1}$ plane for
$M_2=400$ GeV, $|\mu|=300$ GeV, 
$\tan \beta=5$, $\varphi_{A_{\tau}}=0$, $A_{\tau}=250$~GeV 
and $(P_{e^-},P_{e^+})=(-0.8,0.6)$.
\label{fig5}}
\end{figure}

For $|\mu|=300$~GeV and $M_2=400$~GeV, we show in 
Figs.~\ref{fig5}a,b contour lines of 
$\sigma$ and ${\mathcal A}_{\rm CP}$, respectively,  
in the $\varphi_{\mu}$--$\varphi_{M_1}$ plane.
The asymmetry ${\mathcal A}_{\rm CP}$ is very sensitive
to variations of the phases $\varphi_{M_1}$ and $\varphi_{\mu}$.
Even for small phases, e.g.  $\varphi_{\mu},\varphi_{M_1}\approx0.1$,
we have ${\mathcal A}_{\rm CP}\approx15\%$.
%Small values of the phases, especially of $\varphi_{\mu}$,
%are suggested by constraints on
%electron and neutron electric dipole moments (EDMs) \cite{edmsexp} 
%for a typical SUSY scale of the order of a few 100 GeV
%(for a review see, e.g., \cite{edmstheo}).

\begin{figure}[t]
	\setlength{\unitlength}{0.035cm}
%	\fbox{
			\begin{minipage}{0.47\textwidth}
 \begin{picture}(100,230)(0,0)
	 \put(0,0){\includegraphics{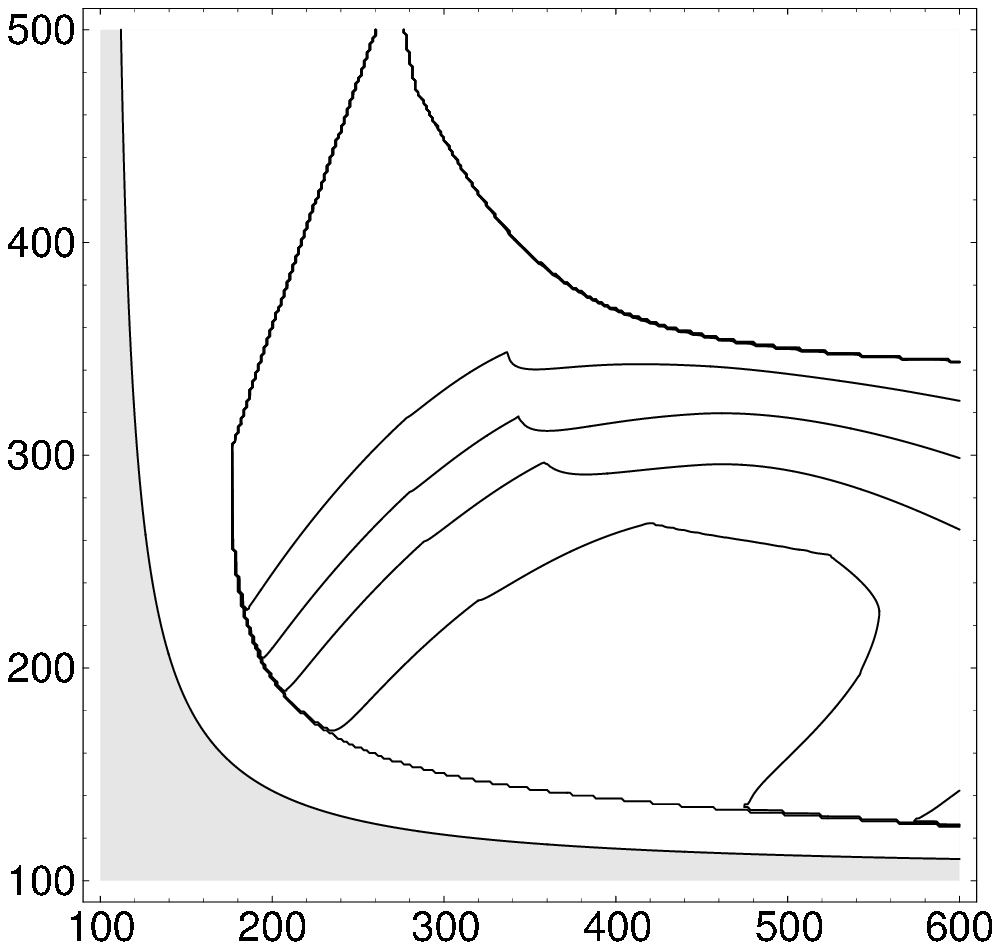}}
	\put(80,210){\fbox{sensitivity $S$}}
	\put(170,-5){$|\mu|~[{\rm GeV}]$}
	\put(0,210){$M_2~[{\rm GeV}]$}
	\put(121,80){\footnotesize{0.15}}
	\put(137,97){\footnotesize{0.3}}
	\put(119,114){\footnotesize{0.5}}
	\put(100,126){\footnotesize{1}}
 \end{picture}
\vspace*{.25cm}
\caption{Contour lines of  $S$ for $\varphi_{A_{\tau}}=0.2\pi$,
$\varphi_{M_1}=\varphi_{\mu}=0$, $A_{\tau}=1$~TeV, $\tan \beta=5$  and 
$(P_{e^-},P_{e^+})=(-0.8,0.6)$.
The blank area outside the area of the contour lines is kinematically
forbidden since here either $\sqrt{s}<m_{\chi_1^0}+m_{\chi_2^0}$ or
$m_{\tilde\tau_1}+m_{\tau}>m_{\chi_2^0}$. The gray area is excluded by
$m_{\chi^{\pm}_1}<104$~GeV.
\label{fig6}}
\end{minipage}
%}
\hspace*{0.5cm}
\begin{minipage}{0.47\textwidth}
 \begin{picture}(100,230)(0,0)
	\put(0,0){\includegraphics{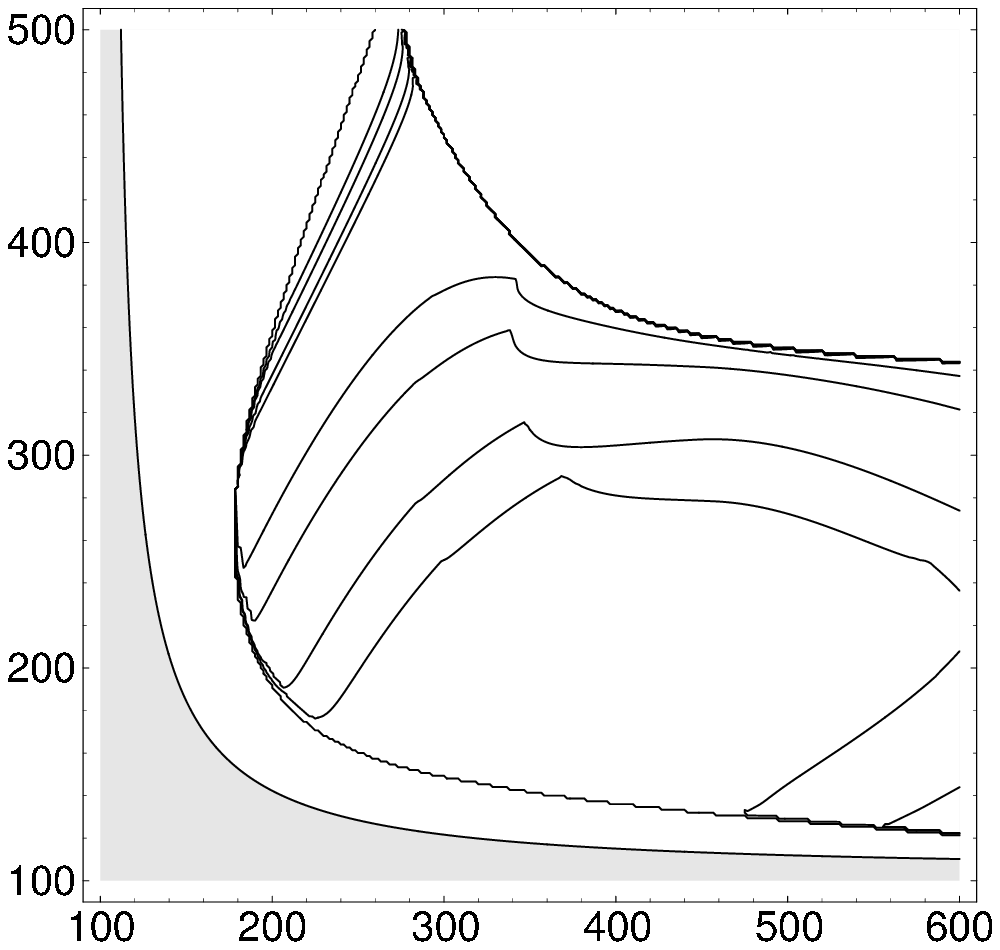}}
	\put(80,210){\fbox{sensitivity $S$}}
	\put(170,-5){$|\mu|~[{\rm GeV}]$}
	\put(0,210){$M_2~[{\rm GeV}]$}
	\put(95,85){\footnotesize{0.1}}
	\put(125,112){\footnotesize{0.15}}
	\put(90,118){\footnotesize{0.3}}
	\put(88,143){\footnotesize{0.5}}
 \end{picture}
\vspace*{.25cm}
\caption{Contour lines of  $S$
for $\varphi_{M_1}=0.2\pi$,
$\varphi_{A_{\tau}}=\varphi_{\mu}=0$,
$A_{\tau}=250$~GeV, $\tan \beta=5$  and
$(P_{e^-},P_{e^+})=(-0.8,0.6)$.
The blank area outside the area of the contour lines is kinematically
forbidden since here either $\sqrt{s}<m_{\chi_1^0}+m_{\chi_2^0}$ or
$m_{\tilde\tau_1}+m_{\tau}>m_{\chi_2^0}$. The gray area is excluded by
$m_{\chi^{\pm}_1}<104$~GeV.
\label{fig7}}
\end{minipage}
\vspace*{0.5cm}
\end{figure}

The polarization of the $\tau$ can be analyzed through its decay 
distributions. 
The sensitivities for measuring the polarization of the $\tau$ 
for the various decay modes are given in \cite{davier}.
The numbers quoted there are for an ideal detector and 
for longitudinal $\tau$ polarization and 
it is expected that the sensitivities for transversely polarized $\tau$ leptons
are somewhat smaller.
Combining informations from all $\tau$ decay modes
a sensitivity of $S=0.35$ \cite{atwood} has been obtained. 
Following \cite{davier}, the relative statistical error of  
$P_2$ (and of $\bar P_2$ analogously)
can be calculated as $\delta P_2= \Delta P_2/|P_2|=
\sigma^s/(S|P_2|\sqrt{N})$, for $\sigma^s$ standard deviations,
and $N=\sigma{\mathcal L}$ events for the integrated luminosity 
$\mathcal L$ and the cross section 
$\sigma = \sigma_P(e^+e^-\to \tilde\chi^0_1\tilde\chi^0_2)\times 
{\rm BR}(\tilde\chi^0_2\to\tilde\tau_1^+\tau^-)$.
Then for ${\mathcal A}_{\rm CP}$~(\ref{stau:asy}),
it follows $\Delta {\mathcal A}_{\rm CP}=\Delta P_2/\sqrt{2}$.
We show in Fig.~\ref{fig6} the contour lines of the sensitivity 
$S=\sqrt{2}/(|{\mathcal A}_{\rm CP}|\sqrt{N})$ which is needed to measure 
${\mathcal A}_{\rm CP}$ at $95\%$ CL ($\sigma^s=2$) for
${\mathcal L}=500~{\rm fb}^{-1}$, for 
$\varphi_{A_{\tau}}=0.2\pi$ and $\varphi_{M_1}=\varphi_{\mu}=0$.
In Fig.~\ref{fig7} we show the contour lines of the sensitivity 
$S$ for $\varphi_{M_1}=0.2\pi$ and $\varphi_{\mu}=\varphi_{A_{\tau}}=0$.
It is interesting to note that in a large region in the $|\mu|$--$M_2$
plane in Figs.~\ref{fig6} and \ref{fig7} we obtain a sensitivity $S<0.35$, 
which means that the asymmetries can be measured at $95\%$ CL.

\subsection{Summary of Section \ref{A CP asymmetry in neutralino production and decay into staus with tau polarization}}

We have defined and analyzed a CP odd asymmetry
${\mathcal A}_{\rm CP}$ of the transverse $\tau$ polarization 
in  neutralino production 
$e^+e^- \to\tilde\chi^0_i \tilde\chi^0_j$
and subsequent two-body decay 
$\tilde\chi^0_i \to \tilde\tau_k^{\pm}  \tau^{\mp}$. 
The asymmetry is sensitive to CP-violating phases of the 
the trilinear scalar coupling parameter  $A_{\tau}$ and
the gaugino and Higgsino mass parameters $M_1$, $\mu$. 
The asymmetry occurs already at tree level and is due to spin effects in the 
neutralino production and decay process. 
In a numerical study for 
$e^+e^- \to\tilde\chi^0_1 \tilde\chi^0_2$
and neutralino decay $\tilde\chi^0_2 \to \tilde\tau_1^{\pm} \tau^{\mp}$
we have shown that the asymmetry can be as large as 60\%.
It can be sizable even for small phases of $\mu$ and $M_1$,
suggested by the experimental limits on EDMs. 
%Depending on the MSSM scenario, the asymmetry should be 
%accessible in future electron-positron linear collider experiments 
%in the 500 GeV range. Longitudinally polarized electron and positron
%beams can considerably enhance both the asymmetry
%and the production cross section.

		\section{T-odd observables in neutralino production and decay
	into a $Z$ boson
	\label{CP observables in neutralino production and decay
	into the Z boson}}

%We study CP sensitive observables in neutralino production 
%$e^+e^- \to\tilde\chi^0_i \tilde\chi^0_j$
%and the subsequent two-body decays of the neutralino
%$\tilde\chi^0_i \to \chi^0_n Z$ and of the $Z$ boson
%$Z \to \ell \bar\ell (q\bar q)$.
%We identify the CP-odd  elements of the $Z$ boson
%density matrix and  propose CP sensitive 
%triple-product asymmetries. 
%We calculate these observables and the cross sections 
%in the Minimal Supersymmetric Standard Model with complex parameters 
%$\mu$ and $M_1$ for  an $e^+e^-$ linear collider with 
%$\sqrt{s}=800$ GeV and longitudinally polarized beams. 
%We show that the asymmetries can reach  $3\%$ 
%for $Z \to \ell \bar\ell $ and
%$18\%$ for $Z \to  q\bar q$
%and discuss the feasibility of measuring these asymmetries.
\begin{figure}[h]
\begin{picture}(5,6.)(-2,.5)
		\put(1,4.7){${\bf p}_{\chi_j^0}$}
   \put(3.4,6){${\bf p}_{e^- }$}
   \put(3.3,2.3){${\bf p}_{e^+}$}
   \put(4.8,4.7){${\bf p}_{\chi_i^0}$}
   \put(7.,5.9){$ {\bf p}_{\chi_n^0}$}
   \put(6.2,3.){${\bf p}_{Z}$}
   \put(9.2,2.3){${\bf p}_{\bar f}$}
   \put(6.4,1.6){$ {\bf p}_{f}$}
\end{picture}
\scalebox{1.9}{
\begin{picture}(0,0)(1.3,-0.25)
\ArrowLine(40,50)(0,50)
\Vertex(40,50){2}
\ArrowLine(55,80)(40,50)
\ArrowLine(25,20)(40,50)
\ArrowLine(40,50)(80,50)
\ArrowLine(80,50)(110,75)
%\ArrowLine(80,50)(100,20)
\Photon(80,50)(100,20){2}{5}
\Vertex(80,50){2}
\ArrowLine(100,20)(125,15)
\ArrowLine(100,20)(85,0)
\Vertex(100,20){2}
\end{picture}}
\caption{\label{shematic picture of Z}
          Schematic picture of the neutralino production
          and decay process.}
\end{figure}
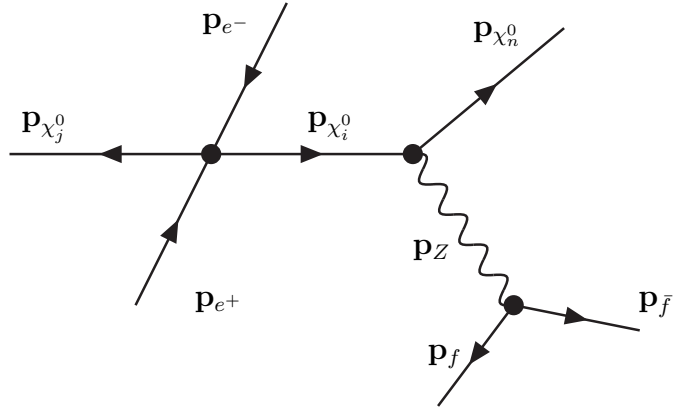

%\subsection{Introduction}

A further possibility to study CP violation in the neutralino
sector is the two-body decay of the neutralino into a Z boson.
Due to spin correlations of the neutralino and the Z boson,
observables can be defined which have not only CP-odd 
contributions from the neutralino production, but 
also from its decay.

We study CP violation in neutralino production
\begin{eqnarray} \label{Z:production}
	e^++e^-&\to&\tilde\chi^0_i+\tilde\chi^0_j; 
	\quad i,j =1,\dots,4,  
\end{eqnarray}
with the subsequent  two-body decay of one neutralino into a $Z$ boson
%(for recent studies see \cite{staudecay,choi2})
\begin{eqnarray} \label{Z:decay_1}
     \tilde\chi^0_i \to \chi^0_n +Z; \quad n<i,
\end{eqnarray}
followed by the decay of the $Z$ boson
	\begin{eqnarray} \label{Z:decay_2}
		Z \to f +\bar f; \quad  f=\ell,q,\quad\ell=e,\mu,\tau, \quad q=c,b.
\end{eqnarray}
For a schematic picture of the neutralino production and decay process 
see Fig.~\ref{shematic picture of Z}.
If CP is violated, the phases $\varphi_{M_1}$ and $\varphi_{\mu}$ 
lead to CP sensitive elements of the $Z$ boson density matrix. 
They involve CP-odd asymmetries ${\mathcal A}_{f}$ in the angular 
distribution of the decay fermions
 \begin{eqnarray}\label{Z:AT1}
	 {\mathcal A}_{f} &=& 
	 \frac{\sigma({\mathcal T}_{f}>0)-\sigma({\mathcal T}_{f}<0)}
	{\sigma({\mathcal T}_{f}>0)+\sigma({\mathcal T}_{f}<0)},
\end{eqnarray}
of the triple product
 \begin{eqnarray}\label{Z:tripleproduct}
	 {\mathcal T}_{f} &=& 
	 {\bf p}_{e^-}\cdot({\bf p}_{f} \times {\bf p}_{\bar f}),
 \end{eqnarray}
and the cross section $\sigma$
of neutralino production (\ref{Z:production}) and
subsequent decay chain (\ref{Z:decay_1})-(\ref{Z:decay_2}).
Due to the correlations between the $\tilde\chi^0_i$ 
polarization and the $Z$ boson polarization, there are 
CP-odd contributions to the $Z$ boson density matrix
and to the asymmetries ${\mathcal A}_{f}$ from the 
production (\ref{Z:production}) and from the 
decay process (\ref{Z:decay_1}).

In  Section (\ref{T odd asymmetries in neutralino 
production and decay into sleptons}) 
we have studied asymmetries for neutralino decay into 
sleptons $\tilde\chi^0_i\to \tilde\ell \ell$.
We have shown that these asymmetries have in contrast only contributions 
from the neutralino  production process, since the neutralino 
decay is a two-body decay into scalars.

Note that if we would replace the triple product 
${\mathcal T}_{f}$ by 
${\mathcal T}_{f}' ={\bf p}_{e^-}\cdot({\bf p}_{\chi_i^0} \times {\bf p}_{Z})$,
and would calculate the corresponding asymmetry,
where the $Z$ boson polarization is summed, all spin correlations and 
thus this asymmetry would vanish identically because 
of the Majorana properties of the neutralinos.

%The triple product ${\mathcal T}_{f}$ (\ref{Z:tripleproduct}) 
%changes sign under time reversal and is thus T-odd. Due to CPT invariance, 
%the corresponding T-odd asymmetries ${\mathcal A}_{f}$ 
%are also CP-odd if final state
%interactions are neglected.
%The final state interactions would also contribute to 
%${\mathcal A}_{f}$. However, they only arise at loop level and 
%are neglected in the present work.

\subsection{Cross section
     \label{Cross section}}

For the calculation of the cross section for the
combined process of neutralino production (\ref{Z:production})
and the subsequent two-body decays
(\ref{Z:decay_1}), (\ref{Z:decay_2}) of $\tilde{\chi}^0_i$
we use the same spin-density matrix formalism as in
\cite{gudineutralino,spinhaber}.
The (unnormalized)  spin-density matrix of the $Z$ boson 
\begin{eqnarray}       \label{Zdensitymatrix}
\rho_{P}(Z)^{\lambda_k\lambda'_k}&=&
|\Delta(\tilde\chi^0_i)|^2~
\sum_{\lambda_i,\lambda'_i}~
\rho_P   (\tilde\chi^0_i)^{\lambda_i \lambda_i'}\;
\rho_{D_1}(\tilde\chi^0_i)_{\lambda_i'\lambda_i}^{\lambda_k\lambda'_k},
\end{eqnarray}
is composed of the neutralino propagator
$\Delta(\tilde\chi^0_i)$, the
spin-density production matrix
%\begin{eqnarray} 
%	\rho_P(\tilde{\chi}^{0}_i)^{\lambda_i \lambda_i'}&=&\sum_{\lambda_j}
%	T_P^{\lambda_i \lambda_j}T_P^{\lambda_i' \lambda_j \ast}
%\end{eqnarray}
$\rho_P(\tilde{\chi}^{0}_i)$,
defined in~(\ref{neut:rhoPdef}), and the decay matrix
%\begin{eqnarray}
%\rho_{D_1}(\tilde\chi^0_i)_{\lambda_i' \lambda_i}^{\lambda_k\lambda'_k} &=&
%\sum_{\lambda_n}T_{D_1,\lambda_i}^{\lambda_n\lambda_k}
%T_{D_1,\lambda_i'}^{\lambda_n\lambda_k'\ast}.
%\end{eqnarray}
$\rho_{D_1}(\tilde\chi^0_i)$,
defined in~(\ref{Z:rhoD1A}).
%With the decay matrix for the $Z$ decay
%\begin{eqnarray}
%	\rho_{D_2}(Z)_{\lambda_k' \lambda_k}&=&
%	\sum_{\lambda_f, \lambda_{\bar f}}
%	T_{D_2,\lambda_k}^{\lambda_f \lambda_{\bar f} }
%	T_{D_2,\lambda_k'}^{\lambda_f \lambda_{\bar f} \ast}
%\end{eqnarray}
The amplitude squared for the complete process
$ e^+e^-\to\tilde\chi^0_i\tilde\chi^0_j$;
$\tilde\chi^0_i\to\tilde\chi^0_n Z $;
$Z \to f \bar f$ 
can now be written
\begin{eqnarray}       \label{Z:amplitude}
|T|^2&=&|\Delta(Z)|^2
	\sum_{\lambda_k,\lambda'_k}~
	\rho_{P}(Z)^{\lambda_k\lambda'_k}\;
	\rho_{D_2}(Z)_{\lambda'_k\lambda_k},
\end{eqnarray}
with the decay matrix $\rho_{D_2}(Z)$ for the $Z$ decay,
defined in~(\ref{Z:rhoD2A}).
Inserting the density matrices 
$\rho_P(\tilde{\chi}^{0}_i)$~(\ref{neut:rhoP}) 
and $\rho_{D_1}(\tilde\chi^0_i)$~(\ref{Z:rhoD1expanded}) 
into~(\ref{Zdensitymatrix}) leads to
\begin{eqnarray}\label{Zdensitymatrixunnorm}
\rho_{P}(Z)^{\lambda_k\lambda'_k}&=&
4~|\Delta(\tilde\chi^0_i)|^2~
\left[
	P  D_1 ~\delta^{\lambda_k\lambda_k'}
	+\Sigma_P^a \,^{c}\Sigma^{a}_{D_1}~(J^{c})^{\lambda_k\lambda_k'}
	+P  \;^{cd}D_1 ~(J^{cd})^{\lambda_k\lambda_k'}
\right],
\end{eqnarray}
summed over $a,c,d$. 
Here the $Z$ production matrix $\rho_{P}(Z)$ 
is decomposed into contributions of scalar (first term), 
vector (second term) and tensor parts (third term).
Inserting then $\rho_{P}(Z)$~(\ref{Zdensitymatrixunnorm}) 
and $\rho_{D_2}(Z)$~(\ref{Z:rhoD2expanded}) 
into~(\ref{Z:amplitude}) leads finally to
\begin{eqnarray} \label{Z:amplitude2}
|T|^2 &=& 4~|\Delta(\tilde{\chi}^{0}_i)|^2~ |\Delta(Z)|^2
		\left[3 P  D_1  D_2 +  
		2\Sigma_P^a \,^c\Sigma_{D_1}^a \, ^cD_2 
		+4P( ^{cd} D_1  ^{cd} D_2-
			{\textstyle \frac{1}{3}}\,^{cc} D_1 \, ^{dd} D_2)\right].
	\nonumber \\
\end{eqnarray}
%summed over $a,c,d$, which is the decomposition of the amplitude 
%squared in its scalar (first term), vector (second term)
%and tensor part (third term). 
The differential cross section is then given by 
\begin{equation}\label{Z:crossection}
	d\sigma=\frac{1}{2 s}|T|^2 
	d{\rm Lips}(s;p_{\chi_j^0 },p_{\chi_n^0},p_{f},p_{\bar f}),
\end{equation}
where $d{\rm Lips}$
%$d{\rm Lips}(s,p_{\chi_j },p_{\chi_n},p_{f},p_{\bar f})$
is the Lorentz invariant phase-space element
defined in~(\ref{Lipsbosonic2}).
%of Appendix \ref{Phase space}.
%More details concerning kinematics and phase space  
%can be found in Appendices \ref{Bosonic decays}
%and \ref{Phase space for bosonic decays}.

\subsection{$Z$ boson polarization
     \label{Z boson matrix}}

The mean polarization of the $Z$ boson is given 
by its $3\times3$ density matrix 
$<\rho(Z)>$ with ${\rm Tr}\{<\rho(Z)>\}=1$.
We obtain $<\rho(Z)>$ in the laboratory system
by integrating~(\ref{Zdensitymatrixunnorm}) 
over the Lorentz invariant phase-space element 
$d{\rm Lips}(s;p_{\chi_j^0},p_{\chi_n^0},p_{Z})$ (\ref{Lipsbosonic1}),
and normalizing by the trace
\begin{equation}\label{Zdensitymatrixnorm}
<\rho(Z)^{\lambda_k\lambda'_k}>=
\frac{\int \rho_{P}(Z)^{\lambda_k\lambda'_k}~d{\rm Lips}}
		{\int {\rm Tr} \{\rho_{P}(Z)^{\lambda_k\lambda'_k}\}~d{\rm Lips}}
={\textstyle \frac{1}{3}}\delta^{\lambda_k\lambda_k'}
+V_c ~(J^{c})^{\lambda_k\lambda_k'}
+T_{cd}  ~(J^{cd})^{\lambda_k\lambda_k'}.\label{defcoef}
\end{equation}
%summed over $c$, $d$. 
The components $V_c$ of the vector polarization and $T_{cd}$ 
of the tensor polarization are given by
%%which is the decomposition of the normalized $Z$ spin-density 
%%matrix in its scalar (first term), vector (second term)
%%and tensor part (third term). In Eq.~(\ref{defcoef}) we have 
%%defined the vector and tensor coefficients as
%\begin{equation}
%V_c=\frac{\int |\Delta(\tilde{\chi}^{0}_i)|^2~\Sigma_P^a
%	\,^{c}\Sigma^{a}_{D_1} ~d{\rm Lips}}
%{3 \int |\Delta(\tilde{\chi}^{0}_i)|^2 ~P  D_1~d{\rm Lips}},\quad
%T_{cd}=T_{dc}=
%\frac{\int |\Delta(\tilde{\chi}^{0}_i)|^2  ~P  \;^{cd}D_1 ~d{\rm Lips}}
%{3 \int |\Delta(\tilde{\chi}^{0}_i)|^2 ~P  D_1~d{\rm Lips}},
%\end{equation}
%with sum over $a$. 
\begin{equation}
V_c=\frac{\int \Sigma_P^a
	\,^{c}\Sigma^{a}_{D_1} ~d{\rm Lips}}
{3 \int 2 ~P  D_1~d{\rm Lips}},\quad
T_{cd}=T_{dc}=
\frac{\int  P  \;^{cd}D_1 ~d{\rm Lips}}
{3 \int  P  D_1~d{\rm Lips}},
\end{equation}
%with sum over $a$, 
where we have used the narrow width approximation
for the neutralino propagator.
The tensor components $T_{12}$ and $T_{23}$ vanish due to
phase-space integration. The density matrix in the helicity basis,
see Appendix~\ref{Spin 1 matrices}, is given  by 
\begin{eqnarray} \label{Z:density1}
	<\rho(Z)^{--}> &= &
	{\textstyle \frac{1}{2}}-V_3+T_{33}, \\
	<\rho(Z)^{00}> &=&-2T_{33},\\
	<\rho(Z)^{-0}> &= &
	{\textstyle \frac{1}{\sqrt{2}}}(V_1+iV_2)-\sqrt{2}\,T_{13},\\
		<\rho(Z)^{-+}> &= & T_{11},\\
	<\rho(Z)^{0+}> &= & {\textstyle \frac{1}{\sqrt{2}}}(V_1+iV_2)
	+\sqrt{2}\,T_{13},\label{Z:density5}
\end{eqnarray}
where we have used $T_{11}+T_{22}+T_{33}=-\frac{1}{2}$
and $T_{12}=T_{23}=0$.

\subsection{T-odd asymmetry
	\label{Z:T-odd asymmetry}}

From~(\ref{Z:amplitude2}) we obtain for the asymmetry~(\ref{Z:AT1})
%\begin{eqnarray} \label{Z:asym}
%	 {\mathcal A}_{f} 
%	 = \frac{\int {\rm Sign}[{\mathcal T}_{f}]
%		 |T|^2 ~d{\rm Lips}}
%           {\int |T|^2 ~d{\rm Lips}}
%%= \frac{\int |\Delta(\tilde{\chi}^{0}_i)|^2 |\Delta(Z)|^2~
%%	{\rm Sign}[{\mathcal T}_{f}]
%%		~2~\Sigma_P^a \, ^c\Sigma_{D_1}^a \,^cD_2 ~d{\rm Lips}}
%%	{\int |\Delta(\tilde{\chi}^{0}_i)|^2 |\Delta(Z)|^2~
%%		3 ~P D_1D_2 ~d{\rm Lips}},
%= \frac{\int |\Delta(\tilde{\chi}^{0}_i)|^2 |\Delta(Z)|^2~
%	{\rm Sign}[{\mathcal T}_{f}]
%		~2~\Sigma_P^a \, ^c\Sigma_{D_1}^a \,^cD_2 ~d{\rm Lips}}
%	{\int |\Delta(\tilde{\chi}^{0}_i)|^2 |\Delta(Z)|^2~
%		3 ~P D_1D_2 ~d{\rm Lips}},
%\end{eqnarray}
%%summed over $a$, $c$,
%%and 
\begin{eqnarray} \label{Z:asym}
	 {\mathcal A}_{f} 
	 = \frac{\int {\rm Sign}[{\mathcal T}_{f}]
		 |T|^2 ~d{\rm Lips}}
           {\int |T|^2 ~d{\rm Lips}}
%= \frac{\int |\Delta(\tilde{\chi}^{0}_i)|^2 |\Delta(Z)|^2~
%	{\rm Sign}[{\mathcal T}_{f}]
%		~2~\Sigma_P^a \, ^c\Sigma_{D_1}^a \,^cD_2 ~d{\rm Lips}}
%	{\int |\Delta(\tilde{\chi}^{0}_i)|^2 |\Delta(Z)|^2~
%		3 ~P D_1D_2 ~d{\rm Lips}},
= \frac{\int {\rm Sign}[{\mathcal T}_{f}]
		~2~\Sigma_P^a \, ^c\Sigma_{D_1}^a \,^cD_2 ~d{\rm Lips}}
	{\int 3 ~P D_1D_2 ~d{\rm Lips}},
\end{eqnarray}
%summed over $a$, $c$,
and $d{\rm Lips}(s;p_{\chi_j^0},p_{\chi^0_n},p_{f},p_{\bar f})$
is the Lorentz invariant phase-space element
defined in~(\ref{Lipsbosonic2}), where we have already used the narrow
width approximation for the propagators.
In the numerator only the vector part of $|T|^2$ 
remains which contains the triple product
%{
%Note that if one would replace the triple product 
%${\mathcal T}_{f}$ by 
%${\mathcal T}_{f} =\vec p_{e^-}\cdot(\vec p_{\chi_i} \times \vec p_{Z})$,
%and would calculate the corresponding asymmetry,
%where the $Z$ boson polarization is summed, all spin correlations and 
%thus this asymmetry would vanish identically because 
%of the Majorana properties of the neutralinos.
%}
${\mathcal T}_{f} = 
 {\bf p}_{e^-}\cdot({\bf p}_{f} \times {\bf p}_{\bar f})$.
In the denominator the vector and tensor parts of $|T|^2$ vanish,
since the complete phase-space integration eliminates the spin
correlations. Due to the correlations 
$\Sigma_P^a \, ^c\Sigma_{D_1}^a$ between
the $\tilde\chi^0_i$ and the $Z$ boson polarization, 
there are CP-odd contributions to the asymmetry ${\mathcal A}_{f}$ 
from both the neutralino production process~(\ref{Z:production}),
and from the neutralino decay process~(\ref{Z:decay_1}).
The contribution from the production is given by the term with $a=2$ 
in~(\ref{Z:asym}) and is proportional to 
the transverse polarization $\Sigma^{2}_P$~(\ref{neut:Sigma2P})
of the neutralino perpendicular to the production plane. 
For the production of a pair of equal neutralinos, 
$e^+e^- \to\tilde\chi^0_i \tilde\chi^0_i$, we have $\Sigma^{2}_P=0$.
The contributions  from the decay, given by the terms with $a=1,3$
in~(\ref{Z:asym}), are proportional to  
\begin{eqnarray}\label{Z:adependence}
	^c\Sigma_{D_1}^a \,^cD_2&\supset&
	-8m_{\chi_n^0}
	%	\frac{8g^4m_{\chi_k}}{\cos^4\theta_W}
	(Im O^{''L}_{ni})(Re O^{''L}_{ni})(R_f^2-L_f^2)
	(t^c_Z\cdot p_{\bar f})
	\epsilon_{\mu\nu\rho\sigma}s^{a,\mu}_{\chi_i^0}p_{\chi_i^0}^{\nu}
	p_Z^{\rho}t^{c,\sigma}_Z,
\end{eqnarray}
see last term of~(\ref{Z:csigmaaD1}), 
which contains the $\epsilon$-tensor.
Thus ${\mathcal A}_{f}$ may be enhanced (reduced) if the contributions 
from production and decay  have the same (opposite) sign.

Note that the contributions from the decay would vanish 
for a two-body decay of the neutralino into a scalar particle,
as discussed in Section (\ref{T odd asymmetries in neutralino 
		production and decay into sleptons}). We have found in 
this case, that only contributions to ${\mathcal A}_{f}$ from the
production remain, which are multiplied by a decay 
factor $\propto (|R|^2-|L|^2)$, and thus  
${\mathcal A}_{f}\propto (|R|^2-|L|^2)/(|R|^2+|L|^2)$,
see (\ref{Amixing}), where $R$ and $L$ are the right and left couplings
of the scalar particle to the neutralino.

For the measurement of ${\mathcal A}_{f}$ 
the charges and the flavors of $f$ and $\bar f$
have to be  distinguished. For $f=e,\mu$ this will be 
possible on an event by event basis. 
For $f=\tau$ it will be possible after taking into account
corrections due to the reconstruction of the $\tau$ 
momentum. For $f=q$ the distinction of the quark 
flavors should be possible by flavor tagging in the 
case $q=b,c$ \cite{flavortaggingatLC,Aubert:2002rg}. 
However, in this case the quark charges will  be distinguished
statistically for a given event sample only.
Note that ${\mathcal A}_{q}$ is always larger than
${\mathcal A}_{\ell}$, due to the dependence of ${\mathcal A}_{f}$ 
on the $Z$-$\bar f$-$f$ couplings,
which follows from~(\ref{Z:D2}), (\ref{Z:cD2}) and (\ref{Z:asym}):
%\cite{oshimo,staudecay}:
\begin{eqnarray} \label{Z:propto}
{\mathcal A}_{f}\propto \frac{R_f^2-L_f^2}{R_f^2+L_f^2}
\quad\Rightarrow
{\mathcal A}_{b(c)}=
\frac{R_{\ell}^2+L_{\ell}^2}
     {R_{\ell}^2-L_{\ell}^2}
\frac{R_{b(c)}^2-L_{b(c)}^2}
     {R_{b(c)}^2+L_{b(c)}^2}~
	  {\mathcal A}_{\ell}\simeq 
	6.3~(4.5)\times{\mathcal A}_{\ell},
\end{eqnarray}
compare also Section~\ref{sferm:T-odd asymmetry}.

The significance for measuring  the asymmetry 
is given by $S_f = |{\mathcal A}_{f}| \sqrt{N}$,
see~(\ref{significanceofAT}).
%The relative statistical error of
%${\mathcal A}_{f}$ is given by $\delta {\mathcal A}_{f} = 
%\Delta {\mathcal A}_{f}/|{\mathcal A}_{f}| = 
%S_f/(|{\mathcal A}_{f}| \sqrt{N})$ \cite{olaf1},
%with $S_f$ standard deviations
%and $N={\mathcal L} \cdot\sigma$ the number of events with 
%${\mathcal L}$ the integrated luminosity and 
%the cross section 
%$\sigma=\sigma_P(e^+e^-\to\tilde\chi^0_i\tilde\chi^0_j) 
%\times{\rm BR}(\tilde{\chi}^0_i \to Z\tilde{\chi}_n^0)\times
%{\rm BR}(Z\to f\bar f)$. Taking $\delta {\mathcal A}_{f}
%=1$ it follows 
%$S_f = |{\mathcal A}_{f}| \sqrt{N}$.
Note that $S_f$ is larger  for $f=b,c$ than for $f=\ell=e,\mu,\tau$ with
$S_{b} \simeq 7.7\times S_{\ell}$ and $S_{c} \simeq 4.9\times S_{\ell}$,
which follows from~(\ref{Z:propto}) and from
${\rm BR}(Z\to b\bar b) \simeq1.5\times{\rm BR}(Z\to\ell\bar\ell)$,
${\rm BR}(Z\to c\bar c) \simeq1.2\times{\rm BR}(Z\to\ell\bar\ell)$ \cite{PDG}.

\subsection{Numerical results
	\label{Z:Numerical results}}

We study the dependence of the  
$Z$ density matrix $<\rho(Z)>$~(\ref{Zdensitymatrixnorm}),
the asymmetry ${\mathcal A}_{\ell} (\ell=e,\mu,\tau)$~(\ref{Z:AT1}), 
and the cross section 
$\sigma=\sigma_P(e^+e^-\to\tilde\chi^0_i\tilde\chi^0_j ) \times
{\rm BR}( \tilde\chi^0_i \to \tilde\chi^0_1 Z)\times
{\rm BR}(Z \to \ell \bar \ell)$
on the MSSM parameters 
$\mu = |\mu| \, e^{ i\,\varphi_{\mu}}$ and 
$M_1 = |M_1| \, e^{ i\,\varphi_{M_1}}$ for $\tan \beta=10$.
In order to reduce the number of parameters, we assume the 
relation $|M_1|=5/3  M_2\tan^2\theta_W $ and  
use the renormalization group equations  for the 
sfermion masses, 
Appendix~\ref{First and second generation sfermion masses}, 
%$m_{\tilde\ell_R}^2 = m_0^2 +0.23 M_2^2
%-m_Z^2\cos 2 \beta \sin^2 \theta_W$, 
%$m_{\tilde\ell_L  }^2 = m_0^2 +0.79 M_2^2
%+m_Z^2\cos 2 \beta(-1/2+ \sin^2 \theta_W)$,
with $m_0=300$ GeV.
For the branching ratio $Z\to\ell\bar\ell$, summed over
$\ell=e,\mu,\tau$, we take ${\rm BR}(Z\to\ell\bar\ell)=0.1$ \cite{PDG}.
The values for ${\mathcal A}_{b,c}$ are given by~(\ref{Z:propto}).
We choose  a center of mass energy of $\sqrt{s} = 800$ GeV
and longitudinally polarized beams with
beam polarizations $(P_{e^-},P_{e^+})=(\pm0.8,\mp0.6)$.

For the calculation of the neutralino widths $\Gamma_{\chi_i^0}$ and 
branching ratios ${\rm BR}( \tilde\chi^0_i \to \tilde\chi^0_1 Z)$,
see Appendix \ref{Neutralino decay widths},
we include the following two-body decays, if kinematically allowed,
\begin{eqnarray}
	\tilde\chi^0_i &\to& \tilde e_{R,L} e,~ 
	\tilde \mu_{R,L}\mu,~
	\tilde\tau_{m}\tau,~
	\tilde\nu_{\ell} \bar\nu_{\ell},~
	\tilde\chi^0_n Z,~
	\tilde\chi^{\mp}_m W^{\pm},~
	\tilde\chi^0_n H_1^0,~
	\ell=e,\mu,\tau, ~ m=1,2,~ n<i
\end{eqnarray}
and neglect three-body decays.
%with $H_1^0$ being the lightest neutral Higgs boson.
%The Higgs parameter is chosen $m_{A}=1$~TeV and thus 
%the decays  $\tilde\chi^0_i \to \tilde\chi^{\pm}_n H^{\mp}$
%into the charged Higgs bosons,
%and the decays $\tilde\chi^0_i \to \tilde\chi^0_n~H_{2,3}^0$
%into  the heavy neutral Higgs bosons are forbidden in our scenarios.
The Higgs parameter is chosen $m_{A}=1$~TeV and in the stau sector, 
we fix the trilinear scalar coupling parameter $A_{\tau}=250$ GeV.

\subsubsection{Production of $\tilde\chi^0_1 \, \tilde\chi^0_2$ }

\begin{figure}
 \begin{picture}(20,20)(0,-2)
	\put(2.2,16.5){\fbox{$\sigma_P(e^+e^- \to\tilde{\chi}^0_1 
			\tilde{\chi}^0_2)$ in fb}}
\put(0,9){\includegraphics{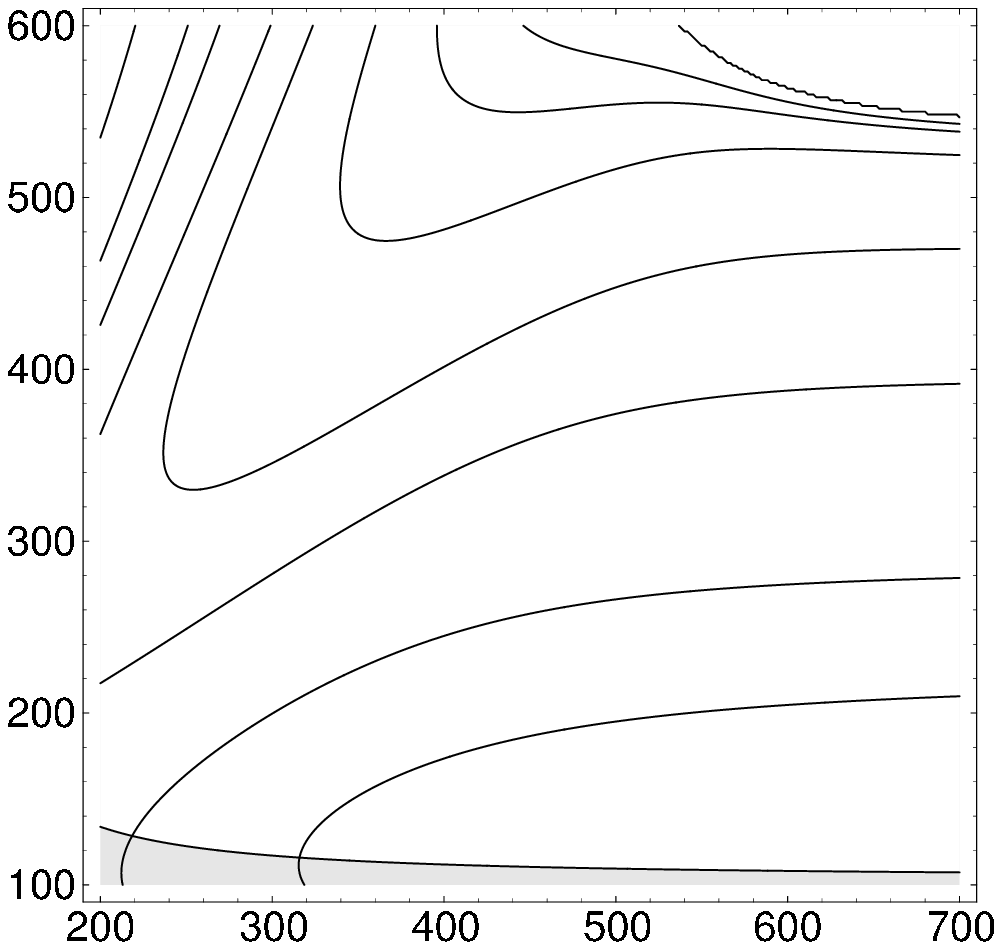}}
\put(5.5,8.8){$|\mu|~[{\rm GeV}]$}
   \put(0,16.3){$M_2~[{\rm GeV}]$ }
	\put(6.5,15.6){\begin{picture}(1,1)(0,0)
			\CArc(0,0)(6,0,380)
			\Text(0,0)[c]{{\scriptsize A}}
		\end{picture}}
	\put(4.3,15.6){ \footnotesize$10$}
	\put(3.3,15.3){ \footnotesize$15$}
	\put(2.6,14.5){ \footnotesize$25$}
	\put(2.8,13.5){ \footnotesize$50$}
	\put(3.5,13.){ \footnotesize$75$}
	\put(4.,11.9){ \footnotesize$100$}
	\put(5.3,11.1){ \footnotesize$110$}
\put(0.5,8.8){Fig.~\ref{plot_12}a}
   \put(8,9){\includegraphics{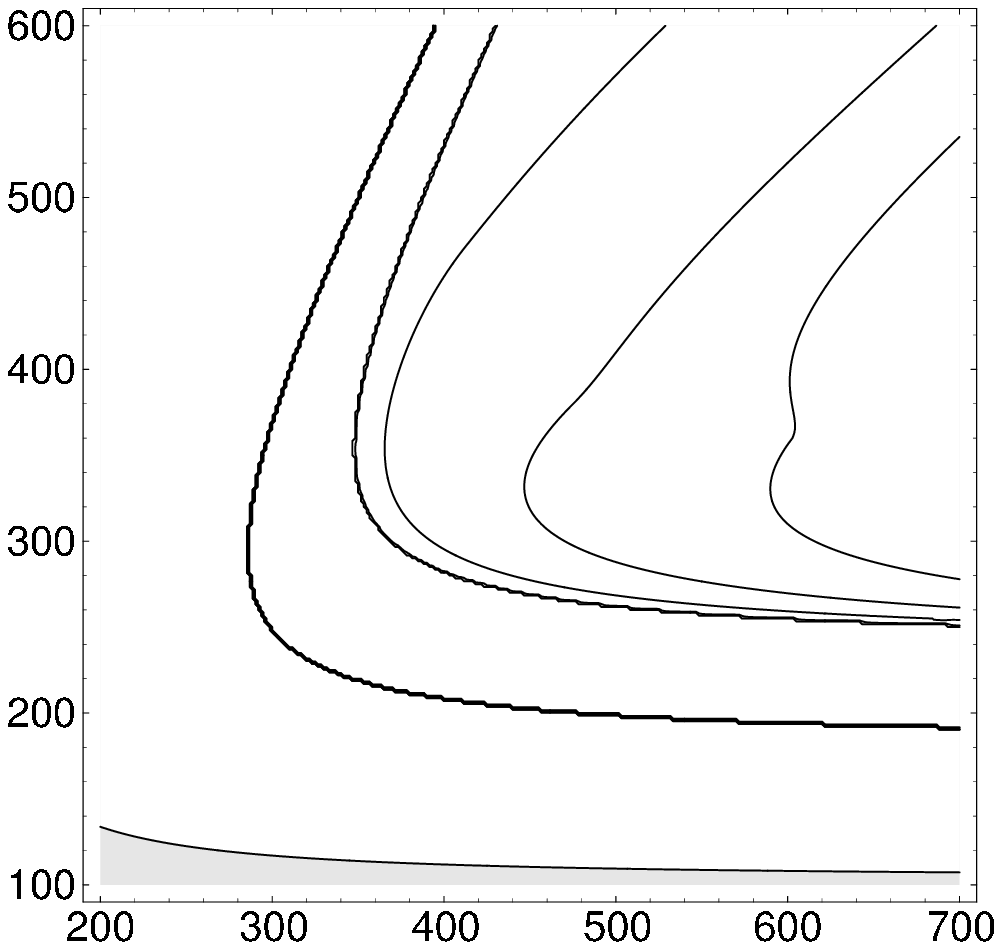}}
	\put(10.2,16.5){\fbox{${\rm BR}(\tilde{\chi}^0_2 \to Z\tilde{\chi}_1^0)$ in \%}}
   \put(13.5,8.8){$|\mu|~[{\rm GeV}]$}
	\put(8,16.3){$M_2~[{\rm GeV}]$}
	\put(9.3,15){\begin{picture}(1,1)(0,0)
			\CArc(0,0)(6,0,380)
			\Text(0,0)[c]{{\scriptsize B}}
	\end{picture}}
	\put(10.4,11.8){\begin{picture}(1,1)(0,0)
			\CArc(0,0)(6,0,380)
			\Text(0,0)[c]{{\scriptsize C}}
	\end{picture}}
	\put(13.6,12.7){ \footnotesize 15}
	\put(12.2,13.1){ \footnotesize 20}
	\put(11.,13.5){ \footnotesize 30}
	\put(11.,11.5){ \footnotesize 100}
\put(8.5,8.8){Fig.~\ref{plot_12}b}
	\put(0,0){\includegraphics{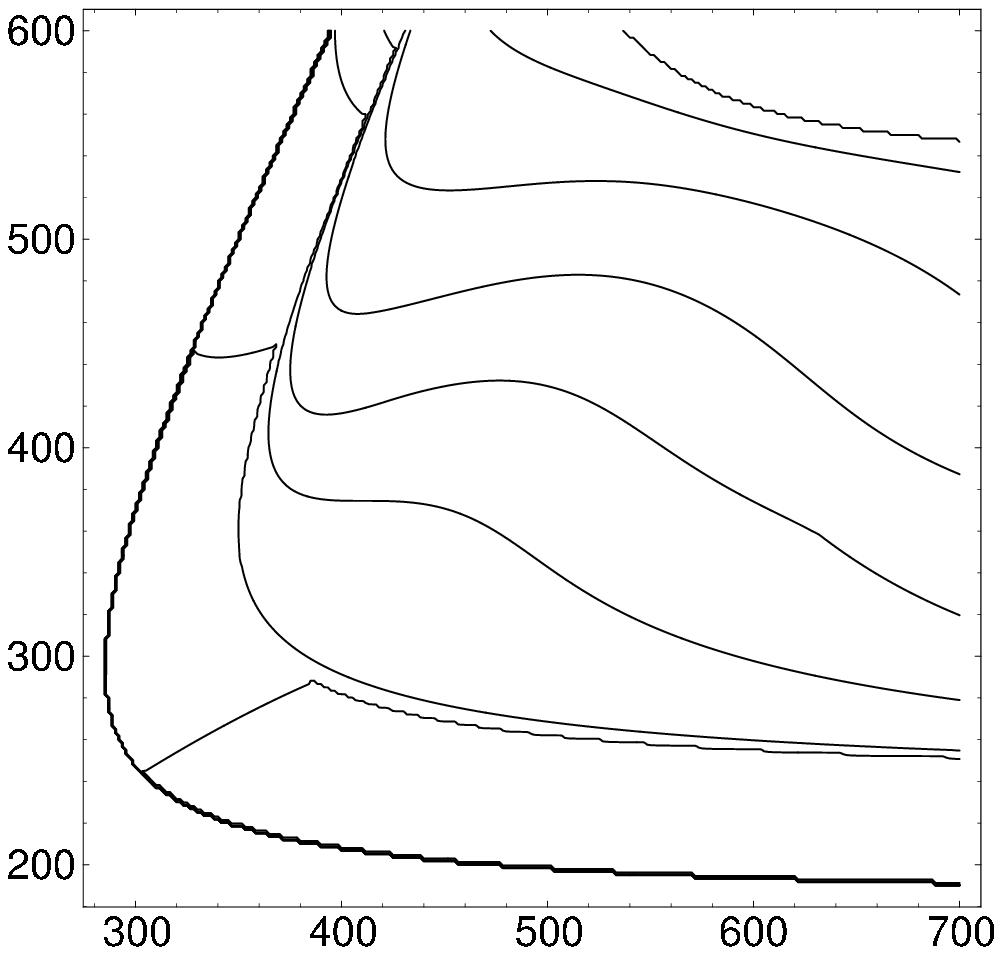}}
	\put(3.5,7.5){\fbox{$\sigma$ in fb}}
	\put(5.5,-0.3){$|\mu|~[{\rm GeV}]$}
	\put(0,7.3){$M_2~[{\rm GeV}]$}
	\put(6.5,6.6){\begin{picture}(1,1)(0,0)
			\CArc(0,0)(6,0,380)
			\Text(0,0)[c]{{\scriptsize A}}
	\end{picture}}
		\put(1.3,6){\begin{picture}(1,1)(0,0)
			\CArc(0,0)(6,0,380)
			\Text(0,0)[c]{{\scriptsize B}}
		\end{picture}}
	\put(4.5,5.9){\footnotesize 0.3}
	\put(4.3,5.4){\footnotesize 0.6}
	\put(3.8,4.7){\footnotesize 0.9}
	\put(3.3,4.){\footnotesize 1.2}
	\put(2.9,3.){\footnotesize 1.5}
	\put(2.1,2.6){\footnotesize 3}
	\put(1.8,1.65){\footnotesize 6}
	\put(2.15,6.1){\scriptsize 1.5}
	\put(1.5,4.15){\footnotesize 3}
	\put(0.5,-0.3){Fig.~\ref{plot_12}c}
   \put(8,0){\includegraphics{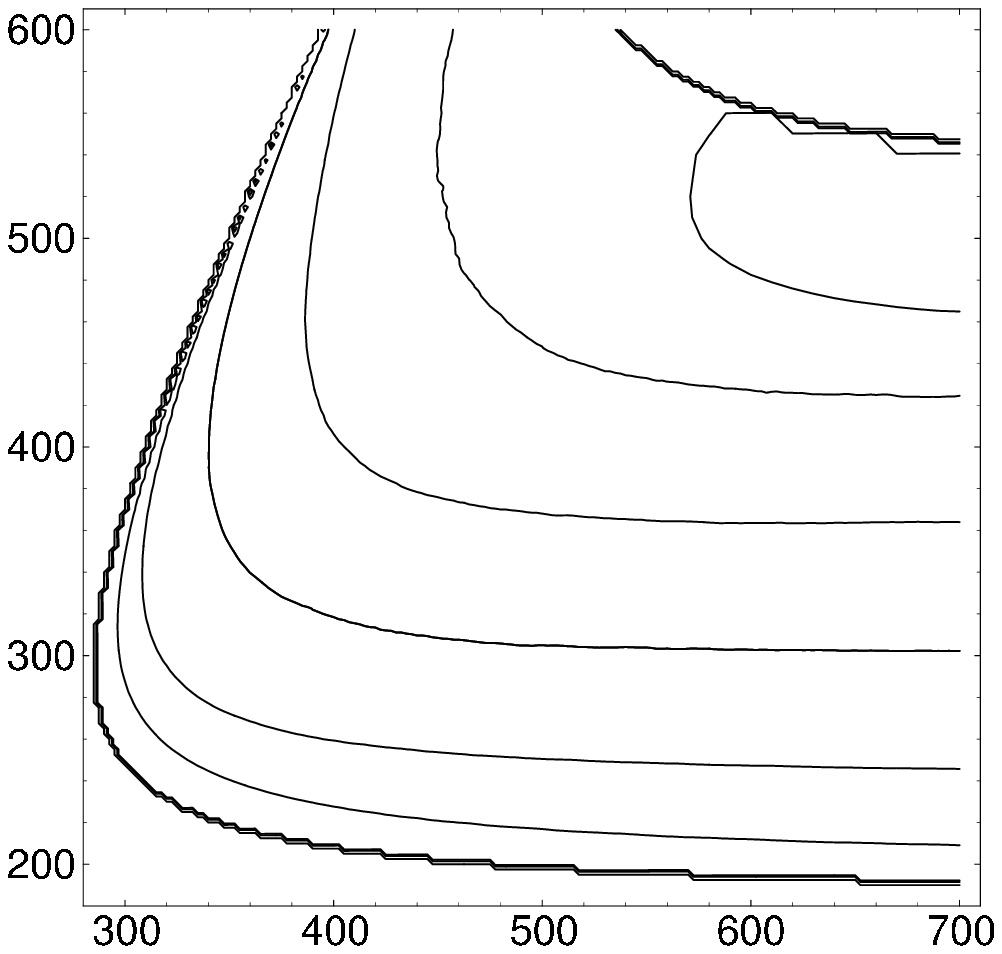}}
	\put(11.5,7.5){\fbox{${\mathcal A}_{\ell}$ in \% }}
	\put(8,7.3){$M_2~[{\rm GeV}]$}
	\put(13.5,-0.3){$|\mu|~[{\rm GeV}]$}
	\put(14.5,6.6){\begin{picture}(1,1)(0,0)
			\CArc(0,0)(6,0,380)
			\Text(0,0)[c]{{\scriptsize A}}
	\end{picture}}
		\put(9.3,6){\begin{picture}(1,1)(0,0)
			\CArc(0,0)(6,0,380)
			\Text(0,0)[c]{{\scriptsize B}}
	\end{picture}}
	\put(13.1,5.5){\footnotesize -1.6}
	\put(12.1,4.8){\footnotesize -1.5}
	\put(11.3,3.7){\footnotesize -1}
	\put(10.6,2.7){\footnotesize 0}
	\put(10.1,1.9){\footnotesize 1}
	\put(10.,1.35){\footnotesize 1.5}
   \put(8.5,-0.3){Fig.~\ref{plot_12}d}
\end{picture}
\vspace*{-1.5cm}
\caption{
	Contour plots for  
	\ref{plot_12}a: $\sigma_P(e^+e^- \to\tilde\chi^0_1\tilde\chi^0_2)$, 
	\ref{plot_12}b: ${\rm BR}(\tilde{\chi}^0_2 \to Z\tilde{\chi}_1^0)$,
	\ref{plot_12}c: $\sigma=\sigma_P(e^+e^-\to\tilde\chi^0_1\tilde\chi^0_2 ) 
	\times{\rm BR}(\tilde{\chi}^0_2 \to Z\tilde{\chi}_1^0)\times
	{\rm BR}(Z\to\ell\bar\ell)$ with ${\rm BR}(Z\to\ell\bar\ell)= 0.1$,
	\ref{plot_12}d: the asymmetry ${\mathcal A}_{\ell}$,
	in the $|\mu|$--$M_2$ plane for $\varphi_{M_1}=0.5\pi $, 
	$\varphi_{\mu}=0$, $\tan \beta=10$, $m_0=300$ GeV,
%	$A_{\tau}=250$ GeV, 
	$\sqrt{s}=800$ GeV and $(P_{e^-},P_{e^+})=(-0.8,0.6)$.
	The area A (B) is kinematically forbidden by
	$m_{\tilde\chi^0_1}+m_{\tilde\chi^0_2}>\sqrt{s}$
	$(m_{Z}+m_{\tilde\chi^0_1}> m_{\tilde\chi^0_2})$.
	In area C of plot \ref{plot_12}b:  
	${\rm BR}(\tilde{\chi}^0_2 \to Z\tilde\chi_1^0)=100 \%$.
	The gray  area is excluded by $m_{\tilde\chi_1^{\pm}}<104 $ GeV. 
	\label{plot_12}}
\end{figure}

In Fig.~\ref{plot_12}a we show the cross section for
$\tilde\chi^0_1 \tilde\chi^0_2$ production in the $|\mu|$--$M_2$
plane for $\varphi_{\mu}=0$ and $\varphi_{M_1}=0.5\pi$. 
For $|\mu|  \gsim 250 $ GeV the left selectron exchange dominates
due to the larger $\tilde\chi^0_2-\tilde e_L$ coupling,
such that the polarization $(P_{e^-},P_{e^+})=(-0.8,0.6)$
enhances the cross section to values of more than $110$~fb. 
The branching ratio  
${\rm BR} (\tilde\chi^0_2 \to Z \tilde\chi^0_1)$,
see Fig.~\ref{plot_12}b, can even be $100\%$ and decreases
with increasing $|\mu|$ and $M_2$, when the two-body decays
into sleptons and/or into the lightest neutral Higgs boson   
are kinematically allowed. 
The cross section 
$\sigma=\sigma_P(e^+e^-\to\tilde\chi^0_1\tilde\chi^0_2 ) 
\times{\rm BR}(\tilde{\chi}^0_2 \to Z\tilde{\chi}_1^0)\times
{\rm BR}(Z\to\ell\bar\ell)$, see Fig.~\ref{plot_12}c,
does however not  exceed $7$~fb, due to 
the small ${\rm BR}(Z\to\ell\bar\ell)=0.1$.
Fig.~\ref{plot_12}d shows the $|\mu|$--$M_2$ dependence 
of the asymmetry ${\mathcal A}_{\ell}$ for 
$\varphi_{M_1}=0.5\pi $ and $\varphi_{\mu}=0$.
The asymmetry $|{\mathcal A}_{\ell}|$ can reach a value of $1.6 \% $.
%On the contour ${\mathcal A}_{\ell}=0$ 
The (positive) contributions
from the production cancel the (negative) contributions from the
decay on the contour ${\mathcal A}_{\ell}=0$.
We also studied the $\varphi_{\mu}$ dependence of 
${\mathcal A}_{\ell}$. In the $|\mu|$--$M_2$ plane for $\varphi_{M_1}=0$ 
and $\varphi_{\mu}=0.5\pi$ we found $|{\mathcal A}_{\ell}|<0.5\%$. 
%------------------------------------------------------------------
%            CHI 1 CHI 2 -- P L O T (S)    1
%-----------------------------------------------------------------

\begin{figure}
\setlength{\unitlength}{1cm}
			\begin{minipage}{0.45\textwidth}
 \begin{picture}(10,6.5)(.5,0.7)
	\put(0,0){\includegraphics{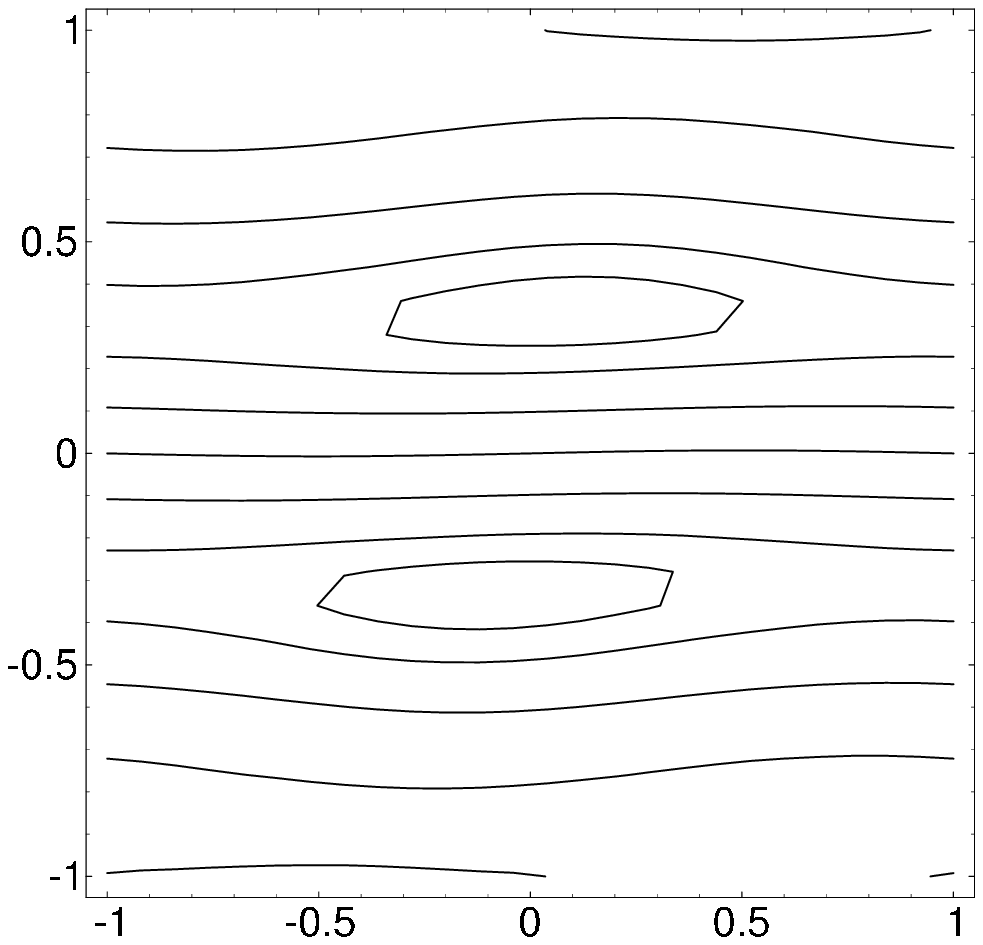}}
	\put(3.3,7.3){\fbox{${\mathcal A}_{\ell}$ in \% }}
	\put(6.5,-.3){$\varphi_{\mu}~[\pi]$}
	\put(0.3,7.3){$ \varphi_{M_1}~[\pi]$ }
		\put(6.2,6.45){\scriptsize 0}
		\put(5.8,6.15){\scriptsize 0}
		\put(5.9,5.6){\scriptsize 0.7}
		\put(5.6,5.2){\scriptsize 1.2}
		\put(4.8,4.7){\scriptsize 1.4}
		\put(5.8,4.5){\scriptsize 1.2}
		\put(6.1,4.15){\scriptsize 0.7}
		\put(6.4,3.8){\scriptsize 0}
		\put(5.7,3.5){\scriptsize -0.7}
		\put(5.4,3.2){\scriptsize -1.2}
		\put(4.3,2.7){\scriptsize -1.4}
		\put(5.3,2.6){\scriptsize -1.2}
		\put(5.6,2.15){\scriptsize -0.7}
		\put(6.0,1.65){\scriptsize 0}
		\put(2.5,0.85){\scriptsize 0}
	\end{picture}
\vspace*{.3cm}
\caption{Contour lines of the asymmetry ${\mathcal A}_{\ell}$
for $e^+e^-\to\tilde\chi^0_1\tilde\chi^0_2; 
	  \tilde{\chi}^0_2 \to Z\tilde{\chi}_1^0;
	Z\to\ell\bar\ell (\ell=e,\mu,\tau)$,	
	in the $\varphi_{\mu}$--$\varphi_{M_1}$ plane
for $M_2=250$ GeV, $|\mu|=400$ GeV, $\tan \beta=10$, $m_0=300$ GeV,
$\sqrt{s}=800$ GeV and $(P_{e^-},P_{e^+})=(-0.8,0.6)$.
\label{plot_3}}
\end{minipage}
\hspace*{0.5cm}
%&&&&&&&&&&&&&&&&&&&&&&&&&&&&&&&&&&&&&&&&&&&&&&&&&&&&&&&&&&&&&&
%                    P L O T  3 
%&&&&&&&&&&&&&&&&&&&&&&&&&&&&&&&&&&&&&&&&&&&&&&&&&&&&&&&&&&&&&&6
%
	\begin{minipage}{0.45\textwidth}
% \begin{picture}(10,6.5)(.3,1.2)
\begin{picture}(10,6.)(.3,1.2)
\put(-1,8){\includegraphics{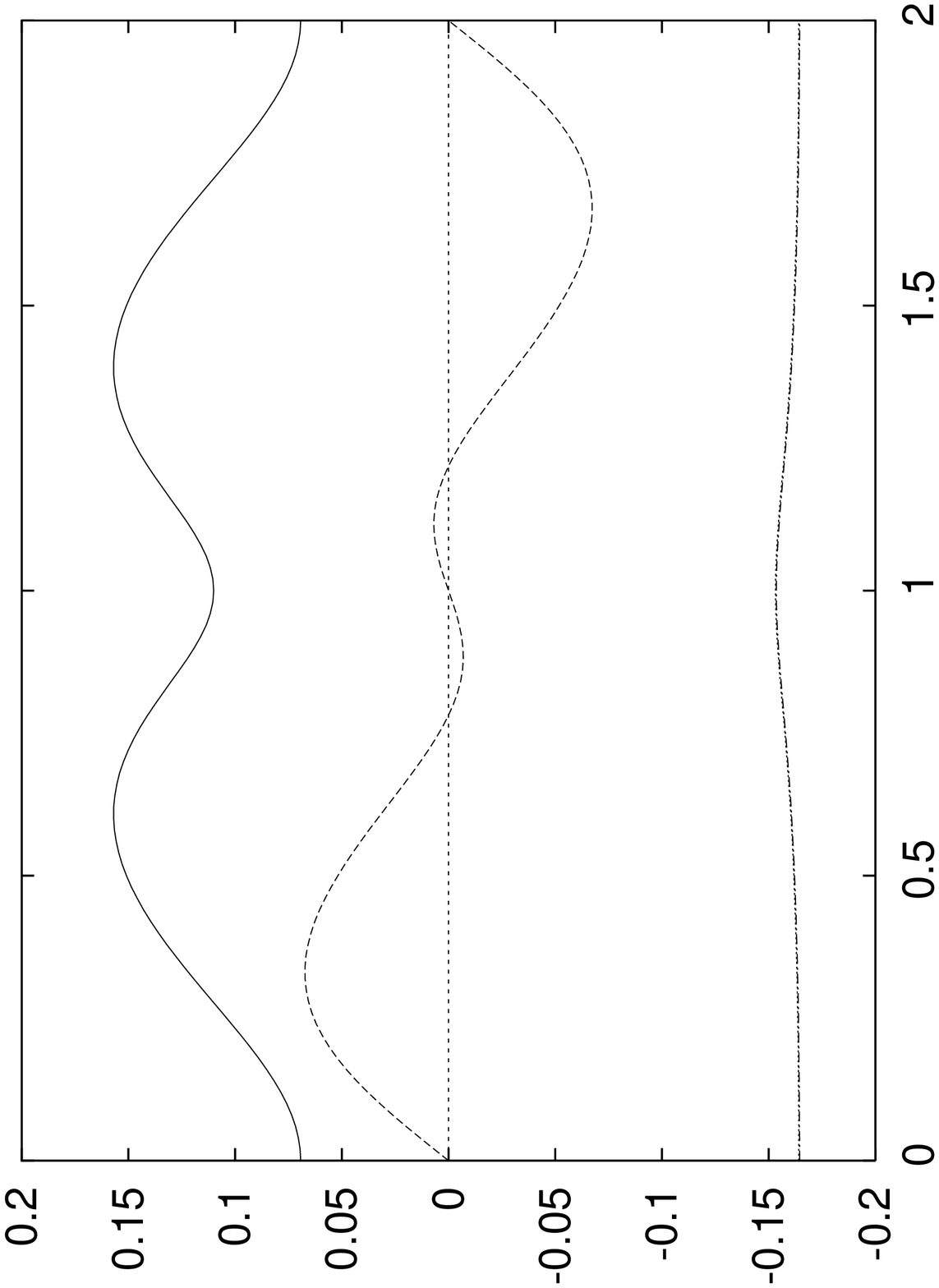}}
%	\put(2.5,7.5){\fbox{components of the $Z$ boson polarization}}
	\put(6.3,-0.3){$\varphi_{M_1}~[\pi]$}
	\put(3.5,5.85){\footnotesize $V_1 $}
	\put(2.6,4.5){\footnotesize $ V_2$}
	\put(1.5,3.35){\footnotesize $ V_3\approx 0$}
	\put(3.3,1.65){\footnotesize $ T_{11}\approx T_{22}$}
\end{picture}
\vspace*{.8cm}
\caption{
%	Dependence on $\varphi_{M_1}$ of the 
	Vector $(V_i)$ and tensor $(T_{ii})$ 
	components of the $Z$ density matrix  
	%$<\rho(Z)>$
	for $e^+e^-\to\tilde\chi^0_1\tilde\chi^0_2;
	\tilde{\chi}^0_2 \to Z\tilde{\chi}_1^0$, 
	for $M_2=250$ GeV,  $|\mu|=400$ GeV, 
	$\varphi_{\mu}=0$, $\tan \beta=10$, $m_0=300$ GeV,
	$\sqrt{s}=800$ GeV 
	and $(P_{e^-},P_{e^+})=(-0.8,0.6)$.
\label{plot_4}}
\end{minipage}
\end{figure}

In Fig.~\ref{plot_3} we show the $\varphi_{\mu}$--$\varphi_{M_1}$ 
dependence of ${\mathcal A}_{\ell}$ for $|\mu|=400$~GeV and 
$M_2=250$~GeV. The asymmetry ${\mathcal A}_{\ell}$ is more sensitive 
to $\varphi_{M_1}$ than to $\varphi_{\mu}$. It is remarkable
that the maximal phases of  $\varphi_{M_1},\varphi_{\mu}=\pm \pi/2$ do
not  lead to the highest values of 
${\mathcal A}_{\ell} \approx \pm 1.4\%$, which are reached for
$(\varphi_{M_1},\varphi_{\mu}) \approx (\pm0.3 \pi,0)$.
The reason for this is that the spin-correlation terms 
$\Sigma_P^a \,^c\Sigma_{D_1}^a \,^cD_2$ in the numerator of 
${\mathcal A}_{f}$~(\ref{Z:asym}), are products of CP-odd and
CP-even factors. The CP-odd (CP-even) factors have a sine-like
(cosine-like) phase dependence. Therefore, the maximum of the CP
asymmetry ${\mathcal A}_{f}$ is shifted from 
$\varphi_{M_1},\varphi_{\mu}= \pm \pi/2$ to a smaller or larger
value.

In the $\varphi_{\mu}$--$\varphi_{M_1}$ region shown 
in Fig.~\ref{plot_3} also the cross section 
$\sigma=\sigma_P(e^+e^-\to\tilde\chi^0_1\tilde\chi^0_2 ) 
\times{\rm BR}(\tilde{\chi}^0_2 \to Z\tilde{\chi}_1^0)\times
{\rm BR}(Z\to\ell\bar\ell)$ with 
${\rm BR}(\tilde{\chi}^0_2 \to Z\tilde{\chi}_1^0)=1$
and ${\rm BR}(Z\to\ell\bar\ell)=0.1$,
is rather insensitive to $\varphi_{\mu}$ and 
varies between $7$~fb ($\varphi_{M_1}=0$) and $14$~fb ($\varphi_{M_1}=\pm\pi$).
The statistical significance for measuring the
asymmetry for the leptonic decay of the $Z$
is given by $S_{\ell} =|{\mathcal A}_{\ell}| 
\sqrt{{\mathcal L}\cdot\sigma}$, 
see  Section~\ref{Z:T-odd asymmetry}.
For ${\mathcal L}= 500$~fb$^{-1}$, we have
$S_{\ell}<1$ in the scenario of Fig.~\ref{plot_3} and thus
${\mathcal A}_{\ell}$ cannot be measured at the 
$68\%$ confidence level $(S_{\ell}=1)$.
For hadronic decays into $b (c)$ quarks,
however, the significance is larger $S_{b(c)}=7.7(4.9)S_{\ell}$, 
as discussed Section~\ref{Z:T-odd asymmetry}. 
For ${\mathcal L}= 500$~fb$^{-1}$ and 
$(\varphi_{M_1},\varphi_{\mu})=(\pm0.3 \pi,0)$ in 
Fig.~\ref{plot_3} we find 
$S_{b(c)}=8(5)$ and thus ${\mathcal A}_{b(c)}$ can be measured.
%However note that we have $S_{\ell}<1$ in this scenario and thus
%${\mathcal A}_{\ell}$ cannot be measured at the 
%$68\%$ confidence level $(S_{\ell}=1)$.
%&&&&&&&&&&&&&&&&&&&&&&&&&&&&&&&&&&&&&&&&&&&&&&&&&&&&&&&&&&&&&&
%                    P L O T  2  
%&&&&&&&&&&&&&&&&&&&&&&&&&&&&&&&&&&&&&&&&&&&&&&&&&&&&&&&&&&&&&&66

In Fig.~\ref{plot_4} we show the 
$\varphi_{M_1}$ dependence of the vector $(V_i)$
$(V_i)$ and tensor $(T_{ii})$ components
of the $Z$ boson polarization. 
The components $T_{11}$, $T_{22}$ and $V_1$ have a CP-even dependence
on $\varphi_{M_1}$. The component $V_2$ is CP-odd and is not only zero
for $\varphi_{M_1}=0$ and $\varphi_{M_1}=\pi$, 
but also for $\varphi_{M_1}\approx(1\pm0.2)\pi$, 
due to the destructive interference of the
contributions from CP violation in production and decay.
The interference of the contributions from the
CP-even effects in production and decay cause the two
maxima of $V_1$. As discussed in 
Appendix~\ref{Neutralino decay into the Z boson},
the tensor components $T_{11}$ and $T_{22}$ are almost 
equal. Compared to  $V_1$ and $V_2$, they have the same order of 
magnitude but their dependence on $\varphi_{M_1}$ is rather weak.
The components $T_{13}, V_3<10^{-6}$ are small, 
and thus the density matrix  $<\rho(Z)>$ is almost symmetric. 
In the CP conserving case, e.g. for
$\varphi_{M_1}=\varphi_{\mu}=0$,
$M_2=250$ GeV,  $|\mu|=400$ GeV, 
$\tan \beta=10$, $m_0=300$ GeV, 
$\sqrt{s}=800$ GeV and  $(P_{e^-},P_{e^+})=(-0.8,0.6)$
it reads
\begin{eqnarray}
<\rho(Z)>  =\left(
        \begin{array}{ccc}
			  0.329  & 0.049 & 0.0003\\
			  0.049  & 0.343 & 0.049 \\
			  0.0003 & 0.049 & 0.329
        \end{array}
	\right).
\end{eqnarray}
In the CP violating case, e.g. for $\varphi_{M_1}=0.5\pi $ 
and the other parameters as above, $<\rho(Z)>$ has 
imaginary parts due to a non-vanishing $V_2$
\begin{eqnarray}
<\rho(Z)>  =\left(
        \begin{array}{ccc}
			  0.324        & 0.107+  0.037i&0.0003\\
			  0.107 -0.037i& 0.352&0.107 + 0.037i \\
			  0.0003       & 0.107 -0.037i&0.324
        \end{array}
	\right).
\end{eqnarray}
Imaginary elements of $<\rho(Z)>$ are thus an indication of CP violation. 
Note that also the CP even diagonal elements are changed for
$ \varphi_{M_1}\ne 0 $ (and also for $ \varphi_{\mu}\ne 0 $). 
This fact has been exploited in \cite{choi2} as a possibility 
to determine the CP violating phases.
The $ \varphi_{M_1}, \varphi_{\mu}$ dependence of the $Z$-density 
matrix elements has also been studied in \cite{reimer}, 
for  $e^+e^-\to\tilde\chi^0_1\tilde\chi^0_3$ 
followed by $\tilde\chi^0_3\to Z\tilde\chi^0_1$.

%&&&&&&&&&&&&&&&&&&&&&&&&&&&&&&&&&&&&&&&&&&&&&&&&&&&&&&&&&&&&&&
%                    P L O T  4 
%&&&&&&&&&&&&&&&&&&&&&&&&&&&&&&&&&&&&&&&&&&&&&&&&&&&&&&&&&&&&&&6
%
\begin{figure}
\setlength{\unitlength}{1cm}
\begin{picture}(10,8)(0,0)
   \put(0,0){\includegraphics{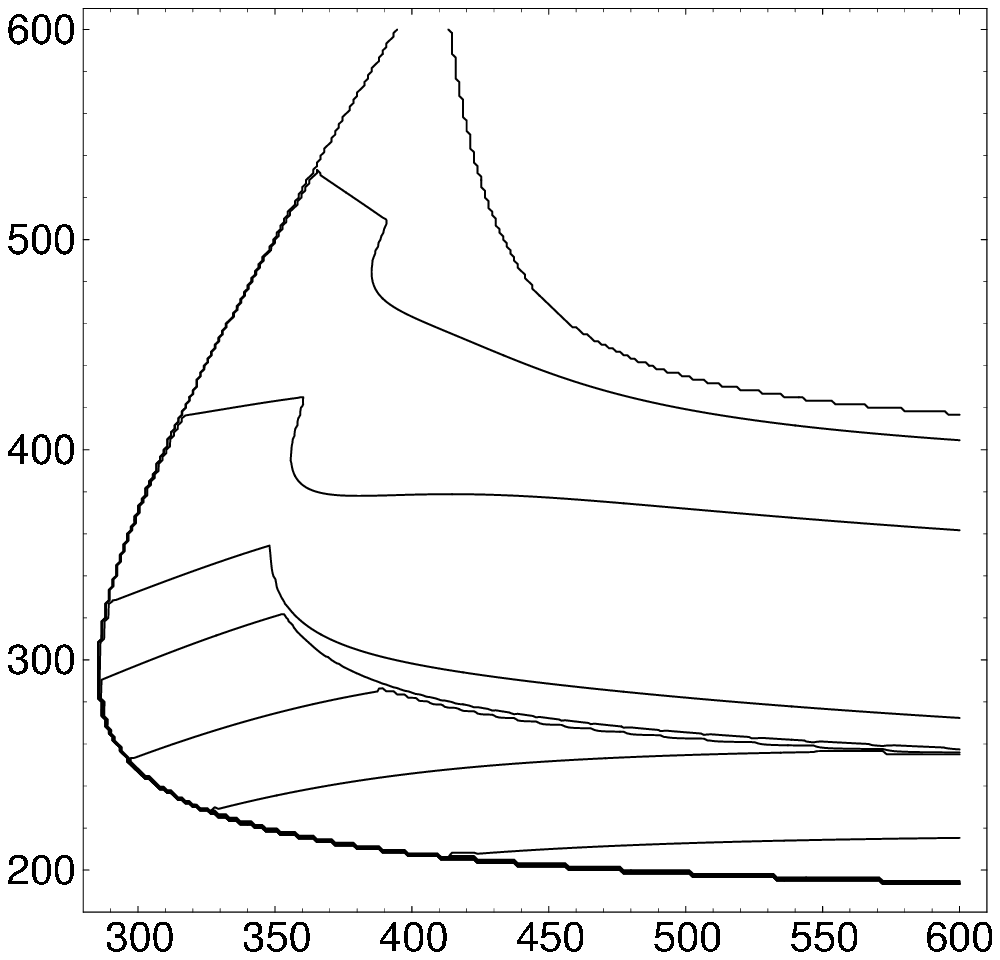}}
	\put(3.5,7.4){\fbox{$\sigma$ in fb}}
	\put(5.5,-0.3){$|\mu|~[{\rm GeV}]$}
	\put(0,7.4){$M_2~[{\rm GeV}]$}
	\put(4.5,1.1){\footnotesize 12}
	\put(3.3,1.3){\footnotesize 9}
	\put(2.,1.7){\footnotesize 6}
	\put(1.4,2.2){\footnotesize 3}
	\put(1.15,2.65){\footnotesize 1.5}
	\put(3.8,2.3){\footnotesize 1.5}
	\put(3.5,3.2){\footnotesize 0.3}
	\put(3.,4.2){\footnotesize 0.03}
  	\put(5.5,6){\begin{picture}(1,1)(0,0)
			\CArc(0,0)(7,0,380)
			\Text(0,0)[c]{{\footnotesize A}}
	\end{picture}}
			\put(1.3,5.5){\begin{picture}(1,1)(0,0)
			\CArc(0,0)(7,0,380)
			\Text(0,0)[c]{{\footnotesize B}}
		\end{picture}}
\put(0.5,-.3){Fig.~\ref{plot_22}a}
	\put(8,0){\includegraphics{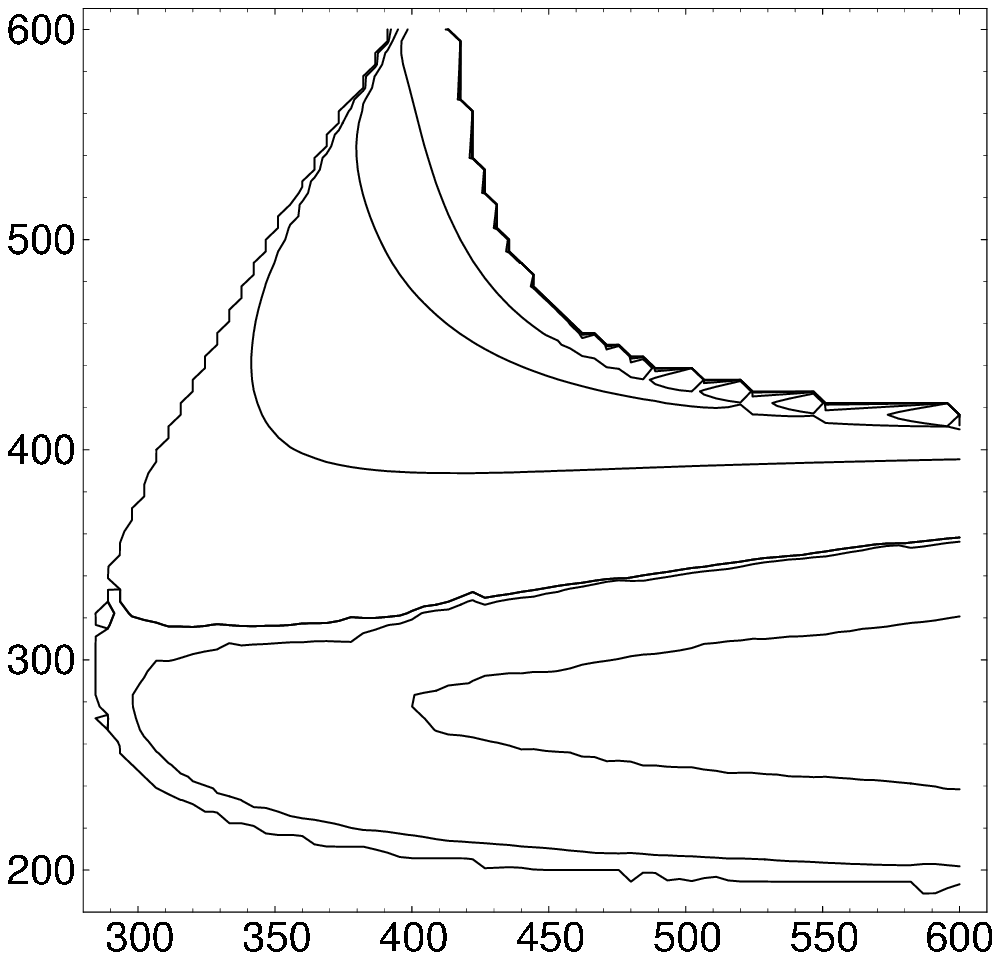}}
	\put(11.,7.4){\fbox{${\mathcal A}_{\ell}$ in \% }}
	\put(13.5,-.3){$|\mu|~[{\rm GeV}]$}
	\put(8,7.4){$M_2~[{\rm GeV}]$}
	\put(11.05,6.0){\footnotesize 3}
	\put(10.75,5.3){\footnotesize 1}
	\put(10.05,3.9){\footnotesize 0.2}
	\put(9.4,2.7){\footnotesize 0}
	\put(9.3,1.9){\footnotesize -0.01}
	\put(11.2,1.9){\footnotesize -0.05}
	  	\put(13.5,6){\begin{picture}(1,1)(0,0)
			\CArc(0,0)(7,0,380)
			\Text(0,0)[c]{{\footnotesize A}}
	\end{picture}}
			\put(9.3,5.5){\begin{picture}(1,1)(0,0)
			\CArc(0,0)(7,0,380)
			\Text(0,0)[c]{{\footnotesize B}}
		\end{picture}}
	\put(8.5,-.3){Fig.~\ref{plot_22}b}
\end{picture}
\vspace*{.5cm}
\caption{
	Contour lines of 
	$\sigma=\sigma_P(e^+e^-\to\tilde\chi^0_2\tilde\chi^0_2 ) 
	\times{\rm BR}(\tilde{\chi}^0_2 \to Z\tilde{\chi}_1^0)\times
	{\rm BR}(Z\to\ell\bar\ell)$ (\ref{plot_22}a),
	and the asymmetry ${\mathcal A}_{\ell}$ (\ref{plot_22}b)
	in the $|\mu|$--$M_2$ plane for $\varphi_{M_1}=0.5\pi $, 
	$\varphi_{\mu}=0$, $\tan \beta=10$, $m_0=300$ GeV,
	$\sqrt{s}=800$ GeV and $(P_{e^-},P_{e^+})=(-0.8,0.6)$.
	The area A (B) is kinematically forbidden by
	$m_{\tilde\chi^0_2}+m_{\tilde\chi^0_2}>\sqrt{s}$
	$(m_{Z}+m_{\tilde\chi^0_1}> m_{\tilde\chi^0_2})$.
	\label{plot_22}}
\end{figure}

\subsubsection{Production of $\tilde\chi^0_2 \, \tilde\chi^0_2$ 
           \label{Productionof22}}

In Fig.~\ref{plot_22}a we show the cross section 
$\sigma=\sigma_P(e^+e^-\to\tilde\chi^0_2\tilde\chi^0_2 ) 
\times{\rm BR}(\tilde{\chi}^0_2 \to Z\tilde\chi_1^0)\times
{\rm BR}(Z\to\ell\bar\ell)$ in the $|\mu|$--$M_2$ plane
for $\varphi_{\mu}=0$ and $\varphi_{M_1}=0.5\pi$.
The production cross section
$\sigma_P(e^+e^-\to\tilde\chi^0_2\tilde\chi^0_2 )$, 
which is not shown, is enhanced by the choice
$(P_{e^-},P_{e^+})=(-0.8,0.6)$ and reaches values up to $130$~fb.
The branching ratio ${\rm BR}(\tilde{\chi}^0_2 \to
Z\tilde{\chi}_1^0)$ can be $100\%$, see Fig.~\ref{plot_12}b,
however, due to ${\rm BR}(Z\to\ell\bar\ell)=0.1$,
$\sigma$ is not larger than $13$~fb,
see Fig.~\ref{plot_22}a. 

If two equal neutralinos are produced,
the  CP sensitive transverse polarization of the neutralinos 
perpendicular to  the production plane vanishes, 
$\Sigma^{2}_P=0 $ in~(\ref{Z:asym}). However, the asymmetry 
${\mathcal A}_{f}$ obtains CP sensitive contributions 
from the  neutralino decay process,
terms with $a=1,3$ in~(\ref{Z:adependence}).
In Fig.~\ref{plot_22}b we show for $\varphi_{M_1}=0.5\pi $ and 
$\varphi_{\mu}=0$ the $|\mu|$ and $M_2$ dependence 
of the asymmetry ${\mathcal A}_{\ell}$, which 
reaches more than  $3 \% $.
Along the contour  ${\mathcal A}_{\ell}=0$ 
in  Fig.~\ref{plot_22}b the contribution to ${\mathcal A}_{\ell}$ 
which is proportional
to $ \Sigma_P^1$, see~\ref{Z:asym}, cancels that 
which is proportional to $ \Sigma_P^3$.
As the largest values of ${\mathcal A}_{\ell}\gsim 0.2\% $ 
and ${\mathcal A}_{q}\gsim 1\% $
lie in a region of the $|\mu|$--$M_2$ plane where $\sigma\lsim0.3$~fb,
it will be difficult to measure ${\mathcal A}_{f}$  
in a statistically significant way.
We also studied the $\varphi_{\mu}$ dependence of 
${\mathcal A}_{\ell}$. In the $|\mu|$--$M_2$ plane for $\varphi_{M_1}=0$ 
and $\varphi_{\mu}=0.5\pi$ we found $|{\mathcal A}_{\ell}|<0.5\%$, 
and thus the influence of $\varphi_{\mu}$ is also small.

\begin{figure}
\begin{minipage}{0.95\textwidth}
\begin{picture}(10,7)(-3.5,0)
   \put(-1,8){\includegraphics{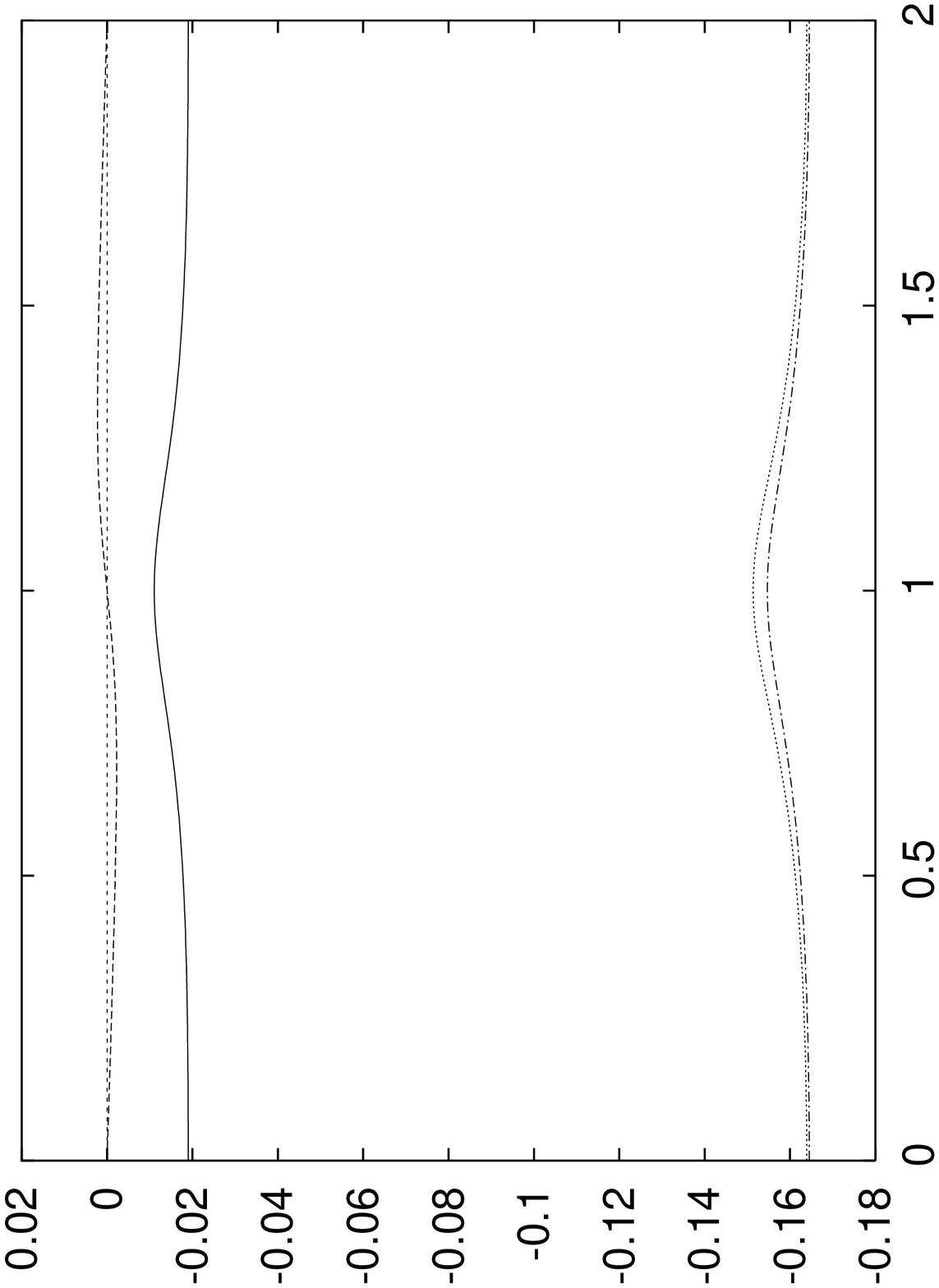}}
%	\put(2.5,7.3){\fbox{ matrix elements}}
	\put(7.8,0.2){$\varphi_{M_1}~[\pi]$}
	\put(3.8,5.4){\footnotesize $V_1 $}
	\put(5.05,6.){\footnotesize $V_3=0 $}
	\put(5.3,6.5){\footnotesize $V_2$}
	\put(3.8,1.7){\footnotesize $ T_{11}$}
	\put(3.8,1.05){\footnotesize $ T_{22}$}
\end{picture}
\vspace*{.3cm}
\caption{
%	Dependence on $\varphi_{M_1}$ of the 
	Vector $(V_i)$ and tensor $(T_{ii})$ 
	components of the $Z$ density matrix  
	for $e^+e^-\to\tilde\chi^0_2\tilde\chi^0_2;
	\tilde{\chi}^0_2 \to Z\tilde{\chi}_1^0$, 
	for $M_2=250$ GeV, $|\mu|=400$ GeV, 
	$\varphi_{\mu}=0$, $\tan \beta=10$, $m_0=300$ GeV,
	$\sqrt{s}=800$ GeV and $(P_{e^-},P_{e^+})=(-0.8,0.6)$.
	\label{plot_6}}
\end{minipage}
\end{figure}

In Fig.~\ref{plot_6} we show the 
$\varphi_{M_1}$ dependence of the vector $(V_i)$ and tensor $(T_{ii})$ 
components of the $Z$ boson polarization.
Because there are only CP sensitive contributions from the  neutralino 
decay process, $V_2$ is only zero at $\varphi_{M_1}=0,\pi$ and
$V_1$ has one maximum at $\varphi_{M_1}=\pi$, 
compared to the components for $\tilde\chi^0_1 \, \tilde\chi^0_2$
production, shown in Fig.~\ref{plot_4}.
In addition, the vector components $V_1$ and 
$V_2$ in Fig.~\ref{plot_6} are much smaller than the 
tensor components $T_{11}\approx T_{22}$.
The smallness of $V_2$ accounts for the
smallness of the asymmetry $|{\mathcal A}_{\ell}|<0.05\%$.
Furthermore, the other components are small, i.e. $T_{13}<10^{-6}$
and $V_3=0$.

%
%&&&&&&&&&&&&&&&&&&&&&&&&&&&&&&&&&&&&&&&&&&&&&&&&&&&&&&&&&&&&&&
%                    P L O T  5 
%&&&&&&&&&&&&&&&&&&&&&&&&&&&&&&&&&&&&&&&&&&&&&&&&&&&&&&&&&&&&&&6
%

\subsubsection{Production of $\tilde\chi^0_1 \, \tilde\chi^0_3$ }

\begin{figure}
\setlength{\unitlength}{1cm}
%\begin{picture}(10,7.8)(-0.5,0)
	\begin{picture}(10,7.5)(0,0)
	\put(0,0){\includegraphics{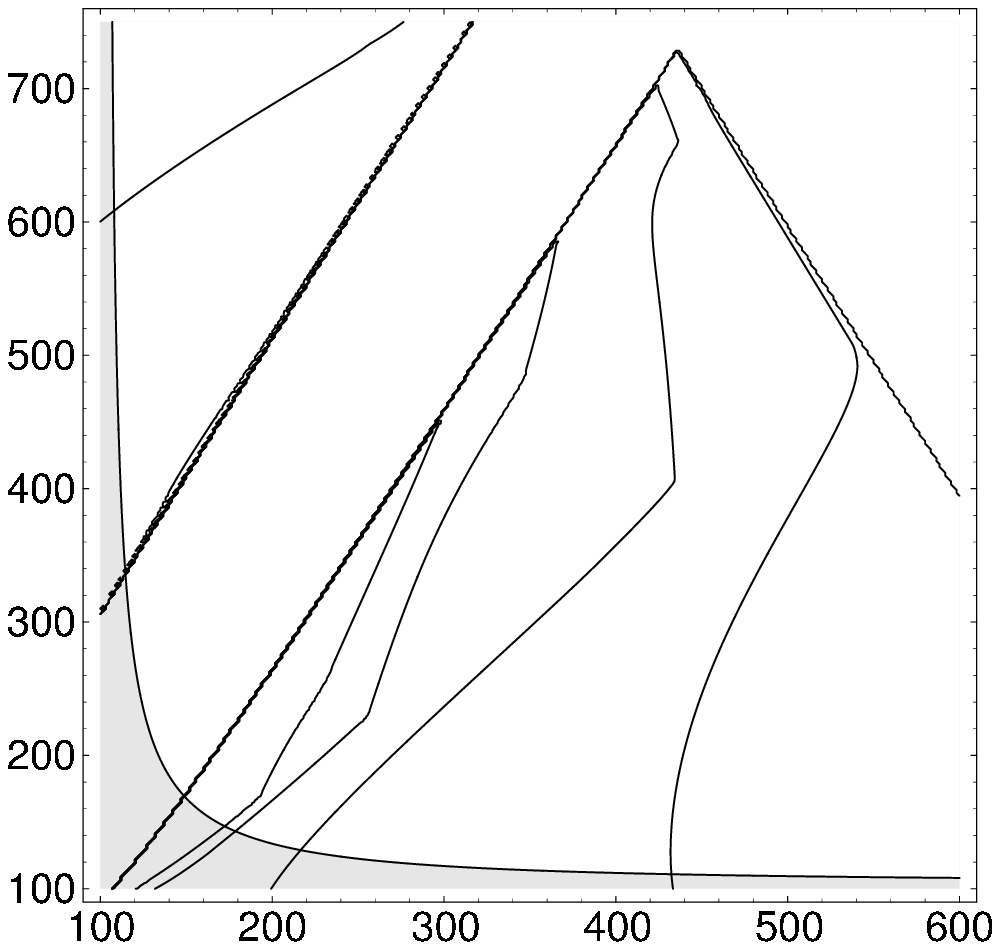}}
	\put(3.5,7.4){\fbox{$\sigma$ in fb}}
	\put(5.5,-0.3){$|\mu|~[{\rm GeV}]$}
	\put(0,7.4){$M_2~[{\rm GeV}]$}
	\put(1.35,6.3){\footnotesize 0.03}
	\put(1.,4.){\footnotesize 0.3}
	\put(1.95,1.8){\scriptsize 3}
	\put(3.05,2.8){\footnotesize 1.5}
	\put(4.85,3.8){\footnotesize 0.3}
	\put(5.1,2.0){\footnotesize 0.03}
  	\put(6.2,6){\begin{picture}(1,1)(0,0)
			\CArc(0,0)(7,0,380)
			\Text(0,0)[c]{{\footnotesize A}}
	\end{picture}}
			\put(2.6,4.5){\begin{picture}(1,1)(0,0)
			\CArc(0,0)(7,0,380)
			\Text(0,0)[c]{{\footnotesize B}}
		\end{picture}}
\put(0.5,-.3){Fig.~\ref{plot_13}a}
	\put(8,0){\includegraphics{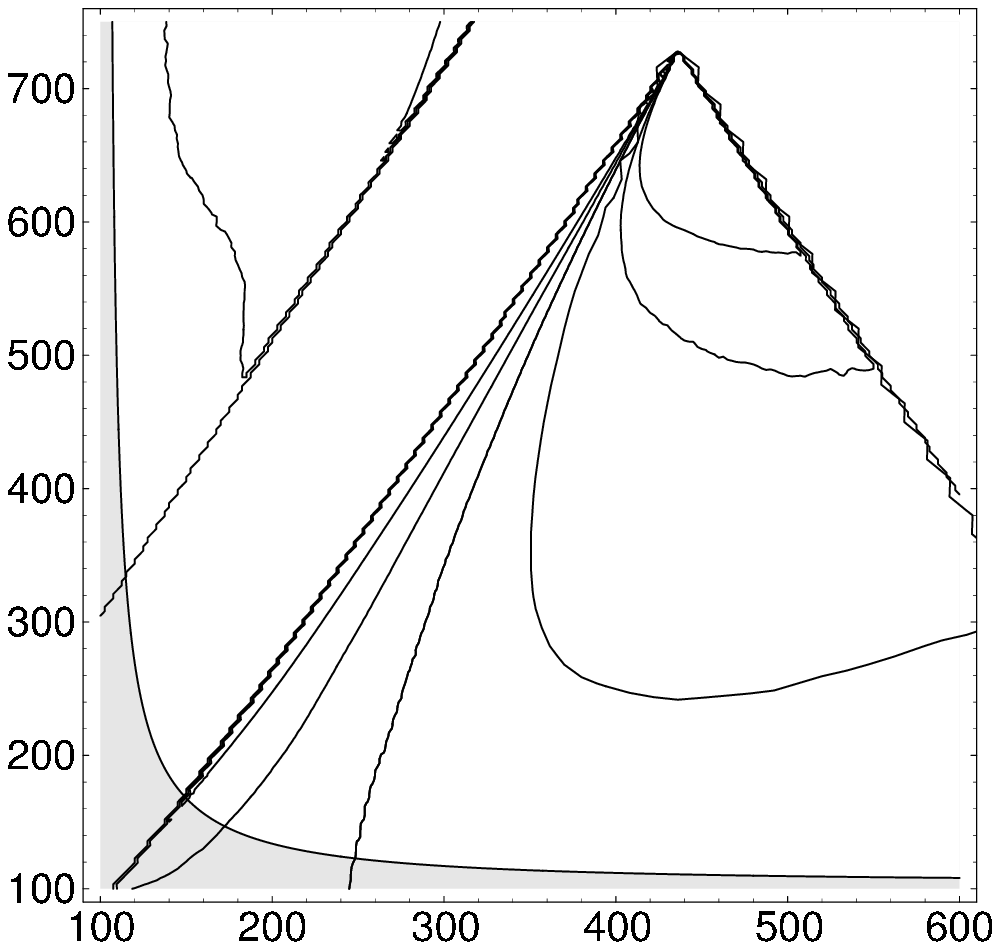}}
	\put(11.,7.4){\fbox{${\mathcal A}_{\ell}$ in \% }}
	\put(13.5,-.3){$|\mu|~[{\rm GeV}]$}
	\put(8,7.4){$M_2~[{\rm GeV}]$}
	\put(10.8,6.55){\footnotesize 1}
	\put(9.4,5.9){\footnotesize 0.5}
	\put(12.75,5.45){\footnotesize -1.3}
	\put(12.9,4.7){\footnotesize -1}
	\put(12.1,2.3){\footnotesize -0.5}
	\put(10.7,1.1){\footnotesize 0}
	\put(10.15,1.6){\footnotesize 0.5}
	\put(10.0,2.){\scriptsize 1}
	  	\put(14.2,6){\begin{picture}(1,1)(0,0)
			\CArc(0,0)(7,0,380)
			\Text(0,0)[c]{{\footnotesize A}}
	\end{picture}}
			\put(10.6,4.5){\begin{picture}(1,1)(0,0)
			\CArc(0,0)(7,0,380)
			\Text(0,0)[c]{{\footnotesize B}}
		\end{picture}}
	\put(8.5,-.3){Fig.~\ref{plot_13}b}
%---------------------------------------
\end{picture}
\vspace*{.4cm}
\caption{
	Contour lines of 
	$\sigma=\sigma_P(e^+e^-\to\tilde\chi^0_1\tilde\chi^0_3) 
	\times{\rm BR}(\tilde{\chi}^0_3 \to Z\tilde{\chi}_1^0)\times
	{\rm BR}(Z\to\ell\bar\ell)$ (\ref{plot_13}a),
	and the asymmetry ${\mathcal A}_{\ell}$ (\ref{plot_13}b)
	in the $|\mu|$--$M_2$ plane for $\varphi_{M_1}=0.5\pi $, 
	$\varphi_{\mu}=0$, $\tan \beta=10$, $m_0=300$ GeV,
	$\sqrt{s}=800$ GeV and $(P_{e^-},P_{e^+})=(0.8,-0.6)$.
	The area A (B) is kinematically forbidden by
	$m_{\tilde\chi^0_1}+m_{\tilde\chi^0_3}>\sqrt{s}$
	$(m_{Z}+m_{\tilde\chi^0_1}> m_{\tilde\chi^0_3})$.
		The gray area is excluded by $m_{\tilde\chi_1^{\pm}}<104$ GeV.
	\label{plot_13}}
\end{figure}

In Fig.~\ref{plot_13}a we show the cross section 
$\sigma=\sigma_P(e^+e^-\to\tilde\chi^0_1\tilde\chi^0_3) 
\times{\rm BR}(\tilde{\chi}^0_3 \to Z\tilde{\chi}_1^0)\times
{\rm BR}(Z\to\ell\bar\ell)$ in the $|\mu|$--$M_2$ plane
for $\varphi_{\mu}=0$ and $\varphi_{M_1}=0.5\pi$.
The production cross section 
$\sigma_P(e^+e^-\to\tilde\chi^0_1\tilde\chi^0_3)$,
which is not shown, is enhanced by the choice 
$(P_{e^-},P_{e^+})=(0.8,-0.6)$ and reaches up to $50$~fb.
The branching ratio ${\rm BR}(\tilde{\chi}^0_3 \to
Z\tilde{\chi}_1^0)$, which is not shown, can be $100\%$,
however, due to ${\rm BR}(Z\to\ell\bar\ell)=0.1$,
the cross section shown in Fig.~\ref{plot_13}a does not exceed $5$~fb.
In Fig.~\ref{plot_13}b we show the $|\mu|$--$M_2$ dependence 
of the asymmetry ${\mathcal A}_{\ell}$.  
The asymmetry $|{\mathcal A}_{\ell}|$ reaches  $1.3 \% $
at its maximum, however in a region, where $\sigma<0.3$~fb,
the asymmetry ${\mathcal A}_{\ell}$ thus cannot be measured. 
%We also studied the $\varphi_{\mu}$ dependence of 
%${\mathcal A}_{\ell}$. 
In the $|\mu|$--$M_2$ plane for $\varphi_{M_1}=0$ 
and $\varphi_{\mu}=0.5\pi$ we found $|{\mathcal A}_{\ell}|<0.7\%$. 
%&&&&&&&&&&&&&&&&&&&&&&&&&&&&&&&&&&&&&&&&&&&&&&&&&&&&&&&&&&&&&&
%                    P L O T  6 
%&&&&&&&&&&&&&&&&&&&&&&&&&&&&&&&&&&&&&&&&&&&&&&&&&&&&&&&&&&&&&&6
%

\vspace{-0.5cm}
\subsubsection{Production of $\tilde\chi^0_2 \, \tilde\chi^0_3$ }

\begin{figure}
\setlength{\unitlength}{1cm}
%	\fbox{
\begin{picture}(10,7.6)(0,0)
   \put(0,0){\includegraphics{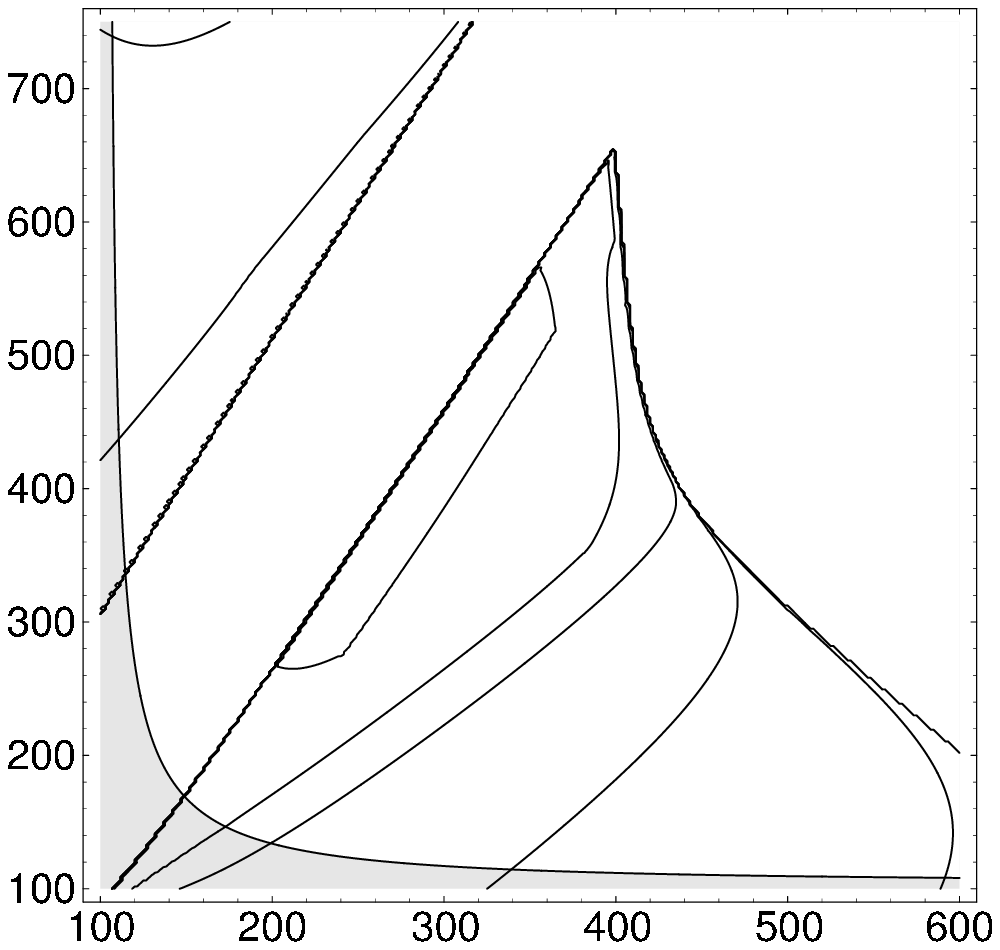}}
	\put(3.5,7.4){\fbox{$\sigma$ in fb}}
	\put(5.5,-0.3){$|\mu|~[{\rm GeV}]$}
	\put(0,7.4){$M_2~[{\rm GeV}]$}
	\put(1.4,6.5){\footnotesize 0.03}
	\put(1.4,5.4){\footnotesize 0.3}
	\put(2.8,3.4){\footnotesize 4.5}
	\put(3.8,3.4){\footnotesize 1.5}
	\put(4.25,2.4){\footnotesize 0.3}
	\put(5.1,2.0){\footnotesize 0.03}
	\put(5.9,1.0){\footnotesize 0.003}
%	\put(.,.){\footnotesize 0.001}
  	\put(5.52,5){\begin{picture}(1,1)(0,0)
			\CArc(0,0)(7,0,380)
			\Text(0,0)[c]{{\footnotesize A}}
	\end{picture}}
			\put(2.7,4.5){\begin{picture}(1,1)(0,0)
			\CArc(0,0)(7,0,380)
			\Text(0,0)[c]{{\footnotesize B}}
		\end{picture}}
\put(0.5,-.3){Fig.~\ref{plot_23}a}
	\put(8,0){\includegraphics{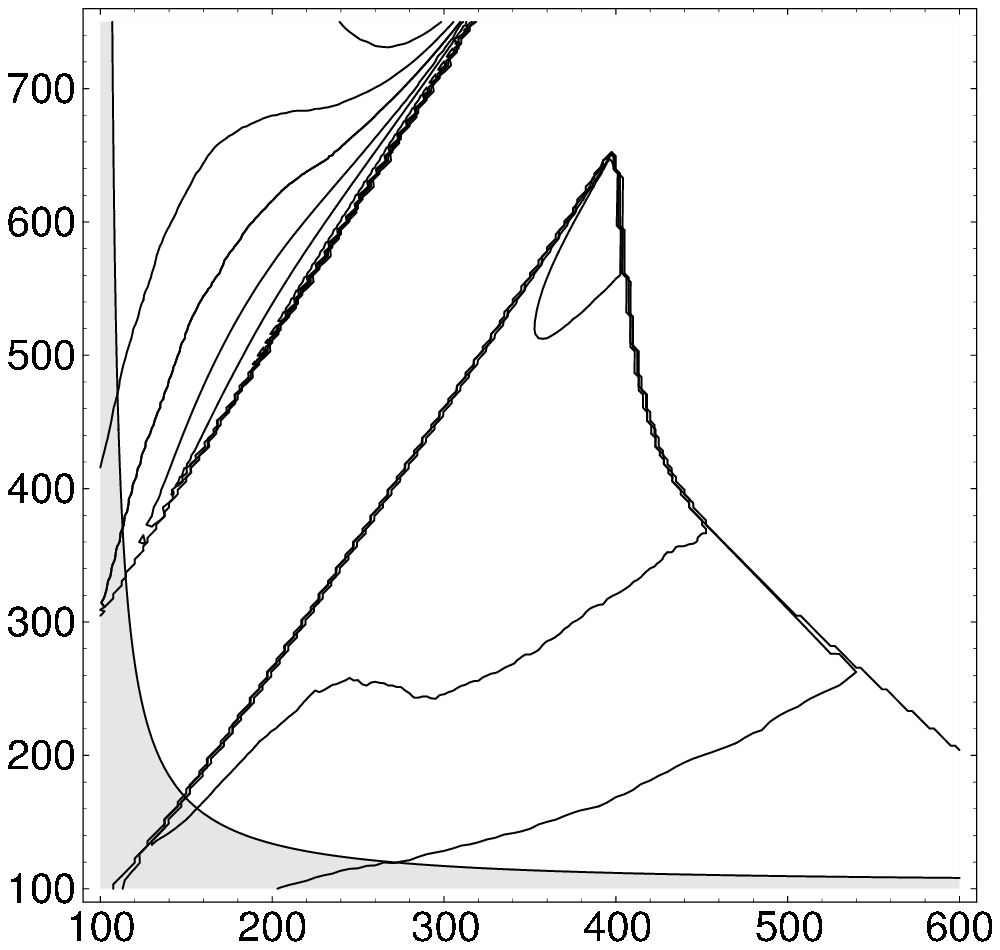}}
	\put(11.,7.4){\fbox{${\mathcal A}_{\ell}$ in \% }}
	\put(13.5,-.3){$|\mu|~[{\rm GeV}]$}
	\put(8,7.4){$M_2~[{\rm GeV}]$}
	\put(10.7,6.75){\scriptsize 1}
	\put(9.3,6.15){\footnotesize 0.5}
	\put(9.58,5.4){\footnotesize 0}
	\put(9.85,4.85){\scriptsize -1}
	\put(9.73,5.15){\scriptsize -.5}
	\put(12.,4.9){\footnotesize -2}
	\put(12.15,2.9){\footnotesize -1}
	\put(13.,1.95){\footnotesize -0.5}
	  	\put(13.8,5){\begin{picture}(1,1)(0,0)
			\CArc(0,0)(7,0,380)
			\Text(0,0)[c]{{\footnotesize A}}
	\end{picture}}
			\put(10.7,4.5){\begin{picture}(1,1)(0,0)
			\CArc(0,0)(7,0,380)
			\Text(0,0)[c]{{\footnotesize B}}
		\end{picture}}
	\put(8.5,-.3){Fig.~\ref{plot_23}b}
%---------------------------------------
\end{picture}
%}
\vspace*{.5cm}
\caption{
	Contour lines of 
	$\sigma=\sigma_P(e^+e^-\to\tilde\chi^0_2\tilde\chi^0_3) 
	\times{\rm BR}(\tilde{\chi}^0_3 \to Z\tilde{\chi}_1^0)\times
	{\rm BR}(Z\to\ell\bar\ell)$ (\ref{plot_23}a),
	and the asymmetry ${\mathcal A}_{\ell}$ (\ref{plot_23}b)
	in the $|\mu|$--$M_2$ plane for $\varphi_{M_1}=0.5\pi $, 
	$\varphi_{\mu}=0$, $\tan \beta=10$, $m_0=300$~GeV,
	$\sqrt{s}=800$~GeV and $(P_{e^-},P_{e^+})=(0.8,-0.6)$.
	The area A (B) is kinematically forbidden by
	$m_{\tilde\chi^0_2}+m_{\tilde\chi^0_3}>\sqrt{s}$
	$(m_{Z}+m_{\tilde\chi^0_1}> m_{\tilde\chi^0_3})$.
		The gray area is excluded by $m_{\tilde\chi_1^{\pm}}<104$~GeV.
	\label{plot_23}}
\end{figure}

For the  process 
$e^+e^-\to\tilde\chi^0_2\tilde\chi^0_3$
we discuss the decay  $\tilde\chi^0_3\to Z\tilde{\chi}_1^0$ of the
heavier neutralino which has a larger kinematically allowed region
than that for $\tilde\chi^0_2\to Z\tilde{\chi}_1^0$. Similar to 
$\tilde\chi^0_1 \, \tilde\chi^0_3$ production and decay, the cross
section $\sigma_P(e^+e^-\to\tilde\chi^0_2\tilde\chi^0_3)$ reaches
values up to $50$~fb for a beam polarization of
$(P_{e^-},P_{e^+})=(0.8,-0.6)$ and that  for the complete
process $\sigma=\sigma_P(e^+e^-\to\tilde\chi^0_2\tilde\chi^0_3) 
\times{\rm BR}(\tilde{\chi}^0_3 \to Z\tilde{\chi}_1^0)\times
{\rm BR}(Z\to\ell\bar\ell)$ attains values up to $5$~fb in the 
investigated regions  of the $|\mu|$--$M_2$ plane in Fig.~\ref{plot_23}a.

The asymmetry ${\mathcal A}_{\ell}$, Fig.~\ref{plot_23}b,
is somewhat larger than that  for  
$\tilde\chi^0_1 \, \tilde\chi^0_3$ production and decay, and
reaches at its maximum 2\%. 
%Although in the respective region the cross
%section is also a bit larger, $\sigma \lsim 4$~fb, 
However, it will be difficult to measure ${\mathcal A}_{\ell}$,
since e.g.  for $|\mu|=380$ GeV, $M_2=560$ GeV and
$(\varphi_{M_1},\varphi_{\mu})=(0.5 \pi,0)$,
we found $S_{\ell}\approx1$, for ${\mathcal L}=500~{\rm fb}^{-1}$.
For the hadronic decays of the $Z$ boson we have $S_{b(c)}\approx 8(5)$
and thus ${\mathcal A}_{b(c)}$ is accessible
for $\tilde\chi^0_1 \, \tilde\chi^0_3$ production.
For $\varphi_{\mu}=0.5\pi$ and $\varphi_{M_1}=0$
we found that $|{\mathcal A}_{\ell}|\lsim1\%$
in regions of the $|\mu|$--$M_2$ plane where $\sigma \lsim 0.5$~fb,
and $|{\mathcal A}_{\ell}|\lsim0.4\%$
in regions where $\sigma \lsim 5$~fb.
%
%&&&&&&&&&&&&&&&&&&&&&&&&&&&&&&&&&&&&&&&&&&&&&&&&&&&&&&&&&&&&&&
%                    P L O T  8 
%&&&&&&&&&&&&&&&&&&&&&&&&&&&&&&&&&&&&&&&&&&&&&&&&&&&&&&&&&&&&&&6
%

\subsection{Summary of Section \ref{CP observables in neutralino production and decay 	into the Z boson}
	\label{Summary and conclusion}}

We have analyzed CP sensitive observables in 
neutralino production $e^+e^- \to\tilde{\chi}^0_i  \tilde{\chi}^0_j$
and the subsequent two-body decay  of one  neutralino
into a $Z$ boson $\tilde\chi^0_i \to \tilde\chi^0_n Z$,
followed by the decay $Z \to \ell \bar\ell $ for $ \ell= e,\mu,\tau$, 
or $Z \to q\bar q$ with $q=c,b$.
The CP sensitive observables are defined by the vector component $V_2$ 
of the $Z$ boson density matrix and the CP asymmetry 
$ {\mathcal A}_{\ell(q)}$, which involves the triple product
${\mathcal T}_{\ell(q)}= {\bf p}_{e^-}\cdot({\bf p}_{\ell(q)} 
\times {\bf p}_{\bar\ell(\bar q)}).$
The tree level contributions to these observables
are due to correlations of the neutralino $\tilde\chi^0_i$ spin and 
the $Z$ boson spin. 
In a numerical study of the MSSM parameter space with 
complex $M_1$ and $\mu$ for  
$\tilde{\chi}^0_1  \tilde{\chi}^0_2$,
$\tilde{\chi}^0_2  \tilde{\chi}^0_2$,
$\tilde{\chi}^0_1  \tilde{\chi}^0_3$ and
$\tilde{\chi}^0_2  \tilde{\chi}^0_3$ production,
we have shown that the asymmetry ${\mathcal A}_{\ell}$ 
can go up to 3\%. For the hadronic decays of the Z boson,  
larger asymmetries  are obtained with
${\mathcal A}_{c(b)} \simeq6.3(4.5)\times {\mathcal A}_{\ell}$.
%By analyzing their statistical errors, we found that the
%asymmetries ${\mathcal A}_{c(b)}$
%could be accessible in future electron positron linear collider 
%experiments in the 500-800 GeV range with high luminosity and 
%longitudinally polarized beams.

	\chapter{CP violation in production and decay of charginos
	\label{CP violation in production and decay of charginos}}

{\Large\bf Overview}\\
\vspace{0.3cm}

We study chargino production with longitudinally 
polarized beams
$e^+~e^- \to\tilde\chi^+_i~\tilde\chi^-_j$
with the subsequent leptonic decay of one chargino
$\tilde\chi^+_i \to \ell^+\tilde\nu_\ell $ 
for $ \ell= e,\mu,\tau$ \cite{charg1}.
This decay mode allows the definition of a CP asymmetry
which is sensitive to the phase $\varphi_{\mu}$
and probes CP violation in the chargino production process.
For chargino decay into a $W$ boson
$\tilde\chi^+_i \to W^+\chi^0_n$
%$W^+ \to c~\bar s $ 
\cite{charg2},
CP observables can be obtained  which are also sensitive to 
$\varphi_{M_1}$.
We present numerical results for the asymmetries, 
$W$ polarizations, cross sections
and branching ratios at a linear electron-positron collider 
with $\sqrt{s}=800$~GeV.
%The asymmetries can go up to $30\%$ and we discuss the event rates 
%which are necessary to observe the asymmetries. 
%Polarized electron and positron beams 
%can significantly enhance the asymmetries and cross sections.

		\section{CP asymmetry in chargino production and decay into a sneutrino
	\label{CP violation in chargino production and decay into the sneutrino}}

%We study CP-odd asymmetries  in chargino production 
%$e^+e^- \to\tilde\chi^{\pm}_1 \tilde\chi^{\mp}_2$
%and the subsequent two-body decay of one chargino
%into a sneutrino. 
%We show that in the Minimal Supersymmetric Standard Model with complex  
%parameter $\mu$ the asymmetries can reach  $30\%$. 
%We discuss the feasibility of measuring these asymmetries
%at a  linear collider with $\sqrt{s}=800$ GeV and 
%longitudinally polarized beams.
%
%\subsection{Introduction}

We study chargino production
\begin{eqnarray} \label{sneut:production}
	e^++e^-&\to&\tilde\chi^+_i+\tilde\chi^-_j; 
	\quad i,j =1,2,  
\end{eqnarray}
with longitudinally polarized beams and
the subsequent  two-body decay of one of the 
charginos into a sneutrino
\begin{eqnarray} \label{sneut:decay_1A}
	\tilde\chi^+_i \to \ell^+ + \tilde\nu_{\ell}; 
	\quad \ell = e,\mu,\tau.
\end{eqnarray}
We define the triple product
 \begin{eqnarray}\label{sneut:tripleproduct1}
	 {\mathcal T}_{\ell} &=& 
	 ({\bf p}_{e^-} \times {\bf p}_{\chi^+_i}) \cdot {\bf p}_{\ell}
 \end{eqnarray}
and the T-odd asymmetry 
\begin{eqnarray}\label{sneut:AT1}
{\mathcal A}_{\ell}^{\rm T} &=& 
	\frac{\sigma({\mathcal T}_{\ell}>0)-\sigma({\mathcal T}_{\ell}<0)}
		{\sigma({\mathcal T}_{\ell}>0)+\sigma({\mathcal T}_{\ell}<0)},
\end{eqnarray}
of the cross section  $\sigma$ for chargino 
production~(\ref{sneut:production}) and decay~(\ref{sneut:decay_1A}).
The asymmetry ${\mathcal A}_{\ell}^{\rm T}$ is not only sensitive
to the phase  $\varphi_{\mu}$, but also to absorptive contributions,
%which could enter via s-channel resonances or final-state interactions.
%In order to eliminate the contributions from  
%the absorptive parts, which do not signal CP violation,
%we study the CP asymmetry
which are eliminated in the CP asymmetry
\begin{equation}\label{sneut:ACP1}
	{\mathcal A}_{\ell} = 
	\frac{1}{2}({\mathcal A}_{\ell}^{\rm T}-\bar{\mathcal A}_{\ell}^{\rm T}),
\end{equation}
where $\bar{\mathcal A}_{\ell}^{\rm T}$ is the CP conjugated asymmetry for
the process 
$e^+e^-\to\tilde\chi^-_i\tilde\chi^+_j; \; 
\tilde\chi^-_i \to  \ell^-\, \bar{\tilde\nu_{\ell}}.$ 
In this context it is interesting to note that in chargino production it
is not possible to construct a triple product and a corresponding
asymmetry by using transversely polarized $e^+$ and $e^-$ beams 
\cite{choichargino,holger}, therefore, one has to rely on the transverse
polarization of the produced chargino.

\subsection{Cross section
     \label{sneut:Cross section}}

For the calculation of the cross section for the
combined process of chargino production (\ref{sneut:production})
and the subsequent two-body decay of
$\tilde\chi^+_i$ (\ref{sneut:decay_1A}),
we use the spin-density matrix formalism as in
\cite{gudichargino,spinhaber}.
The amplitude squared,  
\begin{eqnarray} \label{sneut:amplitude}
|T|^2 &=&
	|\Delta(\tilde\chi^+_i)|^2~
	\sum_{\lambda_i,\lambda'_i}~
	\rho_P   (\tilde\chi^+_i)^{\lambda_i \lambda_i'}\;
	\rho_{D}(\tilde\chi^+_i)_{\lambda_i'\lambda_i},
	\end{eqnarray}
is composed of the (unnormalized) spin-density production matrix
$\rho_P(\tilde\chi^+_i)$, defined in~(\ref{char:rhoP}), 
and the decay matrix
$\rho_{D}(\tilde\chi^+_i)$, defined in~(\ref{sneut:rhoD}),
with the helicity indices 
$\lambda_i$ and $ \lambda_i'$ of the chargino.
Inserting the density matrices 
into~(\ref{sneut:amplitude}) leads to
\begin{eqnarray} \label{sneut:amplitude2}
		|T|^2 &=& 4~|\Delta(\tilde\chi^+_i)|^2~  
			  ( P D +  \Sigma_P^a \Sigma_{D}^a ),
\end{eqnarray}
where we sum over a.
The cross section and distributions
are then obtained by integrating 
$|T|^2$ over the Lorentz invariant phase space 
element $d{\rm Lips}$, defined in~(\ref{Lipsleptonic1}):
\begin{equation}\label{sneut:crossection}
d\sigma=\frac{1}{2 s}|T|^2d{\rm Lips}(s;p_{\chi_j^-},p_{\ell},p_{\tilde\nu_{\ell}}). 
%%		(2\pi)^4 \delta^4(p_1+p_2-\sum_{i=3}^8 p_i) 
%d{\rm Lips}(s,p_{\chi_j },p_{{\ell}_1},p_{\chi_1},p_{{\ell}_2})\label{eq_13},
\end{equation}
%where we use the narrow width approximation for the chargino 
%propagator.

\subsection{CP asymmetries
	\label{sneut:CP asymmetries}}

Inserting the cross section~(\ref{sneut:crossection}) into the definition 
of the asymmetry~(\ref{sneut:AT1}) we obtain 
\begin{eqnarray}\label{sneut:properties}
	{\mathcal A}^{\rm T}_{\ell} 
	 = \frac{\int {\rm Sign}[{\mathcal T_{\ell}}]~
		 |T|^2 ~d{\rm Lips}}
           {\int |T|^2~d{\rm Lips}}
	=  \frac{\int {\rm Sign}[{\mathcal T}_{\ell}]~
	\Sigma_P^2 \Sigma_{D}^2~ d{\rm Lips}}
          {\int  P D ~d{\rm Lips}},
\end{eqnarray}
where we have already used the narrow width approximation for the
chargino propagator. In the numerator of~(\ref{sneut:properties}) 
only the CP sensitive contribution 
$\Sigma_P^2 \Sigma_{D}^2$ from chargino polarization perpendicular to 
the production plane  remains, since only
this term contains the triple product
${\mathcal T}_{\ell}=({\bf p}_{e^-} \times {\bf p}_{\chi^+_i}) 
\cdot {\bf p}_{\ell}$~(\ref{sneut:tripleproduct1}).
In the denominator only the term $P D$ remains,
since all spin correlations $\Sigma_P^a \Sigma_{D}^a$ 
vanish due to  the integration over the complete phase space. 

The coefficient $\Sigma^{2}_P$ is non-zero only for 
production of an unequal pair of charginos,
$e^+e^- \to\tilde\chi^{\pm}_1 \tilde\chi^{\mp}_2$,
and obtains contributions from $Z$-exchange and $Z$-$\tilde \nu$ 
interference only, see~(\ref{charg:sigmaP}). 
The contribution to $\Sigma^{2}_P$ from $Z$-exchange,
see~(\ref{char:ZZ}), is non-zero only for $\varphi_{\mu}\neq 0,\pi$, 
whereas the $Z$-$\tilde \nu$ 
interference term, see~(\ref{char:Zsnu}), obtains also
absorptive contributions due to the finite $Z$-width which do not
signal CP violation. These, however, will be eliminated in the 
asymmetry ${\mathcal A}_{\ell}$~(\ref{sneut:ACP1}).

For chargino decay into a tau sneutrino,
$\tilde\chi^+_i \to \tau^+  \tilde\nu_{\tau}$,
the asymmetry $	{\mathcal A}_{\tau}^{\rm T}\propto
	(|V_{i1}|^2 -Y_{\tau}^2|U_{i2}|^2)/
	 (|V_{i1}|^2 +Y_{\tau}^2|U_{i2}|^2)$
is reduced, which follows from the expressions for $D$
and $\Sigma_{D}^2 $, given in~(\ref{sneut:D_1B}) and~(\ref{sneut:SD_1B}).

%Note that in order to measure ${\mathcal A}_{\ell}$ 
%in the reaction (\ref{sneut:production})
%the momentum of $\tilde\chi^+_i$, i.e. the
%production plane, has to be determined.
%This can be done if the corresponding information from the decay
%of the other chargino  $\tilde\chi^-_j$ on the opposite side is
%also available. This is the case if, for example, the 
%$\tilde\chi^-_j$ decays like 
%$\tilde\chi^-_j \to \tilde\chi_1^- Z^0$,
%$\tilde\chi^-_j \to \tilde\chi^0_1 W^-$ or
%$\tilde\chi^-_j \to \tilde\chi_1^- H_1^0$
%and $Z,W,H_1^0$ decay hadronically,
%$Z^0\to q\,\bar q $, $W^-\to q\,\bar q' $, $H_1^0\to b \,\bar b$.
%If the masses of the charginos and $ \tilde\nu_{\ell}$
%as well as the masses of $H_1^0$ and $ \tilde\chi^0_1$
%are known, then the momentum ${\bf p}_{\chi^-_j}$ can be
%kinematically reconstructed. This is also possible if the
%leptonic decays $Z^0\to \ell^+ \ell^-$, $H_1^0\to \tau^+\tau^-$
%or $\tilde\chi^-_j \to \ell^-\tilde\nu_{\ell}$ are used.
%In order  to predict the expected accuracy of measuring 
%${\mathcal A}_{\ell}$, it is clear that also detailed
%Monte Carlo studies taking into account background and detector
%simulations are necessary. 
%However, this is beyond the scope of the present work.

\subsection{Numerical results
	\label{Numerical results}}

We present numerical results for the  
asymmetries  ${\mathcal A}_{\ell}$ ~(\ref{sneut:ACP1}),
for $\ell=e,\mu$, and the cross sections 
$\sigma=\sigma_P(e^+e^-\to\tilde\chi^+_1\tilde\chi^-_2 ) \times
{\rm BR}( \tilde\chi^+_1 \to \ell^+\tilde\nu_{\ell})$.
We study the dependence of the asymmetries and cross sections
on the MSSM parameters 
$\mu = |\mu| \, e^{ i\,\varphi_{\mu}}$, 
$M_2$ and $\tan \beta$.
We choose a center of mass energy of   $\sqrt{s} = 800$ GeV
and longitudinally polarized beams with
beam polarizations $(P_{e^-},P_{e^+})=(-0.8,+0.6)$,
which enhance $\tilde\nu_{e}$ exchange in the
production process. This results in larger cross sections
and asymmetries.

We study the decays of the lighter 
chargino $\tilde\chi^+_1$. For the calculation of the chargino 
widths $\Gamma_{\chi_1^+}$ and the branching ratios 
${\rm BR}( \tilde\chi^+_1 \to\ell^+ \tilde\nu_{\ell})$ 
%and ${\rm BR}( \tilde\chi^+_1 \to W^+\tilde\chi^0_1 )$ 
we include the following two-body decays, 
\begin{eqnarray}
	\tilde\chi^+_1 &\to& 
	W^+\tilde\chi^0_n,~
	e^+\tilde\nu_{e},~
	\mu^+\tilde\nu_{\mu},~
	\tau^+\tilde\nu_{\tau},~
	\tilde e_{L}^+\nu_{e},~
	\tilde\mu_{L}^+\nu_{\mu},~
	\tilde\tau_{1,2}^+\nu_{\tau},
\end{eqnarray}
and neglect three-body decays.
%including for $\tilde\chi^+_2$ the decays into the $Z$ boson
%and the  lightest neutral Higgs boson
%\begin{eqnarray}
%	\tilde\chi^+_2 &\to&
%	\tilde\chi_1^+ Z^0,~
%	\tilde\chi_1^+ H_1^0.
%\end{eqnarray}
%The Higgs parameter is chosen $m_{A}=1$~TeV and thus 
%the decays  $\tilde\chi^+_i \to \tilde\chi^0_n H^{\mp}$
%into the charged Higgs bosons are forbidden in our scenarios.
In order to reduce the number of parameters, we assume the 
relation $|M_1|=5/3~M_2\tan^2\theta_W $.
For all scenarios we fix the sneutrino and slepton masses, 
$m_{\tilde\nu_{\ell}}=185$~GeV, $\ell =e,\mu,\tau$,
$ m_{\tilde\ell_L}=200$~GeV, $\ell =e,\mu$.
These values are obtained from the renormalization group 
equations~(\ref{mselL}) and~(\ref{msneut}),
%%$m_{\tilde\ell_R}^2 = m_0^2 +0.23 M_2^2
%%-m_Z^2\cos 2 \beta \sin^2 \theta_W$, 
%$m_{\tilde\ell_L  }^2 = m_0^2 +0.79 M_2^2
%+m_Z^2\cos 2 \beta(-1/2+ \sin^2 \theta_W)$ and
%$m_{\tilde\nu_{\ell}}^2 = m_0^2 +0.79 M_2^2 +m_Z^2/2\cos 2 \beta$,
for $M_2=200$~GeV, $m_0=80$~GeV and $\tan\beta=5$.
In the stau sector, see Appendix \ref{Stau mixing}, 
we fix the trilinear scalar coupling
parameter to $A_{\tau}=250$~GeV.
The stau masses are fixed to $m_{\tilde\tau_{1}}=129$~GeV and
$m_{\tilde\tau_{2}}=202$~GeV.

%&&&&&&&&&&&&&&&&&&&&&&&&&&&&&&&&&&&&&&&&&&&&&&&&&&&&&&&&&&&&&&
%                    P L O T  1 
%&&&&&&&&&&&&&&&&&&&&&&&&&&&&&&&&&&&&&&&&&&&&&&&&&&&&&&&&&&&&&&6
%
\begin{figure}[h]
\setlength{\unitlength}{1cm}
\begin{picture}(10,8)(0,0)
   \put(0,0){\includegraphics{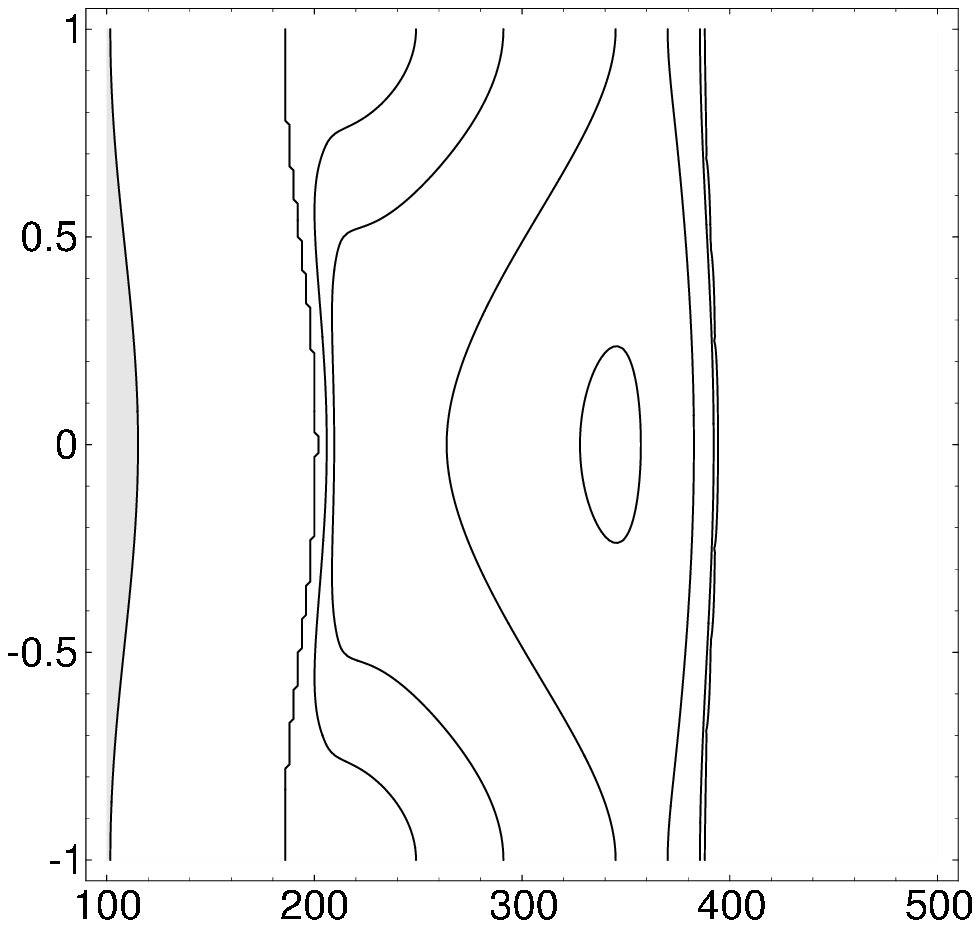}}
	\put(3.5,7.4){\fbox{$\sigma$ in fb}}
	\put(5.5,-0.3){$M_2~[{\rm GeV}]$}
	\put(0,7.4){ $\varphi_{\mu}~[\pi]$}
	\put(2.25,0.8){\footnotesize 0}
	\put(2.5,1.1){\footnotesize 10}
	\put(2.85,1.65){\footnotesize 20}
	\put(3.4,2.3){\footnotesize 40}
	\put(4.35,3.7){\footnotesize 58}
	\put(1.8,3.7){\begin{picture}(1,1)(0,0)
			\CArc(0,0)(7,0,380)
			\Text(0,0)[c]{{\footnotesize A}}
	\end{picture}}
	\put(6.0,3.7){\begin{picture}(1,1)(0,0)
			\CArc(0,0)(7,0,380)
			\Text(0,0)[c]{{\footnotesize B}}
	\end{picture}}
\put(0.5,-.3){Fig.~\ref{sneut:plot_1}a}
	\put(8,0){\includegraphics{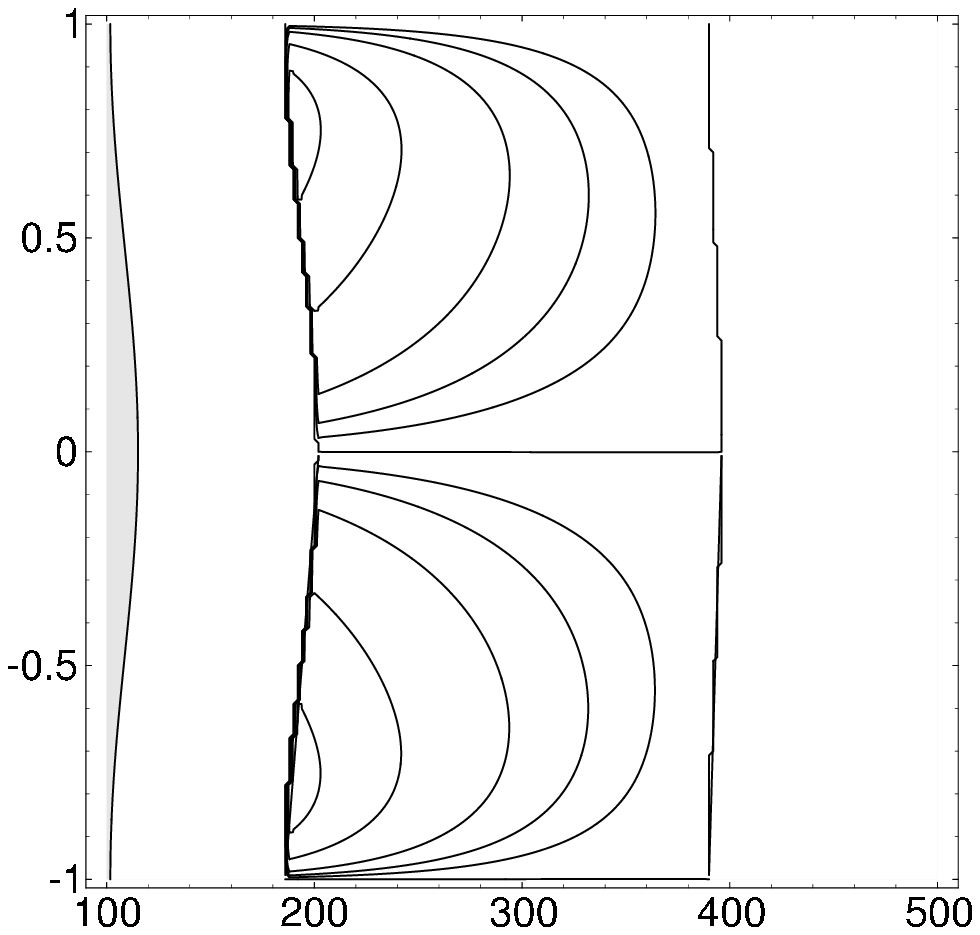}}
	\put(11.,7.4){\fbox{${\mathcal A}_{\ell}$ in \% }}
	\put(13.5,-.3){$M_2~[{\rm GeV}]$}
	\put(8,7.4){ $\varphi_{\mu}~[\pi]$}
	\put(10.4,1.6){\scriptsize 10}
	\put(11.05,1.8){\footnotesize 5}
	\put(11.75,2.2){\footnotesize 2}
	\put(12.25,2.5){\footnotesize 1}
	\put(12.65,2.8){\footnotesize 0.5}
	\put(13.0,1.1){\footnotesize 0}
	\put(10.45,5.8){\scriptsize -10}
	\put(11.05,5.5){\footnotesize -5}
	\put(11.75,5.15){\footnotesize -2}
	\put(12.3,4.9){\footnotesize -1}
	\put(12.5,4.3){\footnotesize -0.5}
	\put(13.0,6.3){\footnotesize 0}
		\put(9.8,3.7){\begin{picture}(1,1)(0,0)
			\CArc(0,0)(7,0,380)
			\Text(0,0)[c]{{\footnotesize A}}
	\end{picture}}
	\put(14.0,3.7){\begin{picture}(1,1)(0,0)
			\CArc(0,0)(7,0,380)
			\Text(0,0)[c]{{\footnotesize B}}
	\end{picture}}
	\put(8.5,-.3){Fig.~\ref{sneut:plot_1}b}
\end{picture}
\vspace*{.5cm}
\caption{
	Contour lines of 
	$\sigma=\sigma_P(e^+e^-\to\tilde\chi^+_1\tilde\chi^-_2) 
	\times {\rm BR}( \tilde\chi^+_1 \to \ell^+\tilde\nu_{\ell})$, 
	summed over $\ell = e,\mu$,
	(\ref{sneut:plot_1}a), and the asymmetry ${\mathcal A}_{\ell}$ 
	for $\ell =e$ or $\mu$ (\ref{sneut:plot_1}b),
	in the $M_2$--$\varphi_{\mu}$ plane for 
	%	$\varphi_{M_1}=0.7\pi $, $\varphi_{\mu}=0$, 
	$|\mu|=400$~GeV, $\tan\beta =5$, 
	$m_{\tilde\nu_{\ell}}=185$~GeV, 
	$\sqrt{s}=800$ GeV and $(P_{e^-},P_{e^+})=(-0.8,0.6)$.
	The gray  area is excluded by $m_{\chi_1^{\pm}}<104 $ GeV.
	The area A is kinematically forbidden by
%	$m_{\tilde\chi^+_1}+m_{\tilde\chi^-_2}>\sqrt{s}$
	$m_{\tilde\nu_{\ell}}+m_{\chi^0_1}> m_{\chi^+_1}$.
	The area B is kinematically forbidden by
	$m_{\chi^+_1}+m_{\chi^-_2}>\sqrt{s}$.
%	$m_{\tilde\nu_{\ell}}+m_{\tilde\chi^0_1}> m_{\tilde\chi^+_1}$.
	\label{sneut:plot_1}}
\end{figure}
In Fig.~\ref{sneut:plot_1}a we show the contour lines of the 
cross section for chargino production and decay
$\sigma=\sigma_P(e^+e^-\to\tilde\chi^+_1\tilde\chi^-_2) 
	\times {\rm BR}( \tilde\chi^+_1 \to \ell^+\tilde\nu_{\ell})$
in the $M_2$--$\varphi_{\mu}$ plane
for $|\mu|=400$~GeV and  $\tan\beta =5$. 
The production cross section 
$\sigma_P(e^+e^-\to\tilde\chi^+_1\tilde\chi^-_2)$
can attain values from $10$~fb to $150$~fb
and ${\rm BR}(\tilde\chi^+_1 \to \ell^+\tilde\nu_{\ell})$,
summed over $\ell = e,\mu$, can be as large as $50\%$. Note that $\sigma$ 
is very sensitive to $\varphi_{\mu}$, which has been exploited 
in \cite{choichargino,choigaiss} to constrain $\cos(\varphi_{\mu})$.

The $M_2$--$\varphi_{\mu} $ dependence of the CP asymmetry
${\mathcal A}_{\ell}$ for $\ell =e$ or $\mu$ is shown 
in Fig.~\ref{sneut:plot_1}b.
The asymmetry can be as large as $10\%$ and it does, however, not attain 
maximal values for $\varphi_{\mu}=\pm0.5\pi$.
The reason is that  ${\mathcal A}_{\ell}$ is proportional to a product of
a CP-odd ($\Sigma_P^2$) and a CP-even factor ($\Sigma_{D}^2$),
see~(\ref{sneut:properties}). The  CP-odd (CP-even) factor has
as sine-like (cosine-like) dependence on  $\varphi_{\mu}$.
Thus the maximum of ${\mathcal A}_{\ell}$ is shifted  
towards $\varphi_{\mu}=\pm\pi$ in Fig.~\ref{sneut:plot_1}b. 
Phases close  to the
CP conserving points, $\varphi_{\mu}= 0,\pm \pi$, are favored by
the experimental upper limits on the EDMs, as discussed in 
Section~\ref{CP violating phases and electric dipole moments}.
%For example in the constrained MSSM, we have
%$|\varphi_{\mu}|\lsim \pi/10$ \cite{edms}.
%However, the restrictions are very model dependent, e.g., 
%if also lepton flavor violating terms are included \cite{BMPW},
%the restrictions may disappear. In order to show the full phase dependence 
%of the asymmetries, we have relaxed the EDM restrictions for this purpose. 

%&&&&&&&&&&&&&&&&&&&&&&&&&&&&&&&&&&&&&&&&&&&&&&&&&&&&&&&&&&&&&&
%                    P L O T  2 
%&&&&&&&&&&&&&&&&&&&&&&&&&&&&&&&&&&&&&&&&&&&&&&&&&&&&&&&&&&&&&&6
%
\begin{figure}[h]
\setlength{\unitlength}{1cm}
\begin{picture}(10,8)(0,0)
   \put(0,0){\includegraphics{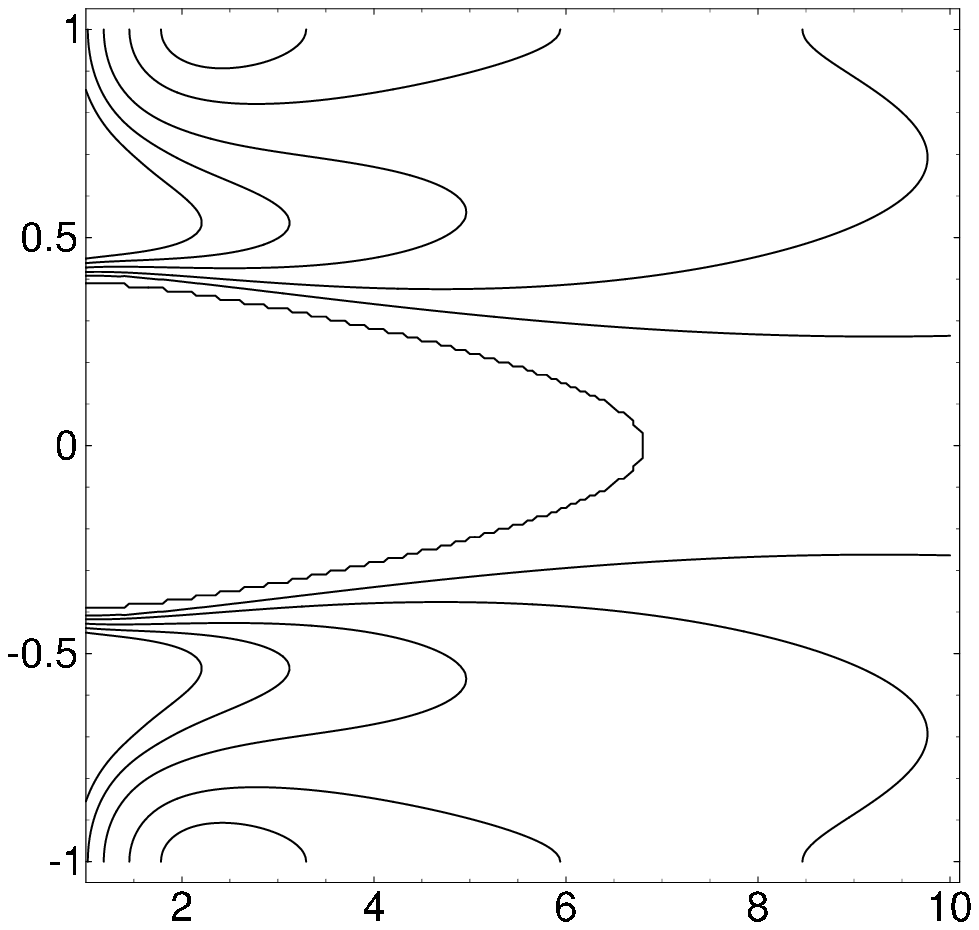}}
	\put(3.5,7.4){\fbox{$\sigma$ in fb}}
	\put(6.3,-0.3){$\tan\beta$}
	\put(0,7.4){ $\varphi_{\mu}~[\pi]$}
	\put(1.1,2.0){\footnotesize 20}
	\put(1.75,2.05){\footnotesize 15}
	\put(3.0,2.0){\footnotesize 10}
	\put(5.6,2.1){\footnotesize 5}
	\put(6.3,3.1){\footnotesize 2}
	\put(1.8,0.8){\footnotesize 2}
	\put(3.1,0.9){\footnotesize 5}
	\put(4.8,3.7){\footnotesize 0}
	\put(1.1,5.3){\footnotesize 20}
	\put(1.7,5.3){\footnotesize 15}
	\put(3.0,5.35){\footnotesize 10}
	\put(5.6,5.2){\footnotesize 5}
	\put(6.3,4.2){\footnotesize 2}
	\put(1.8,6.5){\footnotesize 2}
	\put(3.1,6.4){\footnotesize 5}
		\put(2.0,3.7){\begin{picture}(1,1)(0,0)
			\CArc(0,0)(7,0,380)
			\Text(0,0)[c]{{\footnotesize A}}
	\end{picture}}
\put(0.5,-.3){Fig.~\ref{sneut:plot_2}a}
	\put(8,0){\includegraphics{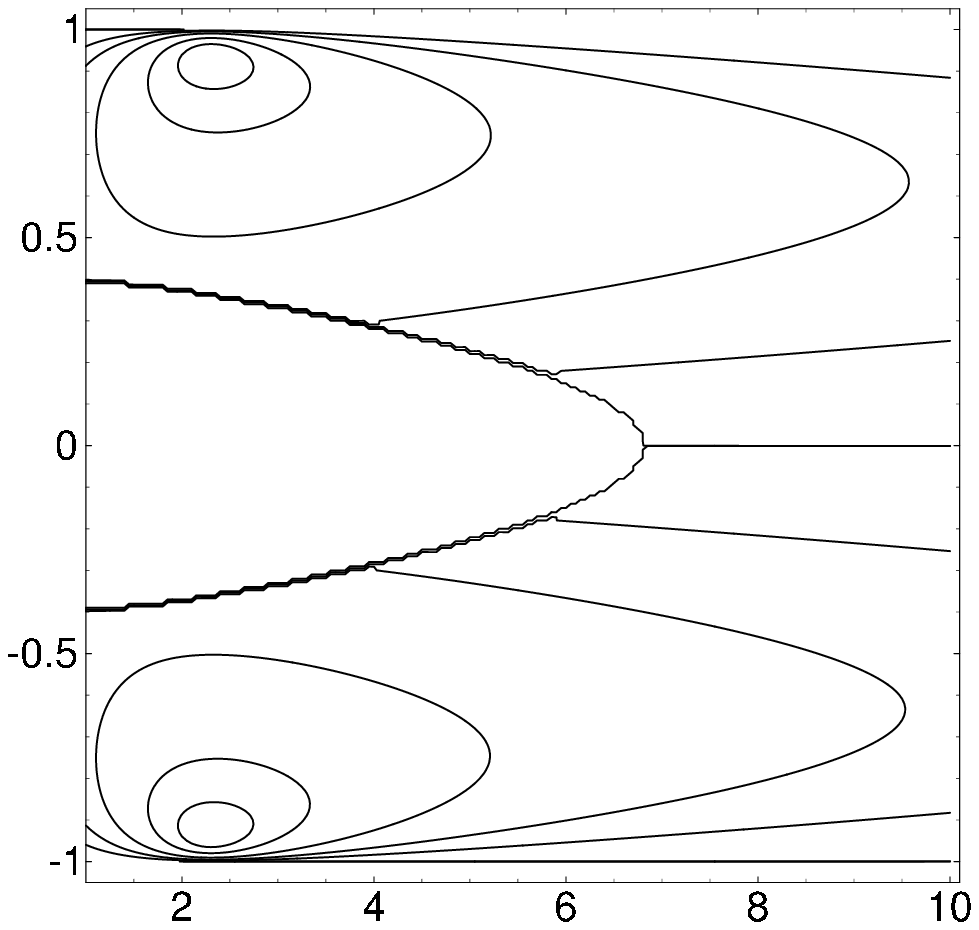}}
	\put(11.,7.4){\fbox{${\mathcal A}_{\ell}$ in \% }}
	\put(14.3,-.3){$\tan\beta$}
	\put(8,7.4){ $\varphi_{\mu}~[\pi]$}
	\put(9.55,1.){\scriptsize 30}
	\put(10.3,1.4){\scriptsize 20}
	\put(11.4,1.9){\footnotesize 10}
	\put(14.2,1.8){\footnotesize 5}
	\put(14.3,2.75){\footnotesize 2.5}
	\put(14.3,1.2){\footnotesize 2.5}
	\put(14.4,3.85){\footnotesize 0}
	\put(14.65,0.85){\footnotesize 0}
	\put(9.5,6.4){\scriptsize -30}
	\put(10.3,6.0){\scriptsize -20}
	\put(11.5,5.6){\footnotesize -10}
	\put(14.2,5.55){\footnotesize -5}
	\put(14.2,4.6){\footnotesize -2.5}
	\put(14.3,6.1){\footnotesize -2.5}
	\put(10.,3.7){\begin{picture}(1,1)(0,0)
			\CArc(0,0)(7,0,380)
			\Text(0,0)[c]{{\footnotesize A}}
	\end{picture}}
	\put(8.5,-.3){Fig.~\ref{sneut:plot_2}b}
\end{picture}
\vspace*{.5cm}
\caption{
	Contour lines of 
	$\sigma=\sigma_P(e^+e^-\to\tilde\chi^+_1\tilde\chi^-_2) 
	\times {\rm BR}( \tilde\chi^+_1 \to \ell^+\tilde\nu_{\ell})$, 
	summed over $\ell = e,\mu$, (\ref{sneut:plot_2}a), 
	and the asymmetry ${\mathcal A}_{\ell}$ for 
	$\ell =e$ or $\mu$ (\ref{sneut:plot_2}b),
	in the $\tan\beta$--$\varphi_{\mu}$ plane for 
	%	$\varphi_{M_1}=0.7\pi $, $\varphi_{\mu}=0$, 
	$M_2=200$~GeV, $|\mu|=400$~GeV, $m_{\tilde\nu_{\ell}}=185$~GeV,
	$\sqrt{s}=800$ GeV and $(P_{e^-},P_{e^+})=(-0.8,0.6)$.
	The area A  is kinematically forbidden by
%	$m_{\tilde\chi^+_1}+m_{\tilde\chi^-_1}>\sqrt{s}$
	$m_{\tilde\nu_{\ell}}+m_{\chi^0_1}> m_{\chi^+_1}$.
	\label{sneut:plot_2}}
\end{figure}
For $M_2=200$ GeV, we show the 
$\tan\beta $--$\varphi_{\mu} $ dependence of 
$\sigma$ and ${\mathcal A}_{\ell}$ in Figs.~\ref{sneut:plot_2}a,b.
%for $\varphi_{M_1}=0.7\pi$ and $\varphi_{\mu}=0$.
%taking  $|\mu|=400$ GeV and $m_0=80$ GeV.
%The production cross section
%$\sigma(e^+e^-\to\tilde\chi^+_1\tilde\chi^-_2)$
%increases with increasing $m_0$ and decreasing $\tan\beta $.
%The branching ratio ${\rm BR}( \tilde\chi^+_1 \to \tilde\chi^0_1W^+)<1$ 
%for $m_0\lsim200$~GeV, as the channels of $\tilde\chi^+_1$
%into sleptons and/or sneutrinos open.
%
%The $\tan\beta $--$\varphi_{\mu}$ dependence of ${\mathcal A}_{\ell}$
%is shown in Fig.~\ref{sneut:plot_2}b.
The asymmetry can reach values up to 30\% and  shows a
strong  $\tan\beta$ dependence and decreases with increasing $\tan\beta$. 
The feasibility of measuring the asymmetry depends also on the cross 
section $\sigma=\sigma_P(e^+e^-\to\tilde\chi^+_1\tilde\chi^-_2)\times
 {\rm BR}(\tilde\chi^+_1 \to \ell^+\tilde\nu_{\ell})$,
Fig.~\ref{sneut:plot_2}a, which attains values up to $20$~fb.

\begin{figure}[h]
\setlength{\unitlength}{1cm}
\begin{picture}(10,8)(0,0)
   \put(0,0){\includegraphics{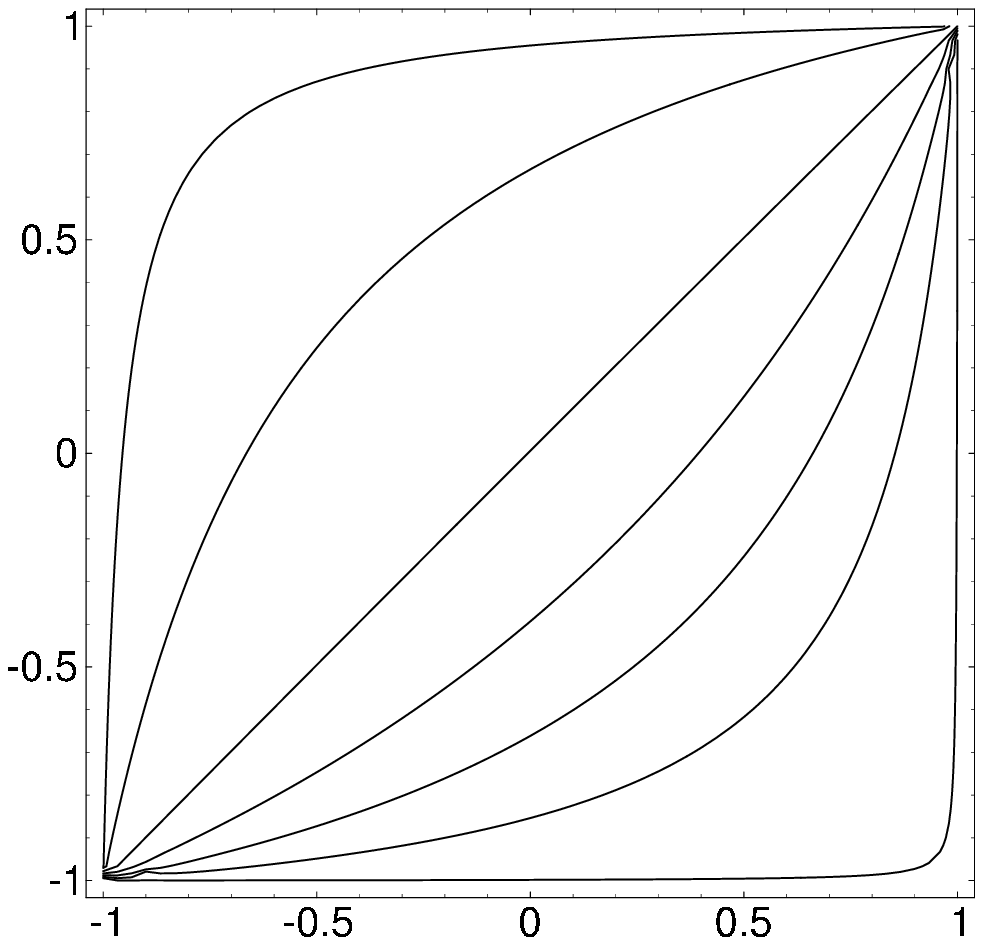}}
	\put(3.5,7.4){\fbox{${\mathcal A}_{\ell}$ in \% }}
	\put(6.5,-0.3){$ P_{e^-}$}
	\put(0.5,7.4){$ P_{e^+} $ }
	\put(1.3,6.2){\footnotesize -7.3}
	\put(2.6,5.1){\footnotesize -7}
	\put(3.6,3.8){\footnotesize -6}
	\put(4.3,3.2){\footnotesize -5}
	\put(4.85,2.7){\footnotesize -4}
	\put(5.45,2.2){\footnotesize -3}
	\put(6.5,1.0){\footnotesize -2}
%	\put(.,.){\footnotesize }
\put(0.5,-.3){Fig.~\ref{sneut:plot_3}a}
	\put(8,0){\includegraphics{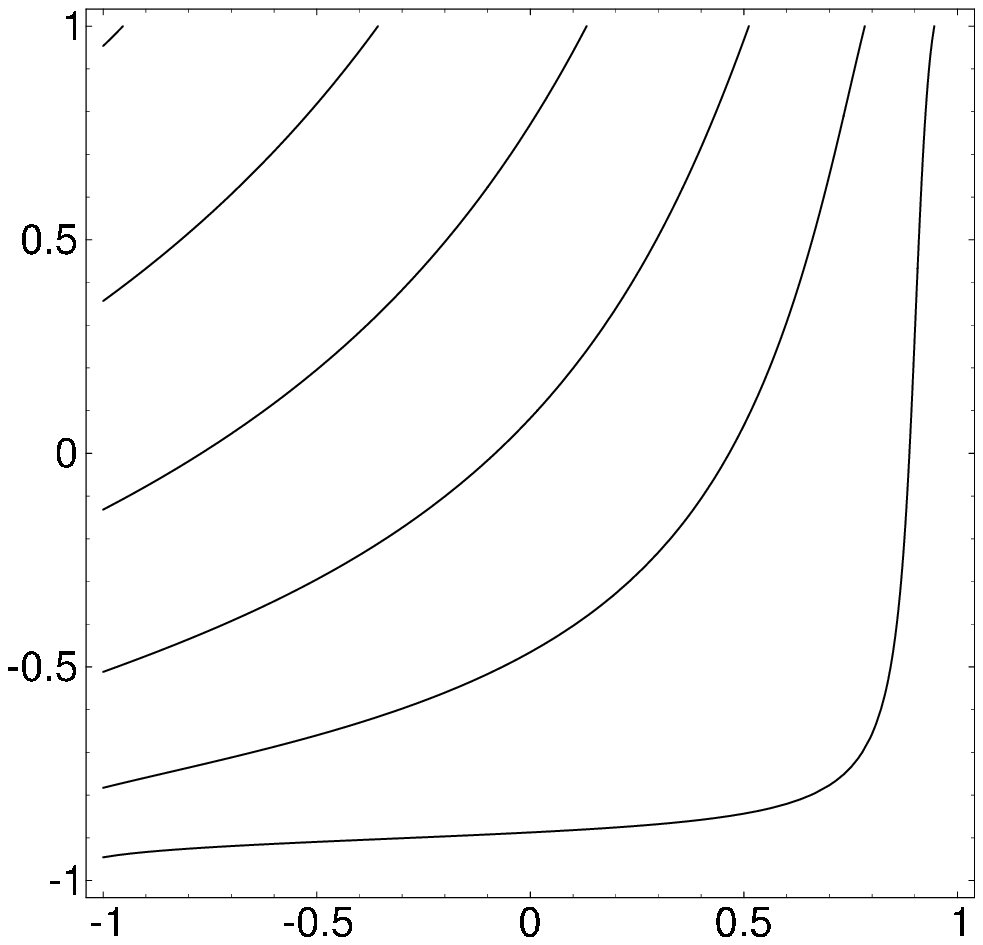}}
	\put(10.3,7.4){\fbox{$S_{\ell} =|{\mathcal A}_{\ell}| 
			\sqrt{2{\mathcal L}\cdot\sigma}$  }}
	\put(14.5,-.3){$ P_{e^-}$ }
	\put(8.5,7.4){$  P_{e^+}$ }
	\put(9.,6.5){\footnotesize 6}
	\put(9.7,5.7){\footnotesize 5}
	\put(10.7,4.8){\footnotesize 4}
	\put(11.7,4.0){\footnotesize 3}
	\put(12.7,3.1){\footnotesize 2}
	\put(14.,1.6){\footnotesize 1}
\put(8.5,-.3){Fig.~\ref{sneut:plot_3}b}
\end{picture}
\vspace*{.5cm}
\caption{
	Contour lines of  
	the asymmetry ${\mathcal A}_{\ell}$ for $\ell =e$ or $\mu$
	(\ref{sneut:plot_3}a),
	and the significance $S_{\ell}$ (\ref{sneut:plot_3}b),
	for $e^+e^-\to\tilde\chi^+_1\tilde\chi^-_2;~ 
	\tilde\chi^+_1 \to \ell^+\tilde\nu_{\ell}$
	in the $ P_{e^-}$--$P_{e^+}$ plane for $\varphi_{\mu}=0.9\pi$,
	$|\mu|=400$~GeV, $M_2=200$~GeV,
	$\tan \beta=5$, $m_{\tilde\nu_{\ell}}=185$~GeV, 
	$\sqrt{s}=800$ GeV and ${\mathcal L}=500~{\rm fb}^{-1}$.
	\label{sneut:plot_3}}
\end{figure}

For the phase $\varphi_{\mu}=0.9\pi$ and $\tan \beta=5$,
we study the beam polarization dependence of ${\mathcal A}_{\ell}$,
which can be strong as shown in Fig.~\ref{sneut:plot_3}a. 
An electron beam polarization $P_{e^-}>0$ and a positron
beam polarization $P_{e^+}<0$ enhance the channels with $\tilde\nu_{e}$
exchange in the chargino production process. 
For e.g. $(P_{e^-},P_{e^+})=(-0.8,0.6)$
the asymmetry can attain $-7\%$, Fig.~\ref{sneut:plot_3}a,
with  $\sigma_P(e^+e^-\to\tilde\chi^+_1\tilde\chi^-_2)\approx10$~fb
and ${\rm BR}(\tilde\chi^+_1 \to \ell^+\tilde\nu_{\ell})\approx50\%$,
summed over $\ell = e,\mu$.
The cross section $\sigma=\sigma_P(e^+e^-\to\tilde\chi^+_1\tilde\chi^-_2)\times
 {\rm BR}(\tilde\chi^+_1 \to \ell^+\tilde\nu_{\ell})$
 ranges between $2.3$~fb for $(P_{e^-},P_{e^+})=(0,0)$ and
$6.8$~fb for $(P_{e^-},P_{e^+})=(-1,1)$.
The statistical significance of ${\mathcal A}_{\ell}$,  given by
$S_{\ell} =|{\mathcal A}_{\ell}| \sqrt{2{\mathcal L}\cdot\sigma}$,
is shown in  Fig.~\ref{sneut:plot_3}b for ${\mathcal L}=500~{\rm fb}^{-1}$.
We have $S_{\ell}\approx 5$ for $(P_{e^-},P_{e^+})=(-0.8,0.6)$,
and thus ${\mathcal A}_{\ell}$ could be accessible at a linear
collider, even for $\varphi_{\mu}=0.9\pi$, by using polarized beams.

%&&&&&&&&&&&&&&&&&&&&&&&&&&&&&&&&&&&&&&&&&&&&&&&&&&&&&&&&&&&&&&
%                    P L O T  3 
%&&&&&&&&&&&&&&&&&&&&&&&&&&&&&&&&&&&&&&&&&&&&&&&&&&&&&&&&&&&&&&6
%
\subsection{Summary of Section 
	\ref{CP violation in chargino production and decay into the sneutrino}
	\label{sneut:Summary}}

We have studied CP violation in chargino production with longitudinally 
polarized beams, $e^+e^- \to\tilde\chi^+_i  \tilde\chi^-_j$,
and subsequent two-body decay  of one  chargino
into the sneutrino $\tilde\chi^+_i \to \ell^+\tilde\nu_{\ell}$.
We have defined the T-odd asymmetries 
$ {\mathcal A}_{\ell}^{\rm T}$ of the triple product 
$({\bf p}_{e^-} \times {\bf p}_{\tilde\chi^+_i}) \cdot {\bf p}_{\ell}$.
The CP-odd asymmetries 
${\mathcal A}_{\ell} = \frac{1}{2}({\mathcal A}_{\ell}^{\rm T}-
	 \bar{\mathcal A}_{\ell}^{\rm T})$,
where $\bar{\mathcal A}_{\ell}^{\rm T}$ denote the CP conjugated
of ${\mathcal A}_{\ell}^{\rm T}$, are sensitive to the phase  
$\varphi_{\mu}$ of the Higgsino mass parameter $\mu$. 
At tree level, the asymmetries have 
large CP sensitive contributions from spin-correlation effects in the 
production of an unequal pair of charginos.
In a numerical discussion for 
$e^+e^- \to\tilde\chi^+_1  \tilde\chi^-_2$ production, we have found
that  ${\mathcal A}_{\ell}$ for $\ell =e$ or $\mu$  can attain values 
up to 30\%. By analyzing the statistical errors, we have shown that,
even for e.g. $\varphi_{\mu}\approx0.9 \pi$, the 
asymmetries could be accessible in future  $e^+e^-$ collider 
experiments in the 800 GeV range with high luminosity and 
longitudinally polarized beams.

		\section{CP violation in chargino production and decay into a $W$
	      boson
			\label{CP observables in chargino production and decay into the W
	      boson}}

%We study CP sensitive observables in chargino production 
%in electron-positron collisions
%%$e^+e^- \to\tilde\chi^+_i \tilde\chi^-_j$
%with subsequent two-body decay of one chargino
%into a $W$ boson.
%%$\tilde\chi^{+}_i \to \chi^0_n W^{+}$ and of the $W$ boson
%%$ W^{+} \to \bar\ell \nu_{\ell} (c\bar s)$.
%We identify the CP odd  elements of the $W$ boson
%density matrix and  propose CP sensitive 
%triple-product asymmetries of the chargino decay products. 
%We calculate the density-matrix elements, the CP asymmetries 
%and the cross sections 
%in the Minimal Supersymmetric Standard Model with complex parameters 
%$\mu$ and $M_1$ for  an $e^+e^-$ linear collider with 
%$\sqrt{s}=800$ GeV and longitudinally polarized beams. 
%The asymmetries can reach  $7\%$ 
%and we discuss the feasibility of measuring these asymmetries.

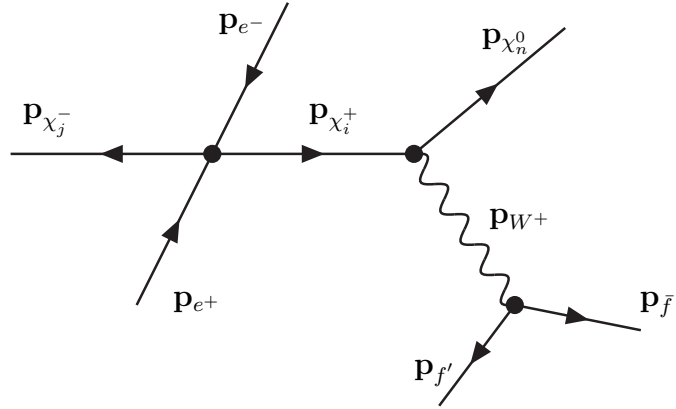
\begin{figure}[h]
\begin{picture}(5,6.)(-2,.5)
	\put(1,4.8){${\bf p}_{\chi^-_j}$}
	\put(3.6,6){${\bf p}_{e^-}$}
	\put(3.,2.3){${\bf p}_{e^+}$}
	\put(4.8,4.8){${\bf p}_{\chi^+_i}$}
	\put(7.1,5.8){${\bf p}_{\chi^0_n}$}
	\put(7.2,3.4){${\bf p}_{W^+}$}
	\put(9.2,2.3){${\bf p}_{\bar f}$}
	\put(6.2,1.4){${\bf p}_{f^{'}}$}
\end{picture}
\scalebox{1.9}{
\begin{picture}(0,0)(1.3,-0.25)
\ArrowLine(40,50)(0,50)
\Vertex(40,50){1.8}
\ArrowLine(55,80)(40,50)
\ArrowLine(25,20)(40,50)
\ArrowLine(40,50)(80,50)
\ArrowLine(80,50)(110,75)
%\ArrowLine(80,50)(100,20)
\Photon(80,50)(100,20){2}{5}
\Vertex(80,50){1.8}
\ArrowLine(100,20)(125,15)
\ArrowLine(100,20)(85,0)
\Vertex(100,20){1.8}
\end{picture}}
\caption{\label{shematic picture of W}
          Schematic picture of the chargino production
          and decay process.}
\end{figure}

%\subsection{Introduction
%	\label{W:Introduction}}

CP violation in the chargino sector can be studied also 
by the two-body decay of the chargino into a $W$ boson.
In contrast to the decay into a sneutrino, the spin 
correlations of chargino and $W$ lead to CP observables,
which are also sensitive to the phase of $M_1$.
Since these observables have in addition contributions 
from the decay, they do not necessarily vanish for
$e^+e^-\to\tilde\chi^{\pm}_1\tilde\chi^{\mp}_1$
production. In this mode CP violation could be established,
even if the production of the heavier pair 
$\tilde\chi^{\pm}_1\tilde\chi^{\mp}_2$
is not yet accessible.

We study chargino production
\begin{eqnarray} \label{W:productionA}
	e^++e^-&\to&\tilde\chi^+_i+\tilde\chi^-_j; 
	\quad i,j =1,2,  
\end{eqnarray}
with longitudinally polarized beams and the subsequent two-body decay
\begin{eqnarray} \label{W:decay_1A}
     \tilde\chi^+_i \to W^+ +\tilde\chi^0_n. 
 \end{eqnarray}
We define the triple product
 \begin{eqnarray}\label{W:tripleproduct1}
	 {\mathcal T}_{I} &=& 
	 {\mathbf p}_{e^-}\cdot({\mathbf p}_{\chi^+_i} \times {\mathbf p}_W)
 \end{eqnarray}
and the T-odd asymmetry
\begin{eqnarray}\label{W:AT1}
	 {\mathcal A}_{I}^{\rm T} &=& 
	 \frac{\sigma({\mathcal T}_{I}>0)-\sigma({\mathcal T}_{I}<0)}
	{\sigma({\mathcal T}_{I}>0)+\sigma({\mathcal T}_{I}<0)},
\end{eqnarray}
with $\sigma$ the cross section of chargino production~(\ref{W:productionA}) 
and decay~(\ref{W:decay_1A}).
The asymmetry ${\mathcal A}_{I}^{\rm T}$ is sensitive
to the CP violating phase $\varphi_{\mu}$. 
%In this context it is interesting to note that 
%asymmetries vanish if they correspond to
%a triple product which contains a  transverse
%polarization vector of the $e^+$ and $e^-$ beams
%\cite{choichargino,holger}.

In order to probe also the phase $\varphi_{M_1}$, which enters
in the chargino decay process~(\ref{W:decay_1A}), 
we consider the subsequent hadronic decay of the $W$ boson
\begin{eqnarray} \label{W:decay_2B}
%		W^+ \to \bar\ell + \nu_{\ell}, \quad\ell=e,\mu,\tau, \\
	W^+ \to  c + \bar s.
%q_u + \bar q_d, \quad q_u=u,c,\quad  q_d=d,s.
\end{eqnarray}
The correlations between the $\tilde\chi^+_i$ 
polarization and the $W$ boson polarization lead 
to CP sensitive elements of the $W$ boson density matrix.
%which we will identify and discuss in detail.
The triple product
 \begin{eqnarray}\label{W:tripleproduct2}
	 {\mathcal T}_{II} &=& 
	 {\mathbf p}_{e^-}\cdot({\mathbf p}_{c} \times {\mathbf p}_{ \bar s}),
\end{eqnarray}
which includes the momenta of the $W$ decay products, and thus
probes the $W$ polarization, defines a second T-odd asymmetry
\begin{eqnarray}\label{W:AT2}
	 {\mathcal A}_{II}^{\rm T} &=& 
	 \frac{\sigma({\mathcal T}_{II}>0)-\sigma({\mathcal T}_{II}<0)}
	{\sigma({\mathcal T}_{II}>0)+\sigma({\mathcal T}_{II}<0)}.
\end{eqnarray}
Here, $\sigma$ is  the cross section of production~(\ref{W:productionA}) 
and decay of the chargino~(\ref{W:decay_1A}) followed by that of the 
$W$ boson~(\ref{W:decay_2B}). Owing to the spin correlations, 
${\mathcal A}_{II}^{\rm T}$ has CP sensitive
contributions from $\varphi_{\mu}$ due to the chargino production 
process~(\ref{W:productionA}) and contributions due to $\varphi_{\mu}$ 
and $\varphi_{M_1}$ from the chargino decay process~(\ref{W:decay_1A}). 

The T-odd asymmetries ${\mathcal A}_{I}^{\rm T}$ and 
${\mathcal A}_{II}^{\rm T}$ have also absorptive contributions
from s-channel resonances or final-state interactions,
%which do not signal CP violation. In order to eliminate these 
%contributions, we study the two CP-odd asymmetries 
which are eliminated in the CP-odd asymmetries
\begin{equation}
{\mathcal A}_{I} = 
\frac{1}{2}({\mathcal A}_{I}^{\rm T}-\bar{\mathcal A}_{I}^{\rm T}),
\quad 
{\mathcal A}_{II} = 
\frac{1}{2}({\mathcal A}_{II}^{\rm T}-\bar{\mathcal A}_{II}^{\rm T}),
\label{W:ACP}
\end{equation}
where $\bar{\mathcal A}_{I,II}^{\rm T}$ are the CP conjugated asymmetries 
for the processes 
$e^+e^-\to\tilde\chi^-_i\tilde\chi^+_j;
\tilde\chi^-_i \to W^- \tilde\chi^0_n$
and
$e^+e^-\to\tilde\chi^-_i\tilde\chi^+_j;
\tilde\chi^-_i \to W^- \tilde\chi^0_n;W^- \to\bar c s$,
respectively.

%\subsection{Cross section
%     \label{W:Cross section}}

\subsection{Spin density matrix of the $W$ boson
     \label{Density matrix}}

For the calculation of the amplitudes squared for the
combined process of chargino production~(\ref{W:productionA})
and the subsequent two-body decays~(\ref{W:decay_1A})
and~(\ref{W:decay_2B}) of $\tilde\chi^+_i$
we use the same spin-density matrix formalism as in 
\cite{gudichargino,spinhaber}. 
The (unnormalized)  spin-density matrix of the $W$ boson 
\begin{eqnarray}       \label{Wdensitymatrix}
\rho_{P}(W^+)^{\lambda_k\lambda'_k}&=&
|\Delta(\tilde\chi^+_i)|^2~
\sum_{\lambda_i,\lambda'_i}~
\rho_P   (\tilde\chi^+_i)^{\lambda_i \lambda_i'}\;
\rho_{D_1}(\tilde\chi^+_i)_{\lambda_i'\lambda_i}^{\lambda_k\lambda'_k},
\end{eqnarray}
is composed of the chargino propagator $\Delta(\tilde\chi^+_i)$,
the spin-density production matrix
$\rho_P(\tilde\chi^+_i)$, defined in~(\ref{char:rhoP}), 
and the decay matrix $\rho_{D_1}(\tilde\chi^+_i)$,
defined in~(\ref{W:rhoD1}).
The amplitude squared for the complete process
$ e^+e^-\to\tilde\chi^+_i\tilde\chi^-_j$;
$\tilde\chi^+_i\to W^+\tilde\chi^0_n $;
$W^+ \to f^{'} \bar f$ 
can now be written
\begin{eqnarray}       \label{W:amplitude}
|T|^2&=&|\Delta(W^+)|^2
	\sum_{\lambda_k,\lambda'_k}~
	\rho_{P}(W^+)^{\lambda_k\lambda'_k}\;
	\rho_{D_2}(W^+)_{\lambda'_k\lambda_k},
\end{eqnarray}
with the decay matrix for $W$ decay $\rho_{D_2}(W^+)$,
defined in~(\ref{W:rhoD2}).
Inserting the density matrices $\rho_P(\tilde\chi^+_i)$~(\ref{char:rhoP})
and $\rho_{D_1}(\tilde\chi^+_i)$~(\ref{W:rhoD1expanded})
into~(\ref{Wdensitymatrix}) gives
\begin{eqnarray}\label{Wdensitymatrixunnorm}
\rho_{P}(W^+)^{\lambda_k\lambda'_k}&=&
4~|\Delta(\tilde\chi^+_i)|^2~[
	(P  D_1 + \Sigma^{a}_{P}\Sigma^{a}_{D_1}) ~\delta^{\lambda_k\lambda_k'}
	+(P \;^{c}D_1+\Sigma_P^a \,^{c}\Sigma^{a}_{D_1})
	~(J^{c})^{\lambda_k\lambda_k'}\nonumber\\
&&+(P  \;^{cd}D_1 +\Sigma^{a}_{P} \;^{cd}\Sigma^{a}_{D_1})
	~(J^{cd})^{\lambda_k\lambda_k'}],
\end{eqnarray}
summed over $a,c,d$. We have thus decomposed the $W$ production
matrix $\rho_{P}(W^+)$ into contributions of scalar (first term),
vector (second term), and tensor parts (third term).
Inserting then $\rho_{P}(W^+)$~(\ref{Wdensitymatrixunnorm}) 
and $\rho_{D_2}(W^+)$~(\ref{W:rhoD2expanded}) 
into~(\ref{W:amplitude}) gives
\begin{eqnarray} \label{W:amplitude2}
|T|^2 &=& 4~|\Delta(\tilde\chi^+_i)|^2~ |\Delta(W^+)|^2
	\{3( P  D_1 + \Sigma^{a}_{P}\Sigma^{a}_{D_1} )D_2 +  
		2(P \;^{c}D_1+\Sigma_P^a \,^c\Sigma_{D_1}^a) \,^cD_2 \nonumber\\
&& \quad \quad  
		+4[(P \;^{cd}D_1  +\Sigma^{a}_{P} \;^{cd}\Sigma^{a}_{D_1} )\,^{cd} D_2-
		{\textstyle \frac{1}{3}}
		(P\;^{cc}D_1 +\Sigma^{a}_{P}\;^{cc}\Sigma^{a}_{D_1})\,^{dd}
		D_2]\}.
\end{eqnarray}
%summed over $a,c,d$, 
%which is the decomposition of the amplitude 
%squared in its scalar (first term), vector (second term)
%and tensor part (third term).

\subsection{$W$ boson polarization
     \label{W boson polarization}}

The mean polarization of the $W$ bosons in the laboratory system 
is given by the $3\times3$ density matrix $<\rho(W^+)>$,
%with ${\rm Tr}\{<\rho(W^+)>\}=1$
obtained by integrating~(\ref{Wdensitymatrixunnorm}) 
over the Lorentz invariant phase-space element 
$d{\rm Lips}(s;p_{\chi^-_j },p_{\chi^0_n},p_{W})$,
%$=
%\frac{1}{(2\pi)^2}~d{\rm Lips}(s,p_{\chi^+_i},p_{\chi^-_j} )
%~d s_{\chi^+_i} \,\sum_{\pm}
%d{\rm Lips}(s_{\chi^+_i},p_{\chi^0_n},p_{W}^{\pm})$,
see~(\ref{Lipsbosonic1}), and normalizing by the trace
\begin{equation}\label{Wdensitymatrixnorm}
<\rho(W^+)^{\lambda_k\lambda'_k}>=
\frac{\int \rho_{P}(W^+)^{\lambda_k\lambda'_k}~d{\rm Lips}}
		{\int {\rm Tr} \{\rho_{P}(W^+)^{\lambda_k\lambda'_k}\}~d{\rm Lips}}
={\textstyle \frac{1}{3}}\delta^{\lambda_k\lambda_k'}
+V_c ~(J^{c})^{\lambda_k\lambda_k'}
+T_{cd}  ~(J^{cd})^{\lambda_k\lambda_k'}.
%\label{defcoef}
\end{equation}
%summed over $c$, $d$. 
The components $V_c$ of the vector polarization and $T_{cd}$ 
of the tensor polarization are 
%which is the decomposition of the normalized $W^+$ spin-density 
%matrix in its scalar (first term), vector (second term)
%and tensor part (third term). In Eq.~(\ref{defcoef}) we have 
%defined the vector and tensor coefficients as:
%\begin{eqnarray}
%	V_c&=&\frac{\int |\Delta(\tilde\chi^+_i)|^2
%		~(P \, ^cD_1 + \Sigma_P^a \,^{c}\Sigma^{a}_{D_1}) 
%		~d{\rm Lips}}
%{3 \int |\Delta(\tilde\chi^+_i)|^2 ~P  D_1~d{\rm Lips}},\\
%T_{cd}&=&T_{dc}=
%\frac{\int |\Delta(\tilde\chi^+_i)|^2  
%	~(P  \;^{cd}D_1+\Sigma^{a}_{P} \;^{cd}\Sigma^{a}_{D_1} )
%		~d{\rm Lips}}
%{3 \int |\Delta(\tilde\chi^+_i)|^2 ~P  D_1~d{\rm Lips}},
%\end{eqnarray}
\begin{eqnarray}
	V_c&=&\frac{\int (P \, ^cD_1 + \Sigma_P^a \,^{c}\Sigma^{a}_{D_1}) 
		~d{\rm Lips}}
{3 \int ~P  D_1~d{\rm Lips}},\\
T_{cd}&=&T_{dc}=
\frac{\int (P  \;^{cd}D_1+\Sigma^{a}_{P} \;^{cd}\Sigma^{a}_{D_1} )
		~d{\rm Lips}}
{3 \int ~P  D_1~d{\rm Lips}},
\end{eqnarray}
%with sum over a. 
where we have already used the narrow width approximation
for the chargino propagator.
%The tensor coefficients $T_{12}$ and $T_{23}$ vanish due to
%phase-space integration.
The density matrix elements in the helicity basis 
(\ref{circularbasis}) are given  by 
\begin{eqnarray} \label{W:density1}
	<\rho(W^+)^{--}> &= &
	{\textstyle \frac{1}{2}}-V_3+T_{33}, \\
	<\rho(W^+)^{00}> &=&-2~T_{33},\\
	<\rho(W^+)^{-0}> &= &
	{\textstyle \frac{1}{\sqrt{2}}}(V_1+iV_2)-\sqrt{2}\,(T_{13}+iT_{23}),\\
		<\rho(W^+)^{-+}> &= & T_{11}-T_{23}+2iT_{12},\\
	<\rho(W^+)^{0+}> &= & {\textstyle \frac{1}{\sqrt{2}}}(V_1+iV_2)
	+\sqrt{2}\,(T_{13}+iT_{23}),\label{W:density5}
\end{eqnarray}
where we have used $T_{11}+T_{22}+T_{33}=-\frac{1}{2}$.
%and $T_{12}=T_{23}=0$.

\subsection{T-odd asymmetries
	\label{W:T-odd asymmetries}}

From~(\ref{Wdensitymatrixunnorm}) we obtain for asymmetry  
${\mathcal A}_{I}^{\rm T}$~(\ref{W:AT1}):
%%\begin{eqnarray} 
%%	 {\mathcal A}_{I}^{\rm T} 
%%	 &=& \frac{\int {\rm Sign}[{\mathcal T}_{I}]
%%		 {\rm Tr} \{\rho_{P}(W^+)^{\lambda_k\lambda'_k}\} d{\rm Lips}}
%%	 {\int {\rm Tr} \{\rho_{P}(W^+)^{\lambda_k\lambda'_k}\} d{\rm Lips}}
%%	 \nonumber \\
%%&=& \frac{\int |\Delta(\tilde\chi^+_i)|^2 ~
%%	{\rm Sign}[{\mathbf p}_{e^-}\cdot({\mathbf p}_{\chi^+_i} 
%%			\times {\mathbf p}_{W})]
%%		\Sigma_P^2 \, \Sigma_{D_1}^2  d{\rm Lips}}
%%	{\int  |\Delta(\tilde\chi^+_i)|^2~ P D_1 d{\rm Lips}},
%% \label{asymI}
%%\end{eqnarray}
\begin{equation} 
	 {\mathcal A}_{I}^{\rm T} 
	 = \frac{\int {\rm Sign}[{\mathcal T}_{I}]
		 {\rm Tr} \{\rho_{P}(W^+)^{\lambda_k\lambda'_k}\}~ d{\rm Lips}}
	 {\int {\rm Tr} \{\rho_{P}(W^+)^{\lambda_k\lambda'_k}\}~ d{\rm Lips}}
	 =
% \frac{\int |\Delta(\tilde\chi^+_i)|^2 ~
%	{\rm Sign}[{\mathcal T}_{I}]
%		\Sigma_P^2 \, \Sigma_{D_1}^2  d{\rm Lips}}
%	{\int  |\Delta(\tilde\chi^+_i)|^2~ P D_1 ~d{\rm Lips}},
	\frac{\int {\rm Sign}[{\mathcal T}_{I}]
		\Sigma_P^2 \, \Sigma_{D_1}^2  d{\rm Lips}}
	{\int  P D_1 ~d{\rm Lips}},
		\label{W:asymI}
\end{equation}
with
$d{\rm Lips}(s;p_{\chi^-_j },p_{\chi^0_n},p_{W})$
given in~(\ref{Lipsbosonic1}),
where we have used the narrow width approximation.
%%$=\frac{1}{(2\pi)^2}~d{\rm Lips}(s,p_{\chi^+_i},p_{\chi^-_j} )
%$$~d s_{\chi^+_i} \,\sum_{\pm}
%$$d{\rm Lips}(s_{\chi^+_i},p_{\chi^0_n},p_{W}^{\pm})$,
In the numerator of~(\ref{W:asymI}), only the spin correlations
$\Sigma_P^2 \, \Sigma_{D_1}^2$ perpendicular to the production plane
remain, since only this term contains the  triple product
${\mathcal T}_{I} ={\mathbf p}_{e^-}\cdot({\mathbf p}_{\chi^+_i} 
\times {\mathbf p}_{W})$. In the denominator only the term $P D_1$ remains,
and all spin correlations vanish due to the integration over the
complete phase space. Note that 
${\mathcal A}_{I}^{\rm T}\propto
\Sigma_{D_1}^2\propto(|O^{R}_{ni}|^2-|O^{L}_{ni}|^2)$,
see~(\ref{W:SigmaaD1}), and thus ${\mathcal A}_{I}^{\rm T}$ may be reduced
for $|O^{R}_{ni}| \approx |O^{L}_{ni}|$.
Moreover, ${\mathcal A}_{I}^{\rm T}$ will be small for
$m_{\chi^+_i}^2-m_{\chi^0_n}^2 \approx 2 m_W^2$,
see~(\ref{W:SigmaaD1}).

For the asymmetry  ${\mathcal A}_{II}^{\rm T}$~(\ref{W:AT2}),
we obtain from~(\ref{W:amplitude2}): 
%%\begin{eqnarray} 
%%	 {\mathcal A}_{II}^{\rm T} 
%%	 &=& \frac{\int {\rm Sign}[{\mathcal T}_{II}]
%%		 |T|^2 d{\rm Lips}}
%%           {\int |T|^2 d{\rm Lips}} = \nonumber \\ 
%% &= &\frac{\int |\Delta(\tilde\chi^+_i)|^2 |\Delta(W^+)|^2~
%%	 {\rm Sign}[{\mathbf p}_{e^-}\cdot({\mathbf p}_{c} 
%%			 \times {\mathbf p}_{\bar s})]
%%%	2(P \, ^cD_1 +\Sigma_P^a \, ^c\Sigma_{D_1}^a )\,^cD_2 d{\rm Lips}}
%%	2\Sigma_P^a \, ^c\Sigma_{D_1}^a \,^cD_2 d{\rm Lips}}
%%	{\int |\Delta(\tilde\chi^+_i)|^2 |\Delta(W^+)|^2~
%%		3 P D_1D_2 d{\rm Lips}}, \label{asymII}
%%\end{eqnarray}
\begin{equation} 
	 {\mathcal A}_{II}^{\rm T} 
	 = \frac{\int {\rm Sign}[{\mathcal T}_{II}]
		 |T|^2 d{\rm Lips}}
           {\int |T|^2 d{\rm Lips}}  
% = \frac{\int |\Delta(\tilde\chi^+_i)|^2 |\Delta(W^+)|^2~
%	 {\rm Sign}[{\mathcal T}_{II}]
%%	2(P \, ^cD_1 +\Sigma_P^a \, ^c\Sigma_{D_1}^a )\,^cD_2 d{\rm Lips}}
%	2\Sigma_P^a \, ^c\Sigma_{D_1}^a \,^cD_2 d{\rm Lips}}
%	{\int |\Delta(\tilde\chi^+_i)|^2 |\Delta(W^+)|^2~
%		3 P D_1D_2 d{\rm Lips}}, 
	 = \frac{\int  {\rm Sign}[{\mathcal T}_{II}]
%	2(P \, ^cD_1 +\Sigma_P^a \, ^c\Sigma_{D_1}^a )\,^cD_2 d{\rm Lips}}
	2\Sigma_P^a \, ^c\Sigma_{D_1}^a \,^cD_2 ~d{\rm Lips}}
	{\int 3 P D_1D_2 ~d{\rm Lips}}, 
		\label{W:asymII}
\end{equation}
%summed over $a$ and $c$, 
with $d{\rm Lips}(s;p_{\chi^-_j },p_{\chi^0_n},p_{f^{'}},p_{\bar f})$, 
defined in~(\ref{Lipsbosonic2}), where we have used the narrow
width approximations for the propagators.
In the numerator only the vector part of $|T|^2$ 
remains because only the vector part contains the triple product
${\mathcal T}_{II}={\mathbf p}_{e^-}\cdot({\mathbf p}_{c} 
			 \times {\mathbf p}_{\bar s}) $.
In the denominator the vector and tensor parts of $|T|^2$ vanish
due to phase-space integration. Owing to the correlations 
$\Sigma_P^a \, ^c\Sigma_{D_1}^a$ between
the $\tilde\chi^+_i$ and the $W$ boson polarization, there are 
contributions to the asymmetry ${\mathcal A}_{II}^{\rm T}$ 
from the chargino production process~(\ref{W:productionA}),
and/or from the chargino decay process~(\ref{W:decay_1A}).
The contribution from the production is given by the term with $a=2$ 
in~(\ref{W:asymII}) and it is proportional to 
the transverse polarization of the chargino perpendicular to the 
production plane $\Sigma^{2}_P$, see
Appendix~\ref{Chargino polarization}. 
For the production of a pair of equal charginos, 
$e^+e^- \to\tilde\chi^+_i \tilde\chi^-_i$, we have $\Sigma^{2}_P=0$.
The contributions  from the decay, given by terms with $a=1,3$
in~(\ref{W:asymII}), are proportional to  
\begin{eqnarray}\label{W:adependence}
	^c\Sigma_{D_1}^a \,^cD_2&\supset&
	-2g^4 m_{\chi_n^0}
	%	\frac{8g^4m_{\tilde\chi_k}}{\cos^4\theta_W}
	Im( O^{R\ast}_{ni}O^{L}_{ni})(t^c_W\cdot p_{\bar f})
	\epsilon_{\mu\nu\rho\sigma}~s^{a,\mu}_{\chi^+_i}~p_{\chi^+_i}^{\nu}
	~p_W^{\rho}~t^{c,\sigma}_W,
\end{eqnarray}
see last term of~(\ref{W:csigmaaD1}), 
which contains the $\epsilon$-tensor. Thus ${\mathcal A}_{II}^{\rm T}$ 
can be enhanced (reduced) if the contributions 
from production and decay  have the same (opposite) sign.
Note that the contributions from the decay  vanish 
for a two-body decay of the chargino into a scalar particle
instead of a $W$ boson.
%
%
%
%With $S_{I}^{\rm T}$ standard deviations, the relative statistical error 
%of ${\mathcal A}_{I}^{\rm T}$ is given by $\delta {\mathcal A}_{I}^{\rm T} = 
%\Delta {\mathcal A}_{I}^{\rm T}/|{\mathcal A}_{I}^{\rm T}| = 
%S_{I}^{\rm T}/(|{\mathcal A}_{I}^{\rm T}| \sqrt{N})$ \cite{olaf1}, where 
%$N={\mathcal L} \cdot\sigma$ is the number of events for the integrated 
%luminosity ${\mathcal L}$ and the cross section 
%$\sigma=\sigma_P(e^+e^-\to\tilde\chi^+_i\tilde\chi^-_j) 
%\times{\rm BR}(\tilde\chi^+_i \to W^+\tilde\chi^0_n)$.
%For the CP asymmetry ${\mathcal A}_{I}$ (\ref{ACP}), we have
%$\Delta {\mathcal A}_{I}=\Delta {\mathcal A}_{I}^{\rm T}/\sqrt{2} $.
%Taking $\delta {\mathcal A}_{I}=1$ it follows 
%$S_{I} = |{\mathcal A}_{I}| \sqrt{2{\mathcal L}\cdot\sigma}$.
%A similar result is obtained for 
%the asymmetry ${\mathcal A}_{II}$ with 
%$S_{II} = |{\mathcal A}_{II}| 
%\sqrt{2{\mathcal L}\cdot\sigma}$ and the cross section
%$\sigma=\sigma_P(e^+e^-\to\tilde\chi^+_i\tilde\chi^-_j) 
%\times{\rm BR}(\tilde\chi^+_i \to W^+\tilde\chi^0_n)\times
%{\rm BR}(W^+\to c\bar s)$.
In order to measure ${\mathcal A}_{I}$ 
the momentum of $\tilde\chi^+_i$, i.e. the
production plane, has to be determined.
This could be accomplished by measuring the 
decay of the other chargino $\tilde\chi^-_j$. 
%as discussed in Section~\ref{sneut:CP asymmetries}.
For the measurement of ${\mathcal A}_{II}$, 
the flavors of the quarks $c$ and $\bar s$ have to be  distinguished,
which will be possible by flavor tagging of the 
$c$-quark \cite{flavortaggingatLC,flavortagginginWdecays}. 
In principle, for the decay  $W \to u ~\bar d$ also an asymmetry similar 
to ${\mathcal A}_{II}$ can be considered, 
if it is possible to distinguish between the 
$u$ and $\bar d$ jet, for instance, by measuring the average charge.
%Also it is clear that detailed
%Monte Carlo studies taking into account background and detector
%simulations are necessary to predict the expected accuracies.
%However, this is beyond the scope of the present work.

\subsection{Numerical results
	\label{W:Numerical results}}

%We present numerical results for the asymmetries  
%${\mathcal A}_{I}$ and ${\mathcal A}_{II}$ (\ref{ACP}),
%the $W$ boson density matrix $<\rho(W^+)>$ (\ref{Wdensitymatrixnorm}),
%and the cross section 
%$\sigma=\sigma_P(e^+e^-\to\tilde\chi^+_i\tilde\chi^-_j ) \times
%{\rm BR}( \tilde\chi^+_i \to W^+\tilde\chi^0_1)\times
%{\rm BR}(W^+ \to c \bar s)$, with the experimental value
%${\rm BR}(W^+\to c \bar s)=0.31$  \cite{PDG}. 
We study the dependence of  
${\mathcal A}_{I}$,  ${\mathcal A}_{II}$,
%~(\ref{W:ACP}), and 
and the density matrix $<\rho(W^+)>$
%~(\ref{Wdensitymatrixnorm}),
on the MSSM parameters 
$\mu = |\mu| \, e^{ i\,\varphi_{\mu}}$, 
$M_1 = |M_1| \, e^{ i\,\varphi_{M_1}}$,
$\tan \beta$ and the universal scalar mass parameter $m_0$. 
%We will allow $\varphi_{M_1}\in[\pi,-\pi ]$, however take into account 
%$|\varphi_{\mu}|\lsim 0.1 \pi$, as suggested  from 
%the EDM analyses \cite{nath,edms}. 
%In order to show the full phase dependence 
%of the asymmetries, however, 
%we will relax the EDM restrictions in some scenarios. 
%We choose  a center  of mass energy of   $\sqrt{s} = 800$ GeV
%and longitudinally polarized beams with
%the polarizations $(P_{e^-},P_{e^+})=(-0.8,+0.6)$.
%This enhances sneutrino exchange in the chargino production process 
%which results in larger cross sections and asymmetries.
The feasibility of measuring the asymmetries depends also on the cross 
sections $\sigma=\sigma_P(e^+e^-\to\tilde\chi^+_i\tilde\chi^-_j ) \times
{\rm BR}(\tilde\chi^+_i \to W^+\tilde\chi^0_1)\times
{\rm BR}(W^+\to c \bar s)$, 
which we will discuss in our scenarios. 
%For the $W$ boson decay we take  the experimental value
%${\rm BR}(W^+\to c \bar s)=0.31$  \cite{PDG}.
We choose  a center  of mass energy of   $\sqrt{s} = 800$~GeV
and longitudinally polarized beams with $(P_{e^-},P_{e^+})=(-0.8,+0.6)$.
This choice enhances sneutrino exchange in the chargino 
production process, which results in larger cross sections 
and asymmetries. For the calculation of the  branching ratios 
${\rm BR}( \tilde\chi^+_i \to W^+\tilde\chi^0_1)$
and widths $\Gamma_{\chi_1^+}$, we include the two-body decays
\begin{eqnarray}\label{W:chardecaysBR}
	\tilde\chi^+_1 &\to& 
	W^+\tilde\chi^0_n,~
	\tilde e_{L}^+\nu_{e},~
	\tilde\mu_{L}^+\nu_{\mu},~
	\tilde\tau_{1,2}^+\nu_{\tau},~
	e^+\tilde\nu_{e},~
	\mu^+\tilde\nu_{\mu},~
	\tau^+\tilde\nu_{\tau},
\end{eqnarray}
and neglect three-body decays. 
%We also include the decays of the heavier chargino $\tilde\chi^+_2$ into 
%the $Z$ boson and into the lightest neutral Higgs boson:
%\begin{eqnarray}
%	\tilde\chi^+_2 &\to&
%	\tilde\chi_1^+ Z^0,~
%	\tilde\chi_1^+ H_1^0.
%\end{eqnarray}
%The Higgs parameter is chosen $m_{A}=1$~TeV and thus 
%the decays  $\tilde\chi^+_i \to \tilde\chi^0_n H^{\mp}$
%into the charged Higgs bosons are forbidden in our scenarios.
For the $W$ boson decay
we take  the experimental value
${\rm BR}(W^+\to c \bar s)=0.31$~\cite{PDG}.
In order to reduce the number of parameters, we assume the 
relation $|M_1|=5/3 \, M_2\tan^2\theta_W $ and  
use the renormalization group equations for the 
slepton and sneutrino masses, see 
Appendix~\ref{First and second generation sfermion masses}.
%%$m_{\tilde\ell_R}^2 = m_0^2 +0.23 M_2^2
%%-m_Z^2\cos 2 \beta \sin^2 \theta_W$, 
%$m_{\tilde\ell_L  }^2 = m_0^2 +0.79 M_2^2
%+m_Z^2\cos 2 \beta(-1/2+ \sin^2 \theta_W)$ and 
%$m_{\tilde\nu_{\ell}  }^2 = m_0^2 +0.79 M_2^2
%+m_Z^2/2\cos 2 \beta$.
In the stau sector, see Appendix \ref{Stau mixing}, 
we fix the trilinear scalar coupling
parameter $A_{\tau}=250$~GeV.

%\newpage

%
\subsubsection{Production of $\tilde\chi^+_1 \, \tilde\chi^-_1$ }

\begin{figure}[t]
\setlength{\unitlength}{1cm}
\begin{picture}(10,8)(0,0)
	\put(0,0){\includegraphics{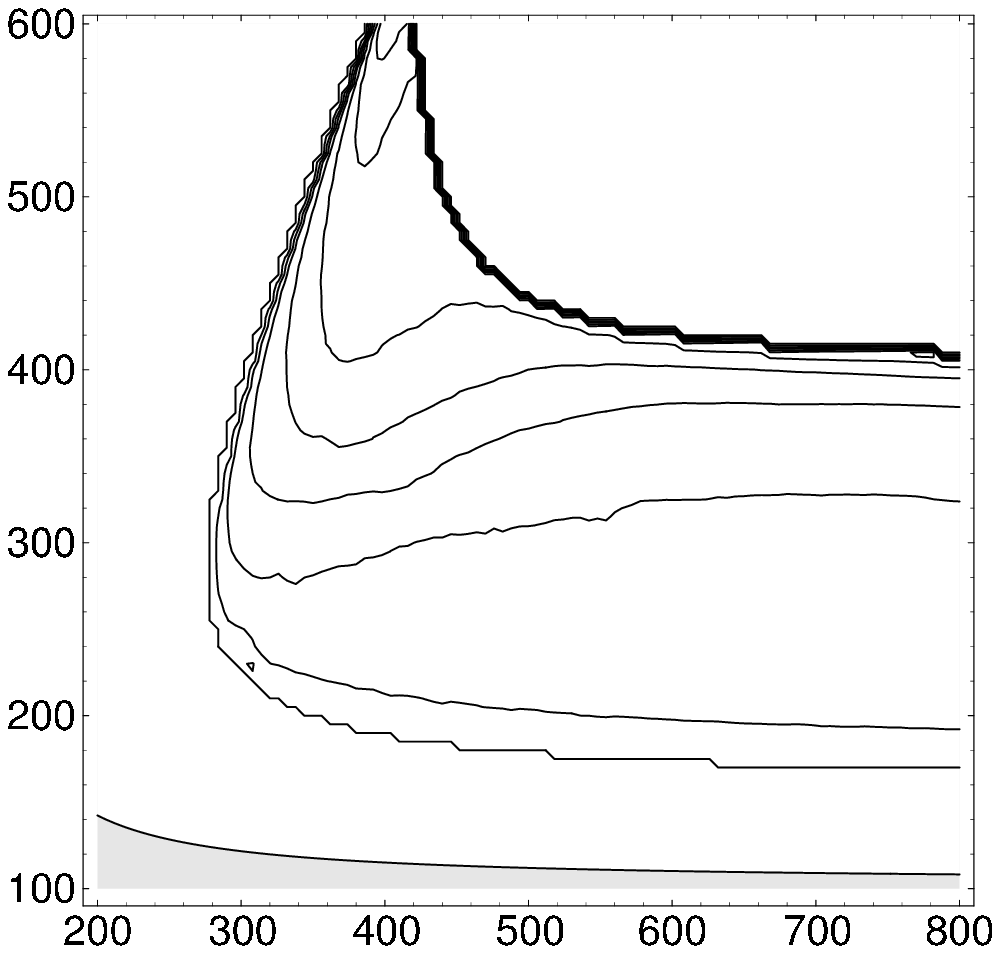}}
		\put(3.,7.4){\fbox{${\mathcal A}_{II}$ in \%}}
		\put(5.5,-0.3){$|\mu|\,[{\rm GeV}]$}
	\put(0,7.4){$M_2\,[{\rm GeV}]$ }
	\put(2.7,6.35){\scriptsize 7}
	\put(2.6,5.55){\footnotesize 6}
	\put(2.8,4.35){\footnotesize 5}
	\put(3.,3.85){\footnotesize 4}
	\put(3.2,3.45){\footnotesize 3}
	\put(3.5,2.9){\footnotesize 2}
	\put(4.,1.65){\footnotesize 1}
	\put(5.5,6){\begin{picture}(1,1)(0,0)
			\CArc(0,0)(7,0,380)
			\Text(0,0)[c]{{\footnotesize A}}
	\end{picture}}
			\put(1.3,5.0){\begin{picture}(1,1)(0,0)
			\CArc(0,0)(7,0,380)
			\Text(0,0)[c]{{\footnotesize B}}
		\end{picture}}
\put(0.5,-.3){Fig.~\ref{W:plot2}a}
\put(8,0){\includegraphics{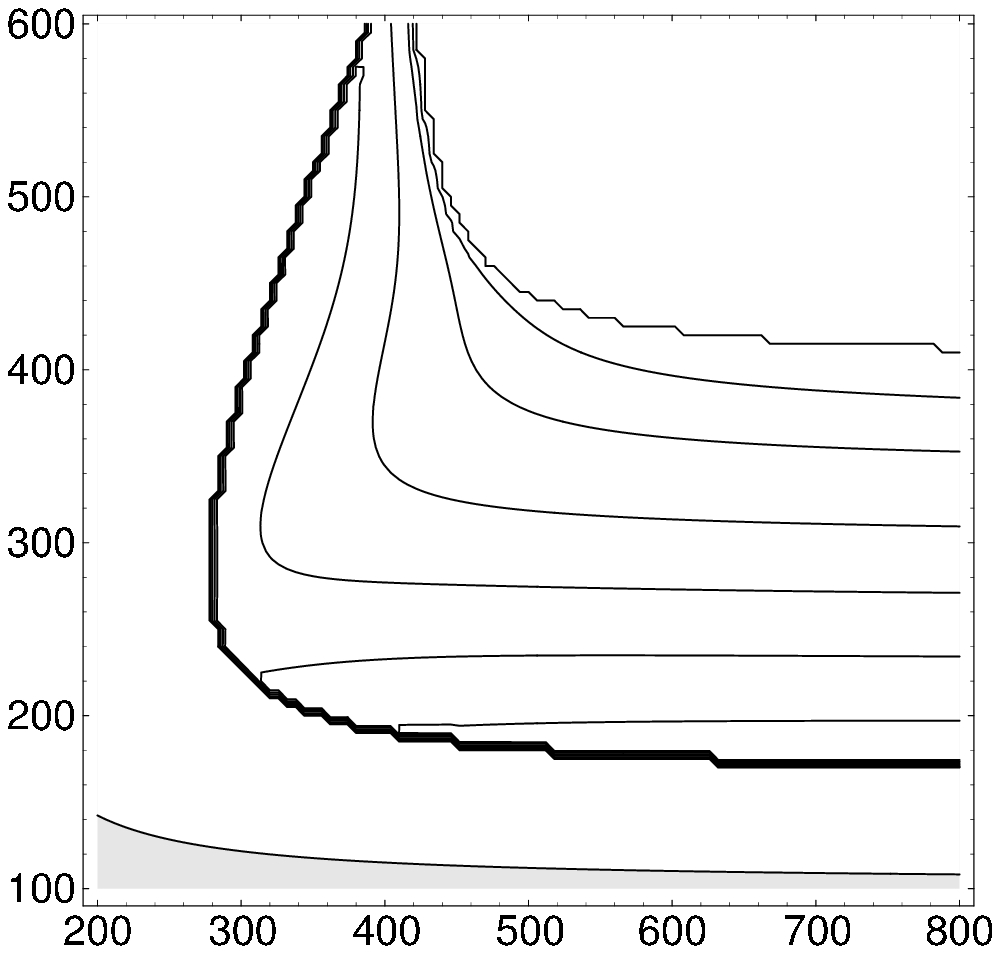}}
	\put(11.5,7.4){\fbox{$\sigma$ in fb}}
	\put(13.5,-.3){$|\mu|\,[{\rm GeV}]$}
	\put(8,7.4){$M_2\,[{\rm GeV}]$ }
	\put(14.0,4.25){\footnotesize 10}
	\put(13.6,3.9){\footnotesize 25}
	\put(12.9,3.4){\footnotesize 50}
	\put(12.5,2.9){\footnotesize 75}
	\put(12.0,2.4){\footnotesize 100}
	\put(11.5,1.9){\footnotesize 125}
	  	\put(13.5,6){\begin{picture}(1,1)(0,0)
			\CArc(0,0)(7,0,380)
			\Text(0,0)[c]{{\footnotesize A}}
	\end{picture}}
			\put(9.3,5.0){\begin{picture}(1,1)(0,0)
			\CArc(0,0)(7,0,380)
			\Text(0,0)[c]{{\footnotesize B}}
		\end{picture}}
	\put(8.5,-.3){Fig.~\ref{W:plot2}b}
\end{picture}
\vspace*{.5cm}
\caption{
	Contour lines of
	the asymmetry ${\mathcal A}_{II}$ (\ref{W:plot2}a)
	and $\sigma=\sigma_P(e^+e^-\to\tilde\chi^+_1\tilde\chi^-_1) 
	\times {\rm BR}( \tilde\chi^+_1 \to W^+\tilde\chi^0_1)
	\times {\rm BR}(W^+ \to c \bar s)$ (\ref{W:plot2}b),
	in the $|\mu|$--$M_2$ plane for 
	$(\varphi_{M_1},\varphi_{\mu})=(0.5\pi,0)$,
%	$\varphi_{M_1}=0.5\pi $, $\varphi_{\mu}=0$, 
	$\tan \beta=5$, $m_0=300$ GeV,
	$\sqrt{s}=800$ GeV and $(P_{e^-},P_{e^+})=(-0.8,0.6)$.
	The area A (B) is kinematically forbidden by
	$m_{\chi^+_1}+m_{\chi^-_1}>\sqrt{s}$
	$(m_{W}+m_{\chi^0_1}> m_{\chi^+_1})$.
	The gray  area is excluded by $m_{\chi_1^{\pm}}<104 $ GeV.
	\label{W:plot2}}
\end{figure}

For the production $e^+e^-\to\tilde\chi^+_i\tilde\chi^-_i $
of a pair of charginos the polarization perpendicular to the production
plane vanishes, and thus ${\mathcal A}_{I}=0$.
However, ${\mathcal A}_{II}$ need not to be zero and is 
sensitive to $\varphi_{\mu}$ and $\varphi_{M_1}$, because this
asymmetry has contributions from the chargino decay process.
%as discussed in Section \ref{T odd asymmetries}.
For $(\varphi_{M_1},\varphi_{\mu})=(0.5\pi,0)$ we show 
in Fig.~\ref{W:plot2}a the $|\mu|$--$M_2$ dependence of 
${\mathcal A}_{II}$, which can reach values of $5\%-7\%$ 
for $M_2 \gsim 400$~GeV. We also studied the $\varphi_{\mu}$ dependence of 
${\mathcal A}_{II}$ in the $|\mu|$--$M_2$ plane. For
$\varphi_{M_1}=0$, $\varphi_{\mu}=0.1\pi(0.5\pi)$ 
and the other parameters as given in the caption of Fig.~\ref{W:plot2},
we find $|{\mathcal A}_{II}|<2\%(7\%)$. 

In Fig.~\ref{W:plot2}b we show the contour lines of the
cross section 
%$\tilde\chi^+_1 \tilde\chi^-_1$ production and decay,
$\sigma=\sigma_P(e^+e^-\to\tilde\chi^+_1\tilde\chi^-_1 ) \times
{\rm BR}( \tilde\chi^+_1 \to W^+\tilde\chi^0_1)\times
{\rm BR}(W^+ \to c \bar s)$ in the $|\mu|$--$M_2$ plane for 
$(\varphi_{M_1},\varphi_{\mu})=(0.5\pi,0)$.
%$\varphi_{\mu}=0$ and $\varphi_{M_1}=0.5\pi$. 
%The choice of polarization $(P_{e^-},P_{e^+})=(-0.8,0.6)$
%enhances the production cross section 
The production cross section 
$\sigma_P(e^+e^-\to\tilde\chi^+_1\tilde\chi^-_1 )$
reaches more than $400$~fb. For our choice
of $m_0=300$~GeV, $ \tilde\chi^+_1 \to W^+\tilde\chi^0_1$
is the only allowed two-body decay  channel.
%------------------------------------------------------------------
%            CHI 1 CHI 1 -- P L O T S   
%-----------------------------------------------------------------
%&&&&&&&&&&&&&&&&&&&&&&&&&&&&&&&&&&&&&&&&&&&&&&&&&&&&&&&&&&&&&&
%                    P L O T  2 
%&&&&&&&&&&&&&&&&&&&&&&&&&&&&&&&&&&&&&&&&&&&&&&&&&&&&&&&&&&&&&&6
%
%
%

%\newpage

\begin{figure}[t]
\setlength{\unitlength}{1cm}
\begin{picture}(10,8)(0,0)
   \put(0,0){\includegraphics{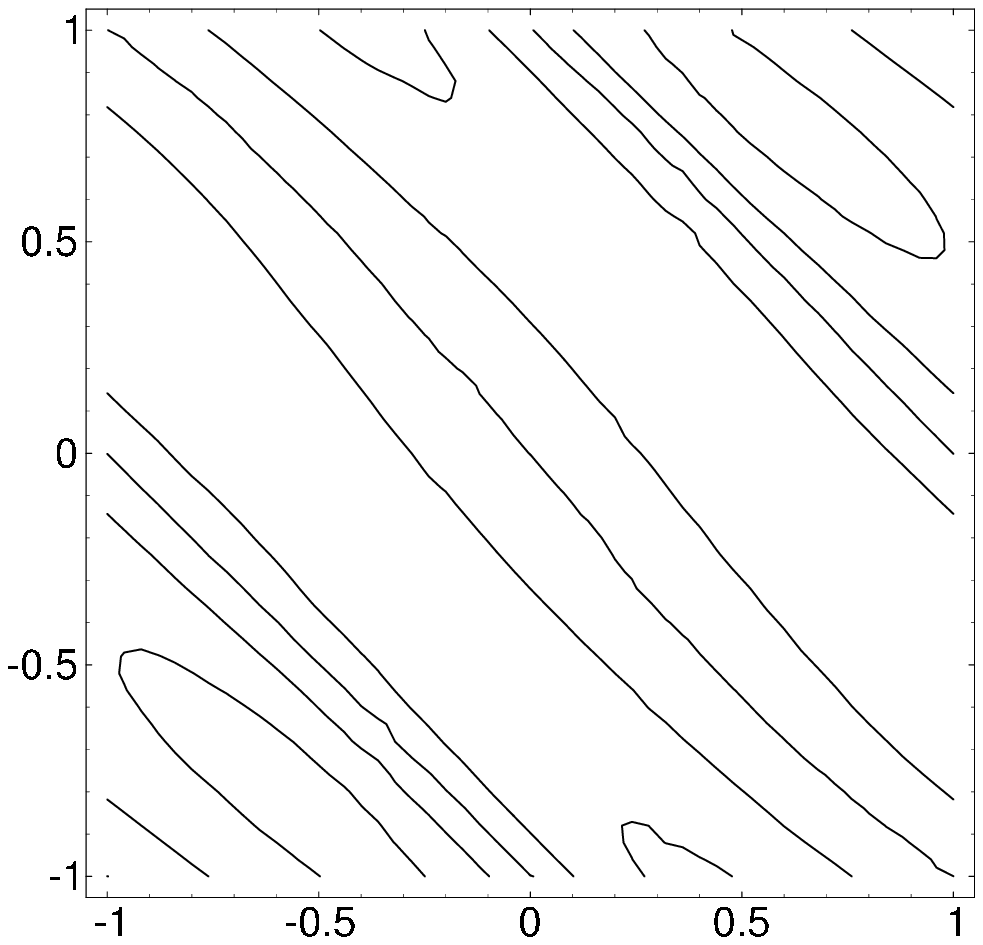}}
	\put(3.5,7.4){\fbox{${\mathcal A}_{II}$ in \% }}
	\put(6.0,-0.3){$\varphi_{\mu}[\pi]$}
	\put(0,7.4){$\varphi_{M_1}[\pi]$ }
	\put(1.1,0.8){\scriptsize 3}
	\put(1.6,1.5){\scriptsize 6}
	\put(2.,2.){\scriptsize 3}
	\put(2.25,2.43){\scriptsize 0}
	\put(2.55,2.6){\scriptsize -3}
	\put(4.4,1.2){\scriptsize -6}
	\put(3.2,3.1){\scriptsize -3}
	\put(4.0,3.7){\scriptsize 0}
	\put(4.35,4.3){\scriptsize 3}
	\put(3.1,6.4){\scriptsize 6}
	\put(5.0,5.0){\scriptsize 3}
	\put(5.3,5.2){\scriptsize 0}
	\put(5.5,5.5){\scriptsize -3}
	\put(5.8,5.8){\scriptsize -6}
	\put(6.6,6.5){\scriptsize -3}
\put(0.5,-.3){Fig.~\ref{W:plot3}a}
	\put(8,0){\includegraphics{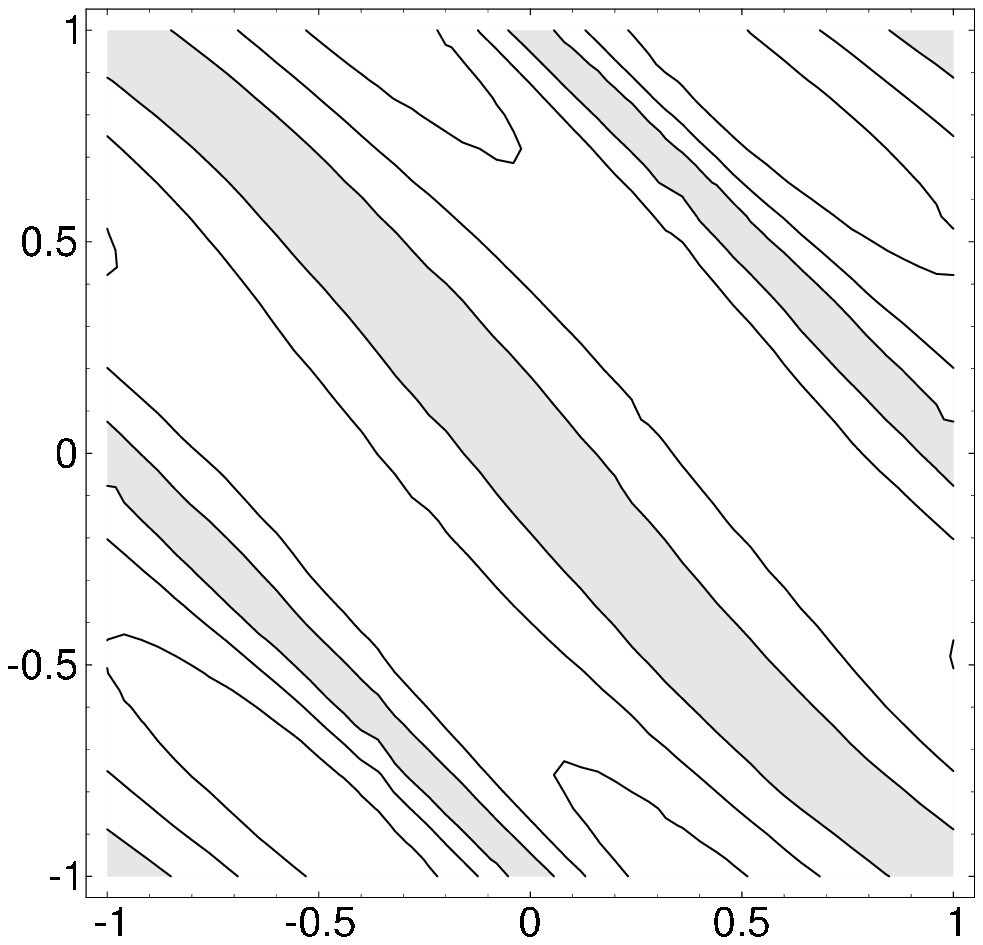}}
	\put(10.3,7.4){\fbox{$S_{II} =|{\mathcal A}_{II}| 
			\sqrt{2{\mathcal L}\cdot\sigma}$  }}
	\put(14.,-.3){$\varphi_{\mu}[\pi] $ }
	\put(8,7.4){$\varphi_{M_1}[\pi]$ }
	\put(9.1,1.36){\scriptsize 10}
	\put(9.6,1.5){\scriptsize 15}
	\put(10.,2.0){\scriptsize 10}
	\put(10.5,2.8){\scriptsize 10}
	\put(10.8,3.3){\scriptsize 10}
	\put(12.3,1.2){\scriptsize 15}
	\put(9.0,5.0){\scriptsize 15}
	\put(12.4,4.5){\scriptsize 10}
	\put(11.3,6.2){\scriptsize 15}
	\put(12.9,4.8){\scriptsize 10}
	\put(14.6,4.7){\scriptsize 10}
	\put(13.2,6.2){\scriptsize 15}
	\put(14.1,5.8){\scriptsize 15}
	\put(14.4,6.1){\scriptsize 10}
\put(8.5,-.3){Fig.~\ref{W:plot3}b}
\end{picture}
\vspace*{.5cm}
\caption{
	Contour lines of 
	and the asymmetry ${\mathcal A}_{II}$ (\ref{W:plot3}a)
	and the significance $S_{II}$ (\ref{W:plot3}b)
	for $e^+e^-\to\tilde\chi^+_1\tilde\chi^-_1;~ 
	\tilde\chi^+_1 \to  W^+\tilde\chi^0_1 ;~W^+ \to c \bar s$,
	in the $\varphi_{\mu}$--$\varphi_{M_1}$ plane  
	for $|\mu|=350$~GeV, $M_2=400$~GeV,
	$\tan \beta=5$, $m_0=300$ GeV,
	$\sqrt{s}=800$ GeV, $(P_{e^-},P_{e^+})=(-0.8,0.6)$
	and ${\mathcal L}=500~{\rm fb}^{-1}$.
	In the gray shaded area of Fig.~\ref{W:plot3}b
	we have $S_{II}<5$.
%	The area A (B) is kinematically forbidden by
%	$m_{\chi^+_1}+m_{\chi^-_1}>\sqrt{s}$
%	$(m_{W}+m_{\chi^0_1}> m_{\chi^+_1})$.
	\label{W:plot3}}
\end{figure}

In Fig.~\ref{W:plot3}a we plot the contour lines   
of ${\mathcal A}_{II}$ for $|\mu|=350$~GeV and $M_2=400$~GeV
in the $\varphi_{\mu}$--$\varphi_{M_1}$ plane.
Fig.~\ref{W:plot3}a shows that ${\mathcal A}_{II}$ is
essentially depending on the sum $\varphi_{\mu}+\varphi_{M_1}$.
%The dependence of ${\mathcal A}_{II}$ on the two phases
%is almost equal, 
However, maximal phases of  
$\varphi_{M_1}=\pm 0.5\pi$ and $\varphi_{\mu}=\pm 0.5\pi$ 
do not  lead to the highest values of 
$|{\mathcal A}_{II}| \gsim  6\%$, which are reached for
$(\varphi_{M_1},\varphi_{\mu}) \approx (\pm0.8 \pi,\pm0.6 \pi)$.
The reason for this is that the spin-correlation terms 
$\Sigma_P^a \,^c\Sigma_{D_1}^a \,^cD_2$ in the numerator of 
${\mathcal A}_{II}$~(\ref{W:asymII}) are products of CP-odd and
CP-even factors. The CP-odd (CP-even) factors have a sine-like
(cosine-like) phase dependence. Therefore, the maximum of the CP
asymmetry ${\mathcal A}_{II}$ may be shifted to smaller or larger
values of the phases. 
In the $\varphi_{\mu}$--$\varphi_{M_1}$ region shown 
in Fig.~\ref{W:plot3}a the cross section 
$\sigma=\sigma_P(e^+e^-\to\tilde\chi^+_1\tilde\chi^-_1) 
\times {\rm BR}( \tilde\chi^+_1 \to W^+\tilde\chi^0_1)
\times {\rm BR}(W^+ \to c \bar s)$
with ${\rm BR}( \tilde\chi^+_1 \to \tilde\chi^0_1 W^+)=1$
does not depend on $\varphi_{M_1}$ and ranges between 74~fb 
for $\varphi_{\mu}=0$ and 66~fb for $\varphi_{\mu}=\pi$.

In Fig.~\ref{W:plot3}b we show the contour lines of the 
%statistical 
significance
%standard deviations
$S_{II} =|{\mathcal A}_{II}| \sqrt{2{\mathcal L}\cdot\sigma}$,
defined in~(\ref{significanceofACP}).
For ${\mathcal L}= 500$~fb$^{-1}$ and for e.g. 
$(\varphi_{M_1},\varphi_{\mu}) \approx ( \pi,0.1 \pi)$
we have $S_{II} \approx 8$ 
and thus ${\mathcal A}_{II}$ should be measured 
%at a confidence level larger than 68\% $(S_{II}=1)$,
even for small $\varphi_{\mu}$. 
%and real $M_1$.
%&&&&&&&&&&&&&&&&&&&&&&&&&&&&&&&&&&&&&&&&&&&&&&&&&&&&&&&&&&&&&&
%                    P L O T  3 
%&&&&&&&&&&&&&&&&&&&&&&&&&&&&&&&&&&&&&&&&&&&&&&&&&&&&&&&&&&&&&&6
%

\newpage

\begin{figure}[t]
\setlength{\unitlength}{1cm}
\begin{picture}(10,8)(0,0)
   \put(0,0){\includegraphics{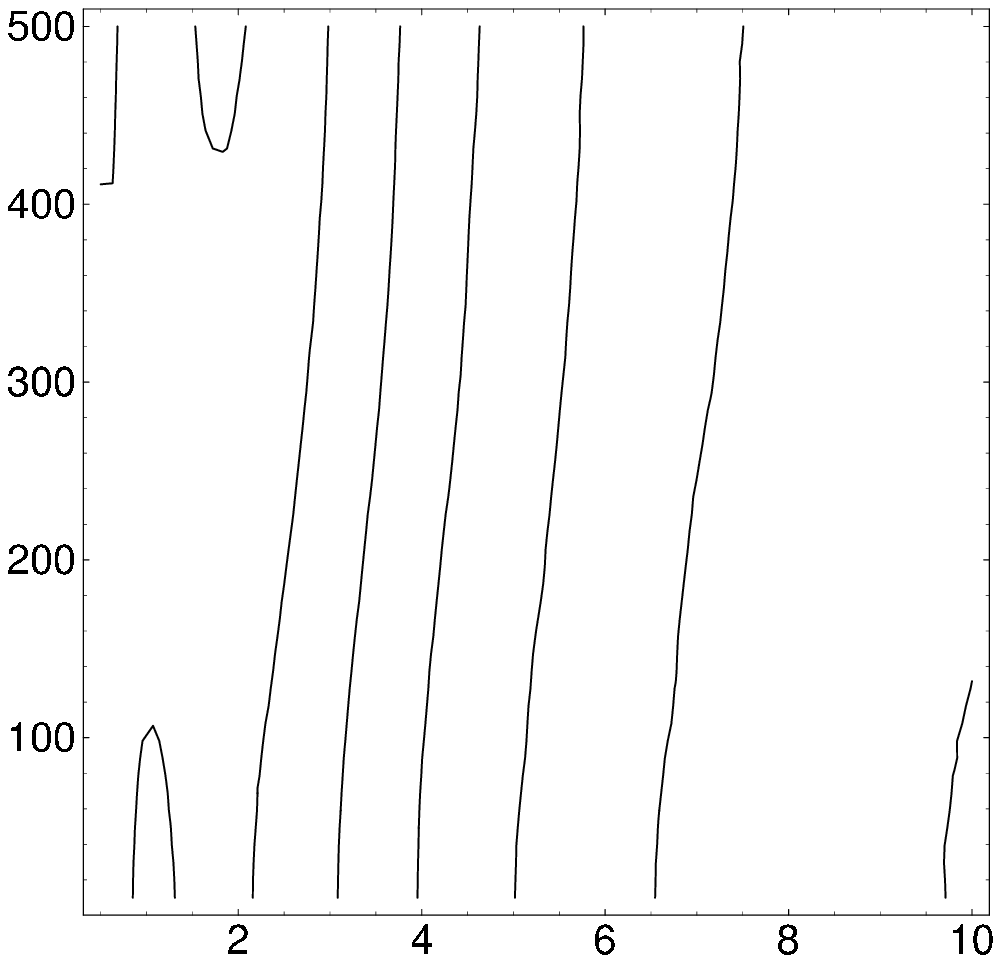}}
	\put(3.,7.4){\fbox{${\mathcal A}_{II}$ in \% }}
	\put(6.3,-0.3){$\tan\beta$}
	\put(0,7.4){$m_0[{\rm GeV}]$}
	\put(0.9,6.){\footnotesize 8}
	\put(1.05,0.9){\footnotesize 9}
	\put(1.5,6.5){\footnotesize 9}
	\put(1.9,4.){\footnotesize 8}
	\put(2.5,4.){\footnotesize 7}
	\put(3.05,4.){\footnotesize 6}
	\put(3.8,4.){\footnotesize 5}
	\put(4.8,4.){\footnotesize 4}
	\put(6.8,1.0){\footnotesize 3}
\put(0.5,-.3){Fig.~\ref{W:plot4}a}
\put(0.5,-.3){Fig.~\ref{W:plot4}a}
	\put(8,0){\includegraphics{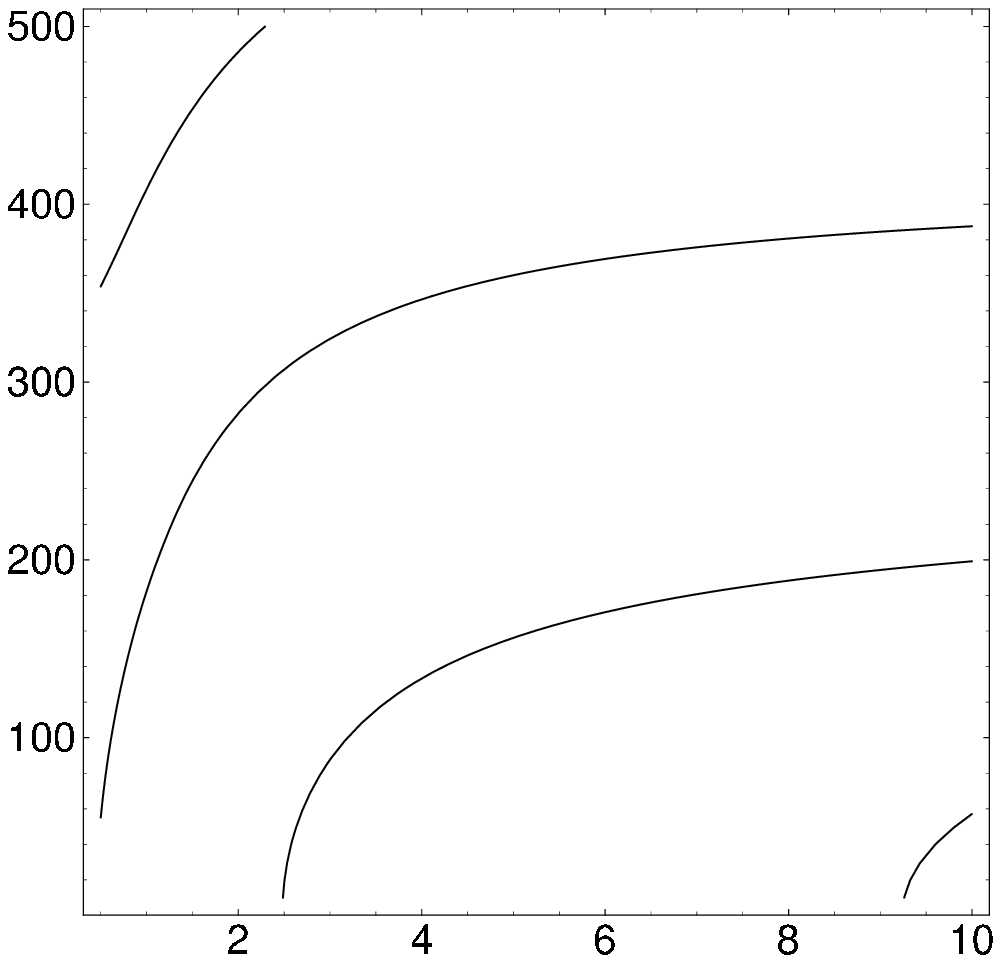}}
	\put(11.5,7.4){\fbox{$\sigma$ in fb}}
	\put(14.3,-.3){$\tan\beta$}
	\put(8,7.4){$m_0[{\rm GeV}]$}
	\put(9.5,6.0){\footnotesize 100}
	\put(11.2,4.5){\footnotesize 80}
	\put(12.3,2.3){\footnotesize 60}
	\put(14.6,0.65){\footnotesize 40}
	\put(8.5,-.3){Fig.~\ref{W:plot4}b}
\end{picture}
\vspace*{.5cm}
\caption{
	Contour lines of 
	the asymmetry ${\mathcal A}_{II}$ (\ref{W:plot4}a)
	and $\sigma=\sigma_P(e^+e^-\to\tilde\chi^+_1\tilde\chi^-_1) 
	\times {\rm BR}( \tilde\chi^+_1 \to W^+\tilde\chi^0_1)
	\times {\rm BR}(W^+ \to c \bar s)$ (\ref{W:plot4}b),
	in the $\tan\beta$--$m_0$ plane for 
	$(\varphi_{M_1},\varphi_{\mu})=(0.7\pi,0)$,
%	$\varphi_{M_1}=0.7\pi $, $\varphi_{\mu}=0$, 
	$M_2=400$~GeV, $|\mu|=350$~GeV,
	$\sqrt{s}=800$ GeV and $(P_{e^-},P_{e^+})=(-0.8,0.6)$.
%	The area A (B) is kinematically forbidden by
%	$m_{\chi^+_1}+m_{\chi^-_1}>\sqrt{s}$
%	$(m_{W}+m_{\chi^0_1}> m_{\chi^+_1})$.
	\label{W:plot4}}
\end{figure}

In Figs.~\ref{W:plot4}a,b we show the 
$\tan\beta $--$m_0$ dependence of ${\mathcal A}_{II}$ and $\sigma$ 
for $(\varphi_{M_1},\varphi_{\mu})=(0.7\pi,0)$.
%for $\varphi_{M_1}=0.7\pi$ and $\varphi_{\mu}=0$.
%$|\mu|=350$ GeV and $M_2=400$ GeV.
The asymmetry is rather insensitive to  $m_0$
and shows strong dependence on  $\tan\beta$ and
decreases with increasing $\tan\beta\gsim 2$.
The production cross section
$\sigma_P(e^+e^-\to\tilde\chi^+_1\tilde\chi^-_1 )$
increases with increasing $m_0$ and decreasing $\tan\beta$.
For $m_0\lsim 200$~GeV, the branching ratio 
${\rm BR}(\tilde\chi^+_1 \to W^+\tilde\chi^0_1)<1$, 
since the decay channels of $\tilde\chi^+_1$ 
into sleptons and/or sneutrinos open,
see~(\ref{W:chardecaysBR}).
%\cite{choigaiss}.
%&&&&&&&&&&&&&&&&&&&&&&&&&&&&&&&&&&&&&&&&&&&&&&&&&&&&&&&&&&&&&&
%                    P L O T  4 
%&&&&&&&&&&&&&&&&&&&&&&&&&&&&&&&&&&&&&&&&&&&&&&&&&&&&&&&&&&&&&&6
%

%\newpage

\begin{figure}[t]
\setlength{\unitlength}{1cm}
\begin{picture}(10,8)(0.2,0)
%   \put(0,0){\special{psfile=./4a.eps angle=0
%			voffset=0 hoffset=0 hscale=70 vscale=70}}
	\put(-1,8){\includegraphics{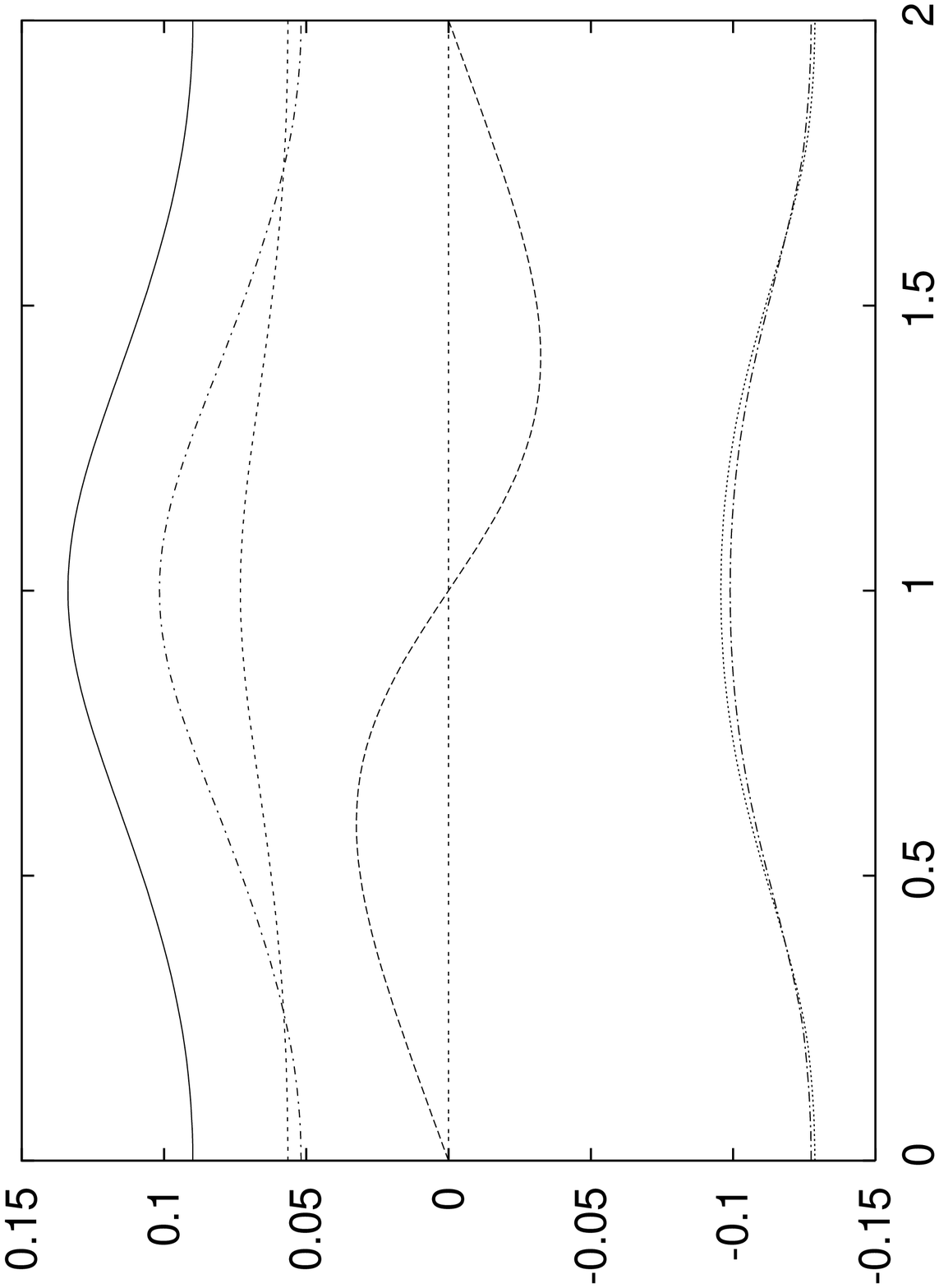}}
	%	\put(14.4,6.1){\scriptsize 10}
	\put(3.2,6.6){\footnotesize $V_1 $}
	\put(3.8,6.1){\footnotesize $ T_{13} $}
	\put(5.7,2.7){\footnotesize $ V_2$}
	\put(3.9,4.85){\footnotesize $ V_3 $}
	\put(1.3,3.2){\footnotesize $T_{12} \approx T_{23}\approx0$}
	\put(3.5,1.9){\footnotesize $ T_{11}\approx T_{22}$}
	\put(1.3,7.4){\fbox{components for $\varphi_{M_1}=\pi$}}
	\put(6.3,-0.3){$\varphi_{\mu}[\pi]$}
%	\put(0,7.4){$\varphi_{M_1}/\pi$ }
\put(0.5,-.3){Fig.~\ref{W:plot5}a}
	\put(7,8){\includegraphics{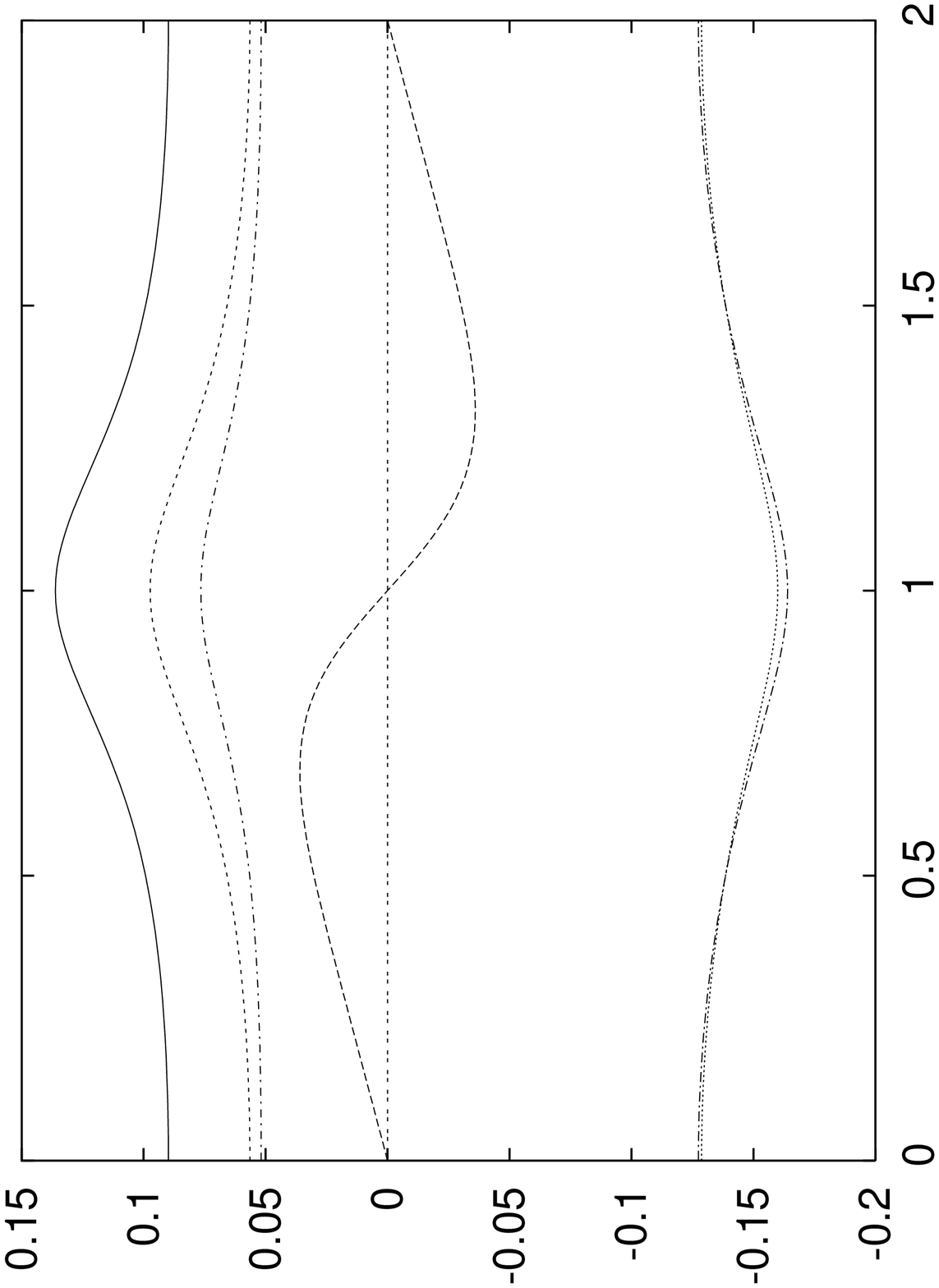}}
	\put(9.3,7.4){\fbox{components for $\varphi_{\mu}=0$ }}
	\put(14.3,-.3){$ \varphi_{M_1}[\pi]$ }
	\put(11.2,6.6){\footnotesize $V_1 $}
	\put(11.8,6.1){\footnotesize $V_3 $}
	\put(13.7,3.35){\footnotesize $ V_2$}
	\put(11.9,5.2){\footnotesize $T_{13} $}
	\put(9.4,3.75){\footnotesize $T_{12} \approx T_{23}\approx0$}
	\put(11.3,1.65){\footnotesize $ T_{11}\approx T_{22}$}
%	\put(8,7.4){$\varphi_{M_1}/\pi$ }
%	\put(14.4,6.1){\scriptsize 10}
\put(8.5,-.3){Fig.~\ref{W:plot5}b}
\end{picture}
\vspace*{.5cm}
\caption{
   Dependence of vector $(V_i)$ and tensor $(T_{ij})$ components
	of the $W^+$ density matrix on $\varphi_{\mu}$  (\ref{W:plot5}a)
	and on $\varphi_{M_1}$ (\ref{W:plot5}b),
	for $e^+e^-\to\tilde\chi^+_1\tilde\chi^-_1;~ 
	\tilde\chi^+_1 \to W^+\tilde\chi^0_1 $,
%	in the $\varphi_{\mu}$--$\varphi_{M_1}$ plane for 
	for $|\mu|=350$~GeV, $M_2=400$~GeV,
	$\tan \beta=5$, $m_0=300$ GeV,
	$\sqrt{s}=800$ GeV and $(P_{e^-},P_{e^+})=(-0.8,0.6)$.
%	and ${\mathcal L}=500~{\rm fb}^{-1}$.
%	In the gray shaded area of Fig.~\ref{W:plot_11phi}b
%	we have $S_{II}<5$.
%	The area A (B) is kinematically forbidden by
%	$m_{\tilde\chi^+_1}+m_{\tilde\chi^-_1}>\sqrt{s}$
%	$(m_{W}+m_{\tilde\chi^0_1}> m_{\tilde\chi^+_1})$.
	\label{W:plot5}}
\end{figure}

In Fig.~\ref{W:plot5}a we show the 
$\varphi_{\mu}$ dependence of the vector $(V_i)$ and tensor $(T_{ij})$ 
components of the density matrix  $<\rho(W^+)>$
for $\varphi_{M_1}=\pi$.
In Fig.~\ref{W:plot5}b we plot their $\varphi_{M_1}$ dependence 
for $\varphi_{\mu}=0$. In both figures, the element $V_2$ is CP-odd,
while $T_{13}$, $T_{11}$, $T_{22}$ and $V_1$, $V_3$ show a CP-even behavior.
%The element $V_2$ is CP odd and is not only zero at 
%$\varphi_{M_1}=0$ and $\varphi_{M_1}=\pi$, 
%but also at $\varphi_{M_1}\approx(1\pm0.2)\pi$, 
%which is due to the destructive interference of the
%contributions from CP violation in production and decay.
%The interference of the contributions from the
%CP even effects in production and decay cause the two
%maxima of $V_1$. 
As discussed in Appendix~\ref{Chargino decay into the W boson},
the components $T_{11}$ and $T_{22}$ are almost 
equal and  have the same order of magnitude as $V_1$ and $V_3$, 
%but their dependence on $\varphi_{M_1}$ is rather weak.
whereas $T_{12},|T_{23}|<10^{-5}$ are small.
%and thus the density matrix  $<\rho(Z)>$ 
%assumes a symmetric shape. 
For CP conserving phases
$(\varphi_{M_1},\varphi_{\mu})=(0,0)$
%$\varphi_{M_1}=\varphi_{\mu}=0$
the density matrix reads
\begin{eqnarray}
	<\rho(W^+)>  =
	\left(
        \begin{array}{ccc}
			   <\rho^{--}> & <\rho^{-0}> & <\rho^{-+}> \\
			   <\rho^{0-}> & <\rho^{00}> & <\rho^{0+}> \\
			   <\rho^{+-}> & <\rho^{+0}> & <\rho^{++}>
        \end{array}
	\right) =
 	 \left(
        \begin{array}{ccc}
			   0.200 & -0.010 & -0.001\\
			  -0.010 &  0.487 &  0.137 \\
			  -0.001 &  0.137 &  0.313
        \end{array}
	  \right),
	  \end{eqnarray}
for $M_2=400$~GeV,  $|\mu|=350$~GeV, 
$\tan \beta=5$, $m_0=300$~GeV, 
$\sqrt{s}=800$~GeV, $(P_{e^-},P_{e^+})=(-0.8,0.6)$
For CP violating phases, e.g.  
$(\varphi_{M_1},\varphi_{\mu})=(0.7\pi,0)$
%$\varphi_{M_1}=0.7\pi $ 
and the other parameters as above, the density matrix has 
imaginary parts due to a non-vanishing $V_2$:
\begin{eqnarray}
<\rho(W^+)>  =\left(
        \begin{array}{ccc}
			  0.219        &-0.010 +0.025i & 0.002\\
			 -0.010 -0.025i& 0.405         & 0.171 + 0.025i \\
			  0.002        & 0.171 -0.025i & 0.376
        \end{array}.
	\right)
\end{eqnarray}
Imaginary parts of the density matrix are thus an 
indication of CP violation. 
%&&&&&&&&&&&&&&&&&&&&&&&&&&&&&&&&&&&&&&&&&&&&&&&&&&&&&&&&&&&&&&
%                    P L O T  5 
%&&&&&&&&&&&&&&&&&&&&&&&&&&&&&&&&&&&&&&&&&&&&&&&&&&&&&&&&&&&&&&6
%
%

%\newpage

%
\subsubsection{Production of $\tilde\chi^+_1 \, \tilde\chi^-_2$ }

\begin{figure}[t]
\setlength{\unitlength}{1cm}
\begin{picture}(10,8)(0,0)
   \put(0,0){\includegraphics{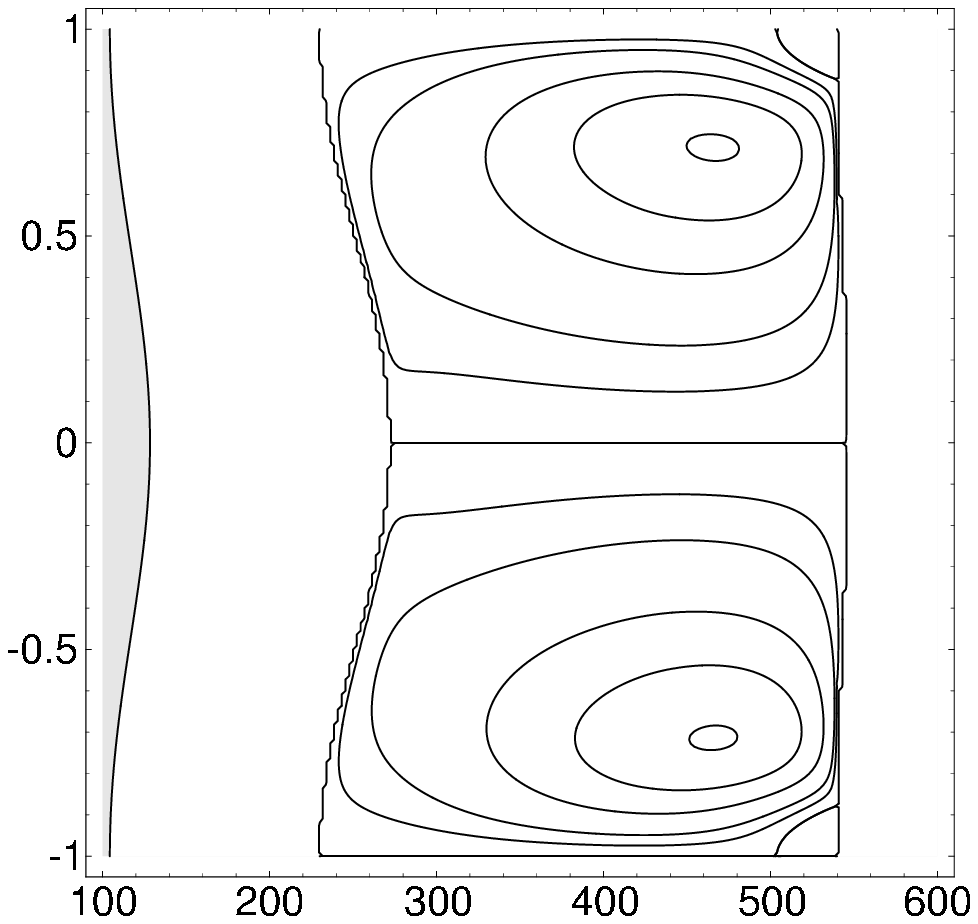}}
	\put(3.,7.4){\fbox{${\mathcal A}_{I}$ in \% }}
	\put(5.5,-0.3){$|\mu|\,[{\rm GeV}]$}
	\put(0,7.4){ $\varphi_{\mu}[\pi]$}
	\put(5.2,5.8){\scriptsize 4}
	\put(4.5,5.6){\scriptsize 3}
	\put(4.15,5.2){\scriptsize 2}
	\put(3.7,4.7){\scriptsize 1}
	\put(3.4,4.3){\scriptsize 0.5}
	\put(3.2,3.8){\scriptsize 0}
	\put(3.4,3.05){\scriptsize -0.5}
	\put(5.2,1.6){\scriptsize -4}
	\put(4.5,1.8){\scriptsize -3}
	\put(4.15,2.1){\scriptsize -2}
	\put(3.7,2.6){\scriptsize -1}
  	\put(6.6,3.7){\begin{picture}(1,1)(0,0)
			\CArc(0,0)(7,0,380)
			\Text(0,0)[c]{{\footnotesize A}}
	\end{picture}}
			\put(2.0,3.7){\begin{picture}(1,1)(0,0)
			\CArc(0,0)(7,0,380)
			\Text(0,0)[c]{{\footnotesize B}}
		\end{picture}}
\put(0.5,-.3){Fig.~\ref{W:plot6}a}
	\put(8,0){\includegraphics{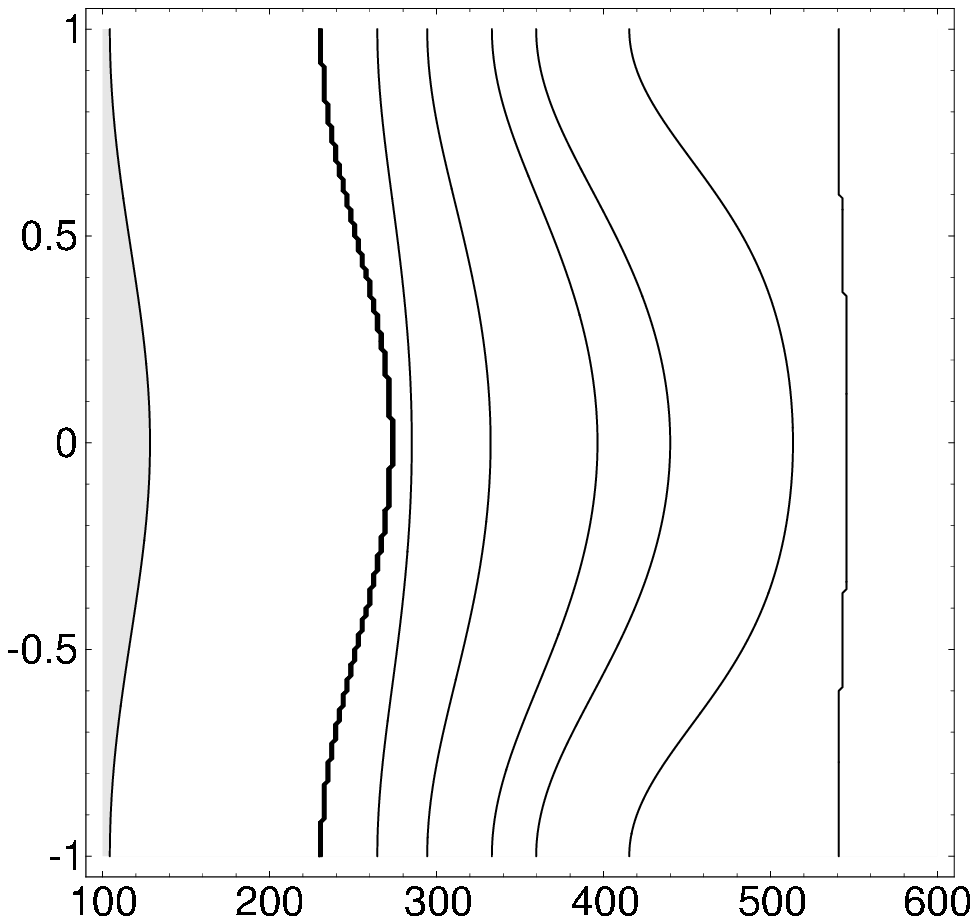}}
	\put(11.5,7.4){\fbox{$\sigma$ in fb}}
	\put(13.5,-.3){$|\mu|\,[{\rm GeV}]$}
	\put(8,7.4){ $\varphi_{\mu}[\pi]$}
	\put(11.2,3.7){\footnotesize 75}
	\put(11.7,3.7){\footnotesize 50}
	\put(12.5,3.7){\footnotesize 25}
	\put(13.0,3.7){\footnotesize 15}
	\put(13.9,3.7){\footnotesize 5}
	\put(14.6,3.7){\begin{picture}(1,1)(0,0)
			\CArc(0,0)(7,0,380)
			\Text(0,0)[c]{{\footnotesize A}}
	\end{picture}}
			\put(10.,3.7){\begin{picture}(1,1)(0,0)
			\CArc(0,0)(7,0,380)
			\Text(0,0)[c]{{\footnotesize B}}
		\end{picture}}
	\put(8.5,-.3){Fig.~\ref{W:plot6}b}
\end{picture}
\vspace*{.5cm}
\caption{
	Contour lines of 
	the asymmetry ${\mathcal A}_{I}$ (\ref{W:plot6}a)
	and $\sigma=\sigma_P(e^+e^-\to\tilde\chi^+_1\tilde\chi^-_2) 
	\times {\rm BR}( \tilde\chi^+_1 \to W^+\tilde\chi^0_1)$ 
	(\ref{W:plot6}b),
	in the $|\mu|$--$\varphi_{\mu}$  plane for $\varphi_{M_1}=0$, 
	$M_2=250$~GeV, $\tan \beta=5$, $m_0=300$ GeV,
	$\sqrt{s}=800$ GeV and $(P_{e^-},P_{e^+})=(-0.8,0.6)$.
	The area A (B) is kinematically forbidden by
	$m_{\chi^+_2}+m_{\chi^-_1}>\sqrt{s}$
	$(m_{W}+m_{\chi^0_1}> m_{\chi^+_1})$.
		The gray  area is excluded by $m_{\chi_1^{\pm}}<104 $ GeV.
	\label{W:plot6}}
\end{figure}

For the production of an unequal pair of charginos,
$e^+e^-\to\tilde\chi^+_1\tilde\chi^-_2 $,
their polarization perpendicular to the production
plane is sensitive to the phase $\varphi_{\mu}$,
which leads to a non-vanishing asymmetry ${\mathcal A}_{I}$~(\ref{W:asymI}).
We will study the decay of the lighter chargino 
$\tilde\chi^+_1\to W^+\tilde\chi^0_1$.
%, since for our choice 
%$m_0=300$~GeV we have  BR$( \tilde\chi^+_1 \to W^+\tilde\chi^0_1)=1$.
%For the decay of $\tilde\chi^+_2$,  the channels into 
%$Z$ and $H$ had to be taken into account, which would reduce 
%BR$(\tilde\chi^+_2\to W^+\tilde\chi^0_1)$.
For $|M_2|=250$~GeV and $\varphi_{M_1}=0$, we show in 
Fig.~\ref{W:plot6}a the  $|\mu|$--$\varphi_{\mu}$
dependence of ${\mathcal A}_{I}$, which attains values up to $4\%$.
Note that ${\mathcal A}_{I}$ is not maximal for 
$\varphi_{\mu}=0.5\pi$, but is rather sensitive for phases 
in the regions $\varphi_{\mu}\in[0.7\pi,\pi]$
and $\varphi_{\mu}\in[-0.7\pi,-\pi]$. As discussed
in Section \ref{CP violating phases and electric dipole moments},
values of $\varphi_{\mu}$ close to the CP conserving points
$\varphi_{\mu}=0, \pm\pi$ are suggested by EDM analyses.
For  $\varphi_{\mu} = 0.9\pi$ and $|\mu|=350$~GeV   
the statistical significance is
$S_{I} =|{\mathcal A}_{I}|\sqrt{2{\mathcal L}\cdot\sigma} \approx 1.5$
with ${\mathcal L}=500~{\rm fb}^{-1}$.
Thus ${\mathcal A}_{I}$ could be measured 
at a confidence level larger than $68\%$ $(S_{I}=1)$.

In Fig.~\ref{W:plot6}b we show contour lines of the 
corresponding cross section 
$\sigma=\sigma_P(e^+e^-\to\tilde\chi^+_1\tilde\chi^-_2) \times
{\rm BR}( \tilde\chi^+_1 \to W^+\tilde\chi^0_1)$
in the $|\mu|$--$\varphi_{\mu}$ plane for the parameters as above.
The cross section shows a CP even behavior, 
which has been used in \cite{choichargino,holger,choigaiss} 
to constrain  $\cos \varphi_{\mu}$.
In our scenario we have considered the decay of the lighter chargino 
$\tilde\chi^+_1\to W^+\tilde\chi^0_1$ 
since for our choice $m_0=300$~GeV we have  
${\rm BR}( \tilde\chi^+_1 \to W^+\tilde\chi^0_1)=1$.
For the decay of $\tilde\chi^+_2$,  one would have to take
into account also the decays into the
$Z$ boson and the lightest neutral Higgs, which would reduce 
${\rm BR}(\tilde\chi^+_2\to W^+\tilde\chi^0_1)\approx 0.2$.
%------------------------------------------------------------------
%            CHI 1 CHI 2 -- P L O T S   
%-----------------------------------------------------------------
%&&&&&&&&&&&&&&&&&&&&&&&&&&&&&&&&&&&&&&&&&&&&&&&&&&&&&&&&&&&&&&
%                    P L O T  6 
%&&&&&&&&&&&&&&&&&&&&&&&&&&&&&&&&&&&&&&&&&&&&&&&&&&&&&&&&&&&&&&6
%

%\newpage

\begin{figure}[t]
\setlength{\unitlength}{1cm}
\begin{picture}(10,8)(0.2,0)
   \put(0,0){\includegraphics{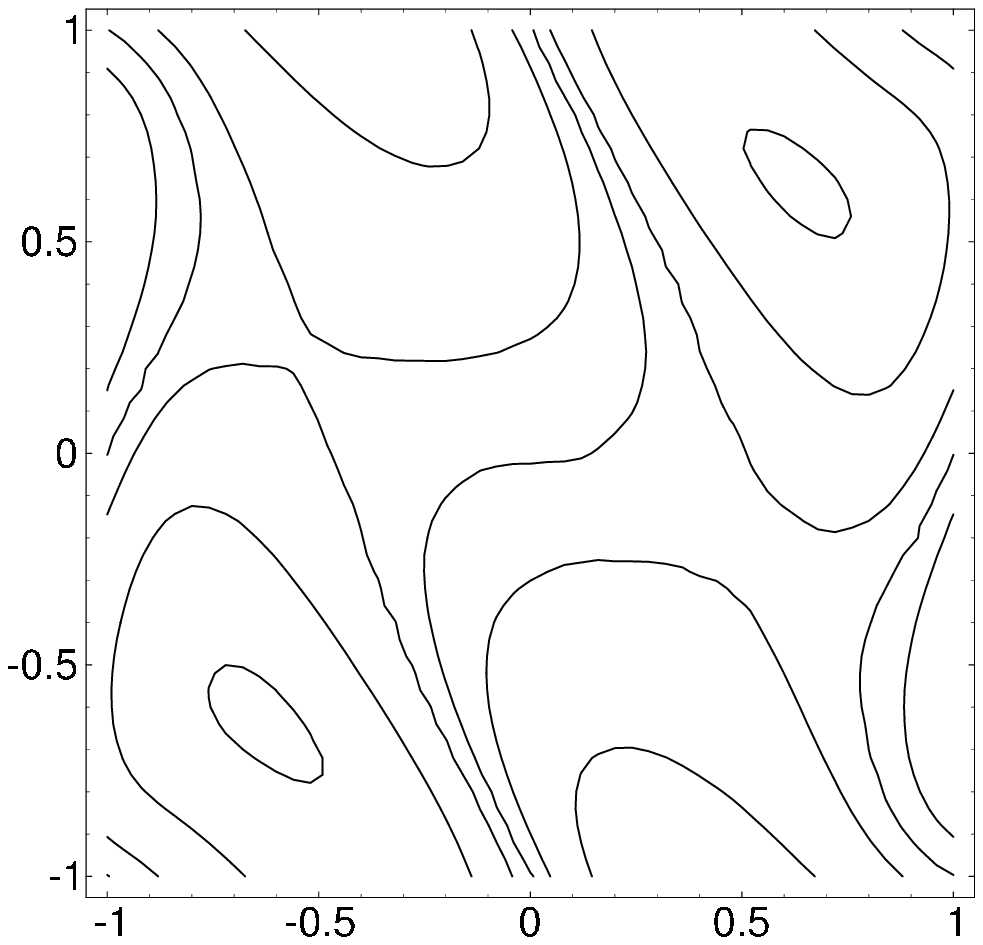}}
	\put(3.5,7.4){\fbox{${\mathcal A}_{II}$ in \% }}
	\put(6.5,-.3){$\varphi_{\mu}[\pi]$}
	\put(0.3,7.3){$ \varphi_{M_1}[\pi]$ }
	\put(1.0,1.0){\scriptsize 0}
	\put(1.2,1.5){\scriptsize 3}
	\put(1.8,1.75){\scriptsize 6.5}
	\put(1.7,4.0){\scriptsize 1}
	\put(1.6,4.8){\scriptsize 0}
	\put(1.,5.3){\scriptsize -1}
	\put(3.15,5.9){\scriptsize 3}
	\put(3.2,4.6){\scriptsize 1}
	\put(3.9,3.8){\scriptsize 0}
	\put(4.4,2.7){\scriptsize -1}
	\put(4.5,1.4){\scriptsize -3}
	\put(6.6,6.4){\scriptsize -1}
	\put(6.2,6.1){\scriptsize -3}
	\put(5.65,5.55){\scriptsize -6.5}
	\put(6.2,4.3){\scriptsize -3}
	\put(6.0,3.4){\scriptsize -1}
	\put(6.1,2.3){\scriptsize 0}
	\put(6.7,2.0){\scriptsize 1}
\put(0.5,-.3){Fig.~\ref{W:plot7}a}
	\put(8,0){\includegraphics{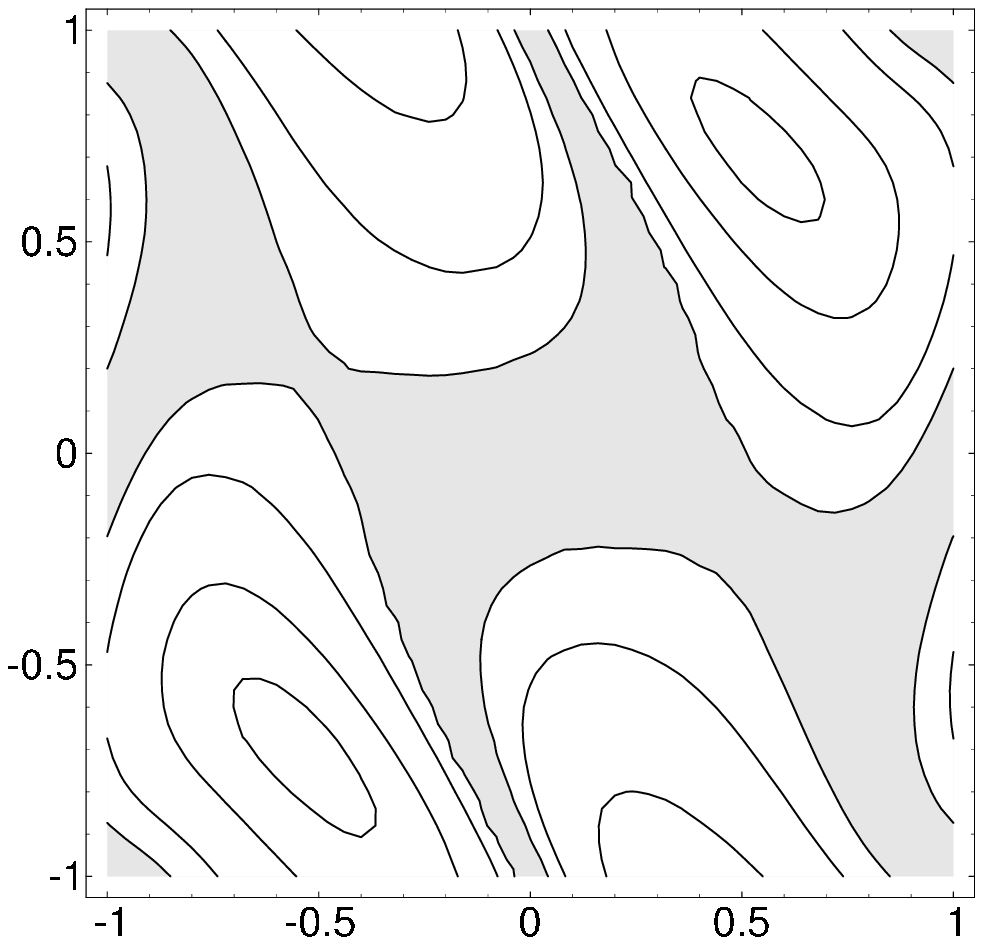}}
	\put(10.3,7.4){\fbox{$S_{II} =|{\mathcal A}_{II}| 
			\sqrt{2{\mathcal L}\cdot\sigma}$  }}
	\put(14.5,-.3){$ \varphi_{\mu}[\pi]$ }
	\put(8.5,7.4){$ \varphi_{M_1}[\pi]$ }
	\put(10.,1.8){\scriptsize 6}
	\put(9.7,2.5){\scriptsize 4}
	\put(9.5,3.3){\scriptsize 2}
	\put(9.0,5.3){\scriptsize 2}
	\put(11.2,5.2){\scriptsize 2}
	\put(11.0,6.3){\scriptsize 4}
	\put(13.3,5.9){\scriptsize 6}
	\put(14.,4.8){\scriptsize 4}
	\put(14.2,4.0){\scriptsize 2}
	\put(12.3,2.1){\scriptsize 2}
	\put(12.6,1.){\scriptsize 4}
	\put(14.75,1.9){\scriptsize 2}
\put(8.5,-.3){Fig.~\ref{W:plot7}b}
\end{picture}
\vspace*{.5cm}
\caption{
	Contour lines of 
	and the asymmetry ${\mathcal A}_{II}$ (\ref{W:plot7}a)
	and the significance $S_{II}$ (\ref{W:plot7}b)
	for $e^+e^-\to\tilde\chi^+_1\tilde\chi^-_2;~ 
	\tilde\chi^+_1 \to W^+\tilde\chi^0_1 ;~W^+ \to c \bar s$,
	in the $\varphi_{\mu} $--$\varphi_{M_1}$ plane for 
	$|\mu|=350$~GeV, $M_2=250$~GeV, $\tan \beta=5$, 
	$m_0=300$ GeV, $\sqrt{s}=800$ GeV,
	$(P_{e^-},P_{e^+})=(-0.8,0.6)$
	and ${\mathcal L}=500~{\rm fb}^{-1}$.
	In the gray shaded area of Fig.~\ref{W:plot7}b
	we have $S_{II}<1$.
	\label{W:plot7}}
\end{figure}

The asymmetry ${\mathcal A}_{II}$  is also sensitive to the 
phase  $\varphi_{M_1}$. 
We show the $\varphi_{\mu}$--$\varphi_{M_1}$ dependence
of  ${\mathcal A}_{II}$, choosing the parameters as above, 
in Fig.~\ref{W:plot7}a. In Fig.~\ref{W:plot7}b we show the contour 
lines of the significance
$S_{II} =|{\mathcal A}_{II}| \sqrt{2{\mathcal L}\cdot\sigma}$
for ${\mathcal L}= 500$ fb$^{-1}$. For 
$(\varphi_{M_1},\varphi_{\mu}) \approx ( \pi,0.1 \pi)$
we have $S_{II} \approx 2.4$ and thus ${\mathcal A}_{II}$ could be  
accessible even for small phases by using polarized beams.
%&&&&&&&&&&&&&&&&&&&&&&&&&&&&&&&&&&&&&&&&&&&&&&&&&&&&&&&&&&&&&&
%                    P L O T  7 
%&&&&&&&&&&&&&&&&&&&&&&&&&&&&&&&&&&&&&&&&&&&&&&&&&&&&&&&&&&&&&&6
%

\subsection{Summary of Section \ref{CP observables in chargino 
		production and decay into the W boson}
	\label{W:Summary}}

We have analyzed CP sensitive observables in 
chargino production, $e^+e^- \to\tilde\chi^+_i  \tilde\chi^-_j$,
with subsequent two-body decay, $\tilde\chi^+_i \to W^+\chi^0_n$.
We have defined the CP asymmetry $ {\mathcal A}_{I}$ of the triple product 
${\mathbf p}_{e^-}\cdot({\mathbf p}_{\tilde\chi^+_i} \times {\mathbf p}_{W})$.
In the MSSM with complex parameters $\mu$ and $M_1$,
we have shown that ${\mathcal A}_{I}$ can reach $4\%$ 
and that even for $\varphi_{\mu}\approx0.9 \pi$ the asymmetry could be 
accessible in the process $e^+e^- \to\tilde\chi^+_1  \tilde\chi^-_2$.
Further we have analyzed the CP sensitive density-matrix
elements of the $W$ boson. 
%The CP-odd vector component $V_2$ 
%is sensitive to $\varphi_{\mu}$ and also to $ \varphi_{M_1}$.
The phase $ \varphi_{M_1}$ enters in the decay 
$\tilde\chi^+_i \to W^+\chi^0_n $
due to correlations of the chargino and the $W$ boson spins,
which can be probed via the hadronic decay 
$W^+ \to c \bar s$. Moreover the triple product 
${\mathbf p}_{e^-}\cdot({\mathbf p}_{c} \times {\mathbf p}_{ \bar s})$
defines the CP asymmetry $ {\mathcal A}_{II}$, 
which can be as large as $7\%$ for 
$\tilde\chi^+_1  \tilde\chi^-_1$ or
$\tilde\chi^+_1  \tilde\chi^-_2$ production. 
By analyzing the statistical errors of ${\mathcal A}_{I}$ and 
${\mathcal A}_{II}$ we found that 
the phases $\varphi_{\mu}$ and $ \varphi_{M_1}$
could be strongly constrained at a future  $e^+e^-$ collider 
with $\sqrt{s}= 800$~GeV, high luminosity and 
longitudinally polarized beams.

	\chapter{CP violation in sfermion decays
  \label{CP violation in sfermion decays}}

%
%{\Large\bf Overview}\\
%\vspace{0.3cm}
%
%
%We study sfermion decays 
%$\tilde f \to f~\tilde\chi^0_1~ \ell~\bar \ell$ or 
%$\tilde f \to f~\tilde\chi^0_1~ q~\bar q$, 
%and define T-odd asymmetries which are based on triple products 
%of the outgoing fermion momenta. 
%%We study this asymmetry in the MSSM with complex parameters.
%The leading contribution to the asymmetries stem from the decay chain
%$\tilde f\to f~\tilde\chi^0_j\to f~\tilde\chi^0_1~Z\to 
%f~\tilde\chi^0_1~\ell~ \bar \ell~~ (f~\tilde\chi^0_1~q ~\bar q)$
%%for which the asymmetries are proportional to the imaginary
%%part of a product of the $\tilde\chi^0_j$-$Z$-$\tilde\chi^0_1$
%%couplings. 
%and are sensitive ot CP phases of $M_1$ and $\mu$.
%%for which we obtain analytic
%%formulae for the amplitude squared. 
%
%In the numerical study we estimate the event rates 
%necessary to measure the asymmetries,
%which can reach $3 \%$ for leptonic decays
%$\tilde \tau_1\to \tau~\tilde\chi^0_1~ \ell~\bar \ell$, and 
%$20 \%$ for semi-leptonic decays like
%$\tilde \tau_1\to \tau~\tilde\chi^0_1 ~b~\bar b$.
%\vspace{0.3cm}

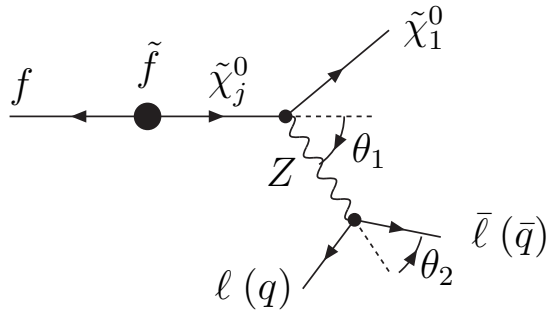
\begin{figure}	
	\setlength{\unitlength}{0.035cm}
%\fbox{
\scalebox{1.3}{
%\scalebox{1.0}{
\begin{picture}(190,95)(-85,-20)
%\begin{picture}(190,95)(0,0)
\ArrowLine(40,50)(0,50)
\put(115,73){$\tilde{\chi}_1^0$}
\Vertex(40,50){4}
%\put(102,97){$\tilde l_1 $}
\put(37,61){$\tilde f $}
\ArrowLine(40,50)(80,50)
\put(0,56){$ f $}
\ArrowLine(80,50)(110,75)
\put(58,58){$ \tilde{\chi}_j^0 $}
\DashLine(80,50)(105,50){1.5}
\ArrowArcn(80,50)(17,0,305)
\put(100,37){$ \theta_1$}
\Photon(80,50)(100,20){2}{5}
\put(75,30){$ Z $}
\Vertex(80,50){2}
\ArrowLine(100,20)(125,15)
%\ArrowLine(100,20)(125,15)
\put(135,12){$\bar \ell~(\bar q)$}
\ArrowArc(100,20)(20,310,346)
\put(120,3){$ \theta_2$}
\ArrowLine(100,20)(85,0)
\put(60,-3){$ \ell ~(q) $}
\DashLine(100,20)(110,5){1.5}
\Vertex(100,20){2}
\end{picture}
}
%}
\caption{Schematic picture of the sfermion decay process.
%	$\tilde f \to f \tilde\chi^0_j,~
%\tilde\chi^0_j \to Z\tilde\chi^0_1,~
%	Z \to \ell \bar\ell ~(q \bar q)$ 
%in the $\tilde f$ rest frame.
\label{picstaudec}}
\end{figure}	

%\section{Introduction}

%We study the decay chain of subsequent two-body decays of a sfermion
For the sfermion decays
\begin{eqnarray}\label{eq:decaychain}
			\tilde f &\to& f+ \tilde\chi^0_j; \quad
	\tilde\chi^0_j \to Z +\tilde\chi^0_1; \quad
						Z \to \ell+\bar \ell \quad \ell =e,\mu,\tau, \\
\tilde f &\to& f+ \tilde\chi^0_j; \quad
	\tilde\chi^0_j \to Z +\tilde\chi^0_1; \quad
						Z \to q+\bar q, \quad q = b,c,
\end{eqnarray}
%with $\ell=e,\mu,\tau$ and $q$ denotes a quark.
%The process is shown schematically in Fig.~(\ref{picstaudec}).
schematically shown in Fig.~(\ref{picstaudec}),
the triple product of the momenta of the outgoing leptons  
\begin{equation}\label{eq:tripleprodlep}
	{\mathcal T}_{\ell}=
	{\bf p}_{f}\cdot({\bf p}_{\ell}\times{\bf p}_{\bar \ell}),
\end{equation}
and that of the outgoing quarks,
\begin{equation}\label{eq:tripleprodquark}
	{\mathcal T}_q=
	{\bf p}_{f}\cdot({\bf p}_{q}\times{\bf p}_{\bar q}),
\end{equation}
define T-odd asymmetries 
\begin{equation}\label{eq:Toddasym}
	{\mathcal A}^{\rm T}_{\ell,q}=
	\frac{\Gamma({\mathcal T}_{\ell,q}>0)-\Gamma({\mathcal T}_{\ell,q}<0)}
	{\Gamma({\mathcal T}_{\ell,q}>0)+\Gamma({\mathcal T}_{\ell,q}<0)}
%= \frac{\int {\rm Sign}[{\mathcal T}_{\ell,q}>0]\;|
%	T_{\ell,q}|^2d{\rm Lips}}
%	{\int |T_{\ell,q}|^2\,d{\rm Lips}},
\end{equation}
%where $T_{\ell,q}$ is the matrix element
of the partial sfermion decay width $\Gamma$
%$\Gamma_{\ell,q}$
for the process~(\ref{eq:decaychain}). The asymmetries \cite{Bartl:2003ck}
%triple products~(\ref{eq:tripleprodlep})
%and~(\ref{eq:tripleprodquark})
are sensitive to correlations between the $\tilde\chi^0_j$ polarization 
and the $Z$ boson polarization, which are encoded in the momenta
of the final leptons or quarks.
The correlations thus would vanish if a scalar particle in 
place of the $Z$ boson is exchanged.
%Final state interactions may also contribute to 
%${\mathcal A}^{\rm T}_{\ell,q}$, however, they arise
%only at one loop level. 
The tree-level contribution to the asymmetries~(\ref{eq:Toddasym})
%triple products~(\ref{eq:tripleprodlep})
%and~(\ref{eq:tripleprodquark}) 
%are proportional to the imaginary
%part of the $\tilde\chi^0_j$-$Z$-$\tilde\chi^0_1$ coupling squared
%and are sensitive to the phases of $M_1$ and $\mu$.
are proportional to the imaginary part of a product of the 
$\tilde\chi^0_j$-$Z$-$\tilde\chi^0_1$ couplings. 
However, they are not  sensitive to the phase 
$\varphi_{A_f}$ of the trilinear scalar coupling parameter $A_f$, 
since the decay $\tilde f\to f\tilde\chi^0_j$~(\ref{eq:decaychain}) 
is a two-body decay of a scalar particle.
As an observable in the process~(\ref{eq:decaychain})
which is sensitive to $\varphi_{A_f}$, one would have to measure the
transverse polarization of the fermion $f$, which is possible for 
$f=\tau, t$ \cite{nojiri}.
Also three-body decays of the sfermion $\tilde f$
can be studied alternatively \cite{Bartl:2002hi}.

In the numerical study we estimate the event rates 
necessary to measure the asymmetries,
which can reach $3 \%$ for leptonic decays
$\tilde \tau_1\to \tau~\tilde\chi^0_1~ \ell~\bar \ell$, and 
$20 \%$ for semi-leptonic decays like
$\tilde \tau_1\to \tau~\tilde\chi^0_1 ~b~\bar b$.

The triple product~(\ref{eq:tripleprodlep}) was 
proposed in \cite{OshimoSfermion} and the size of the asymmetry was calculated
for the decay $\tilde\mu\to\mu~\tilde\chi^0_2\to
\tilde\chi^0_1~\mu~\ell~\bar\ell$, however, for a
specific final state configuration only.
We extend the work of \cite{OshimoSfermion}
by calculating the asymmetries~(\ref{eq:Toddasym}) 
in the entire phase space.

\section{Sfermion decay width}

For the calculation of the amplitude squared of the 
subsequent two-body decays of the sfermion~(\ref{eq:decaychain}), 
we use the spin-density matrix formalism of \cite{spinhaber}: 
\begin{equation}\label{eq:matrixelement}
	|T|^2=|\Delta(\tilde\chi^0_j)|^2~|\Delta(Z)|^2
	\sum_{\lambda_j,\lambda'_j,\lambda_k,\lambda'_k}~
	\rho_{D_1}(\tilde f)_{\lambda_j\lambda'_j}~
	\rho_{D_2}(\tilde\chi^0_j)^{\lambda'_j\lambda_j}_{\lambda_k\lambda'_k}~
	\rho_{D_3}(Z)^{\lambda'_k\lambda_k}.
\end{equation}
The amplitude squared is composed of 
the propagators $\Delta(\tilde\chi^0_j), \Delta(Z)$, the unnormalized 
spin density matrices 
$\rho_{D_1}(\tilde f)$, $\rho_{D_2}(\tilde\chi^0_j)$ and 
$\rho_{D_3}(Z)$, with the helicity indices $\lambda_j,\lambda'_j$ 
of the neutralino and/or the helicity indices $\lambda_k,\lambda'_{k}$ 
of the $Z$ boson. Introducing a set of spin four-vectors 
$s^a_{\chi_j^0},$ $a=1,2,3,$ for the neutralino $\tilde\chi^0_j$, 
see Appendix~(\ref{sfermion:spinchi}), the density matrices can be 
expanded in terms of the Pauli matrices
\begin{eqnarray} \label{eq:rhoD1}
\rho_{D_1}(\tilde f)_{\lambda_j\lambda'_j}&=&\delta_{\lambda_j\lambda'_j}~D_1+
	\sigma^a_{\lambda_j\lambda'_j}~\Sigma^a_{D_1}, \\
\rho_{D_2}(\tilde\chi^0_j)^{\lambda'_j,\lambda_j}_{\lambda_k,\lambda'_k}&=&
	\left[\delta_{\lambda'_j\lambda_j}~D_2^{\mu\nu}+\sigma^b_{\lambda'_j\lambda_j}~
		\Sigma^{b\,\mu\nu}_{D_2}\right] \varepsilon^{\lambda_k\ast}_{\mu}
		\varepsilon^{\lambda'_k}_{\nu},\label{eq:rhoD2}\\
\rho_{D_3}(Z)^{\lambda'_k\lambda_k}&=&
	D_3^{\rho\sigma}~\varepsilon^{\lambda'_k\ast}_{\sigma}
	\varepsilon_{\rho}^{\lambda_k}.\label{eq:rhoD3}
\end{eqnarray}
The polarization vectors 
$\varepsilon^{\lambda_k}_{\mu}$ of the $Z$ boson
obey $p^{\mu}_Z\varepsilon^{\lambda_k}_{\mu}=0$ and the completeness relation
$\sum_{\lambda_k} \varepsilon^{\lambda_k\ast}_{\mu}
\varepsilon^{\lambda_k}_{\nu}= -g_{\mu\nu}+p_{Z \mu}p_{Z \nu}/m_Z^2$.
The expansion coefficients of the density 
matrices~(\ref{eq:rhoD1})-(\ref{eq:rhoD3}) are
\begin{eqnarray} 
D_1 &=& (|a^{\tilde f}_{kj}|^2+|b^{\tilde f}_{kj}|^2)(p_{f'} 
		\cdot p_{\chi_j^0}), \label{eq:D1} \\ [2mm]
\Sigma^a_{D_1} &=& \pm m_{\chi_j^0} (|a^{\tilde f}_{kj}|^2-|b^{\tilde f}_{kj}|^2)
		(p_{f'}\cdot s^a_{\chi_j^0}), \label{eq:sigmaD1} \\ [2mm]
{D_2}_{\rho\sigma}&=& \frac{4~g^2}{\cos^2\theta_W}\Big\{
		 g_{\rho\sigma}\left[2~{\rm Re}
		(O''^L_{1j} {O''^R_{1j}}^{\ast})
		m_{\chi_1^0} m_{\chi_j^0} -(|O''^L_{1j}|^2+|O''^R_{1j}|^2)
		(p_{\chi_1^0}\cdot p_{\chi_j^0})\right] \nonumber \\ [2mm]
&&{}+ (|O''^L_{1j}|^2+|O''^R_{1j}|^2)(p_{\chi_j^0\rho}p_{\chi_1^0\sigma}+
	{p_{\chi_j^0}}_{\sigma}p_{\chi_1^0\rho})\Big\}, \label{eq:D2} \\ [2mm]
\Sigma^{a}_{D_2\rho\sigma} &=& \frac{4~i~g^2}{\cos^2\theta_W}\Big\{ 
	2~m_{\chi_1^0} 
	{\rm Im}(O''^L_{1j}{O''^R_{1j}}^{\ast})(p_{\chi_j^0\rho}~
	s^a_{\chi_j^0\sigma}-p_{\chi_j^0\sigma}~
	s^a_{\chi_j^0\rho})  \nonumber \\ [2mm]
&&{} -\varepsilon_{\rho\sigma\mu\nu}~
	p_{\chi_1^0}^{\mu}s^{a\nu}_{\chi_j^0}m_{\chi_j^0}(|O''^L_{1j}|^2+
	|O''^R_{1j}|^2){}\nonumber\\ [2mm]
&&{}+2~\varepsilon_{\rho\sigma\mu\nu}~
	p_{\chi_j^0}^{\mu}s^{a\nu}_{\chi_j^0}m_{\chi_1^0}{\rm Re}
	(O''^L_{1j}{O''^R_{1j}}^{\ast})\Big\},\label{eq:S2}\\ [2mm]
D_3^{\rho\sigma}&=&\frac{2~g^2}{\cos^2\theta_W}\Big\{-~g^{\rho\sigma}
	(L_f^2+R_f^2)(p_{ f}\cdot p_{\bar f})
	+(p^{\rho}_{f}~p^{\sigma}_{\bar f}+p^{\rho}_{\bar f}~p^{\sigma}_{f})
	(L_f^2+R_f^2) \nonumber\\[2mm]
	&&{} -i~(R_f^2-L_f^2)~\varepsilon^{\rho\sigma\mu\nu}~
	p_{f\mu}~p_{\bar f \nu}\Big\},\label{eq:D3}
\end{eqnarray}
with $\varepsilon_{0123}=1$ 
and the couplings as defined in Appendix~\ref{Lagrangian}.
The negative sign in~(\ref{eq:sigmaD1}) holds for the decay of a 
negatively charged sfermion.
In~(\ref{eq:D1}) and~(\ref{eq:sigmaD1}), $f'$ denotes the fermion
from the first decay $\tilde f \to f'~\tilde\chi^0_j$ 
in~(\ref{eq:decaychain}).
Inserting the density matrices~(\ref{eq:rhoD1})-(\ref{eq:rhoD3})
in~(\ref{eq:matrixelement}), we obtain for the amplitude squared
\begin{equation}\label{eq:matrixelement2}
	|T|^2=2~|\Delta(\tilde\chi^0_j)|^2~|\Delta(Z)|^2~
	\lbrace D_1~D_{2 \rho\sigma}+
	\Sigma^a_{D_1}~\Sigma^a_{D_2\rho\sigma}\rbrace D_3^{\rho\sigma}.
\end{equation}
The $\tilde f$ decay width for the decay 
chain~(\ref{eq:decaychain}) is then given by 
\begin{equation}\label{eq:width}
\Gamma(\tilde f \to f'~\tilde\chi^0_1~f~\bar f)=
	\frac{1}{2 m_{\tilde f}}\int|T|^2
	d{\rm Lips}(m^2_{\tilde f};p_{f'},p_{\chi_1^0},p_{\bar f},p_{f}),
\end{equation}
with the phase-space element $d{\rm Lips}$ defined in 
Appendix~\ref{Phase space for sfermion decays}.

\section{T-odd asymmetry
			\label{sferm:T-odd asymmetry}}

In the following we present in some detail the calculation of the
T-odd asymmetry~(\ref{eq:Toddasym}) for the slepton decays
$\tilde \ell\to \ell ~\tilde\chi^0_j\to \ell~ \tilde\chi^0_1~
Z\to \ell~\tilde\chi^0_1~f~\bar f$.
The replacements that must be made to obtain the asymmetry
for $\tilde q$ decays are obvious.
From~(\ref{eq:Toddasym}) and~(\ref{eq:matrixelement2}) we find
%for the asymmetry
%\begin{equation}\label{eq:Adependence}
%{\mathcal A}^{\rm T}_{\ell,q}=
%\frac{\int |\Delta (\tilde\chi^0_j)|^2 |\Delta (Z)|^2
%	{\rm Sign}[{\mathcal T}_{\ell,q}]
%	\Sigma^a_{D_1}~\Sigma^a_{D_2\rho\sigma}D_3^{\rho\sigma}
%d{\rm Lips}  }
%{\int |\Delta (\tilde\chi^0_j)|^2 |\Delta (Z)|^2
%	D_1~D_{2 \rho\sigma} D_3^{\rho\sigma}
%d{\rm Lips} },
%\end{equation}
\begin{equation}\label{eq:Adependence}
{\mathcal A}^{\rm T}_{\ell,q}=
\frac{\int {\rm Sign}[{\mathcal T}_{\ell,q}]
	\Sigma^a_{D_1}~\Sigma^a_{D_2\rho\sigma}D_3^{\rho\sigma}
d{\rm Lips}  }
{\int D_1~D_{2 \rho\sigma} D_3^{\rho\sigma}
d{\rm Lips} },
\end{equation}
where we have already used the narrow width approximation
for the propagators. In the numerator we have used 
%$ \int |\Delta (\tilde\chi^0_j)|^2 |\Delta (Z)|^2
%{\rm Sign}[{\mathcal T}_{\ell,q}] 
%D_1~D_{2 \rho\sigma} \\D_3^{\rho\sigma}
%d{\rm Lips}=0 $
$ \int {\rm Sign}[{\mathcal T}_{\ell,q}] 
D_1~D_{2 \rho\sigma} D_3^{\rho\sigma}
d{\rm Lips}=0 $
and in the denominator 
%we have used
%$\int |\Delta (\tilde\chi^0_j)|^2 |\Delta (Z)|^2 
%\Sigma^a_{D_1}~\Sigma^a_{D_2\rho\sigma}D_3^{\rho\sigma}
%\\d{\rm Lips}=0$.
$\int \Sigma^a_{D_1}~\Sigma^a_{D_2\rho\sigma}D_3^{\rho\sigma}
d{\rm Lips}=0$.
%As it can be seen from~(\ref{eq:Adependence}), 
%The asymmetry ${\mathcal A}^{\rm T}_{\ell,q}$ is proportional to the
%spin correlation terms 
%$	\Sigma^a_{D_1}~\Sigma^a_{D_2\rho\sigma}D_3^{\rho\sigma}$.
Among the spin correlation terms 
$	\Sigma^a_{D_1}~\Sigma^a_{D_2\rho\sigma}D_3^{\rho\sigma}$
only those contribute to 
${\mathcal A}^{\rm T}_{\ell,q}$, which are proportional to the  
triple product ${\mathcal T}_{\ell,q}$
%~(\ref{eq:tripleprodlep}), 
%(\ref{eq:tripleprodquark}):
%The triple product is included in the product of the first 
%term of~(\ref{eq:S2}) and the last term of~(\ref{eq:D3}):
%These are the terms
\begin{equation}\label{eq:relpart}
\Sigma^{a}_{D_2\rho\sigma}D_3^{\rho\sigma}\supset
	32~m_{\chi_1} {\rm Im}(O''^L_{1j}{O''^R_{1j}}^{\ast})
	(R_f^2-L_f^2)
	\varepsilon^{\rho\sigma\mu\nu}~
	p_{\chi_j^0\rho}~s^a_{\chi_j^0\sigma}~p_{f\mu}~p_{\bar f\nu},
\end{equation}
see first term of~(\ref{eq:S2}) and the last term of~(\ref{eq:D3}).
From the explicit representations of the neutralino spin
vector~(\ref{sfermion:spinchi}) and the lepton momentum 
vector~(\ref{sfermion:chi}), we find  in the sfermion rest frame
$(p_\ell \cdot s^a_{\chi_j^0})=0$ for $a=1,2$ so that 
$\Sigma^{1,2}_{D_1}=0$ in~(\ref{eq:sigmaD1}). Thus only  
$\Sigma^{3}_{D_2\rho\sigma}$ %~(\ref{eq:S2}) 
contributes and the momentum dependent part
%In the sfermion rest frame, the term with the $\varepsilon$-tensor 
of~(\ref{eq:relpart}) can be written as
\begin{equation}\label{eq:epsilonpara}
	\varepsilon^{\rho\sigma\mu\nu}~
	p_{\chi_j^0\rho}~s^3_{\chi_j^0\sigma}~p_{f\mu}~p_{\bar f \nu}=
	m_{\chi_j^0}~{\bf \hat p}_{\ell}\cdot({\bf p}_{f}\times {\bf
			p}_{\bar f}),
	%=m_{\chi_j^0}~|{\bf p}_Z|~|{\bf p}_{\bar f}|
%\sin\theta_1\sin\theta_2\sin\phi_2,
\end{equation}
with ${\bf \hat p} = {\bf p}/|{\bf p}|$.
%Since $0\leq\theta_1,\theta_2\leq \pi$ the
%sign of the correlation 
%${\bf p}_{\ell}\cdot({\bf p}_{f}\times {\bf p}_{\bar f})$ is given
%by the sign of $\sin\phi_2$.
%Thus ${\mathcal T}_{\ell,q}>0$ corresponds to an integration
%$\int^{\pi}_{0} {\rm d}\phi_2$, while ${\mathcal T}_{\ell,q}<0$
%corresponds to an integration $\int^{2\pi}_{\pi} {\rm d}\phi_2$.
%We therefore integrate in~(\ref{eq:Adependence}) over the entire 
%phase space except for $\phi_2$:
%\begin{equation}\label{eq:toddasymangle}
%{\mathcal A}^{\rm T}_f=
%\frac{\left[\int^{\pi}_0\frac{{\rm d}\Gamma}{{\rm d}\phi_2}
%		-\int^{2\pi}_{\pi}\frac{{\rm d}\Gamma}{{\rm d}\phi_2}\right]
%{\rm d}\phi_2}
%{\left[\int^{\pi}_0\frac{{\rm d}\Gamma}{{\rm d}\phi_2}
%		+\int^{2\pi}_{\pi}\frac{{\rm d}\Gamma}{{\rm d}\phi_2}\right]
%	{\rm d}\phi_2}~.
%\end{equation}

The dependence of ${\mathcal A}^{\rm T}_f$ 
on the $\tilde \ell_k$-$\ell$-$\tilde\chi^0_j$ couplings  
$a^{\tilde \ell}_{kj},b^{\tilde \ell}_{kj}$, 
on the $Z$-$\bar f$-$f$ couplings $L_f,R_f$ and on the 
$Z$-$\tilde \chi_1^0$-$\tilde \chi_j^0$ couplings $O''^{L,R}_{1j}$ 
follows  from~(\ref{eq:Adependence}) and~(\ref{eq:relpart}):
%is given by
\begin{equation}\label{eq:prop1}
{\mathcal A}_f^{\rm T}\propto 
	\frac{|a^{\tilde \ell}_{kj}|^2-|b^{\tilde \ell}_{kj}|^2}
	{|a^{\tilde \ell}_{kj}|^2+|b^{\tilde \ell}_{kj}|^2}~
	\frac{L_f^2-R_f^2}{L_f^2+R_f^2}~
	{\rm Im}(O''^L_{1j}{O''^R_{1j}}^{\ast}).
\end{equation}
%which follows  from~(\ref{eq:Adependence}) and~(\ref{eq:relpart}).
Due to the first factor
$\frac{|a^{\tilde \ell}_{kj}|^2-|b^{\tilde \ell}_{kj}|^2}
{|a^{\tilde \ell}_{kj}|^2+|b^{\tilde \ell}_{kj}|^2}$,
the asymmetry  will be strongly suppressed for  
$|a^{\tilde\ell}_{kj}| \approx |b^{\tilde\ell}_{kj}|$
and maximally enhanced for vanishing mixing 
in the slepton sector.
%$\frac{|a^{\tilde \ell}_{kj}|^2-|b^{\tilde \ell}_{kj}|^2}
%{|a^{\tilde \ell}_{kj}|^2+|b^{\tilde \ell}_{kj}|^2}\approx \pm1 $.
Due to the second factor
$\frac{L_f^2-R_f^2}{L_f^2+R_f^2}$,
the asymmetry ${\mathcal A}_{b(c)}^{\rm T}$ for hadronic decays 
of the $Z$ boson  is larger than the asymmetry ${\mathcal A}_\ell^{\rm T}$
for leptonic decays:
%${\mathcal A}_\ell^{\rm T}$ 
\begin{equation}\label{eq:prop2}
{\mathcal A}_{b(c)}^{\rm T}=
	\frac{L_{\ell}^2+ R_{\ell}^2} {L_{\ell}^2-R_{\ell}^2}
	\frac{L_{b(c)}^2- R_{b(c)}^2} {L_{b(c)}^2+R_{b(c)}^2}~
{\mathcal A}_{\ell}^{\rm T}\simeq 6.3~(4.5)
\times{\mathcal A}_{\ell}^{\rm T}.\nonumber
\end{equation}
%Note that the r.h.s of~(\ref{eq:relpart}) and therefore the asymmetry
%in~(\ref{eq:Toddasym}) vanish for $m_{\chi_1}\to 0$, which is related 
%to the fact that it is possible to redefine the Weyl spinor
%$\chi_1\to e^{i\alpha}\chi_1$ in this limit.
For the measurement of ${\mathcal A}_{f}^T$ 
the charges and the flavors of $f$ and $\bar f$
have to be  distinguished. For $f=e,\mu$ this will be 
possible on an event by event basis. 
For $f=\tau$ one has to take into account
corrections due to the reconstruction of the $\tau$ 
momentum. For $f=b,c$ the distinction of the quark 
flavors should be possible by flavor tagging 
\cite{flavortaggingatLC,Aubert:2002rg}. 
However, in this case the quark charges will  be distinguished
statistically for a given event sample only.
%Note that the asymmetry ${\mathcal A}_{f}^{\rm T}$ could in 
%principle also be studied in $ \tilde\chi^0_j$ three-body 
%decays if the two-body decays are kinematically forbidden,
%which will however be treated elsewhere \cite{karldipl}.

\section{Numerical results
         \label{sferm:numerics}}
		
We assume that $\tilde\tau_1$ is the lightest sfermion
and study the decay chain  
$\tilde \tau_1\to \tau\tilde\chi^0_j;\;
\tilde\chi^0_j\to \tilde\chi^0_1Z;\;
Z\to \ell \bar \ell$
for the two cases
$\tilde\tau_1\to\tau\tilde \chi^0_2 $ and
$\tilde\tau_1\to\tau\tilde \chi^0_3 $ separately.
We present numerical results for the T-odd asymmetry
${\mathcal A}_{\ell}^{\rm T}$~(\ref{eq:Toddasym})
and the branching ratios 
%for the $\tilde\tau_1$ decay chain
${\rm BR}(\tau_1 \to \tau~\tilde\chi^0_1~\ell~\bar \ell) 
:={\rm BR}(\tilde \tau_1 \to \tau~\tilde\chi^0_j)\times
{\rm BR}(\tilde\chi^0_j\to Z\tilde\chi^0_1)\times
{\rm BR}(Z \to \ell~\bar \ell)$.
The size of the asymmetry ${\mathcal A}_{b,c}^{\rm T}$ 
for hadronic decays may be obtained from~(\ref{eq:prop2}).

The relevant MSSM parameters are 
$|\mu|, \varphi_{\mu}, |M_1|, \varphi_{M_1}, M_2, \tan\beta,
|A_{\tau}|, \varphi_{A_{\tau}}, m_{\tilde\tau_1}$ and $m_{\tilde\tau_2}$. 
We fix $\tan\beta =10$, $|A_{\tau}|=1$~TeV, $\varphi_{A_{\tau}}=0$, 
$m_{\tilde\tau_1}=300$~GeV, $ m_{\tilde\tau_2}=800$~GeV
and use the relation $|M_1|=5/3\, M_2\tan^2\theta_W$
in order to reduce the number of parameters.

For the calculation of the branching ratios 
${\rm BR}(\tilde \tau_1 \to \tau~\tilde\chi^0_j)$ and
${\rm BR}(\tilde\chi^0_j\to Z\tilde\chi^0_1)$,
we include the decays
\begin{eqnarray}
	\tilde\tau_1 &\to& 
		\tau~\tilde\chi^0_j,~
		\tilde\chi^-_j\nu_{\tau},\\
\tilde\chi^0_j &\to& 
	Z\tilde\chi^0_1,~
	\tilde\chi^{\mp}_m W^{\pm},~
	\tilde\chi^0_n H_1^0,~ m=1,2,~ n<j.
\end{eqnarray}
%$\tilde\tau_1 \to \tilde\chi^-_j\nu_{\tau},
%\tilde\chi^0_j\to H_1^0 ~\tilde\chi^0_1,W^{\pm}~\tilde\chi^{\mp}_1$
%in addition to 
%$\tilde\tau_1 \to  \tilde\chi^0_j~\tau, 
%\tilde\chi^0_j \to Z ~\tilde\chi^0_1$.
%where $H_1^0$ is the lightest neutral Higgs boson.
We fix the Higgs mass parameter $m_{A}=800$~GeV 
so that the decays of the neutralino into charged Higgs bosons 
$\tilde\chi^0_j \to \tilde\chi^{\pm}_m H^{\mp}$,
as well as decays into heavy neutral Higgs bosons 
$\tilde\chi^0_j \to \tilde\chi^0_n~H_{2,3}^0$
are forbidden. The decays via sleptons 
%$\tilde\tau_1\to\tau~\ell~\tilde \ell$
$\tilde\chi^0_j \to \ell~\tilde \ell$
are forbidden due to our assumption that
$\tilde\tau_1$ is the lightest sfermion.

\subsection{Decay chain via $\tilde \tau_1\to \tau~\tilde\chi^0_2$}

\begin{table}[t]
 \label{tab:chi2}
\caption{
		Masses 
		%	of $\tilde\chi^0_i$ 
		and widths 
%	$\Gamma_{\chi^0_2}$, $\Gamma_{\tilde\tau_1}$
		for various combinations 
		of $\varphi_{\mu}$ and $\varphi_{M_1}$, for $|\mu|=300$~GeV, 
		$M_2=280$~GeV,  $\tan\beta =10$, $A_{\tau}=1$~TeV, 
		$m_{\tilde\tau_1}=300$~GeV, $ m_{\tilde\tau_2}=800$~GeV. 
%		for $M_{\tilde E} > M_{\tilde L}$.
}
\begin{center}
\begin{tabular}{|c|c|l|c|c|} \hline
$\varphi_{\mu}$ & $\varphi_{M_1}$ &
$m_{\chi^0_1},m_{\chi^0_2},m_{\chi^0_3},m_{\chi^0_4}  ~[\rm GeV]$  &
$\Gamma_{\chi^0_2}~[\rm MeV]$ & $\Gamma_{\tilde\tau_1}~[\rm MeV]$\\ \hline\hline
0 &            0    &$135,\;\;234,\;\;306,\;\;358$& $4.06$ & $527$ \\
0 & $\frac{\pi}{2}$ &$137,\;\;233,\;\;308,\;\;357$& $1.79$ & $550$ \\
0 & $ \pi   $       &$138,\;\;231,\;\;309,\;\;356$& $0.09$ & $573$ \\ \hline
$\frac{\pi}{2}$ & 0&$137,\;\;239,\;\;307,\;\;353$&$5.43$&$487$  \\
$\frac{\pi}{2}$ &$\frac{\pi}{2}$&$138,\;\;238,\;\;309,\;\;352$&$2.89$&$511$ \\
$\frac{\pi}{2}$ & $\pi$&$137,\;\;237,\;\;311,\;\;351$&$1.49$&$529$  \\\hline
$\pi$ &       0         & $138,\;\;245,\;\;309,\;\;347$&$7.25$&$448$ \\
$\pi$ & $\frac{\pi}{2}$ & $137,\;\;244,\;\;311,\;\;346$&$5.78$&$466$ \\
$\pi$ & $\pi$           & $136,\;\;243,\;\;313,\;\;345$&$4.32$&$484$ \\
\hline
 \end{tabular}
\end{center}
% \label{tab:chi2}
\end{table}

\begin{figure}[b]
\setlength{\unitlength}{0.035cm}
\begin{picture}(120,220)(12,0)
\put(10,0){\includegraphics{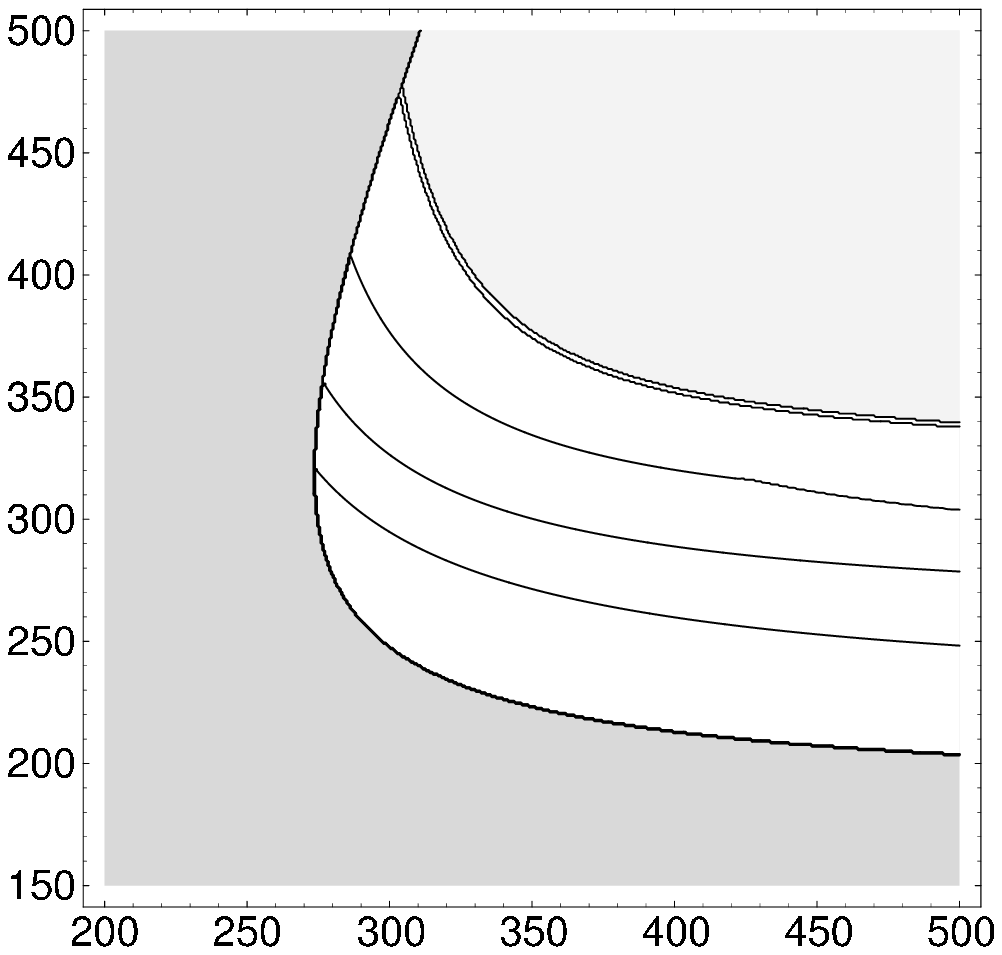}}
	\put(165,-8){$|\mu|~[{\rm GeV}]$}
	\put(10,210){$M_2 [{\rm GeV}]$}
	\put(70,210){\fbox{${\rm BR}(\tilde\tau_1 \;\to \;\tau \;
			\tilde\chi^0_1\;\ell \;\bar \ell)
%		,\;\;\ell = e, \mu, \tau \; \; \; 
		$ in \% }}
	\put(110,117){{\footnotesize $1$}}
	\put(135,93){{\footnotesize $2 $}}
	\put(160,75){{\footnotesize $2.5 $}}
	\put(30,-8){Fig.~\ref{stau2a}a}
\put(240,0){\includegraphics{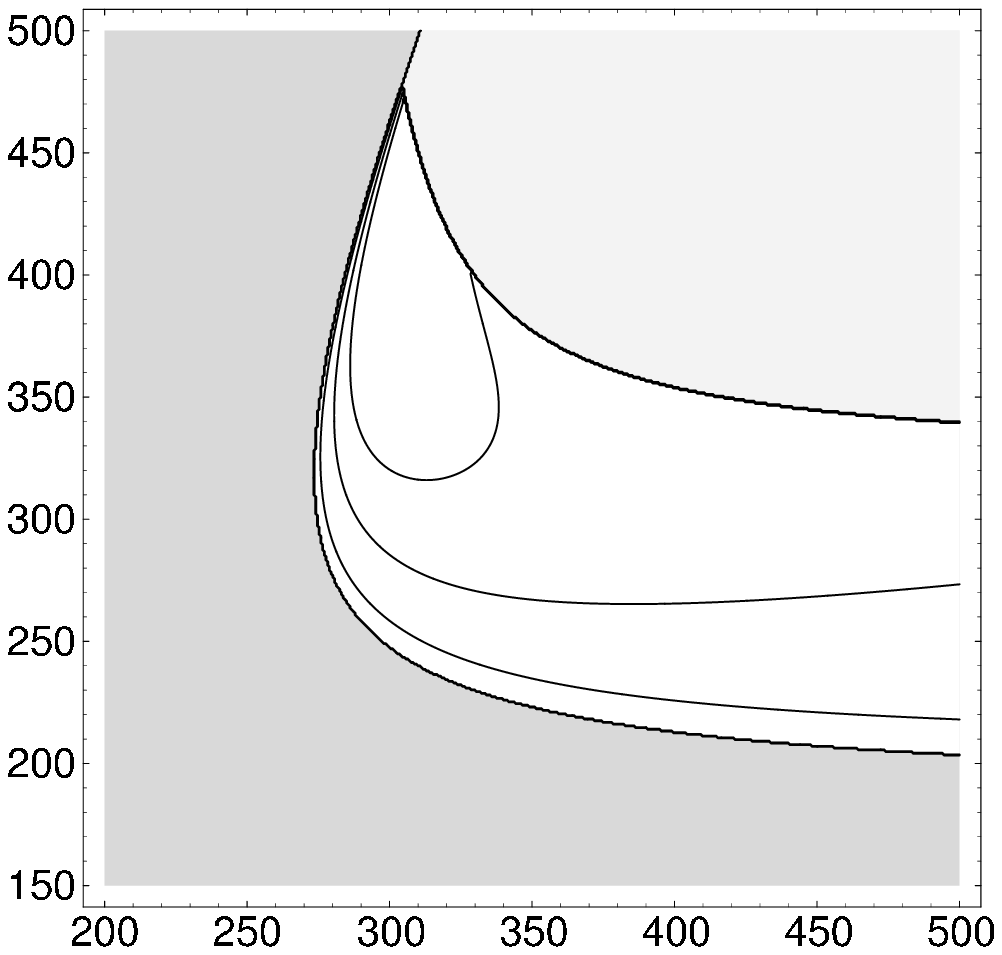}}
   \put(395,-8){$|\mu|~[{\rm GeV}]$}
	\put(240,210){$M_2 [{\rm GeV}]$}
	\put(320,210){\fbox{${\mathcal A}^{\rm T}_\ell$ in \% }}
   \put(323,106){{\footnotesize$3 $}}
   \put(350,80){{\footnotesize$2.5 $}}
	\put(380,60){{\footnotesize$1 $}}
	\put(260,-8){Fig.~\ref{stau2a}b }
\end{picture}
\vspace*{.3cm}
 \caption{Contour lines of the branching ratio 
	for $\tilde\tau_1\to \tilde\chi^0_1\tau \ell\bar\ell$ 
	and asymmetry ${\mathcal A}^{\rm T}_\ell$ in the $\mu$--$M_2$ plane for
	 $\varphi_{M_1}=\pi/2$ and $ \varphi_{\mu}=0$, taking
	$\tan\beta =10$, $A_{\tau}=1$~TeV, 
	$m_{\tilde\tau_1}=300$~GeV, $ m_{\tilde\tau_2}=800$~GeV
	for $M_{\tilde E} > M_{\tilde L}$. The gray areas
	are kinematically forbidden by 
	$m_{\tilde\tau_1}< m_{\chi^0_2}+m_{\tau}$ (light gray) or 
	$m_{\chi^0_2}<m_{\chi^0_1}+m_Z$ (dark gray).
\label{stau2a}}
\end{figure}

We study ${\mathcal A}_\ell^{\rm T}$ for the decay chain
$\tilde \tau_1\to \tau\tilde\chi^0_2;\;
\tilde\chi^0_2\to \tilde\chi^0_1Z;\;
Z\to \ell \bar \ell$ for $\ell=e,\mu,\tau$.
In Fig.~\ref{stau2a}a we show the contour
lines for the branching ratio 
${\rm BR}(\tau_1 \to \tau~\tilde\chi^0_1~\ell~\bar \ell) 
={\rm BR}(\tilde \tau_1 \to \tau~\tilde\chi^0_2)\times
{\rm BR}(\tilde\chi^0_2\to Z\tilde\chi^0_1)\times
{\rm BR}(Z \to \ell~\bar \ell)$, summed over 
$\ell=e,\mu,\tau$, in the $M_2$-$|\mu|$ plane
for $\varphi_{M_1}=\pi/2$ and $ \varphi_{\mu}=0$.
%We choose $M_{\tilde E} > M_{\tilde L}$ since  
%then\footnote{
%	We use the usual notation 
%	$M_{\tilde E}\equiv M_{R\tilde\tau}$, 
%	$M_{\tilde L}\equiv M_{L\tilde\tau}$, see~(\ref{eq:mll}) 
%	and~(\ref{eq:mrr}).
%	The ordering $M_{\tilde E} > M_{\tilde L}$ is suggested in 
%	some scenarios with non-universal scalar mass parameters at the GUT scale  
%	\cite{Baer:2000cb}. Furthermore, in~(\ref{eq:mll}) and (\ref{eq:mrr}) 
%	one could have $M_{\tilde\tau_{LL}}< M_{\tilde\tau_{RR}}$ 
%	in extended models with additional D-terms \cite{Hesselbach:2001ri}.
%} 
The $\tilde \tau_1$-$\tau$-$\tilde\chi^0_2$ coupling 
$|a^{\tilde\tau}_{12}|$ is larger, which implies a larger  
${\rm BR}(\tilde \tau_1 \to \tau~\tilde\chi^0_2)$ 
if we choose $M_{\tilde E} > M_{\tilde L}$.
We use the usual notation 
$M_{\tilde E}\equiv M_{R\tilde\tau}$, 
$M_{\tilde L}\equiv M_{L\tilde\tau}$, see~(\ref{eq:mll}) 
and~(\ref{eq:mrr}).
The ordering $M_{\tilde E} > M_{\tilde L}$ is suggested in 
some scenarios with non-universal scalar mass parameters at the GUT scale  
\cite{Baer:2000cb}. Furthermore, in~(\ref{eq:mll}) and (\ref{eq:mrr}) 
one could have $M_{\tilde\tau_{RR}}>M_{\tilde\tau_{LL}}$ 
in extended models with additional D-terms \cite{Hesselbach:2001ri}.

than for $M_{\tilde E} < M_{\tilde L}$. 
In a large region of the parameter
space ${\rm BR}(\tilde\chi^0_2\to Z\tilde\chi^0_1) = 1$.
The asymmetry ${\mathcal A}^{\rm T}_\ell$ is shown in Fig.~\ref{stau2a}b.
The dependence of ${\mathcal A}^{\rm T}_\ell$ on $M_2$ and $|\mu|$ 
is dominantly determined by 
${\rm Im}(O''^L_{12}{O''^R_{12}}^{\ast})$,
as expected from~(\ref{eq:prop1}).

\begin{figure}[t]
\setlength{\unitlength}{0.035cm}	
\begin{picture}(120,220)(12,0)
\put(10,0){\includegraphics{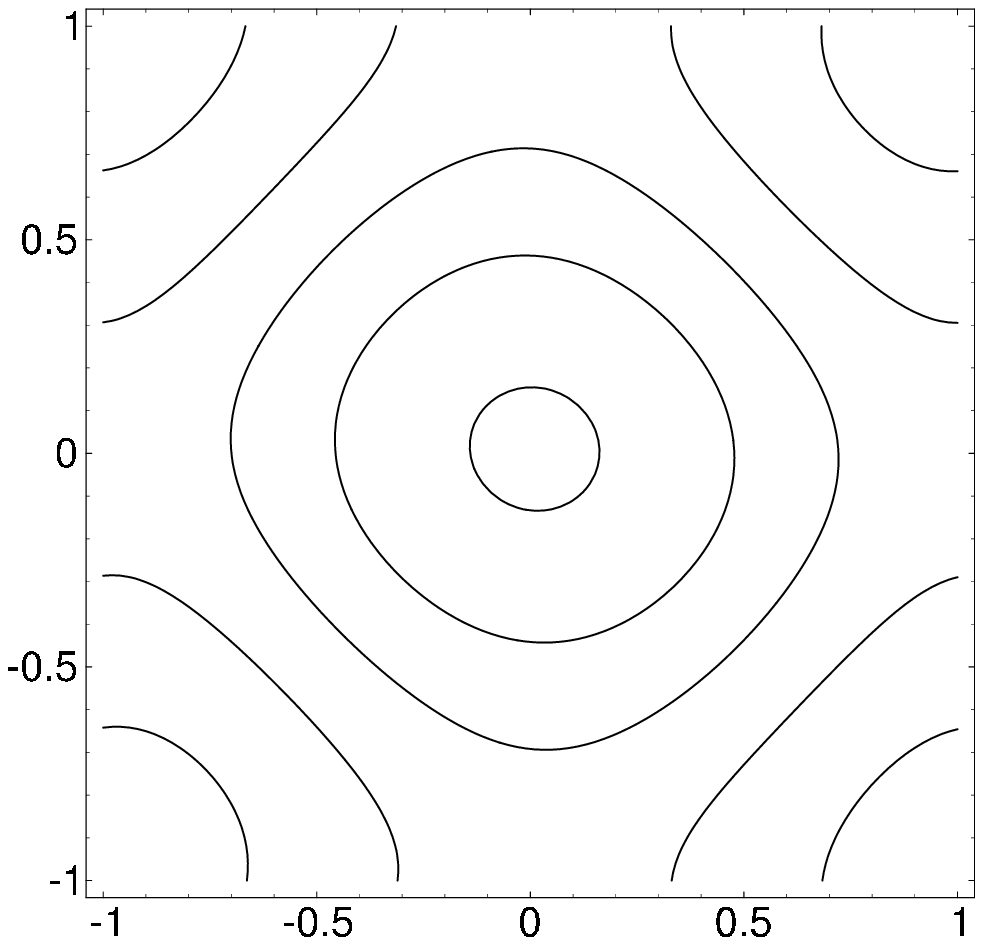}}
   \put(185,-8){$\varphi_{\mu}[\pi] $}
	\put(10,210){$\varphi_{ M_1}[\pi]$}
	\put(70,210){\fbox{${\rm BR}(\tilde\tau_1 \;\to \;\tau  \;
			\tilde\chi^0_1\;\ell \; \bar \ell)
%		,\;\;\ell = e, \mu, \tau \; \; \; 
		$ in \% }}
	\put(118,102){{\footnotesize $2.8 $}}
	\put(139,81){{\footnotesize $2.6 $}}
	\put(153,70){{\footnotesize$2.4 $}}
	\put(167,59){{\footnotesize$2.2 $}}
	\put(176,37){{\footnotesize$2.0 $}}
        \put(120,45){$\otimes$}
        \put(162,148){$\bullet$}
	\put(30,-8){Fig.~\ref{stau2b}a}
\put(240,0){\includegraphics{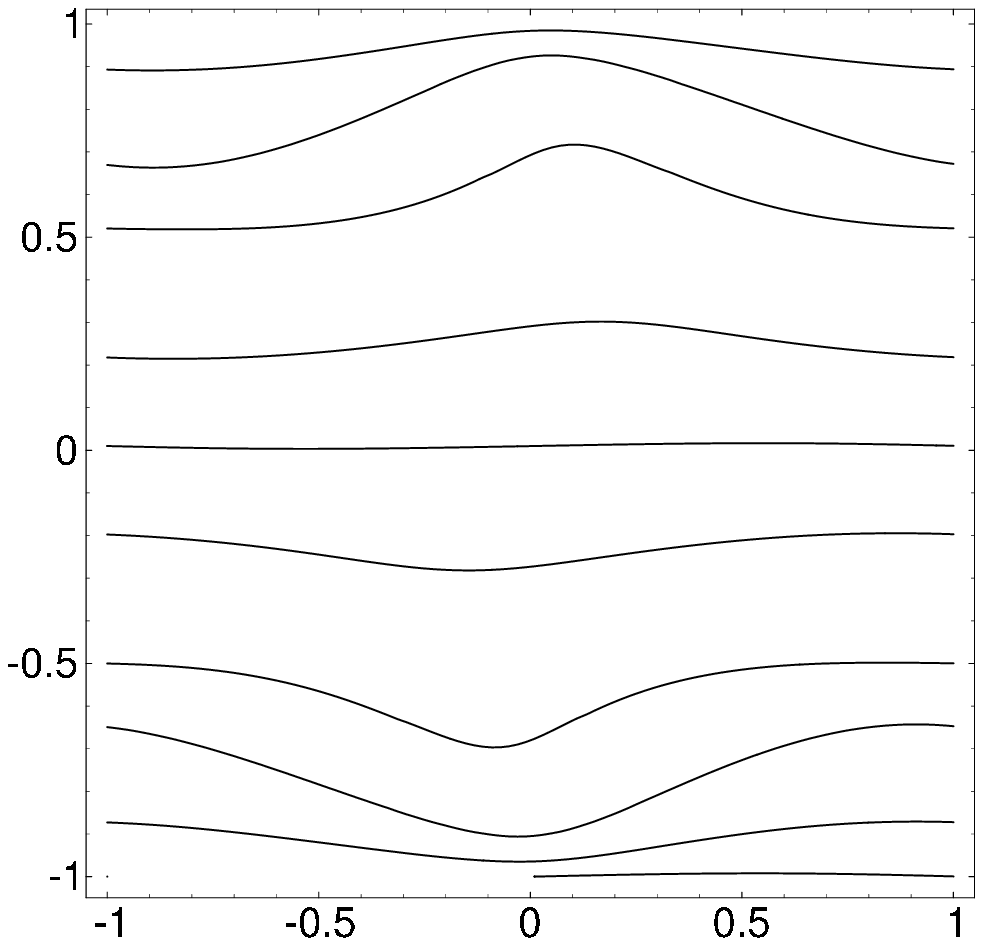}}
   \put(415,-8){$\varphi_{\mu}[\pi] $}
	\put(240,210){$\varphi_{ M_1}[\pi]$}
	\put(320,210){\fbox{${\mathcal A}^{\rm T}_\ell$ in \% }}
   \put(270,186){{\footnotesize $1.5 $}}
   \put(280,166){{\footnotesize $3 $}}
	\put(293,154){{\footnotesize $3 $}}
	\put(300,130){{\footnotesize $1.5 $}}
   \put(330,110){{\footnotesize $0 $}}
	\put(340,87){{\footnotesize $-1.5 $}}
	\put(353,56){{\footnotesize $-3 $}}
   \put(360,37){{\footnotesize $-3 $}}
	\put(395,33){{\footnotesize $-1.5 $}}
        \put(347,45){$\otimes$}
        \put(394,148){$\bullet$}
	\put(260,-8){Fig.~\ref{stau2b}b }
\end{picture}
\vspace*{.3cm}
\caption{Contour lines of the branching ratio 
	for $\tilde\tau_1\to \tilde\chi^0_1\tau \ell\bar\ell$ 
	and asymmetry ${\mathcal A}^{\rm T}_\ell$ in the 
	$\varphi_{\mu} $--$\varphi_{ M_1}$ plane 
	for $|\mu|=300$~GeV, $M_2=280$~GeV,
	taking $\tan\beta =10$, $A_{\tau}=1$~TeV, 
	$m_{\tilde\tau_1}=300$~GeV, $ m_{\tilde\tau_2}=800$~GeV
	for $M_{\tilde E} > M_{\tilde L}$.
%The points denoted by $\bullet$ and $\otimes$, respectively, 
%are for the theoretical estimate 
%of the necessary number of produced $\T_1$'s.
\label{stau2b}}
\end{figure}

In Fig.~\ref{stau2b} we show the $\varphi_{M_1}$ and 
$\varphi_{\mu}$ dependence 
of ${\rm BR}(\tilde\tau_1 \to \tau~\chi^0_1~\ell~\bar \ell)$
and of ${\mathcal A}_{\rm T}^\ell$ 
for $|\mu|=300$~GeV and $M_2=280$~GeV.
For these parameters, we also give in Table \ref{tab:chi2} 
the neutralino masses 
%of $\tilde\chi^0_i, i=1,\dots,4$ 
and the total neutralino and stau widths 
%$\Gamma_{\chi^0_2}$, $\Gamma_{\tilde\tau_1}$ 
for various phase combinations.
%The value of ${\mathcal A}^{\rm T}_\ell$ depends stronger on $\varphi_{M_1}$,
%which also determines the sign of ${\mathcal A}^{\rm T}_\ell$,
%than on $\varphi_{\mu}$.
Note that maximal phases 
$\varphi_{\mu},\varphi_{M_1}=\pm\pi/2$ do not lead to 
the highest value of ${\mathcal A}^{\rm T}_\ell$,
since the asymmetry is proportional to a product of CP even $(\Sigma^3_{D_1})$
and CP odd terms $(\Sigma^3_{D_2\rho\sigma}D_3^{\rho\sigma})$,
see~(\ref{eq:Adependence}).
We give a lower bound on the number $N$ of $\tilde\tau_1$'s 
to be produced at a linear collider, in order to 
measure ${\mathcal A}_{\ell }^{\rm T}$ at 1$\sigma$.
%From the relative statistical error of the asymmetry, see ~(\ref{errorofA}), 
We estimate $N=[({\mathcal A}_{\ell }^{\rm T})^2\times {\rm BR}]^{-1}$
from the relative statistical error of the asymmetry, 
see ~(\ref{errorofA}), with  
${\rm BR}={\rm BR}(\tau_1 \to \tau~\tilde\chi^0_1~\ell~\bar \ell)$.
%The relevant quantity to decide whether 
%${\mathcal A}^{\rm T}_\ell$ is observable at 1$\sigma$ is  
%$[({\mathcal A}_{\ell }^{\rm T})^2\times {\rm BR}]^{-1}$, where 
%${\rm BR}$ is the total decay branching ratio. 
%The number of produced $\tilde\tau_1$'s should then be 
%larger than $[({\mathcal A}_{\ell }^{\rm T})^2\times {\rm BR}]^{-1}$.
%As an example we take 
For the point 
$\varphi_{\mu}=\pi/2$ and $\varphi_{M_1} = \pi/2$,
marked by $\bullet$ in Fig.~\ref{stau2b},
${\rm BR} \approx 2.5\%$ and $|{\mathcal A}^{\rm T}_\ell|\approx 3\% $, 
so that $N\approx 4.4\times10^{5}$.
%implies that $[({\mathcal A}_{\ell}^{\rm T})^2\times {\rm BR}]^{-1}
%\approx 4.4\times10^{5}$.
For the decay $\tilde \tau_1\to b\bar b\tilde\chi^0_1\tau$, however,
${\rm BR} \approx 3.6 \%$ and 
$|{\mathcal A}^{\rm T}_b|\approx 19 \%$, 
so that only $N \approx 7.7\times 10^{2}$ $\tilde\tau_1$'s are
needed. We obtain almost the same results for
%comparison we consider 
smaller CP phases
$\varphi_{\mu}=0$ and $\varphi_{M_1}= -0.3 \pi$,
marked by $\otimes$ in Fig.~\ref{stau2b}. 
%we obtain almost the same results for $N$.
%$[({\mathcal A}_{\ell,b}^{\rm T})^2\times {\rm BR}]^{-1}$.
In these two examples ${\mathcal A}^{\rm T}_{\ell,q}$
should be measurable at an $e^+ e^-$ linear collider  with 
$\sqrt{s}=800$~GeV and an integrated luminosity of $500~{\rm fb}^{-1}$.
%for $m_{\tilde\tau_1}=300$~GeV.
% schon gesagt in intro
%It is clear that detailed Monte Carlo studies taking into
%account background and detector simulations are necessary to get a more 
%precise prediction of $N$.
%However, this is beyond the scope of the present work.

%\vspace{2cm}
%----------------------------------------------------------
%     S F E R M I O N    plot   -1-
%---------------------------------------------------------

%----------------------------------------------------------
%     S F E R M I O N    plot   -2-
%---------------------------------------------------------

\subsection{Decay chain via $\tilde \tau_1\to \tau~\tilde\chi^0_3$
\label{decaychain3}}

\begin{figure}[H]
\setlength{\unitlength}{0.035cm}
%\fbox{
\begin{picture}(120,220)(12,0)
\put(10,0){\includegraphics{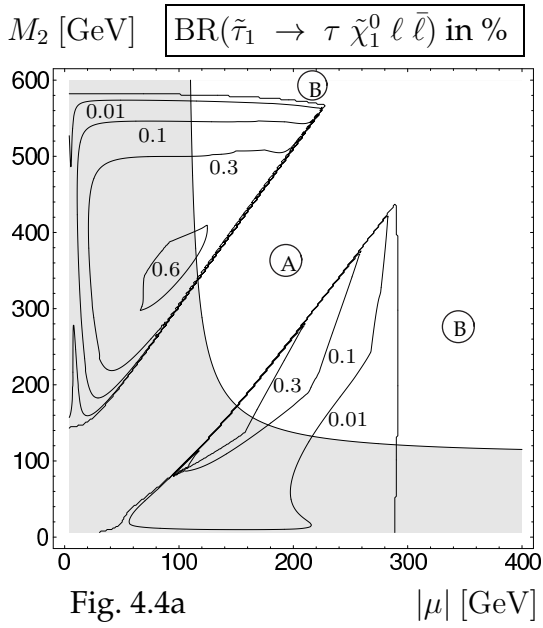}}
	\put(165,-8){$|\mu|~[{\rm GeV}]$}
	\put(10,210){$M_2~[{\rm GeV}] $}
	\put(70,210){\fbox{${\rm BR}(\tilde{\tau}_1 \;\to \;\tau  \;
			\tilde{\chi}^0_1 \;\ell  \;\bar \ell)
%		,\;\;\ell = e, \mu, \tau \; \; \; 
$ in \% }}
\CArc(125,191)(5.5,0,380)
\Text(127,191)[c]{{\scriptsize B}}
\CArc(115,125)(6,0,380)
\Text(117,125)[c]{{\scriptsize A}}
\CArc(180,100)(6,0,380)
\Text(182,100)[c]{{\scriptsize B}}
		\put(65,120){{\scriptsize $0.6 $}}
		\put(87,159){{\scriptsize $0.3 $}}
		\put(60,171){{\scriptsize $0.1 $}}
		\put(40,180){{\scriptsize $0.01 $}}
		\put(111,76){{\scriptsize$0.3 $}}
		\put(132,87){{\scriptsize$0.1 $}}
		\put(132,63){{\scriptsize$0.01 $}}
		\put(34,-8){Fig.~\ref{stau3a}a}
\put(240,0){\includegraphics{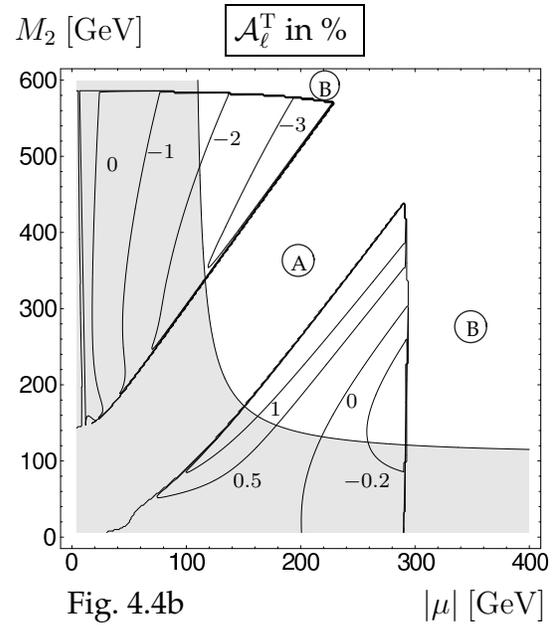}}
   \put(395,-8){$|\mu| ~[{\rm GeV}]$}
	\put(240,210){$M_2~[{\rm GeV}]$}
	\put(320,210){\fbox{${\mathcal A}^{\rm T}_\ell$ in \% }}
\CArc(355,191)(5.5,0,380)
\Text(358,191)[c]{{\scriptsize B}}
\CArc(345,125)(6,0,380)
\Text(348,125)[c]{{\scriptsize A}}
\CArc(410,100)(6,0,380)
\Text(414,100)[c]{{\scriptsize B}}
	\put(275,160){{\scriptsize $0$}}
   \put(290,165){{\scriptsize $-1$}}
	\put(315,170){{\scriptsize $-2 $}}
	\put(340,175){{\scriptsize $-3$}}
		\put(337,67){{\scriptsize $1 $}}
		\put(323,40){{\scriptsize $ 0.5$}}
		\put(366,70){{\scriptsize $0 $}}
		\put(365,40){{\scriptsize$-0.2 $}}
%		\put(295,27){$|a^{\tilde\tau}_{13}|=|b^{\tilde\tau}_{13}|$}
		\put(260,-8){Fig.~\ref{stau3a}b }
 \end{picture}
 %} 
\vspace*{.3cm}
\caption{Contour lines of the branching ratio 
	for $\tilde\tau_1\to \tilde\chi^0_1\tau \ell\bar\ell$ 
	and asymmetry ${\mathcal A}^{\rm T}_\ell$ in the 
	$|\mu|$--$M_2$ plane for $\varphi_{M_1}=\pi/2$,  
	$\varphi_{\mu}=0$, $\tan\beta =10$, $A_{\tau}=1$~TeV, 
	$m_{\tilde\tau_1}=300$~GeV, $ m_{\tilde\tau_2}=800$~GeV 
	and $M_{\tilde E} < M_{\tilde L}$. The area A (B) is 
	kinematically forbidden by  
	$m_{\tilde\chi_3^0}<m_{\tilde\chi_1^0}+m_Z$
	$(m_{\tilde\tau_1}< m_{\tilde\chi_3^0}+m_{\tau})$.  
	The gray area is excluded by $m_{\chi^{\pm}_1}<104$~GeV.
\label{stau3a}}
\end{figure}

We discuss the decay chain 
$\tilde \tau_1\to \tau\tilde\chi^0_3;\;
\tilde\chi^0_3\to \tilde\chi^0_1Z;\;
Z\to \ell \bar \ell$ for $\ell=e,\mu,\tau$.
The decay  $\tilde \tau_1 \to \tau~\tilde\chi^0_3$ 
can be distinguished from 
$\tilde \tau_1 \to \tau~\tilde\chi^0_2$  by the $\tau$ energy.
In Fig.~\ref{stau3a}a we show the contour
lines of 
${\rm BR}(\tau_1 \to \tau~\tilde\chi^0_1~\ell~\bar \ell) 
={\rm BR}(\tilde \tau_1 \to \tau~\tilde\chi^0_3)\times
{\rm BR}(\tilde\chi^0_3\to Z\tilde\chi^0_1)\times
{\rm BR}(Z \to \ell~\bar \ell)$
in the $|\mu|$--$M_2$ plane for $\varphi_{M_1}=\pi/2$ 
and $ \varphi_{\mu}=0$.
We choose $M_{\tilde E} < M_{\tilde L}$ since the 
$\tilde \tau_1$-$\tau$-$\tilde\chi^0_3$ coupling 
$|a^{\tilde\tau}_{13}|$ is larger, 
thus ${\rm BR}(\tilde \tau_1 \to \tau~\tilde\chi^0_3)$
is larger than for $M_{\tilde E} > M_{\tilde L}$.
However, the total branching ratio is smaller than for the previous
decay chain due to the small 
${\rm BR}(\tilde \tau_1 \to \tau~\tilde\chi^0_3)<.75(0.05)$
in the upper (lower) part of Fig.~\ref{stau3a}a.

The asymmetry ${\mathcal A}^{\rm T}_\ell$, shown   
in Fig.~\ref{stau3a}b, vanishes on contours where 
either $|a^{\tilde\tau}_{13}|=|b^{\tilde\tau}_{13}|$
or ${\rm Im}(O''^L_{13}{O''^R_{13}}^{\ast})=0$, see~(\ref{eq:prop1}). 
Along the contour ${\mathcal A}^{\rm T}_\ell=0$
in the lower part of  Fig.~\ref{stau3a}b we have 
$|a^{\tilde\tau}_{13}|=|b^{\tilde\tau}_{13}|$, whereas
along the contour line 0 in the upper part of Fig.~\ref{stau3a}b we have 
${\rm Im}(O''^L_{13}{O''^R_{13}}^{\ast})=0$.
Furthermore, there is a sign change of 
${\rm Im}(O''^L_{13}{O''^R_{13}}^{\ast})$
between the upper and the lower part of Fig.~\ref{stau3a}b (area A).
The first factor in (\ref{eq:prop1}) increases for 
$|\mu|/M_2\to 0$, since the gaugino component
of $\tilde\chi^0_3$ gets enhanced,
resulting in $|b^{\tilde\tau}_{13}|/|a^{\tilde\tau}_{13}|\to 0$.

\begin{figure}[H]
	\setlength{\unitlength}{0.035cm}
\begin{picture}(120,220)(12,0)
\put(10,0){\includegraphics{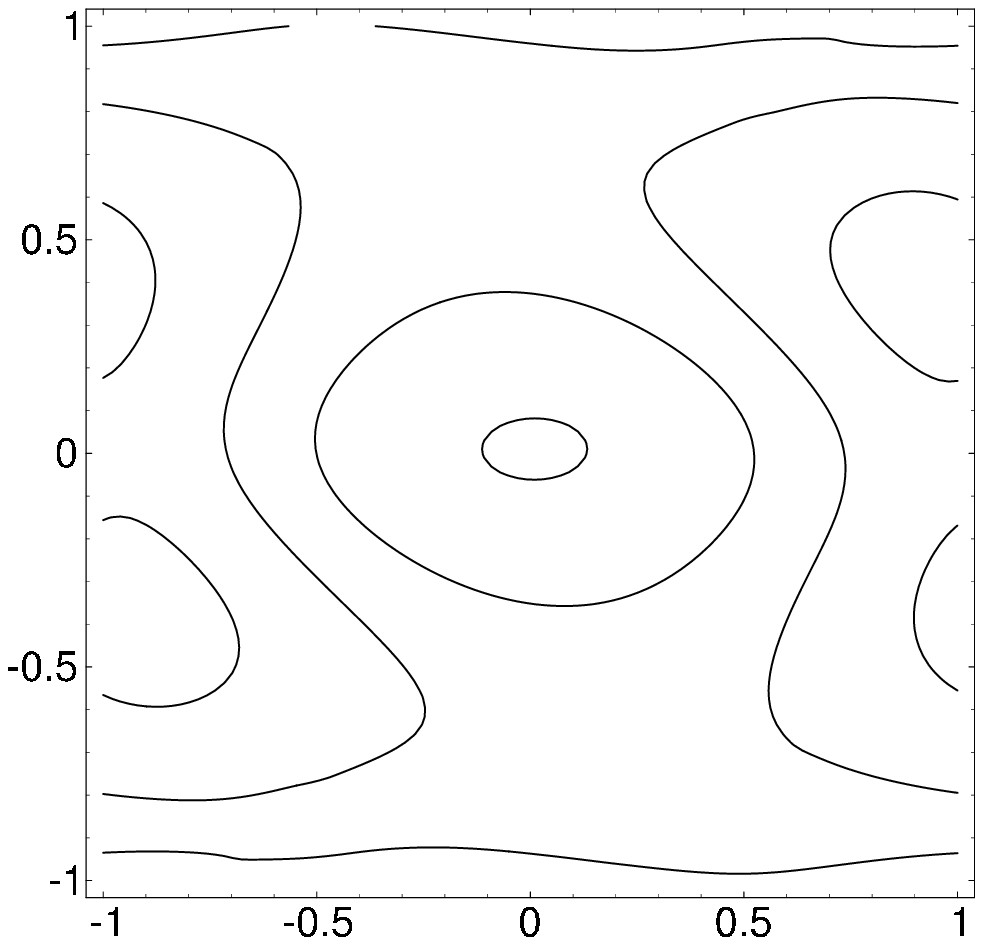}}
	\put(185,-8){$\varphi_{\mu}~[\pi] $}
	\put(10,210){$\varphi_{ M_1}~[\pi]$}
	\put(70,210){\fbox{${\rm BR}(\tilde{\tau}_1 \;\to \;\tau  \;
				\tilde{\chi}^0_1 \;\ell  \; \bar \ell)
%		,\;\;\ell = e, \mu, \tau \; \; \; 
		$ in \% }}
	\put(117,104){{\footnotesize $0.2 $}}
	\put(133,80){{\footnotesize $0.4 $}}
	\put(157,65){{\footnotesize $0.6 $}}
	\put(187,63){{\footnotesize $0.8 $}}
	\put(130,27){{\footnotesize $0.8 $}}
	\put(192,135){{\footnotesize $0.8 $}}
	\put(40,57){{\footnotesize $0.8 $}}
	\put(40,120){{\footnotesize $0.8 $}}
	\put(78,160){{\footnotesize $0.6 $}}
	\put(30,-8){Fig.~\ref{stau3b}a}
\put(240,0){\includegraphics{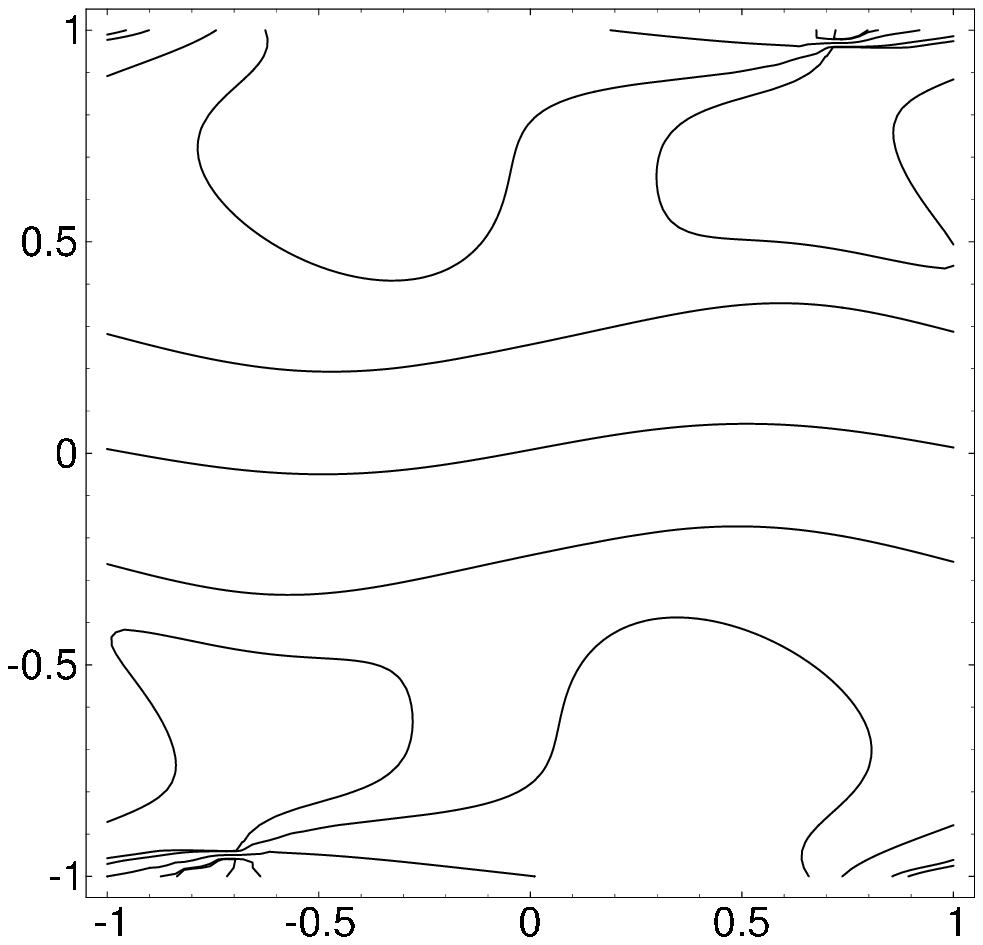}}
   \put(415,-8){$\varphi_{\mu}~[\pi] $}
	\put(240,210){$\varphi_{ M_1}~[\pi]$}
	\put(320,210){\fbox{${\mathcal A}^{\rm T}_{\ell}$ in \% }}
        \put(294,127){{\footnotesize $-3 $}}
	\put(291,160){{\footnotesize $-3 $}}
	\put(400,153){{\footnotesize $-3.3 $}}
		\put(330,108){{\footnotesize $0 $}}
		\put(350,78){{\footnotesize $3 $}}
		\put(385,63){{\footnotesize $ 3 $}}
		\put(300,57){{\footnotesize $3.3 $}}
		\put(260,-8){Fig.~\ref{stau3b}b }
	\end{picture}
	\vspace*{.3cm}
\caption{Contour lines of the branching ratio 
	for $\tilde\tau_1\to \tilde\chi^0_1\tau \ell\bar\ell$ 
	and asymmetry ${\mathcal A}^{\rm T}_{\ell}$ in the 
	$\varphi_{\mu} $--$\varphi_{ M_1}$ plane for
	$|\mu|=150$~GeV, $M_2=450$~GeV, 
	$\tan\beta =10$, $A_{\tau}=1$~TeV, 
	$m_{\tilde\tau_1}=300$~GeV, $ m_{\tilde\tau_2}=800$~GeV
	and $M_{\tilde E} < M_{\tilde L}$.
	\label{stau3b}}
\end{figure}

In Fig.~\ref{stau3b} we show 
%for $|\mu|=150$~GeV and $M_2=450$~GeV
the $\varphi_{M_1}$, $\varphi_{\mu}$ dependence of 
${\rm BR}(\tau_1 \to \tau~\chi^0_1~\ell~\bar \ell)$
and ${\mathcal A}^{\rm T}_\ell$,
for $|\mu|=150$~GeV and $M_2=450$~GeV.
Two points of level crossing appear at  
$(\varphi_{M_1}, \varphi_{\mu}) \approx (\pm 0.95\pi,\pm 0.7\pi)$
in  Fig.~\ref{stau3b}b.
%For these parameters, but with various phase combinations, we give 
Neutralino masses and the neutralino and stau widths are given
in Table \ref{tab:chi3}. 
%Note that maximal CP violating phases 
%$\varphi_{\mu},\varphi_{M_1}=\pm\pi/2$ do not necessarily 
%lead to the highest value of ${\mathcal A}^{\rm T}_\ell$. 
%due to the complex interplay of the phases in 
%${\rm Im}(O''^L_{13}{O''^R_{13}}^{\ast})$.
%The value of ${\mathcal A}^{\rm T}_\ell$ depends stronger on 
%$\varphi_{M_1}$, which also determines the sign of 
%${\mathcal A}^{\rm T}_\ell$, than on $\varphi_{\mu}$.
Comparing Fig.~\ref{stau2b}b and Fig.~\ref{stau3b}b, one can see 
the common and  strong $\varphi_{M_1}$ dependence of the asymmetries. 
In a good approximation 
${\rm Sign}[{\mathcal A}^{\rm T}_\ell]\approx {\rm Sign}[\varphi_{M_1}]$ 
in Fig.~\ref{stau2b}b and  
${\rm Sign}[{\mathcal A}^{\rm T}_\ell]\approx -{\rm Sign}[\varphi_{M_1}]$ 
in Fig.~\ref{stau3b}b, due to the  different phase dependence of 
${\rm Im}(O''^L_{12}{O''^R_{12}}^{\ast})$ and
${\rm Im}(O''^L_{13}{O''^R_{13}}^{\ast})$.
%Moreover, two points of level crossing appear at  
%$(\varphi_{M_1}, \varphi_{\mu}) \approx (\pm 0.95\pi,\pm 0.7\pi)$
%in  Fig.~\ref{stau3b}b.

%
\begin{table}[H]
\caption{
	Masses 
	%of $\tilde\chi^0_i$ 
	and widths 
%$\Gamma_{\tilde\chi^0_3}$, $\Gamma_{\tilde\tau_1}$
	for various combinations of $\varphi_{\mu}$ and 
	$\varphi_{M_1}$, for $|\mu|=150$~GeV, $M_2=450$~GeV,
	$\tan\beta =10$, $A_{\tau}=1$~TeV, 
	$m_{\tilde\tau_1}=300$~GeV, $ m_{\tilde\tau_2}=800$~GeV.
%	for $M_{\tilde E} < M_{\tilde L}$.
\label{tab:chi3}
}
\begin{center}
\begin{tabular}{|c|c|l|c|c|} \hline
$\varphi_{\mu}$ & $\varphi_{M_1}$ &
$m_{\chi^0_1},m_{\chi^0_2},m_{\chi^0_3},m_{\chi^0_4} 
~[\rm GeV]$  &
$\Gamma_{\chi^0_3}~[\rm MeV]$ & $\Gamma_{\tilde\tau_1}~[\rm MeV]$\\ \hline\hline
0 &            0    &$128,\;\;156,\;\;238,\;\;467$&$59.0$ &$ 362$ \\
0 & $\frac{\pi}{2}$ &$132,\;\;153,\;\;238,\;\;466$&$68.2$ &$ 359$ \\
0 & $ \pi   $       &$141,\;\;145,\;\;238,\;\;466$&$75.5$ &$ 356$ \\ \hline
$\frac{\pi}{2}$ &           0     &$131,\;\;158,\;\;237,\;\;466$&$41.5$&$356$\\
$\frac{\pi}{2}$ & $\frac{\pi}{2}$ &$136,\;\;154,\;\;237,\;\;466$&$49.4$&$353$  \\
$\frac{\pi}{2}$ & $\pi$           &$142,\;\;145,\;\;240,\;\;465$&$73.8$&$360$  \\\hline
$\pi$ &       0         & $135,\;\;159,\;\;236,\;\;465$&$27.7$&$ 351$ \\
$\pi$ & $\frac{\pi}{2}$ & $137,\;\;154,\;\;239,\;\;465$&$47.5$&$ 357$ \\
$\pi$ & $\pi$           & $143,\;\;144,\;\;242,\;\;464$&$71.0$&$ 364$ \\
\hline
 \end{tabular}
\end{center}
\end{table}

%----------------------------------------------------------
%     S F E R M I O N    plot   -3-
%---------------------------------------------------------
%
%----------------------------------------------------------
%     S F E R M I O N    plot   -4-
%---------------------------------------------------------

\section{Summary of Chapter \ref{CP violation in sfermion decays}}
   
We have considered a T-odd asymmetry in the
sequential decay of a sfermion
$\tilde f \to f'~\tilde\chi^0_j\to 
f'~\tilde\chi^0_1~Z\to f'~\tilde\chi^0_1~f \bar f$.
The asymmetry is sensitive to the phases in the neutralino sector.
In a numerical study for stau decay
$\tilde\tau_1\to\tau\tilde\chi^0_1~f \bar f$, we have shown that
the asymmetry can be of the order
of $3\%$ for leptonic final states
$\tau~\tilde\chi^0_1~\bar \ell\ell$,
and is larger by a factor 6.3 for the semi-leptonic final state 
$\tau~\tilde\chi^0_1~\bar b b$.
The number of produced $\tilde \tau$'s which are necessary 
to observe the asymmetry is at least of the order $10^{5}$
for leptonic final states, and $10^{3}$
for semi-leptonic final states,
such that the phases in the neutralino sector may be accessible 
at future collider experiments.

	\chapter{Summary and conclusions}
\label{Conclusions}

%\item \textbf{Introduction}

\section{Summary}

%\textbf{The importance of CP violating phases in SUSY models}
%The only source of CP violation in the standard model (SM) is
%given by one phase in the Kobayashi-Maskawa matrix.
%However, this phase alone cannot account for the observed baryon
%asymmetry of the universe, and further
%sources of CP violation have to be introduced.

In supersymmetric (SUSY) extensions of the Standard Model (SM),
several parameters can be complex.
In the neutralino sector of the Minimal extended Supersymmetric 
Standard Model (MSSM), 
these are the Higgsino mass parameter $\mu$ and the gaugino mass
parameter $M_1$. In addition, in the sfermion sector also the
trilinear scalar coupling parameter $A_f$ can be complex.
%The phases $\varphi_{\mu}$, $\varphi_{M_1}$ and $\varphi_{A_f}$ 
The imaginary parts 
of these parameters imply CP violating effects in the 
production and decay of SUSY particles.
%at future colliders.

In this thesis we have analyzed the implications of  
$\varphi_{\mu}$, $\varphi_{M_1}$ and $\varphi_{A_{\tau}}$ in neutralino 
and chargino production and decay in electron-positron collisions. For
the decays of sfermions we have analyzed the effects of 
$\varphi_{\mu}$ and $\varphi_{M_1}$. We have analyzed T-odd and  CP-odd 
asymmetries of triple products of particle momenta 
or spins. Such  asymmetries are non-zero only if CP is violated. 
Their measurements at future colliders allow a determination of the phases, 
in particular also their signs.

The asymmetries involve angular distributions 
of the neutralino, chargino and sfermion decay products. 
The tree-level calculations in the spin-density formalism 
include the complete spin correlations between production and decay.
%, which has been accounted for in the spin density matrix formalism. 
Modular FORTRAN codes have been programmed
for numerical analyses.

\newpage

%\vspace{0.4cm}
%\textbf{The processes studied and results}

For neutralino production,  
$e^+ e^-\to\tilde\chi^0_i~\tilde\chi^0_j$ ,
we can summarize as follows:

\begin{itemize}

	\item
For the leptonic two-body decay chain of one of the neutralinos
$\tilde\chi^0_i \to \tilde\ell  \ell_1$,
$ \tilde\ell \to \tilde\chi^0_1  \ell_2$ for
$ \ell= e,\mu,\tau$, we have analyzed the asymmetries of two
triple products 
${\mathcal T}_{I} = ({\bf p}_{e^-} \times {\bf p}_{\tilde\chi_i^0})
\cdot {\bf p}_{\ell_1}$
and
${\mathcal T}_{II} = ({\bf p}_{e^-} \times {\bf p}_{\ell_2})
\cdot {\bf p}_{\ell_1}$,
which are sensitive to $\varphi_{\mu}$ and 
$\varphi_{M_1}$. In a numerical study for 
$e^+e^- \to\tilde\chi^0_1 \tilde\chi^0_2$
and subsequent neutralino decay into a right slepton
$\tilde\chi^0_2 \to \tilde\ell_R \ell$
we have shown that the asymmetry ${\mathcal A}_{I}$ can be 
as large as $25\%$. The asymmetry ${\mathcal A}_{II}$, which does not require
the identification of the neutralino momentum, can reach $10\%$. 
Asymmetries of the same order are obtained
for the processes 
$e^+e^- \to\tilde\chi^0_1 \tilde\chi^0_3$ and
$e^+e^- \to\tilde\chi^0_2 \tilde\chi^0_3$.

\item
For the two-body decay of one  neutralino
into a stau-tau pair, $\tilde\chi^0_i \to \tilde\tau_k  \tau$,
we have analyzed the asymmetry of the triple product 
${\mathcal T}_{\tau}={\bf s}_{\tau}\cdot({\bf p}_{\tau}\times {\bf
		p}_{e^-})$, 
which includes the transverse $\tau$ polarization ${\bf s}_{\tau}$.
The asymmetry is sensitive to the phases
$\varphi_{\mu}$, $\varphi_{M_1}$ and $\varphi_{A_{\tau}}$
and can attain values up to 
$60\%$ for $e^+e^- \to\tilde\chi^0_1 \tilde\chi^0_2$.

\item
For the neutralino decay into a $Z$ boson, 
$\tilde\chi^0_i \to Z\tilde\chi^0_n$,
followed by the decay $Z \to \ell \bar\ell $ 
for $ \ell= e,\mu,\tau$, or $Z \to q\bar q$ for $q=c,b$,
we have defined and analyzed the asymmetry 
$ {\mathcal A}_{\ell(q)}$ of the triple product
${\mathcal T}_{\ell(q)}= {\bf p}_{e^-}\cdot({\bf p}_{\ell(q)} 
	\times {\bf p}_{\bar\ell(\bar q)})$, which is sensitive to 
$\varphi_{\mu}$ and $\varphi_{M_1}$. We have also identified the CP 
sensitive elements of the $Z$ spin-density matrix.
For $\tilde{\chi}^0_1  \tilde{\chi}^0_2$,
$\tilde{\chi}^0_2  \tilde{\chi}^0_2$,
$\tilde{\chi}^0_1  \tilde{\chi}^0_3$ and
$\tilde{\chi}^0_2  \tilde{\chi}^0_3$ production,
the asymmetry ${\mathcal A}_{\ell}$ can go up to $3\%$. 
For the hadronic decays of the $Z$ boson,  
larger asymmetries  
${\mathcal A}_{c(b)} \simeq6.3(4.5)\times {\mathcal A}_{\ell}$
are obtained.

\end{itemize}

The results for chargino production,
$e^+ e^-\to\tilde\chi^{\pm}_i~\tilde\chi^{\mp}_j$,
can be summarized as follows:

\begin{itemize}

\item
For the two-body decay  of one of the charginos
into a sneutrino $\tilde\chi^+_i \to \ell^+\tilde\nu_{\ell}$,
the  asymmetry of the triple product 
$({\mathbf p}_{e^-} \times {\mathbf p}_{\tilde\chi^+_i}) \cdot 
{\mathbf p}_{\ell}$ is sensitive to 
$\varphi_{\mu}$ and can be as large as $30\%$
for $e^+ e^-\to\tilde\chi^{\pm}_1~\tilde\chi^{\mp}_2$.

\item
For the decay $\tilde\chi^+_i \to W^+\chi^0_n$ the triple product 
${\mathcal T}_{I} = {\mathbf p}_{e^-}\cdot({\mathbf p}_{\tilde\chi^+_i} 
	\times {\mathbf p}_{W})$
defines the asymmetry $ {\mathcal A}_{I}$, which is sensitive 
to $\varphi_{\mu}$. The asymmetry ${\mathcal A}_{I}$ can reach $4\%$ 
for $e^+ e^-\to\tilde\chi^{\pm}_1~\tilde\chi^{\mp}_2$ production. 
Further, we have analyzed the CP sensitive elements of the 
$W$ boson spin-density matrix. 
%The CP odd vector component $V_2$ 
%is sensitive to $\varphi_{\mu}$ and also to $ \varphi_{M_1}$.
The phase $ \varphi_{M_1}$ enters in the decay 
$\tilde\chi^+_i \to W^+\chi^0_n $
due to correlations of the chargino and the $W$ boson polarizations,
which can be probed via the hadronic decay 
$W^+ \to c \bar s$. Moreover the triple product 
${\mathcal T}_{II} = {\mathbf p}_{e^-}\cdot({\mathbf p}_{c} 
	\times {\mathbf p}_{ \bar s})$
defines the $\varphi_{\mu}$ and $\varphi_{M_1}$ sensitive 
asymmetry ${\mathcal A}_{II}$, which can be as large as $7\%$ for 
$\tilde\chi^+_1  \tilde\chi^-_1$ or
$\tilde\chi^+_1  \tilde\chi^-_2$ production.

\end{itemize}

We note in addition that if the neutralinos or charginos 
decay  into scalar particles, like sleptons or sneutrinos,
the asymmetries probe CP violation in the production process only. 
The asymmetries are then caused by the  neutralino or chargino polarizations 
perpendicular to the production plane, which are non-vanishing only for
the production of a non-diagonal pair of neutralinos 
$e^+ e^-\to\tilde\chi^0_i~\tilde\chi^0_j, i\neq j$
or charginos
$e^+ e^-\to\tilde\chi^{\pm}_1~\tilde\chi^{\mp}_2$.
If the neutralinos and charginos decay into particles with spin, 
like $\tau$ or $W$ and $Z$ bosons, we have found that 
also the CP phases which enter in the decay can be probed,
in addition to the CP contributions from the production. 
This is due to the spin correlations between production and decay.

For the  two-body decay chain of a sfermion 
$\tilde f \to f \tilde\chi^0_j\to 
f \tilde\chi^0_1~Z\to f'\tilde\chi^0_1~l ~\bar l~(f' \tilde\chi^0_1~q~
	\bar q)$
we have obtained the following results:

\begin{itemize}
	
\item 
We have defined the asymmetries ${\mathcal A}_{\ell,q}$
of the triple products of the momenta of the outgoing leptons
$	{\mathcal T}_{\ell}=
	{\bf p}_{f}\cdot({\bf p}_{\ell}\times{\bf p}_{\bar \ell})$
	or quarks
${\mathcal T}_q=
{\bf p}_{f}\cdot({\bf p}_{q}\times{\bf p}_{\bar q})$,
which are sensitive to $\varphi_{\mu}$ and $\varphi_{M_1}$.
In a numerical study of stau decay
$\tilde\tau_1\to\tau\chi^0_1~f \bar f$ we found that the
asymmetry ${\mathcal A}_{\ell}$ can be of the order
of $3\%$ percent for leptonic final states.
The number of produced $\tilde \tau$'s necessary to observe
${\mathcal A}_{\ell}$ is at least of the
order $10^{5}$, which may be accessible at
future collider experiments.
For the semi-leptonic final state 
$\tilde\tau_1\to\tau~\tilde\chi^0_1~ b~\bar b$ the asymmetry
${\mathcal A}_{b}$ is larger by a factor $6.3$ 
and could be measured if the  number of produced $\tilde \tau$'s 
is of the order $10^{3}$.

\item 
The asymmetries are not sensitive to the phase $\varphi_{A_f}$, since 
the decay $\tilde f \to f~\tilde\chi^0_j$ is a two-body decay of
a scalar particle. As an observable sensitive to
$\varphi_{A_f}$, one would have to measure the transverse polarization
of the fermion~$f$, or study asymmetries in three-body decays of the 
sfermion $\tilde f$.

\end{itemize}

%\vspace{0.4cm}

\section{Conclusions}

We have shown that in all the processes studied, triple-product 
asymmetries can be defined which are sensitive to the CP phases.
Especially promising for measuring the phases are leptonic 
decays of charginos and neutralinos, where the asymmetries can 
be as large as $30\%$. For neutralino decays into a stau-tau pair,
the asymmetry of the $\tau$ polarization may even reach $60\%$.
The asymmetries of neutralino and chargino decays into
a $Z$ or $W$ boson may be as large as $18\%$ and $7\%$, 
respectively. For the neutralino and chargino production processes, 
we have found that longitudinal polarization of both beams 
can significantly enhance the asymmetries and cross sections.

By analyzing the statistical significances, we have shown that
the asymmetries
%, and thus the MSSM phases
%$\varphi_{\mu}$, $\varphi_{M_1}$ and $\varphi_{A_{\tau}}$,
are accessible in future electron-positron linear 
collider experiments in the $500$~GeV - $800$~GeV range with 
high luminosity and polarized beams. The MSSM phases
$\varphi_{\mu}$, $\varphi_{M_1}$ and $\varphi_{A_{\tau}}$
could thus be strongly constrained.
%For the neutralino and chargino production processes, 
%we have found that longitudinal polarization of both beams 
%can significantly enhance the asymmetries and cross sections.

%As a closing paragraph we can say that the MSSM phases
%$\varphi_{\mu}$, $\varphi_{M_1}$ and $\varphi_{A_{\tau}}$
%could be strongly constrained in future linear 
%electron-positron collider experiments in the $500$~GeV - $800$~GeV range
%with high luminosity and longitudinally polarized beams.

%\vspace{0.4cm}
%\textbf{Outlook}

\section{Outlook}

%In this work we have analyzed CP-asymmetries of triple products 
%of the momenta of the decay products of one of the 
%neutralinos or charginos. 
In further investigations our analysis of 
CP-odd asymmetries in the production and decay of neutralinos and 
charginos should be extended to electron-positron collisions with 
transverse beam polarizations. For chargino production the asymmetries
of triple products including the transverse beam polarization
are only non-vanishing if the decay of both particles 
is included.
%\cite{choichargino,holger}.

Moreover, triple products which
involve the momenta of the decay products of both 
neutralinos (or charginos) could be sensitive 
to correlations between their polarizations.
These spin-spin correlations would yield additional 
information on the CP phases.
%It has to be clarified, whether additional information on the CP
%phases could be obtained this way. 

\begin{appendix}
	\chapter{Basics of the Minimal Supersymmetric Standard Model}

The  Minimal Supersymmetric 
Standard Model (MSSM) is characterized by the following properties:
\begin{itemize}
	\item{
		A minimal gauge group: $SU(3)\times SU(2)\times U(1)$.
	}
\item{
	A minimal content of fields: 3 generations of leptons and
	quarks, 12 gauge bosons, 2 Higgs doublets, and their super
	partners, see Table~\ref{MSSMspectrum}.
	}
\item{
		Explicit SUSY breaking parametrized by soft breaking terms.
	}
\item{
		R-parity conservation.
	}
\end{itemize}

\begin{table}
\caption{Particle spectrum of the  MSSM \label{MSSMspectrum}}
\begin{center}
\begin{tabular}{|c||cl|cl|} \hline
	 & \multicolumn{4}{c|}{SUSY-partners}\\ \cline{2-5}
	 \rb{SM particles} & \multicolumn{2}{c|}{weak eigenstates} &
  \multicolumn{2}{c|}{mass eigenstates}\\ \hline \hline
$q_u=u,c,t$ &  & & & \\ 
$q_d=d,s,b$ & \rb{$\tilde q_L,\tilde q_R$} & \rb{squarks} & \rb{$\tilde q_1,
	\tilde q_2$} & \rb{squarks}\\ \hline
$\ell=e,\mu,\tau$ & $\tilde\ell_L,\tilde\ell_R$ & sleptons & 
$\tilde\ell_1,\tilde{\ell}_2$ & sleptons\\
$\nu=\nu_e,\nu_{\mu},\nu_{\tau}$ & $\tilde\nu_{\ell}$ & sneutrinos &
$\tilde\nu_{\ell}$ & sneutrinos\\ \hline \hline
$g$ & $\tilde{g}$ & gluino & $\tilde{g}$ & gluino\\ \hline
$W^{\pm}$ & $\tilde W^{\pm}$ & wino &  & \\
$(H_1^+,H_2^-)$ & $\tilde H_1^+,\tilde H_2^-$ & Higgsinos &
\rb{$\tilde\chi^{\pm}_{1,2}$} & \rb{charginos} \\\hline
%$(H_1^+)$ & $\tilde H_1^+$ & Higgsino & $\tilde\chi^{\pm}_{1,2}$ 
%& charginos \\
%$(H^-_2)$ & $\tilde H^-_2$ & Higgsino & & \\ \hline
$\gamma$ &  $\tilde \gamma$ & photino & & \\
$Z^0$ & $\tilde Z^0$ & zino & $\tilde\chi^0_{1,\ldots,4}$ 
& neutralinos \\
$H^0_1,(H^0_2) $ & $\tilde H^0_1,\tilde H^0_2$ & Higgsinos & & \\
\hline
%$H^0_1$ & $\tilde H^0_1$ & Higgsino & \raisebox{1.5ex}[1.5ex]
%{$\tilde\chi^0_{1,\ldots,4}$} & \raisebox{1.5ex}[1.5ex]{neutralinos}\\
%$(H^0_2)$ & $\tilde H^0_2$ & Higgsino & & \\ \hline
\end{tabular}
\end{center}
\end{table}

In the following we give a short account on
the relevant Lagrangians and couplings,
and mixing of neutralinos, charginos and
staus.
%For detailed reviews, see e.g., \cite{Sohnius,Haber-Kane}.

In terms of superfields, the field content and the parameter 
content of the MSSM is well visible, and the Lagrangian 
can be written in an elegant and short way.
We thus give its electroweak part in terms of superfields 
in Section~\ref{The MSSM Lagrangian in terms of superfields}, 
where we also define the parameters.

For calculations of interaction amplitudes, however, it is 
more convenient to expand the MSSM Lagrangian in component 
fields, which is albeit a more complex procedure 
with a longish result. We thus restrict our discussions 
on the neutralino, chargino and stau sector, and give the mass 
matrices in Section~\ref{Mass mixing matrices}.
In  Section \ref{Lagrangian} we give the relevant parts of the 
Lagrangian.

For detailed reviews of the MSSM and its Lagrangians, see e.g., 
\cite{Sohnius,Haber-Kane,Simonsen}.
For a detailed study of CP violating sources in the MSSM, 
see \cite{garfielddiss}.

%\section{Particle spectrum
%	\label{Properties and particle spectrum}}
%
%The MSSM is characterized by the following properties:
%\begin{itemize}
%	\item{
%		A minimal gauge group: $SU(3)\times SU(2)\times U(1)$.
%	}
%\item{
%	A minimal content of fields: 3 generations of leptons and
%	quarks, 12 gauge bosons, 2 Higgs doublets, and their super
%	partners, see Table~\ref{MSSMspectrum}.
%	}
%\item{
%		Explicit SUSY breaking parametrized by soft breaking terms.
%	}
%\item{
%		Exact discrete R-parity.
%	}
%	
%\end{itemize}

\section{MSSM Lagrangian in terms of superfields
	\label{The MSSM Lagrangian in terms of superfields}}

We give the electroweak (EW) part of the MSSM Lagrangian in the
superfield formalism, which  consists of a supersymmetric
part and terms which break SUSY softly 
\begin{equation}
	{\mathcal L}_{EW}={\mathcal L}_{SUSY}+{\mathcal L}_{soft}.
\end{equation}

\subsection{Superfield content
	\label{The superfield content}}

The left-handed lepton superfields are arranged in $SU(2)$ doublets
and the right-handed ones in $SU(2)$ singlets,
\begin{equation}\label{lepton super fields}
\hat L =
		\left( \begin{array}{c}
					\hat \nu_l (x,\theta,\bar\theta) \\
					\hat l (x,\theta,\bar\theta)
				\end{array}
			\right)_L, \qquad   
\hat R =\hat l_R (x,\theta,\bar\theta),
\end{equation}
where the generation indices of $e$, $\mu$ and $\tau$ have been suppressed.
The superfields are functions on the superspace,
with spacetime coordinates $x^{\mu}$ and anticommuting 
Grassmann variables $\theta_{\alpha}$.
The lepton superfield contains both bosonic and
fermionic degrees of freedom. It can be expanded 
\begin{eqnarray}
\hat L &=& \tilde L(x) 
	+ i\theta \sigma^{\mu} \bar\theta\partial_{\mu}~ \tilde L(x) 
	- \frac{1}{4} \theta \theta \bar\theta \bar\theta
	\partial^{\mu} \partial_{\mu} ~\tilde L(x)
		~+\nonumber \\ &&
		+\sqrt{2} \theta ~L^{(2)}(x)
		+ \frac{i}{\sqrt{2}}\theta \theta \bar\theta \bar\sigma^{\mu}
		\partial_{\mu} ~ L^{(2)}(x) + \theta\theta~ F_L(x),
		\label{leptonfieldL}
\end{eqnarray}
in terms of the spin-$0$ slepton field $\tilde L$, the spin-$1/2$ Weyl field
$L^{(2)}$ and a spin-$0$ auxiliary field $ F_L$,
which however can be removed by the Euler-Lagrange
equations. The right-handed field $\hat R$ is expanded 
similarly to $\hat L$.

There are two doublets of chiral Higgs superfields 
\begin{equation}
\hat H_1  =
		\left( \begin{array}{c}
					\hat H_1^1 (x,\theta,\bar\theta) \\
					\hat H_1^2 (x,\theta,\bar\theta)
				\end{array}
			\right), \qquad
	\hat H_2 =
		\left( \begin{array}{c}
					\hat H_2^1 (x,\theta,\bar\theta) \\
					\hat H_2^2 (x,\theta,\bar\theta)
				\end{array}
			\right),
\end{equation}
where the upper index denotes the $SU(2)$ index.
The component field expansions of the Higgs fields
is similar to that of the lepton field $\hat L$~(\ref{leptonfieldL}).

The $U(1)$ and $SU(2)$ gauge vector superfields
are respectively given by
\begin{equation}
\hat V' = {\rm \bf Y }   ~\hat v' (x,\theta,\bar\theta), \qquad
\hat V  = {\rm \bf T }^a ~\hat V^a(x,\theta,\bar\theta), 
\end{equation}
with sum over $a=1,2,3$ and
${\rm \bf Y}$ and ${\rm \bf T}^a$ are the
$U(1)$ and $SU(2)$ generators. 
The gauge vector superfield contains 
a bosonic (spin $1$)  gauge field $V^a_{\mu}$,
and a fermionic (spin $1/2$) gaugino Weyl field $\lambda^a$
\begin{eqnarray}
\hat V^a (x,\theta,\bar\theta) &=&
			-~\theta \sigma^{\mu} \bar \theta~  V^a_{\mu}(x)
			+ i \theta \theta \bar\theta~\bar \lambda^a (x) 
			- i \bar\theta\bar\theta\theta ~\lambda^a (x)
			+ \frac{1}{2}\theta\theta\bar\theta\bar\theta ~D^a(x).
\end{eqnarray}
The auxiliary spin-$1/2$ field $D^a$ can be removed 
by the Euler-Lagrange equations.
The  gauge field $\hat v'$ is expanded 
similarly to $\hat V^a$.

\subsection{The supersymmetric Lagrangian
	\label{The supersymmetric Lagrangian}}

The supersymmetric Lagrangian contains the kinetic terms for
chiral and vector superfields, as well as a Higgs part
\begin{equation}
	{\mathcal L}_{SUSY}=  {\mathcal L }_{Lepton} +
						{\mathcal L }_{Gauge} + {\mathcal L }_{Higgs}.
\end{equation}
The lepton and gauge parts are given by
\begin{eqnarray}
{\mathcal L }_{Lepton} &=& \int d^4 \theta ~ [\hat L^{\dagger} 
	~e^{2g \hat V + g' \hat V'} \hat L + 
	\hat R^{\dagger}~ e^{2g \hat V + g' \hat V'} \hat R ], \\
{\mathcal L }_{Gauge} &=& 	\frac{1}{4} \int d^4 \theta ~
[ W^{a \alpha} W^a_{\alpha} +  W'^{ \alpha} W'_{\alpha}] ~
			\delta^2(\bar\theta) + h.c.,
\end{eqnarray}
where $g$ and $g'$ are the $SU(2)$ and $U(1)$ gauge couplings.
The $SU(2)$ and $U(1)$  field strengths are defined by
\begin{eqnarray}
 W_{\alpha} = - \frac{1}{8g}\bar D\bar D 
			e^{-2g \hat V} D_{\alpha}
			e^{ 2g \hat V},   \qquad
W'_{\alpha} = - \frac{1}{4} D D 
				\bar  D_{\alpha} \hat V', 
\end{eqnarray}
where $D$ are covariant derivatives.
The Higgs part
\begin{eqnarray}
{\mathcal L }_{Higgs} &=&   
\int d^4 \theta ~\Big [\hat H_1^{\dagger}
	e^{2g \hat V}+g' \hat V' \hat H_1 + \hat H_2^{\dagger}
	e^{2g \hat V} + g' \hat V' \hat H_2 
%	\nonumber\\ &&
		+ W \delta^2(\bar\theta) + \bar W \delta^2(\theta ) \Big{]}
\end{eqnarray}
contains the superpotential 
\begin{equation}
W = W_H + W_Y. 
\end{equation}
The Higgs (H) and Yukawa (Y) parts are
\begin{eqnarray}
W_H = \mu~\epsilon^{ij}\hat H_1^i \hat H_2^j, \qquad 
W_Y = \epsilon^{ i j} [ f \hat H_1^i \hat L^j \hat R +
						 f_1 \hat H_1^i \hat Q^j \hat D +
						 f_2 \hat H_2^j\hat Q^i \hat U],
\end{eqnarray}
with $\mu$ the Higgsino mass parameter, $f_i$ the Yukawa couplings,
with the generation index suppressed, and 
the antisymmetric tensor $\epsilon^{11}=\epsilon^{22}=0$,
$\epsilon^{12}=-\epsilon^{21}=1$.
%\begin{equation}
%\epsilon =  \left( \begin{array}{rr}
%	                 0&1 \\ -1 & 0 
%						\end{array} \right). 
%\end{equation}
Note that in order to be renormalizable, the superpotential 
can only be cubic in the superfields at its maximum.

By construction, the Lagrangians given in this section 
are gauge invariant and invariant under supersymmetry 
transformations. In addition, they are R-parity conserving,
if trilinear terms in $W|_{\theta=0}$ are disregarded.
The Higgsino mass parameter $\mu$ and the Yukawa couplings
$f$ can have physical CP-phases.

\subsection{The soft SUSY breaking Lagrangian
	\label{The soft SUSY breaking Lagrangian}}

The most general Lagrangian, which breaks SUSY softly,
i.e., which does not lead to quadratical divergencies,
can be classified in scalar mass terms (S) and  
gaugino mass terms (G) \cite{Girardello}:
\begin{equation}
{\mathcal L }_{soft} =  
		{\mathcal L }_{S} +   {\mathcal L }_{G}.
\end{equation}
Note that trilinear scalar interaction terms
are not R-invariant and will thus be neglected.
The scalar mass term reads (neglecting squark fields)
\begin{eqnarray}
	{\mathcal L }_{S}&= &
	-\int d^4 \theta ~ \Big{[} M_L^2 \hat L^{\dagger}\hat L 
		+ m_R^2 \hat R^{\dagger} \hat R
		+ m_1^2 \hat H_1^{\dagger}\hat H_1 
		+ m_2^2 \hat H_2^{\dagger}\hat H_2 
		+\nonumber\\ &&
		+ m_3^2 \epsilon^{ij} ( \hat H_1^i \hat H_2^j + h.c.) \Big{]}
	\delta^4(\theta, \bar\theta), 
\end{eqnarray}
with the abbreviation
%\begin{equation}
$M_L^2 \hat L^{\dagger}\hat L = m^2_{\tilde\nu} \hat\nu^{\dagger}
+ m_L^2 \hat l _L^{\dagger}\hat l _L.$
%\end{equation}
The gaugino mass term 
\begin{equation}
	{\mathcal L }_{G} = \frac{1}{2}\int d^4 \theta ~\Big{[}(
			M_1 W'^{\alpha} W'_{\alpha} + M_2 W^{a \alpha} W^a_{\alpha} )
		+ h.c. \Big{] }~\delta^4(\theta, \bar\theta) 
\end{equation}
includes the $U(1)$ and  $SU(2)$ gaugino mass parameters
$M_1$ and $M_2$, respectively. By redefining the fields,
one parameter, usually $M_2$, can be made real, while $M_1$ 
can have a CP violating phase.

\section{MSSM Lagrangian in component fields 
     \label{Lagrangian}}

The MSSM Lagrangians relevant for 
chargino production and decay are
%(in our notation and conventions we follow closely 
\cite{Haber-Kane}:
\begin{eqnarray}
{\cal L}_{Z^0 \ell \bar \ell} &=&
		-\frac{g}{\cos\theta_W} Z_{\mu}\bar \ell\gamma^{\mu}[L_{\ell}P_L+
		R_{\ell} P_R]\ell,\\
{\cal L}_{\gamma \tilde\chi^+_j \tilde\chi_i^-} &=&
	- e A_{\mu} \bar{\tilde\chi}^+_i \gamma^{\mu} \tilde\chi_j^+
		\delta_{ij},\quad e>0, \\
{\cal L}_{Z^0\tilde\chi_j^+\tilde\chi_i^-} &=&
		 \frac{g}{\cos\theta_W}Z_{\mu}\bar{\tilde\chi}^+_i\gamma^{\mu}
		[O_{ij}^{'L} P_L+O_{ij}^{'R} P_R]\tilde\chi_j^+,\\
{\cal L}_{W^{-}\tilde\chi^+_i\tilde{\chi}^{0}_k}&=&
	 g W_{\mu}^{-}\bar{\tilde{\chi}}^{0}_k\gamma^{\mu}
	[O_{ki}^L P_L+O_{ki}^R P_R]\tilde\chi^+_i +\mbox{h.c.},\\
{\cal L}_{\ell \tilde\nu\tilde\chi^+_i} &=&
	- g U_{i1}^{*} \bar{\tilde\chi}^+_i P_{L} \nu \tilde\ell_L^*
	- g V_{i1}^{*} \bar{\tilde\chi}_i^{+C} P_L \ell 
				 \tilde{\nu}^{*}+\mbox{h.c.},\quad \ell=e,\mu,\\
 {\cal L}_{\tau \tilde\nu_{\tau}\tilde\chi^+_i} &=&
%- g U_{i1}^{*} \bar{\tilde\chi}^+_i P_{L} \nu \tilde\ell_L^*
- g \bar{\tilde\chi}_i^{+C}(V_{i1}^{*} P_L-Y_{\tau}U_{i2} P_R) \tau 
\tilde{\nu}^{*}_{\tau} +\mbox{h.c.},
\end{eqnarray}
and the terms relevant also for neutralino production and decay are \cite{Haber-Kane}:
\begin{eqnarray}
{\cal L}_{Z^0\tilde{\chi}^0_m\tilde{\chi}^0_n} &=&
	\frac{1}{2}\frac{g}{\cos\theta_W}Z_{\mu}\bar{\tilde{\chi}}^0_m\gamma^{\mu}
	[O_{mn}^{''L} P_L+O_{mn}^{''R} P_R]\tilde{\chi}^0_n,\\
{\cal L}_{\ell \tilde{\ell}\tilde\chi^{0}_k} &=&
	  g f_{\ell k}^{L} \bar{\ell} P_R \tilde\chi^0_k \tilde{\ell}_L
	+ g f_{\ell k}^{R} \bar{\ell} P_L \tilde\chi^0_k 
   \tilde{\ell}_R+\mbox{h.c.},\\
{\cal L}_{\nu \tilde{\nu}\tilde{\chi}^{0}_k} &=&
	 g f_{\nu k}^{L} \bar{\nu} P_R \tilde{\chi}^0_k \tilde{\nu}_L+
	\mbox{h.c.},
\end{eqnarray}
with the couplings
\begin{eqnarray}
 L_{\ell}&=&T_{3\ell}-e_{\ell}\sin^2\theta_W, \quad
 R_{\ell}\;=\;-e_{\ell}\sin^2\theta_W,\\
% & &\nonumber\\
O_{ij}^{'L}&=&-V_{i1} V_{j1}^{*}-\frac{1}{2} V_{i2} V_{j2}^{*}+
		\delta_{ij} \sin^2\theta_W,\\
O_{ij}^{'R}&=&-U_{i1}^{*} U_{j1}-\frac{1}{2} U_{i2}^{*} U_{j2}+
		\delta_{ij} \sin^2\theta_W,\\
O_{ki}^L&=&-1/\sqrt{2}\Big( \cos\beta N_{k4}-\sin\beta N_{k3}
		\Big)V_{i2}^{*}
		+\Big( \sin\theta_W N_{k1}+\cos\theta_W N_{k2} \Big) V_{i1}^{*},\\
O_{ki}^R&=&+1/\sqrt{2}\Big( \sin\beta N^{*}_{k4}+\cos\beta
		N^{*}_{k3}\Big) U_{i2}
		+\Big( \sin\theta_W N^{*}_{k1}+\cos\theta_W N^{*}_{k2} \Big)U_{i1},\\
O_{mn}^{''L} &=& -\frac{1}{2} (N_{m3}N_{n3}^*-N_{m4}N_{n4}^*)\cos2\beta
		-\frac{1}{2} (N_{m3}N_{n4}^*+N_{m4}N_{n3}^*)\sin2\beta,\\
O_{mn}^{''R} &=& -O_{mn}^{''L*}, \\
%& &\nonumber\\
f_{\ell k}^L &=& -\sqrt{2}\bigg[\frac{1}{\cos
		\theta_W}(T_{3\ell}-e_{\ell}\sin^2\theta_W)N_{k2}+
	e_{\ell}\sin \theta_W N_{k1}\bigg],\label{sel:L}\\
f_{\ell k}^R  &=&-\sqrt{2}e_{\ell} \sin \theta_W\Big[\tan 
	\theta_W N_{k2}^*-N_{k1}^*\Big],\label{sel:R}\\
f_{\nu k}^L &=& -\sqrt{2}\frac{1}{\cos\theta_W} 
		T_{3\nu}N_{k2},
\end{eqnarray}
with $i,j=1,2$ and $k,m,n=1,\ldots,4$.
%Here $P_{L, R}=\frac{1}{2}(1\mp \gamma_5)$, 
%$g$ is the weak coupling
%constant ($g=e/\sin\theta_W$), 
The charge $e_\ell$ and the third component of the weak isospin $T_{3 \ell}$ 
of each fermion is given in Tab.~(\ref{tabelectriccharge}).
%lepton $\ell$. Furthermore, $\tan\beta=\frac{v_2}{v_1}$ where 
%$v_{1,2}$ are the vacuum expectation values of the two 
%neutral Higgs fields. The chargino-mass eigenstates 
%$\tilde\chi^+_i={\chi_i^+ \choose \bar\chi_i^-}$
%are defined by $\chi^{+}_i=V_{i1}w^{+}+V_{i2} h^{+}$ and 
%$\chi_j^-=U_{j1}w^{-}+U_{j2} h^{-}$ with $w^{\pm}$ and $h^{\pm}$
%the two-component spinor fields of the W-ino and the charged
%Higgsinos, respectively. The complex unitary $2\times 2$ matrices
%$U_{mn}$ and $V_{mn}$ diagonalize the chargino mass 
%matrix $X_{\alpha\beta}$, $U_{m \alpha}^* X_{\alpha\beta}V_{\beta
%n}^{-1}= m_{\tilde{\chi}^+_i}\delta_{mn}$, with $ m_{\tilde{\chi}^+_i}>0$. 
The $\tau$-Yukawa coupling is given by  
$Y_{\tau}= m_{\tau}/(\sqrt{2}m_W\cos\beta)$. 
%with $\tan\beta=\frac{v_2}{v_1}$, where $v_{1,2}$ are the vacuum 
%expectation values of the two neutral Higgs fields. 

%The complex unitary $4\times 4$ matrix $N_{ij}$
%diagonalizes the neutral gaugino-Higgsino mass matrix $Y_{\alpha\beta}$, 
%$N_{i \alpha}^*Y_{\alpha\beta}N_{\beta k}^{\dagger}=
%m_{\tilde\chi^0_i}\delta_{ik}$, with $ m_{\tilde\chi^0_i}>0$,
%in the  neutralino basis 
%$\tilde{\gamma}, \tilde{Z}, \tilde{H}^0_a, \tilde{H}^0_b$.
\begin{table}[t]
\caption{Electric charge $e_\ell$ and weak isospin $T_{3 \ell}$ of fermions}
\label{tabelectriccharge}
\begin{center}
\begin{tabular}{|c|ccccc|}
\hline
\quad & \quad d \quad & \quad u \quad 
& \quad$e_L$\quad & \quad$e_R$\quad 
& \quad$\nu$\quad\\\hline 
$e_\ell$ & $-\frac{1}{3}$ & $+\frac{2}{3}$ & $-1$ & $-1$ & 0 \\
$T_{3 \ell}$ & $-\frac{1}{2}$ & $+\frac{1}{2}$ & $-\frac{1}{2}$ & 0 &
$\frac{1}{2}$\\\hline
\end{tabular}
%\caption{Electric charge $e_\ell$ and weak isospin $T_{3 \ell}$ of fermions}
%\label{tabelectriccharge}
\end{center}
\end{table}

For the neutralino decay into staus 
$\tilde \chi^0_i \to \tilde \tau_k \tau$, we take stau mixing into
account and write for the Lagrangian \cite{thomasstau}:
\begin{eqnarray}
{\mathcal L}_{\tau\tilde{\tau} \chi_i }&=&  g\tilde \tau_k \bar \tau
(a^{\tilde \tau}_{ki} P_R+b^{\tilde \tau}_{ki} P_L)\chi^0_i + {\rm h.c.}~,
 \quad k = 1,2; \; i=1,\dots,4, \label{eq:LagStauchi} 
\end{eqnarray}
with
\begin{eqnarray}
a_{kj}^{\tilde \tau}&=&
({\mathcal R}^{\tilde \tau}_{kn})^{\ast}{\mathcal A}^\tau_{jn},\quad 
b_{kj}^{\tilde \tau}=
({\mathcal R}^{\tilde \tau}_{kn})^{\ast}
{\mathcal B}^\tau_{jn},\quad
\ell_{kj}^{\tilde \tau}=
({\mathcal R}^{\tilde \tau}_{kn})^{\ast}
{\mathcal O}^\tau_{jn},
\quad (n=L,R), 
\label{eq:coupl1}\\
%\end{equation}
%
%\begin{equation}
{\mathcal A}^{\tau}_j&=&\left(\begin{array}{ccc}
f^{L}_{\tau j}\\[2mm]
h^{R}_{\tau j} \end{array}\right),\qquad 
{\mathcal B}^{\tau}_j=\left(\begin{array}{ccc}
h^{L}_{\tau j}\\[2mm]
f^{R}_{\tau j} \end{array}\right),\qquad
{\mathcal O}^{\tau}_j=\left(\begin{array}{ccc}
-U_{j1} \\[2mm]
Y_{\tau} U_{j2}\end{array}\right),
\label{eq:coupl2} \\
%\end{equation}
%
%
%\begin{eqnarray}
h^{L}_{\tau j}&=& (h^{R}_{\tau j})^{\ast}
=-Y_{\tau}( N_{j3}^{\ast}\cos\beta+N_{j4}^{\ast}\sin\beta), \\ 
Y_{\tau}&=& m_{\tau}/(\sqrt{2}m_W \cos\beta), \label{eq:coupl4}
\end{eqnarray}
with ${\mathcal R}^{\tilde \tau}_{kn}$ given in~(\ref{eq:rtau})
and $f^{L,R}_{\tau j}$ given in~(\ref{sel:L}),
(\ref{sel:R}).

\section{Mass matrices
\label{Mass mixing matrices}}

\subsection{Neutralino mass matrix
	\label{Neutralino mixing}}

The complex symmetric mass matrix
of the neutral gauginos and  Higgsinos in the basis
$(\tilde\gamma,\tilde Z, \tilde H^0_a, \tilde H^0_b)$ is given by
\begin{eqnarray}\label{neutralinomassmatrix}
Y &=&
\left(\begin{array}{cccc}
M_2\sin^2\theta_W+M_1\cos^2\theta_W & 
(M_2-M_1)\sin\theta_W\cos\theta_W & 0 & 0 \\
(M_2-M_1)\sin\theta_W\cos\theta_W & 
M_2\cos^2\theta_W+M_1\sin^2\theta_W & m_Z  & 0 \\
0 & m_Z & \mu\sin2\beta & -\mu\cos2\beta \\
0 & 0 & -\mu\cos2\beta & -\mu\sin2\beta 
\end{array}\right),\nonumber\\
& & 
\end{eqnarray}
with $\mu$ the Higgsino mass parameter and
$\tan\beta=\frac{v_2}{v_1}$, where 
$v_{1,2}$ are the vacuum expectation values of the two 
neutral Higgs fields.
The mass matrix $Y_{\alpha\beta}$~(\ref{neutralinomassmatrix}) can be
diagonalized by a complex, unitary $4\times 4$ matrix $N_{ij}$
\cite{Haber-Kane},
%diagonalizes the neutral gaugino-Higgsino mass matrix $Y_{\alpha\beta}$, 
\begin{equation}
N_{i \alpha}^*~Y_{\alpha\beta}~N_{\beta k}^{\dagger}=
m_{\tilde\chi^0_i}~\delta_{ik}, 
\end{equation}
with the neutralino masses
$ m_{\tilde\chi^0_i}>0$. Then the weak eigenstates 
$(\tilde\gamma,\tilde Z, \tilde H^0_a, \tilde H^0_b)$
mix to the neutralino mass eigenstates
$(\tilde\chi^0_1,\tilde\chi^0_2,\tilde\chi^0_3,\tilde\chi^0_4)$.

The diagonalization of the neutralino matrix is achieved with
the singular value decomposition \cite{garfielddiss}. 
Let $Z$ be a complex $n\times n$ matrix, then:
\begin{itemize}

\item{
	The matrices $Z Z^{\dagger}$ and $Z^{\dagger} Z$ are selfadjoint
	and have the same real
	eigenvalues  $\lambda_i \ge 0$.
}

\item{
The eigenvectors  $\hat e_i$ connected to the  eigenvalues $\lambda_i$
built an orthonormal system. If some of the  $\lambda_i=0$,
the eigenvectors can be completed to built an orthonormal system. 
}

\item{
	Let $\hat e_i$ be an eigenvector of $Z^{\dagger} Z$
	with eigenvalue $\lambda_i \ne 0$, that is  
	$Z^{\dagger} Z~\hat e_i =\lambda_i~ \hat e_i$,
	then the vectors $\hat e_i':= \frac{1}{\sqrt{\lambda_i}}Z ~\hat e_i$
	also built an orthonormal system with
	$Z^{\dagger} ~\hat e_i'=\sqrt{\lambda_i}~ \hat e_i$.
}
\end{itemize}
Thus $Z$ can be decomposed into its singular values:
$ \tilde Z = (\hat e_1' \dots \hat e_2' )^{\dagger}
Z (\hat e_1 \dots \hat e_2 )=diag(\sqrt{\lambda_i})$.

\subsection{Chargino mass matrix
	\label{Chargino mixing}}

The complex chargino mass matrix is given by
\begin{equation}
X=\left(\begin{array}{cc}
M_2 & m_W\sqrt{2}\sin\beta\\
m_W\sqrt{2}\cos\beta & \mu
\end{array}\right).
%\hspace*{.5cm},
%\label{1N}
\end{equation}
It can be diagonalized by two complex unitary $2\times 2$ matrices
$U_{mn}$ and $V_{mn}$ \cite{Haber-Kane}, 
%diagonalize the chargino mass matrix $X_{\alpha\beta}$, 
\begin{equation}
U_{m \alpha}^* ~X_{\alpha\beta}~V_{\beta
	n}^{-1}= m_{\tilde\chi^+_i}~\delta_{mn}, 
\end{equation}
with the chargino masses
$ m_{\tilde\chi^+_i}>0$.
The chargino-mass eigenstates 
$\tilde\chi^+_i={\chi_i^+ \choose \bar\chi_i^-}$
are defined by $\chi^{+}_i=V_{i1}w^{+}+V_{i2} h^{+}$ and 
$\chi_j^-=U_{j1}w^{-}+U_{j2} h^{-}$ with $w^{\pm}$ and $h^{\pm}$
the two-component spinor fields of the W-ino and the charged
Higgsinos, respectively.
We diagonalize the chargino mass matrix with the
singular value decomposition, see Section~\ref{Neutralino mixing}.

\subsection{Stau mass matrix 
	\label{Stau mixing}}

The masses and couplings of the $\tau$-sleptons follow from the 
hermitian $2 \times 2$  $\tilde\tau_L - \tilde \tau_R$ mass 
matrix \cite{thomasstau}:
\begin{equation}
{\mathcal{L}}_M^{\tilde \tau}= -(\tilde \tau_L^{\ast},\, \tilde \tau_R^{\ast})
\left(\begin{array}{ccc}
M_{\tilde \tau_{LL}}^2 & e^{-i\varphi_{\tilde \tau}}|M_{\tilde \tau_{LR}}^2|\\[5mm]
e^{i\varphi_{\tilde \tau}}|M_{\tilde \tau_{LR}}^2| & M_{\tilde \tau_{RR}}^2
\end{array}\right)\left(
\begin{array}{ccc}
\tilde \tau_L\\[5mm]
\tilde \tau_R \end{array}\right),
\label{eq:mm}
\end{equation}
with
\begin{eqnarray}
%M_{\tilde \tau_{LL}}^2 & = & m_{\tilde\ell_L}^2 + m_{\tau}^2 ,\\[3mm]
%M_{\tilde \tau_{RR}}^2 & = & m_{\tilde\ell_R}^2 + m_{\tau}^2 ,\\[3mm]
M_{\tilde\tau_{LL}}^2 & = & M_{\tilde L}^2+(-\frac{1}{2}+\sin^2\theta_W)
\cos2\beta \ m_Z^2+m_{\tau}^2,\label{eq:mll}\\[3mm]
M_{\tilde\tau_{RR}}^2 & = & M_{\tilde E}^2-\sin^2\theta_W\cos2\beta \
m_Z^2+m_{\tau}^2, \label{eq:mrr}\\[3mm]
M_{\tilde \tau_{RL}}^2 & = & (M_{\tilde\tau_{LR}}^2)^{\ast}=
  m_{\tau}(A_{\tau}-\mu^{\ast}  
 \tan\beta), \label{eq:mlr}
\end{eqnarray}
\begin{equation}
\varphi_{\tilde \tau}  = \arg\lbrack A_{\tau}-\mu^{\ast}\tan\beta\rbrack,
\label{eq:phtau}
\end{equation}
where $A_{\tau}$ is the complex trilinear scalar coupling parameter
and $M_{\tilde L}$, $M_{\tilde E}$ are the other soft
SUSY--breaking parameters of the $\tilde\tau_i$ system. 
In order to reduce the number of MSSM parameters, we will
often use the renormalization group equations \cite{hall},
$M_{\tilde L}^2=m_0^2 +0.79 M_2^2$ and
$M_{\tilde E}^2=m_0^2 +0.23 M_2^2$.
The $\tilde \tau$ mass eigenstates are 
$(\tilde\tau_1, \tilde \tau_2)=(\tilde \tau_L, \tilde \tau_R)
{{\mathcal R}^{\tilde \tau}}^{T}$, 
with 
\begin{equation}
	{\mathcal R}^{\tilde \tau}
	 =\left( \begin{array}{ccc}
	e^{i\varphi_{\tilde \tau}}\cos\theta_{\tilde \tau} & 
	\sin\theta_{\tilde \tau}\\[5mm]
	-\sin\theta_{\tilde \tau} & 
	e^{-i\varphi_{\tilde \tau}}\cos\theta_{\tilde \tau}
\end{array}\right),
\label{eq:rtau}
\end{equation}
with the mixing angle
\begin{equation}
\cos\theta_{\tilde \tau}=
	\frac{-|M_{\tilde \tau_{LR}}^2|}{\sqrt{|M_{\tilde \tau _{LR}}^2|^2+
	(m_{\tilde \tau_1}^2-M_{\tilde \tau_{LL}}^2)^2}},\quad
\sin\theta_{\tilde \tau}=\frac{M_{\tilde \tau_{LL}}^2-m_{\tilde \tau_1}^2}
	{\sqrt{|M_{\tilde \tau_{LR}}^2|^2+(m_{\tilde \tau_1}^2-M_{\tilde
					\tau_{LL}}^2)^2}},
	\label{eq:thtau}
\end{equation}
and the mass eigenvalues
\begin{equation}
	m_{\tilde \tau_{1,2}}^2 = \frac{1}{2}\left[(M_{\tilde \tau_{LL}}^2+M_{\tilde \tau_{RR}}^2)\mp 
	\sqrt{(M_{\tilde \tau_{LL}}^2 - M_{\tilde \tau_{RR}}^2)^2
		+4|M_{\tilde \tau_{LR}}^2|^2}\;\right].
\label{eq:m12}
\end{equation}

\subsection{First and second generation sfermion masses
			\label{First and second generation sfermion masses}}

The off-diagonal terms of the sfermion mass matrices are
proportional to the fermion mass. For fermions of the 
first and second generation, whose masses are small compared 
to SUSY masses, their mass matrices are diagonal to a good approximation.
For these sfermions we will assume the approximate solutions to the
renormalization group equations \cite{hall}:
\begin{equation}
	m^2_{\tilde{f}_{L,R}}=m_f^2+m_0^2+C(\tilde{f})M_2^2\pm 
	m_Z^2\cos2\beta(T_{3f}-e_f\sin^2\theta_W). \label{1W}
\end{equation}
where 
$T_{3f}$ is the third component of weak isospin,
$m_0$ is the common scalar mass parameter at the GUT-scale,
and $C(\tilde{f})$ depends on the sfermion 
\begin{eqnarray}
	C(\tilde{\ell}_L)\approx 0.79, & \quad & C(\tilde{\ell}_R)\approx 0.23,\\
	C(\tilde{q}_L)\approx 10.8,    & \quad & C(\tilde{q}_R)\approx 10.1.
\end{eqnarray}
For the slepton and sneutrino masses we have
\begin{eqnarray}
	m^2_{\tilde\ell _R} & = & 
	m^2_0+0.23M^2_2-m_Z^2\cos2\beta\,\sin^2\theta_W, \label{mselR}\\
	m^2_{\tilde\ell_L} & = & 
	m^2_0+0.79M^2_2+m_Z^2\cos2\beta(-\frac{1}{2}+\sin^2\theta_W),\label{mselL}\\
	m^2_{\tilde\nu_{\ell}} & = & 
	m^2_0+0.79M^2_2+\frac{1}{2}m_Z^2\cos2\beta. \label{msneut}
\end{eqnarray}

	\chapter{Kinematics and phase space}

\section{Spherical trigonometry}

For the parametrization of the phase space one often needs
the following relations from spherical trigonometry.
Consider the following spherical triangle with sides
$a, b, c $ and angles $A, B, C $.
%The tip of the triangle lies in the origin of 
%the coordinate system.

\begin{center}
%	\fbox{
\setlength{\unitlength}{1cm}
	\begin{picture}(8,8)(0,0)
\put(-3.5,-6){\includegraphics{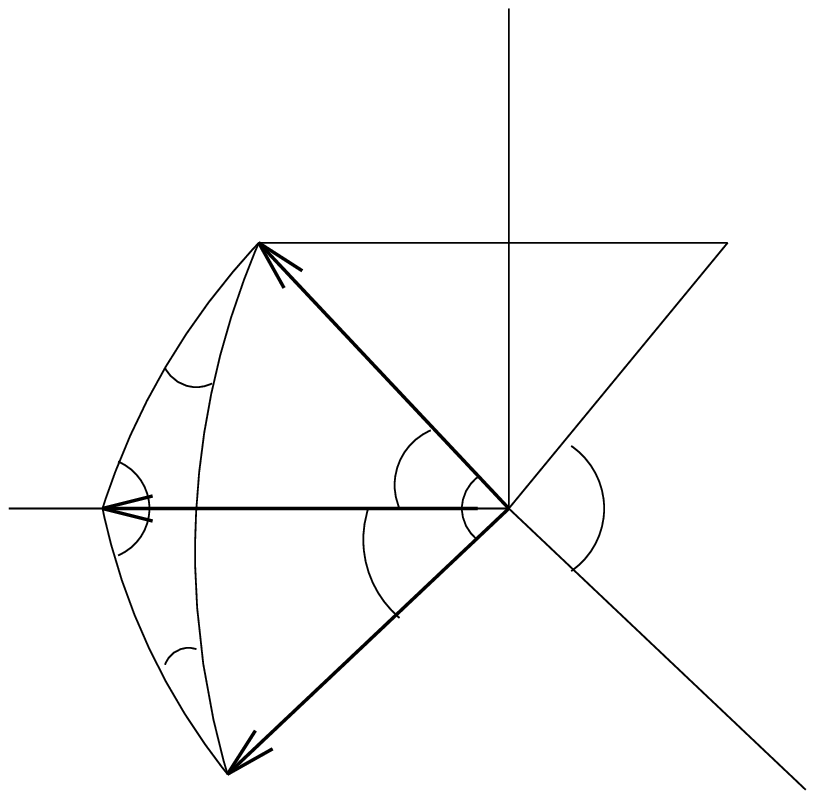}}
\put(1,.5){$x$}
\put(7.4,2.8){$y$}
\put(3.2,7.1){$z$}
\put(3.4,1.9){$A$}
\put(3.7,6.5){$\hat{e}_1$}
\put(.9,5.4){$\hat{e}_2$}
\put(5.9,5){$\hat{e}_3$}
\put(3.3,3.6){$a$}
\put(3.1,4.4){$c$}
\put(3.9,4.1){$b$}
\put(2.5,5.75){$B$}
\put(4.2,5.7){$C$}
\put(2.3,6.7){$A$}
\put(2.7,6.7){\vector(3,-2){.4}}
\end{picture}
%}
\end{center}

The unit vectors of the triangle sides are given by
\begin{eqnarray}
	\hat e_1 & = & (0, 0, 1), \label{anhD_0a}\\
	\hat e_2 & = & (\sin c, 0, \cos c),\label{anhD_0b}\\
	\hat e_3 & = & (\sin b\cos A, \sin b\sin A, \cos b). \label{anhD_0c}
\end{eqnarray}
In the following we give formulas relating
the sides and the angles of the triangle \cite{Byckling,gudidiss}:
\begin{itemize}
\item law of sines
\begin{equation}
	\frac{\sin a}{\sin A}=\frac{\sin b}{\sin B}=\frac{\sin c}{\sin C}. 
	\label{anhD_3}
\end{equation}
\item law of cosines for sides (cosine theorem)
\begin{equation}
	\cos a=\cos b\cos c+\sin b\sin c\cos A. \label{anhD_4}
\end{equation}
Similar formulae for the other sides may be obtained by cyclical 
permutations.
\item law of cosines for angles
\begin{equation}
	\cos A=-\cos B \cos C +\sin B \sin C \cos a, \label{anhD_5}
\end{equation}
etc, cyclically.
\item products of functions from sides and angles
\begin{eqnarray}
	\sin a \cos B&=&\cos b \sin c-\sin b \cos c \cos A, \label{anhD_6a}\\
	\sin a \cos C&=&\cos c \sin b-\sin c \cos b \cos A. \label{anhD_6b}
\end{eqnarray}
The products  $\sin b \cos C$ and $\sin c \cos A$ are obtained
from (\ref{anhD_6a}) by cyclical permutations.
The products $\sin b \cos A$ and $\sin c \cos B$
are obtained from (\ref{anhD_6b}) by cyclical permutations.
\begin{eqnarray}
	\sin A \cos b&=&\cos B \sin C + \sin B \cos C \cos a, \label{anhD_7a}\\
	\sin A \cos c&=&\cos C \sin B + \sin C \cos B \cos a. \label{anhD_7b}
\end{eqnarray}
The products  $\sin B \cos c$ and $\sin C \cos a$ are obtained
from (\ref{anhD_7a}) by cyclical permutations.
The products $\sin B \cos a$ and $\sin C \cos b$  are obtained
from (\ref{anhD_7b}) by cyclical permutations.
\end{itemize}

\section{Kinematics of neutralino/chargino production and decay
  \label{Kinematics of neutralino/chargino production}}

\subsection{Momenta and spin vectors of the production process
  \label{Prod}}

We choose a coordinate frame in the laboratory system
(center of mass system)
such that the momentum of the neutralino $\tilde\chi^0_j$ 
or chargino $\tilde\chi^-_j$, denoted by ${\bf p}_{\chi_j}$, 
points in the $z$-direction (in our definitions we follow 
closely \cite{gudineutralino,gudichargino}). 
The scattering  angle is 
$\theta \angle ({\mathbf p}_{e^-},{\mathbf p}_{\chi_j})$ and 
the azimuth $\phi$ can be chosen zero. The momenta are 
   \begin{eqnarray}
	&&p_{e^-}^{\mu} = E_b(1,-\sin\theta,0, \cos\theta),\quad
     p_{e^+}^{\mu} = E_b(1, \sin\theta,0,-\cos\theta),\\
 &&  p_{\chi_i}^{\mu} = (E_{\chi_i},0,0,-q),\quad
     p_{\chi_j}^{\mu} = (E_{\chi_j},0,0, q),
   \end{eqnarray}
with the beam energy $E_b=\sqrt{s}/2$ and
\begin{eqnarray}
 &&  E_{\chi_i} =\frac{s+m_{\chi_i}^2-m_{\chi_j}^2}{2 \sqrt{s}},\quad
	  E_{\chi_j} =\frac{s+m_{\chi_j}^2-m_{\chi_i}^2}{2 \sqrt{s}},\quad
      q =\frac{\lambda^{\frac{1}{2}}
             (s,m_{\chi_i}^2,m_{\chi_j}^2)}{2 \sqrt{s}}, 
\end{eqnarray}
%where $m_{\chi^+_i}, m_{\chi^-_j}$ are the masses of the charginos and 
where $\lambda(x,y,z) = x^2+y^2+z^2-2(xy+xz+yz)$.
For the description of the polarization of the neutralino 
$\tilde\chi^0_i$ or chargino  $\tilde\chi^+_i$
we choose three spin vectors
\begin{eqnarray}
	&&  s^{1,\,\mu}_{\chi_i}=(0,-1,0,0),\quad
		 s^{2,\,\mu}_{\chi_i}=(0,0,1,0),\quad
		 s^{3,\,\mu}_{\chi_i}=
	  \frac{1}{m_{\chi_i}}(q,0,0,-E_{\chi_i}).
	 \label{spinvec}
\end{eqnarray} 
Together with   
%the unit momentum vector  
$p_{\chi_i}^{\mu}/m_{\chi_i}$ they form an orthonormal set
\begin{eqnarray}
&&s^a_{\chi_i}\cdot s^b_{\chi_i}=-\delta^{ab}, \quad
s^a_{\chi_i}\cdot p_{\chi_i}=0.
\end{eqnarray}

\subsection{Momenta and spin vectors of leptonic decays
	  \label{Leptonic  decays}}

If the neutralino or chargino decays into a lepton,
$\tilde\chi^0_i\to \ell_1 \tilde\ell$ or
$\tilde\chi^+_i\to \ell_1 \tilde\nu_{\ell}$,
in short $\tilde\chi_i\to \ell_1 \tilde\xi$,
it is suitable to parametrize in terms of the angle
$\theta_1=\angle ({\mathbf p}_{\ell_1} ,{\mathbf p}_{\chi_i})$:
 \begin{eqnarray}
 &&  p^{\mu}_{\ell_1} = (                        E_{\ell_1},
            -|{\mathbf p}_{\ell_1}| \sin \theta_1 \cos \phi_1,
             |{\mathbf p}_{\ell_1}| \sin \theta_1 \sin \phi_1,
				-|{\mathbf p}_{\ell_1}| \cos \theta_1), \\
&& E_{\ell_1} = |{\mathbf p}_{\ell_1}| = 
	 \frac{m_{\chi_i}^2-m_{\tilde\xi}^2}{2(E_{\chi_i}-q \cos
			 \theta_{1})}. 
	 	   \label{energy4}
 \end{eqnarray}
For the subsequent slepton decay 
$\tilde\ell \to \ell_2\tilde\chi^0_1$
we define 
$\theta_2=\angle ({\mathbf p}_{\ell_2} ,{\mathbf p}_{\chi_i})$ and
write
 \begin{eqnarray}
&& p^{\mu}_{\ell_2} = (                        E_{\ell_2},
            -|{\mathbf p}_{\ell_2}| \sin \theta_2 \cos \phi_2,
             |{\mathbf p}_{\ell_2}| \sin \theta_2 \sin \phi_2,
				-|{\mathbf p}_{\ell_2}| \cos \theta_2),\\
&& E_{\ell_2} = |{\mathbf p}_{\ell_2}| =
	  \frac{m_{\tilde\ell}^2-m_{\chi^0_1}^2 }
	  {2(E_{\tilde\ell}-|{\mathbf p}_{\chi_i}-
			  {\mathbf p}_{\ell_1}| \cos\theta_{D_2})},
	 	   \label{energy6}
 \end{eqnarray}
with  $\theta_2=\angle ({\mathbf p}_{\ell_{2}} ,{\mathbf p}_{\chi_i})$,
the decay angles 
$\theta_{D_2} \angle ({\mathbf p}_{\tilde\ell},{\mathbf p}_{\ell_2})$,
$\theta_{D_1} \angle ({\mathbf p}_{\chi_i},{\mathbf p}_{\tilde\ell})$ and
 \begin{equation}
\cos\theta_{D_2}=\cos\theta_{D_1}\cos\theta_2-
\sin\theta_{D_1}\sin\theta_2\cos(\phi_2-\phi_1), \quad
\cos\theta_{D_1}=\frac{{\mathbf p}_{\chi_i} 
	({\mathbf p}_{\chi_i}-{\mathbf p}_{\ell_1})}
{ |{\mathbf p}_{\chi_i}|~ |{\mathbf p}_{\chi_i}-{\mathbf p}_{\ell_1}|}.
  \end{equation}
If the neutralino decays into a stau,
$ \tilde\chi^0_i \to \tilde\tau_m \tau$, $m=1,2$ ,
the $\tau$ spin vectors are chosen by
\begin{equation} \label{stau:polvec}
s^1_{\tau}=\left(0,\frac{{\bf s}_2\times{\bf s}_3}
	{|{\bf s}_2\times{\bf s}_3|}\right),\quad
s^2_{\tau}=\left(0, \frac{{\bf p}_{\tau}\times{\bf p}_{e^-}}
	{|{\bf p}_{\tau}\times{\bf p}_{e^-}|}\right),\quad
s^3_{\tau}=\frac{1}{m_{\tau}} \left(|{\bf p}_{\tau}|, 
		\frac{E_{\tau}}{|{\bf p}_{\tau}|}{\bf p}_{\tau} \right).
\end{equation}

\subsection{Phase space for leptonic decays
	  \label{Phase space for leptonic decays}}

For neutralino/chargino 
production $e^+e^-\to\tilde\chi_i\tilde\chi_j$
and subsequent leptonic decay 
$\tilde\chi^0_i\to \ell_1 \tilde\ell$ or
$\tilde\chi^+_i\to \ell_1 \tilde\nu_{\ell}$,
in short $\tilde\chi_i\to \ell_1 \tilde\xi$,
the Lorentz invariant phase-space element
can be decomposed into two-body  phase-space elements \cite{Byckling}:
%\begin{eqnarray}
%&&d{\rm Lips}(s,p_{\chi_j },p_{\ell_1},p_{\chi_1},p_{\ell_2}) =
%	 \nonumber \\ 
%&&\frac{1}{(2\pi)^2}~d{\rm Lips}(s,p_{\chi_i},p_{\chi_j} )
% ~d s_{\chi_i} ~d{\rm Lips}(s_{\chi_i},p_{\tilde\xi},p_{\ell_1})
% ~d s_{\tilde\xi}~d{\rm Lips}(s_{\tilde\xi},p_{\chi_1},p_{\ell_2}),
% \label{Lipsleptonic}
%\end{eqnarray}
\begin{eqnarray}\label{Lipsleptonic1}
d{\rm Lips}(s;p_{\chi_j },p_{\ell_1},p_{\tilde\xi}) =
%	 \nonumber \\ 
\frac{1}{2\pi}~d{\rm Lips}(s;p_{\chi_i},p_{\chi_j} )
 ~d s_{\chi_i}~d{\rm Lips}(s_{\chi_i};p_{\ell_1},p_{\tilde\xi}).
% ~d s_{\tilde\xi}~d{\rm Lips}(s_{\tilde\xi},p_{\chi_1},p_{\ell_2}),
 \label{Lipsleptonic}
\end{eqnarray}
For $\tilde\xi=\tilde\ell$, we can include the subsequent selectron decay
$\tilde\ell \to \ell_2\tilde\chi^0_1$ and have for the complete 
process $e^+e^-\to\tilde\chi_i^0\tilde\chi_j^0$; 
$\tilde\chi^0_i\to \ell_1 \tilde\ell$;
$\tilde\ell \to \ell_2\tilde\chi^0_1$:
\begin{eqnarray}\label{Lipsleptonic2}
d{\rm Lips}(s;p_{\chi_j },p_{\ell_1},p_{\ell_2},p_{\chi_1^0}) =
 \frac{1}{2\pi}~d{\rm Lips}(s;p_{\chi_j },p_{\ell_1},p_{\tilde\ell})
%\frac{1}{(2\pi)^2}~d{\rm Lips}(s,p_{\chi_i},p_{\chi_j} )
% ~d s_{\chi_i} ~d{\rm Lips}(s_{\chi_i},p_{\tilde\xi},p_{\ell_1})
 ~d s_{\tilde\ell}~d{\rm Lips}(s_{\tilde\ell};p_{\ell_2},p_{\chi_1^0},).
\end{eqnarray}
The several parts of the phase space elements are
\begin{eqnarray}
	d{\rm Lips}(s;p_{\chi_i },p_{\chi_j })&=&
	\frac{q}{8\pi\sqrt{s}}~\sin\theta~ d\theta, \\
	d{\rm Lips}(s_{\chi_i};p_{\tilde\xi},p_{\ell_1})&=&
	\frac{1}{2(2\pi)^2}~
	\frac{|{\mathbf p}_{\ell_1}|^2}{m_{\chi_i}^2-m_{\tilde\xi}^2}
	~d\Omega_1,\\
	d{\rm Lips}(s_{\tilde\ell};p_{\ell_2},p_{\chi_1^0})&=&
	\frac{1}{2(2\pi)^2}~\frac{|{\mathbf p}_{\ell_2}|^2}
			{m_{\tilde\xi}^2-m_{\chi_1}^2} ~d\Omega_2,
\end{eqnarray}
with $s_{\chi_i}=p^2_{\chi_i}$, $s_{\tilde\xi}=p^2_{\tilde\xi}$ and 
$ d\Omega_i=\sin\theta_i~ d\theta_i~ d\phi_i$.
We use the narrow width approximation for the propagators
$\int|\Delta(\tilde\chi_i)|^2 $ $ d s_{\chi_i} = 
\frac{\pi}{m_{\chi_i}\Gamma_{\chi_i}}, ~
\int|\Delta(\tilde\ell)|^2 d s_{\tilde\ell} = 
\frac{\pi}{m_{\tilde\ell}\Gamma_{\tilde\ell}}$.
The approximation is justified for
$(\Gamma_{\chi_i}/m_{\chi_i})^2\ll1$,
and $(\Gamma_{\tilde\ell}/m_{\tilde\ell})^2\ll1$,
which holds in our case with 
$\Gamma_{\chi_i}\lsim {\mathcal O}(1 {\rm GeV}) $
and $\Gamma_{\tilde\ell}\lsim {\mathcal O}(1 {\rm GeV}) $.

\subsection{Energy distributions of the decay leptons
	  \label{Energy distributions of the decay leptons}}

For neutralino production 
$e^+e^-\to\tilde\chi_i^0\tilde\chi_j^0$
and the subsequent leptonic decay 
$\tilde\chi_i^0\to \ell_1 \tilde\ell$,
and $\tilde\ell \to \ell_2\tilde\chi^0_1$,
the two decay leptons $\ell_1$ and $\ell_2$
can be distinguished by their different energy distributions. 
The  energy distribution of lepton $\ell_1$ in the laboratory system
has the form of a box with the endpoints
\begin{eqnarray}
  E_{\ell_1,min,max} &=&
  \frac{m_{\chi_i^0}^2-m_{\tilde\ell}^2}{2(E_{\chi_i^0} \pm q)},
\end{eqnarray}
with $q$ the neutralino momentum. The energy distribution 
of the second lepton  $\ell_2$ is obtained  by integrating over 
the energy $E_{\tilde\ell}$ of the decaying slepton 
 \begin{eqnarray}
	 \frac{1}{\sigma} \frac{d \sigma}{dE_{\ell_2}}
	 &=& 
	 \frac{m_{\tilde\ell}^2~m_{\chi_i^0}^2}
	 {q[m_{\chi_i^0}^2-m_{\tilde\ell}^2]
		 [m_{\tilde\ell}^2-m_{\chi_1^0}^2]}  \times
 \left\{ \begin{array}{c@{\quad;\quad}c@{\quad\leq E_{\ell_2}\leq\quad}c}
 ln  \frac{\displaystyle E_{\ell_2}}{\displaystyle A}   &A&a \\
 ln  \frac{\displaystyle   a}{\displaystyle A}   &a&b \\
 ln  \frac{\displaystyle   B}{\displaystyle E_{\ell_2}} &b&B \\
 \end{array}  
 \right.
 \end{eqnarray}
with
 \begin{eqnarray}
 A,B &=& \frac{m_{\tilde\ell}^2-m_{\chi_1^0}^2}
              {2m_{\tilde\ell}^2}
        \left( E_{\tilde\ell,max} \mp 
                 \sqrt{E_{\tilde\ell,max}^2-m_{\tilde\ell}^2} 
        \right)        \\
 a,b &=& \frac{m_{\tilde\ell}^2-m_{\chi_1^0}^2}
              {2m_{\tilde\ell}^2}
        \left( E_{\tilde\ell,min} \mp 
                 \sqrt{E_{\tilde\ell,min}^2-m_{\tilde\ell}^2} 
        \right)\\
    E_{\tilde{\ell},max,min} &=& 
             \frac{E_{\chi_i^0}(m_{\chi_i^0}^2+m_{\tilde\ell}^2) 
   \pm (m_{\chi_i^0}^2-m_{\tilde\ell}^2)\sqrt{E_{\chi_i^0}^2-m_{\chi_i^0}^2}}
                      {2 m_{\chi_i^0}^2}.
 \end{eqnarray}

\subsection{Momenta and spin vectors of bosonic decays
	  \label{Bosonic decays}}

For the bosonic two-body decays of 
neutralino 
$\tilde\chi^0_i\to Z^0\tilde\chi_n^0$
or chargino
$\tilde\chi^+_i\to W^+\tilde\chi_n^0$,
in short
$\tilde\chi_i\to B\tilde\chi_n^0$,
we define the decay angle between neutralino
(or chargino) and the boson $B$ as  
$\theta_{1} \angle ({\mathbf p}_{\chi_i},{\mathbf p}_B)$.
The angle is constrained  by $\sin\theta^{\rm max}_{1}= q'/q$
for $q>q'$, where 
$q'=\lambda^{\frac{1}{2}}(m^2_{\chi_i},m^2_B,m^2_{\chi^0_n})/2m_B$
is the neutralino (chargino) momentum if the boson $B$ is 
produced at rest. In this case there are two solutions 
\begin{eqnarray}	\label{momentumB}
| {\mathbf p}^{\pm}_B|&=& \frac{
(m^2_{\chi_i}+m^2_B-m^2_{\chi^0_n})q\cos\theta_{1}\pm
E_{\chi_i}\sqrt{\lambda(m^2_{\chi_i},m^2_B,m^2_{\chi^0_n})-
	 4q^2~m^2_B~(1-\cos^2\theta_{1})}}
	{2q^2 (1-\cos^2\theta_{1})+2 m^2_{\chi_i}}.\nonumber \\
&&
\end{eqnarray}
For $q'>q$, the angle $\theta_{1}$ is not 
constrained and only the physical solution 
$ |{\mathbf p}^+_B|$ is left. 
We parametrize the momenta of the decay 
$B\to f \bar f$ in the laboratory system as
 \begin{eqnarray}
   p_{B}^{\pm,\,\mu} &=& (                        E_{B}^{\pm},
            -|{\mathbf p}_{B}^{\pm}| \sin \theta_{1} \cos \phi_{1},
             |{\mathbf p}_{B}^{\pm}| \sin \theta_{1} \sin \phi_{1},
				 -|{\mathbf p}_{B}^{\pm}| \cos \theta_{1}), \\
 p_{\bar f}^{\, \mu} &=& (                        E_{\bar f},
            -|{\mathbf p}_{\bar f}| \sin \theta_{2} \cos \phi_{2},
             |{\mathbf p}_{\bar f}| \sin \theta_{2} \sin \phi_{2},
				-|{\mathbf p}_{\bar f}| \cos \theta_{2}),\\
 E^{\mu}_{\bar f} &=& |{\mathbf p}_{\bar f}| = 
\frac{m_B^2}{2(E_{B}^{\pm}-|{\mathbf p}_{B}^{\pm}|\cos\theta_{D_2})},
 \end{eqnarray}
with $\theta_{2} \angle ({\mathbf p}_{\chi_i},{\mathbf p}_{\bar f})$
and the decay angle 
$\theta_{D_2} \angle ({\mathbf p}_{B},{\mathbf p}_{\bar f})$ given by
 \begin{equation}
\cos\theta_{D_2}=\cos\theta_{1}\cos\theta_{2}+
\sin\theta_{1}\sin\theta_{2}\cos(\phi_{2}-\phi_{1}). 
  \end{equation}
The three spin vectors $t^{c}_B$ of the  boson $B=Z^0,W^+$
are in the laboratory system
\begin{eqnarray}\label{defoft}
t^{1,\,\mu}_B=\left(0,\frac{ {\mathbf t}^2_B
\times {\mathbf t}_B^3}{|{\mathbf t}_B^2\times {\mathbf t}^3_B|}\right),\quad
t^{2,\,\mu}_B=\left(0,
\frac{{\mathbf p}_{e^-}\times {\mathbf p}_B}{|{\mathbf p}_{e^-}
\times {\mathbf p}_B|}\right),\quad
t^{3,\,\mu}_B=\frac{1}{m_B}
\left(|{\mathbf p}_B|, E_B \frac{ {\mathbf p}_B}{|{\mathbf p}_B|} \right).
\nonumber \\&&
  \end{eqnarray}
Together with $p_B^{\mu}/m_B$ they form an orthonormal set.
The polarization four-vectors 
$\varepsilon^{\lambda_k}$ 
for helicities $\lambda_k=-1,0,+1$ of the boson
are defined by
\begin{equation}\label{circularbasis}
	\varepsilon^-={\textstyle \frac{1}{\sqrt 2}}(t^1_B-i t^2_B);
	\quad \varepsilon^0=t^3_B; \quad
	\varepsilon^+=-{\textstyle \frac{1}{\sqrt 2}}(t^1_B+i t^2_B).
\end{equation}

\subsection{Phase space for bosonic decays
	  \label{Phase space for bosonic decays}}

For neutralino/chargino 
production $e^+e^-\to\tilde\chi_i\tilde\chi_j$ and subsequent decay 
of the neutralino
$\tilde\chi^0_i\to Z^0\tilde\chi_n^0$
or chargino
$\tilde\chi^+_i\to W^+\tilde\chi_n^0$,
in short
$\tilde\chi_i\to B\tilde\chi_n^0$,
the Lorentz invariant phase-space element
%(\ref{production}) and the decay 
%chain (\ref{decay_1})-(\ref{decay_2}) 
can be decomposed into two-body  phase-space elements
\cite{Byckling}:
\begin{eqnarray}\label{Lipsbosonic1}
d{\rm Lips}(s;p_{\chi_j },p_{\chi^0_n},p_{B}) =
%	 \nonumber \\ 
\frac{1}{2\pi}~d{\rm Lips}(s;p_{\chi_i},p_{\chi_j} )
~d s_{\chi_i} ~\sum_{\pm}d{\rm Lips}(s_{\chi_i};p_{\chi^0_n},p_B^{\pm}).
 \end{eqnarray}
If we include the subsequent decay $B\to f \bar f$ we have
%\begin{eqnarray}
%&&d{\rm Lips}(s,p_{\chi_j },p_{\chi^0_n},p_{f},p_{\bar f}) =
%	 \nonumber \\ 
%&&\frac{1}{(2\pi)^2}~d{\rm Lips}(s,p_{\chi_i},p_{\chi_j} )
%~d s_{\chi_i} ~\sum_{\pm}d{\rm Lips}(s_{\chi_i},p_{\chi^0_n},p_B^{\pm})
% ~d s_B~d{\rm Lips}(s_B,p_f,p_{\bar f}),\label{Lipsbosonic}
% \end{eqnarray}
\begin{eqnarray}\label{Lipsbosonic2}
d{\rm Lips}(s;p_{\chi_j },p_{\chi^0_n},p_{f},p_{\bar f}) =
%	 \nonumber \\ 
  \frac{1}{2\pi}~d{\rm Lips}(s;p_{\chi_j },p_{\chi^0_n},p_{B})
	 ~d s_B~d{\rm Lips}(s_B;p_f,p_{\bar f}).
 \end{eqnarray}
The several parts of the phase space elements are given by
 \begin{eqnarray}
	d{\rm Lips}(s;p_{\chi_i },p_{\chi_j })&=&
	\frac{q}{8\pi\sqrt{s}}~\sin\theta~ d\theta, \\
	d{\rm Lips}(s_{\chi_i};p_{\chi^0_n},p_B^{\pm})&=&
\frac{1}{2(2\pi)^2}~
\frac{|{\mathbf p}_B^{\pm}|^2}{2|E_B^{\pm}~q\cos\theta_1-
	E_{\chi_i}~|{\mathbf p}^{\pm}_B||}~d\Omega_1,\\
	d{\rm Lips}(s_B;p_f,p_{\bar f})&=&
\frac{1}{2(2\pi)^2}~\frac{|{\mathbf p}_{\bar f}|^2}{m_B^2}
	~d\Omega_2,
\end{eqnarray}
with $s_{\chi_i}=p^2_{\chi_i}$, $s_B=p^2_B$ and 
$ d\Omega_i=\sin\theta_i~ d\theta_i~ d\phi_i$.
We use the narrow width approximation for the propagators
$\int|\Delta(\tilde\chi_i)|^2 $ $ d s_{\chi_i} = 
\frac{\pi}{m_{\chi_i}\Gamma_{\chi_i}}, ~
\int|\Delta(B)|^2 d s_B = 
\frac{\pi}{m_B\Gamma_B}$.
The approximation is justified for
$(\Gamma_{\chi_i}/m_{\chi_i})^2\ll1$,
which holds in our case with 
$\Gamma_{\chi_i}\lsim {\mathcal O}(1 {\rm GeV}) $.

\section{Kinematics of sfermion decays
  \label{Kinematics of sfermion decays}}

\subsection{Momenta and spin vectors 
  \label{Momenta and spin vectors}}

We consider the slepton decay chain
$\tilde \ell\to \ell \tilde\chi^0_j$,
$\tilde\chi^0_j \to \tilde\chi^0_1Z$,
$Z\to f\bar f$.
The substitutions which must be made for similar decay 
chains of a squark are obvious.
%\subsection{Momentum and spin vectors
%	  \label{Momentum and spin vectors}}
%
The momenta in the slepton rest frame are
\begin{eqnarray}
	p^{\mu}_Z &=&(E_Z,0,0,|{\bf p}_Z^{\pm}|),\\ 
	p^{\mu}_{\chi^0_j} &=&|{\bf p}_{\chi^0_j}|
	(E_{\chi^0_j}/|{\bf p}_{\chi^0_j}|,\sin\theta_1,0,\cos\theta_1),\\
	p^{\mu}_{\bar f} &=& |{\bf p}_{\bar f}|(E_{\bar f}/|{\bf p}_{\bar f}
		|,\sin\theta_2\cos\phi_2,\sin\theta_2\sin\phi_2,\cos\theta_2),
	\label{sfermion:chi}\\
	|{\bf p}_{\chi^0_j}|&=&\frac{m^2_{\tilde \ell}-
	m^2_{\chi^0_j}}{2~m_{\tilde \ell}},\qquad
	|{\bf p}_{\bar f}|=\frac{m^2_Z}{2(E_Z-|{\bf p}_Z|\cos\theta_2)}.
\end{eqnarray}
There are two solutions for $|{\bf p}_Z^{\pm}|$, see
(\ref{momentumB}), if the decay angle 
$\theta_1=\angle ({\mathbf p}_Z ,{\mathbf p}_{\chi_i})$
is constrained by
\begin{equation}
	\sin\theta^{\rm max}_1 =
	\frac{m_{\tilde \ell}}{m_Z}
	\frac{\lambda^{\frac{1}{2}}(m^2_{\chi_j},m^2_Z,m^2_{\chi_1})}
	{(m^2_{\tilde \ell}-m^2_{\chi_j})}\leq 1.
\end{equation}
%there are two solutions for $|{\bf p}_Z^{\pm}|$, see  (\ref{momentumB}).
The $\tilde\chi^0_j$ spin vectors in the $\tilde \ell$ rest frame are 
\begin{eqnarray} \label{sfermion:spinchi}
s^{1,\, \mu}_{\chi^0_j}=\left(0,\frac{{\bf s}^2_{\chi^0_j}
			\times{\bf s}^3_{\chi^0_j}}
	{|{\bf s}^2_{\chi^0_j}\times{\bf s}^3_{\chi^0_j}|}\right),~%\quad
s^{2,\, \mu}_{\chi^0_j}=\left(0,
	\frac{{\bf p}_{\chi^0_j}\times{\bf p}_Z}{|{\bf p}_{\chi^0_j}
	\times{\bf p}_Z|}\right),~%\quad
s^{3,\, \mu}_{\chi^0_j}=\frac{1}{m_{\chi^0_j}}
	\left(|{\bf p}_{\chi^0_j}|, E_{\chi^0_j} 
		\;\frac{{\bf p}_{\chi^0_j}}{ |{\bf p}_{\chi^0_j}|} \right).
	\nonumber \\
	&&
\end{eqnarray}
Together with $p^{\mu}_{\chi^0_j}/m_{\chi^0_j}$ they
form an orthonormal set.

\subsection{Phase space for sfermion decays
	  \label{Phase space for sfermion decays}}

The Lorentz invariant phase-space element 
for the decay chain $\tilde \ell\to \ell \tilde\chi^0_j$,
$\tilde\chi^0_j \to \tilde\chi^0_1Z$, $Z\to f\bar f$
can be written in the rest frame of $\tilde \ell$ as
\begin{eqnarray}
&&	d{\rm Lips}(m^2_{\tilde\ell};p_{\ell},p_{\chi^0_1},p_{\bar f},p_{f}) =
	\nonumber\\
&&	\frac{1}{(2\pi)^2}~d{\rm Lips}(m^2_{\tilde \ell};p_{\ell},p_{\chi^0_j})
	\,ds_{\chi^0_j}  \sum_{\pm}
	d{\rm Lips}(s_{\chi^0_j};p_{\chi^0_1},p_Z^{\pm})
	\,d s_{Z}\,d{\rm Lips}(s_{Z};p_{\bar f},p_{f}),
\end{eqnarray}
\begin{eqnarray}
	d{\rm Lips}(m^2_{\tilde \ell};p_\ell,p_{\chi^0_j})&=&\frac{1}{8(2\pi)^2}
	\left(1-\frac{m^2_{\chi^0_j}}{m^2_{\tilde \ell}}\right)~ d\Omega,\\
	d{\rm Lips}(s_{\chi^0_j};p_{\chi^0_1},p^{\pm}_Z)&=&\frac{1}{4(2\pi)^2}~
	\frac{|{\bf p}^{\pm}_Z|^2}{|E_Z^{\pm}~|{\bf p}_{\chi^0_j}|\cos\theta_1-
	E_{\chi^0_j}~|{\bf p}^{\pm}_Z||}~ d\Omega_1, \\
	d{\rm Lips}(s_{Z};p_{\bar f},p_{f})&=&\frac{1}{8(2\pi)^2}
	\frac{m^2_Z}{(E_Z^{\pm}-|{\bf p}^{\pm}_Z|\cos\theta_2)^2}~ d\Omega_2,
\end{eqnarray}
with $s_{\chi^0_j}=p^2_{\chi^0_j}$, $s_{Z}=p^2_Z$ and
$ d\Omega_i=\sin\theta_i~ d\theta_i~ d \phi_i$.
We use the narrow width approximation for the propagators
$\int|\Delta(\tilde\chi_j^0)|^2 $ $ d s_{\chi_j^0} = 
\frac{\pi}{m_{\chi_j^0}\Gamma_{\chi_j^0}}, ~
\int|\Delta(Z)|^2 d s_Z = 
\frac{\pi}{m_Z\Gamma_Z}$.
The approximation is justified for
$(\Gamma_{\chi_j^0}/m_{\chi_j^0})^2\ll1$,
which holds in our case with 
$\Gamma_{\chi_j^0}\lsim {\mathcal O}(1 {\rm GeV}) $.

	\chapter{Spin-density matrices for neutralino production and decay 
	\label{Neutralino production and decay matrices}}

We give the analytic formulae for the 
%differential cross section
squared amplitudes for neutralino production, 
$e^+~e^-\to\tilde\chi^0_i~\tilde\chi^0_j$,
with longitudinally polarized beams, 
and for different subsequent two-body decay chains 
of one neutralino.
%In~(\ref{Appendix:neutproduction})  
%Momentum and helicity are denoted by $p$ and $\lambda$, respectively.
%In order to calculate the squared amplitudes 
%for such processes of neutralino 
%production and decay,
%$e^+\,e^- \to\tilde\chi^0_i \, \tilde\chi^0_j$ 
%followed by a two-body decay chain of the  
%neutralino~$\tilde\chi^0_i$, 
We use the spin density matrix formalism 
as in \cite{gudineutralino,spinhaber,gudidiss}.
The amplitude squared can then be written
\begin{eqnarray}    \label{neut:amplitude}
|T|^2 &=& |\Delta(\tilde\chi^0_i)|^2~
	\sum_{\lambda_i\lambda_i'}~  
	\rho_P(\tilde\chi^0_i)^{\lambda_i \lambda_i'}~
	\rho_D(\tilde\chi^0_i)_{\lambda_i'\lambda_i},
\end{eqnarray}
with $\rho_P(\tilde\chi^0_i)$  the spin density production matrix 
of neutralino $ \tilde\chi^0_i$, the propagator 
$ \Delta(\tilde\chi^0_i)=i/[s_{\chi_i^0}-m_{\chi_i^0}^2
	+im_{\chi_i^0}\Gamma_{\chi_i^0}]$ and
the neutralino decay matrix $\rho_D(\tilde\chi^0_i)$.
%The helicities of the neutralino $\tilde\chi^0_i$ are denoted by 
%$\lambda_i$ and $\lambda_i'$. 

\section{Neutralino production
  \label{Neutralino production}}

\begin{figure}[h]
\begin{minipage}[h]{3cm}
\begin{center}
{\setlength{\unitlength}{1cm}
\begin{picture}(2.5,2.5)
\put(1,0){\includegraphics{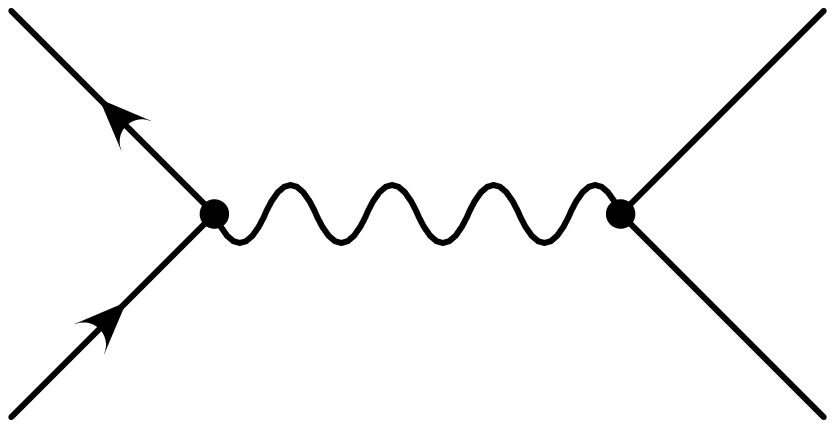}}
\put(0,-.6){\small $e^{-}$}
\put(4.5,-.6){\small $\tilde{\chi}^{0}_i$}
\put(0,1.7){\small $e^{+}$}
\put(4.5,1.7){\small $\tilde{\chi}^{0}_j$}
\put(2.1,1){\small $Z^0$}
\end{picture}}
\end{center}
\end{minipage}
\hspace{3cm}
\vspace{.8cm}
\begin{minipage}[h]{2.5cm}
\begin{center}
{\setlength{\unitlength}{1cm}
\begin{picture}(2.5,2.5)
\put(0,0){\includegraphics{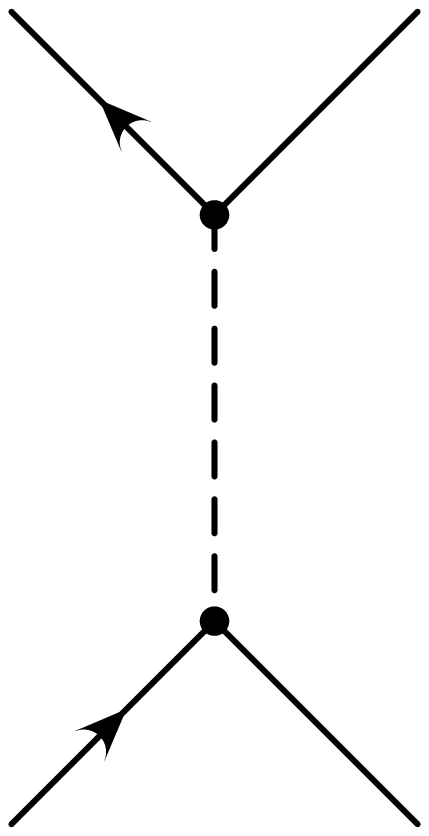}}
\put(-.2,-1.1){\small $e^{-}$}
\put(2.5,-1.1){\small $\tilde{\chi}^{0}_i$}
\put(-.2,2.4){\small $e^{+}$}
\put(2.5,2.4){\small $\tilde{\chi}^{0}_j$}
\put(.5,.5){\small $\tilde{e}_{L,R}$}
\end{picture}}
\end{center}
\end{minipage}
\hspace{3cm}
\vspace{.8cm}
\begin{minipage}[h]{2.5cm}
\begin{center}
{\setlength{\unitlength}{1cm}
\begin{picture}(2.5,2)
\put(0,-.5){\includegraphics{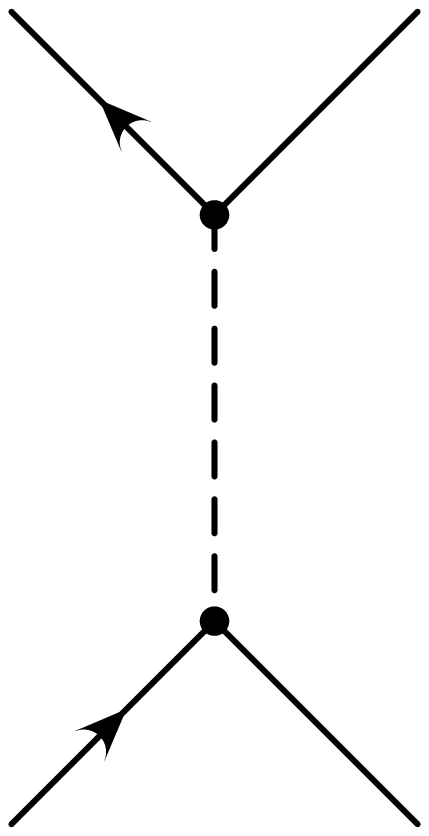}}
\put(-1.4,-1.4){\small $e^{-}$}
\put(-1.4,2.3){\small $e^{+}$}
\put(1.3,2.3){\small $\tilde{\chi}^{0}_i$}
\put(1.3,-1.4){\small $\tilde{\chi}^{0}_j$}
\put(-.8,.2){\small $\tilde{e}_{L,R}$}
\end{picture}}
\end{center}
\end{minipage}
%\begin{center}
%{\parbox{9cm}{\small  
%Fig.~B.3: Feynman--Diagram f\"ur 
%$e^+ e^-\to \tilde{\chi}^0_{i} \tilde{\chi}^0_j$ \la{abb4_3}}}
%\end{center}
\caption{Feynman diagrams for neutralino production
	\label{Feynman diagramms for neutralino production}}
\end{figure}

For the production of neutralinos
\begin{eqnarray} \label{Appendix:neutproduction}
	e^++e^-&\to&\tilde\chi^0_i(p_{\chi_i^0}, \lambda_i)+
	            \tilde\chi^0_j(p_{\chi_j^0}, \lambda_j), 
\end{eqnarray}
with momentum $p$ and helicity $\lambda$, the unnormalized
spin-density matrix of neutralino $\tilde\chi^0_i$ is defined as
\begin{eqnarray} \label{neut:rhoPdef}
	\rho_P(\tilde\chi^0_i)^{\lambda_i \lambda_i'}&=&\sum_{\lambda_j}
	T_P^{\lambda_i \lambda_j}T_P^{\lambda_i' \lambda_j \ast}.
\end{eqnarray}
The helicity amplitudes are \cite{gudineutralino,gudidiss}: 
\begin{eqnarray}
%& &\mbox{\hspace*{-.5cm}}
T_P^{\lambda_i\lambda_j}(s,Z) &=&\frac{g^2}{\cos^2\theta_W}
		\Delta^s(Z)\bar{v}(p_{e^+})\gamma^{\mu}
		(L_{e} P_L+R_{e}P_R)u(p_{e^-}) 
	 \nonumber\\& &
%	 \mbox{\hspace*{-.5cm}}
%	\phantom{T_P^{\lambda_i\lambda_j}(s)
%=-\frac{g^2}{\cos^2\Theta_W}\Delta(Z)}
\times\bar u(p_{\chi^0_j},\lambda_j)\gamma_{\mu} 
	(O^{''L}_{ji} P_L+O^{''R}_{ji} P_R) v (p_{\chi^0_i},\lambda_i),
		\label{neut:T1}\\
%& &\mbox{\hspace*{-.5cm}}
T_P^{\lambda_i\lambda_j}(t,\tilde e_{L})&=&
		-g^2 f^L_{e i}f^{L*}_{e j}\Delta^t(\tilde e_L)
		\bar v (p_{e^+}) P_R   v (p_{\chi^0_i},\lambda_i)
		\bar u (p_{\chi^0_j},\lambda_j)P_L u(p_{e^-}),\label{neut:T2}\\
%& &\mbox{\hspace*{-.5cm}}
T_P^{\lambda_i\lambda_j}(t,\tilde e_{R})&=&
		-g^2 f^R_{ e i}f^{R*}_{e j}\Delta^t(\tilde e_R)
		\bar v (p_{e^+}) P_L  v (p_{\chi^0_i},\lambda_i)
		\bar u (p_{\chi^0_j},\lambda_j)P_R u(p_{e^-}),\label{neut:T3}\\
%& &\mbox{\hspace*{-.5cm}}
T_P^{\lambda_i\lambda_j}(u,\tilde e_L)&=&
		g^2 f^{L*}_{e i} f^L_{e j}\Delta^u(\tilde e_L)
		\bar{v}(p_{e^+}) P_R v (p_{\chi^0_j},\lambda_j)
		\bar u(p_{\chi^0_i},\lambda_i)P_L u(p_{e^-}),\label{neut:T4}\\
%& &\mbox{\hspace*{-.5cm}}
T_P^{\lambda_i\lambda_j}(u,\tilde e_R)&=&
		g^2 f^{R*}_{e i} f^R_{e j}\Delta^u(\tilde e_R)
		\bar v(p_{e^+}) P_L v(p_{\chi^0_j},\lambda_j)
		\bar u(p_{\chi^0_i},\lambda_i)P_R u(p_{e^-}).\label{neut:T5}
\end{eqnarray}
The propagators are 
    \begin{equation}\label{neut:propagators}
        \Delta^s(Z)=\frac{i}{s-m^2_Z+im_Z\Gamma_Z},\quad
                  \Delta^{t}(\tilde{e}_{R,L})=
            \frac{i}{t-m^2_{\tilde{e}_{R,L}}},\quad
              \Delta^u (\tilde{e}_{R,L})=
             \frac{i}{u-m^2_{\tilde{e}_{R,L}}},
         \end{equation}
with 
$s=(p_{e^-}+p_{e^+})^2$, $t=(p_{e^-}-p_{\chi^0_j})^2$ and 
$u=(p_{e^-}-p_{\chi^0_i})^2$.
The Feynman diagrams are shown in 
Fig.~\ref{Feynman diagramms for neutralino production}.

%We give the analytical formula for 
%the neutralino production matrix  $\rho_P(\tilde\chi^0_i)$. 
For the polarization of the decaying neutralino $\tilde\chi^0_i$
with momentum $p_{\chi_i^0}$ we have introduced 
three space like spin vectors $s^a_{\chi_i^0}$~(\ref{spinvec}).
%which together with $p_{\chi_i^0}^{\mu}/m_{\chi_i^0}$
%form an orthonormal set:  
%$s^a_{\chi_i^0}\cdot s^b_{\chi_i^0}=-\delta^{ab}$, 
%$s^a_{\chi_i^0}\cdot p_{\chi_i^0}=0$.
Then the neutralino production matrix~(\ref{neut:rhoPdef})
can be expanded in terms of the Pauli matrices,
see Appendix~\ref{Bouchiat-Michel formulae for spin 1/2 particles}:
\begin{eqnarray} \label{neut:rhoP}
  \rho_P(\tilde\chi^0_i)^{\lambda_i \lambda_i'} &=&
      2\big(\delta_{\lambda_i \lambda_i'} P + 
        \sigma^a_{\lambda_i \lambda_i'}
		\Sigma_P^a\big),   
\end{eqnarray}
where we sum over $a$ and the factor 2 
%in~(\ref{neut:rhoP}) 
is due to the summation of the helicities of the second 
neutralino~$\tilde\chi^0_j$, whose decay will not be considered.
With our choice of the spin vectors, 
$\Sigma^3_P/P$ is the longitudinal polarization of 
neutralino~$\tilde\chi^0_i$,
$\Sigma^1_P/P$ is the transverse polarization in the 
production plane and $\Sigma^2_P/P$ is the polarization
perpendicular to the production plane. 
Only if there are non-vanishing CP phases $\varphi_{M_1}$
and/or $\varphi_{\mu}$ in the neutralino sector, and
only if two different neutralinos are produced,
$e^+~e^-\to\tilde\chi^0_i~\tilde\chi^0_j$, $i \neq j$,
the polarization $\Sigma^2_P/P$ perpendicular to the 
production plane is non-zero. Thus it is a probe for
CP violation in the production of an unequal pair of neutralinos.
Note that $\Sigma^2_P$ also gets contributions from
the finite $Z$ width, which however do not signal CP violation.
%which is however a higher order effect and will be small. 

We give the analytical 
formulae for $P$ and $\Sigma_P^1,\Sigma_P^2,\Sigma_P^3$ 
in the laboratory system \cite{gudidiss} in the following sections. 
Lorentz invariant expressions for these functions 
can be found in \cite{gudineutralino,gudidiss}.

\subsection{Neutralino polarization independent quantities 
     \label{Neutralino polarization independent quantities}}

The coefficient $P$ is independent of the neutralino polarization. 
It can be composed into contributions from the 
 different production channels
\begin{equation}
	P=P(Z Z)+P(Z \tilde{e}_R)+P(\tilde{e}_R \tilde{e}_R)+
	P(Z \tilde{e}_L)+P(\tilde{e}_L \tilde{e}_L),\label{eq_15}
\end{equation}
with
\begin{eqnarray}
P(Z Z)&=&
	2 \frac{g^4}{\cos^4\theta_W}|\Delta^s(Z)|^2
	(R_{e}^2 c_R + L_{e}^2 c_L) E_b^2 
\nonumber\\& &\times
	\Big\{ |O^{''R}_{ij}|^2 (E_{\chi^0_i} E_{\chi^0_j} + 
	q^2\cos^2\theta)-[(Re O^{''R}_{ij})^2 -(Im O^{''R}_{ij})^2]
	m_{\chi^0_i} m_{\chi^0_j}\Big\}\label{eq_16},\label{P_1}\\
P(Z \tilde{e}_R) & = &
	\frac{g^4}{\cos^2 \theta_W} R_{e} c_R E_b^2
	Re\Big\{\Delta^s(Z)
\nonumber\\& & \times
	\Big[ -(\Delta^{t*}(\tilde{e}_R) f^{R*}_{e i} 
	f^{R}_{e j} O^{''R*}_{ij}
	+\Delta^{u*}(\tilde{e}_R) f^R_{e i} f^{R*}_{e j}
	O^{''R}_{ij}) m_{\chi^0_i} m_{\chi^0_j}
\nonumber\\& &
	+(\Delta^{t*}(\tilde{e}_R) f^{R*}_{e i} f^{R}_{e j} O^{''R}_{ij}
	+\Delta^{u*}(\tilde{e}_R) f^R_{e i} f^{R*}_{e j} O^{''R*}_{ij})
(E_{\chi^0_i} E_{\chi^0_j}+q^2\cos^2\theta)
\nonumber\\& & 
	-(\Delta^{t*}(\tilde{e}_R) f^{R*}_{e i} f^{R}_{e j} O^{''R}_{ij}
	-\Delta^{u*}(\tilde{e}_R) f^R_{e i} f^{R*}_{e j} O^{''R*}_{ij})
	2 E_b q \cos\theta
\Big]\Big\},\label{P_2}\\
P(\tilde{e}_R \tilde{e}_R)&=&
	\frac{g^4}{4} c_R  E_b^2 
	\Big\{|f^R_{e i}|^2 |f_{e j}^R|^2  \times \nonumber\\& &
		%		\mbox{\hspace*{-.5cm}}
	\Big[ (|\Delta^t(\tilde{e}_R)|^2 +|\Delta^u(\tilde{e}_R)|^2)
		(E_{\chi^0_i} E_{\chi^0_j}+q^2 \cos^2\theta)
\nonumber\\& &
			-(|\Delta^t(\tilde{e}_R)|^2-|\Delta^u(\tilde{e}_R)|^2)
	2 E_b q \cos\theta\Big]
\nonumber\\ &&
%& &\mbox{\hspace*{-.5cm}}
	-Re\{(f^{R*}_{e i})^2 (f^R_{e j})^2
     \Delta^u(\tilde{e}_R)\Delta^{t*}(\tilde{e}_R)\}
	2 m_{\chi^0_i} m_{\chi^0_j}\Big\}.\label{P_3}
\end{eqnarray}
To obtain the quantities
$P(Z\tilde{e}_L),P(\tilde{e}_L \tilde{e}_L)$ one
has to exchange in (\ref{P_2}) and (\ref{P_3})
\begin{eqnarray}\nonumber
     &&\Delta^{t}(\tilde{e}_R)\to\Delta^{t}(\tilde{e}_L),\quad
	  \Delta^{u}(\tilde{e}_R)\to\Delta^{u}(\tilde{e}_L),\quad
%	  P^3_{-}\to P^3_{+},\quad P^3_{+}\to P^3_{-}\\
	 c_R \to c_L\\ 
	  &&R_{e}\to L_{e},\quad
         O^{''R}_{ij}\to O^{''L}_{ij},\quad
           f_{e i}^R\to f_{e i}^L,\quad
			  f_{e j}^R\to f_{e j}^L. \label{exchange}
\end{eqnarray}

The longitudinal beam polarizations are included in the weighting factors  
\begin{equation}
c_L =(1-P_{e^-})(1+P_{e^+}), \quad c_R= (1+P_{e^-})(1-P_{e^+}).
\end{equation}
%The longitudinal beam polarization of $e^{-} (e^{+})$ are denoted by
%$P_{-}^3 (P_{+}^3)$, respectively.
Generally the contributions from the exchange  
of $\tilde{e}_{R}$ ($\tilde{e}_{L}$) 
is enhanced and that of $\tilde{e}_{L}$ ($\tilde{e}_{R}$) is
suppressed for $P_{e^-} > 0 , P_{e^+} < 0 ~(P_{e^-}<0,P_{e^+}>0)$.

\subsection{Neutralino polarization 
     \label{Neutralino polarization}}

The coefficients $\Sigma^a_P$, which describe the polarization of the 
neutralino $\tilde{\chi}^0_i$, decompose into
\begin{equation}\label{neut:Sigma2P}
\Sigma_P^a=
     \Sigma_P^a(ZZ)
   + \Sigma_P^a(Z\tilde{e}_R)
	+ \Sigma_P^a(\tilde{e}_R \tilde{e}_R)
	+ \Sigma_P^a(Z\tilde{e}_L)
   + \Sigma_P^a(\tilde{e}_L \tilde{e}_L).
\end{equation}
\begin{itemize}
	\item
The contributions to the transverse 
%$\tilde\chi^0_i$ 
polarization in the production plane are
\end{itemize}
\begin{eqnarray}
\Sigma_P^1(ZZ)&=&
	2 \frac{g^4}{\cos^4\theta_W} |\Delta^s(Z)|^2 E_b^2 \sin\theta
	(R_{e}^2c_R - L_{e}^2 c_L )
%	[(1-P^3_-P^3_+)(R_{e}^2-L_{e}^2) + (P^3_- - P^3_+)(R_{e}^2+L_{e}^2)] 
	\nonumber\\ & & \times
	\Big[ |O^{''R}_{ij}|^2 m_{\chi^0_i} E_{\chi^0_j}
	-[(Re O^{''R}_{ij})^2 -(Im O^{''R}_{ij})^2]  m_{\chi^0_j}
	E_{\chi^0_i}\Big],\label{s1_zz} \\
%------------------------------------------------------------------------
\Sigma_P^1(Z\tilde{e}_R) &=& 
	\frac{-g^4}{\cos^2\theta_W} R_{e} c_R E_b^2 \sin\theta
%	(1+P^3_-)(1-P^3_+) 
	\nonumber\\ & & \times
	\Big[- Re\big\{\Delta^s(Z)
	\big[f^{R}_{e i}f^{R*}_{e j} O^{''R*}_{ij} \Delta^{u*}(\tilde{e}_R)
	+f^{R*}_{e i}f^{R}_{e j} O^{''R}_{ij} \Delta^{t*}(\tilde{e}_R)\big]
	m_{\chi^0_i} E_{\chi^0_j}\big\}
	\nonumber\\ & & 
	-Re\big\{\Delta^s(Z) \big[f^{R}_{e i}f^{R*}_{e j} O^{''R*}_{ij}
	\Delta^{u*}(\tilde{e}_R) -f^{R*}_{e i}f^{R}_{e j} O^{''R}_{ij} 
	\Delta^{t*}(\tilde{e}_R)\big]  m_{\chi^0_i} q \cos\theta\big\}
	\nonumber\\ & &
	+Re\big\{\Delta^s(Z) \big[f^{R}_{e i}f^{R*}_{e j} O^{''R}_{ij} 
	\Delta^{u*}(\tilde{e}_R) + f^{R*}_{e i}f^{R}_{e j} O^{''R*}_{ij} 
	\Delta^{t*}(\tilde{e}_R)\big]  m_{\chi^0_j} E_{\chi^0_i}\big\} 
	\Big],\label{s1_zl} \\
%------------------------------------------------------------------------
\Sigma_P^1(\tilde{e}_R \tilde{e}_R)&=&
	\frac{g^4}{4} c_R  E_b^2 \sin\theta
	\Big\{ |f^R_{e i}|^2 |f^R_{e j}|^2 
	\nonumber\\ & & \times
	\Big[ (|\Delta^t (\tilde{e}_R)|^2+|\Delta^u (\tilde{e}_R)|^2)
	 m_{\chi^0_i} E_{\chi^0_j} -(|\Delta^t (\tilde{e}_R)|^2-|\Delta^u (\tilde{e}_R)|^2)
	m_{\chi^0_i} q \cos\theta \Big]
	\nonumber \\ & & 
	-2 Re\{ (f^{R*}_{e i})^2 (f^{R}_{e j})^2
	\Delta^{u}(\tilde{e}_R)\Delta^{t*}(\tilde{e}_R)\} 
	m_{\chi^0_j} E_{\chi^0_i} \Big\}.
\label{s1_ll}
\end{eqnarray}
To obtain the expressions for 
$\Sigma_P^1(Z \tilde{e}_L)$ and
$\Sigma_P^1(\tilde{e}_L \tilde{e}_L)$
one has to apply the exchanges (\ref{exchange})
in (\ref{s1_zl}) and (\ref{s1_ll}) and to change the overall
sign of the right hand side of (\ref{s1_zl}) and (\ref{s1_ll}).

\begin{itemize}
	\item
The contributions to the transverse 
$\tilde\chi^0_i$ 
polarization perpendicular to the production plane are
\end{itemize}
\begin{eqnarray}
\Sigma_P^2 (ZZ)&=&
             \frac{4g^4}{\cos^4\theta_W}  |\Delta^s(Z)|^2 (R_e^2 c_R-L_e^2 c_L) 
%		[(1-P^3_-P^3_+)(L_{e}^2-R_{e}^2) - (P^3_- - P^3_+)(L_{e}^2+R_{e}^2)] 
%		\nonumber\\  & & \label{s2_1} \times 
		\, m_{\chi^0_j} q E_b^2 
                 \sin\theta Re(O^{''R}_{ij}) Im(O^{''R}_{ij}),\\
\Sigma_P^2(Z \tilde{e}_R) &=& 
	 \frac{g^4}{\cos^2\theta_W} R_e c_R  m_{\chi^0_j}
          E_b^2 q \sin\theta \nonumber\\
          & & \times \, Im\Big\{\Delta^s(Z)
            \big[f^{R}_{e i}f^{R*}_{e j} O^{''R}_{ij} 
           \Delta^{u*}(\tilde{e}_R) -f^{R*}_{e i}f^{R}_{e j}
            O^{''R*}_{ij} \Delta^{t*}(\tilde{e}_R) \big]\Big\}, \label{s2_2}\\
\Sigma_P^2(\tilde{e}_R \tilde{e}_R)&=& 
	  -\frac{g^4}{2} c_R  m_{\chi^0_j} E_b^2 q \sin\theta 
%	  \nonumber\\ & & \times 
	  \,Im\Big\{(f^{R*}_{e i})^2 (f^{R}_{e j})^2
          \Delta^u (\tilde{e}_R)\Delta^{t*}
			 (\tilde{e}_R)\Big\}.\label{s2_3}
     \end{eqnarray}
To obtain the expressions for 
$\Sigma_P^2(Z \tilde{e}_L)$ and
$\Sigma_P^2(\tilde{e}_L \tilde{e}_L)$
one has to apply the exchanges (\ref{exchange})
in (\ref{s2_2}) and (\ref{s2_3}).
\begin{itemize}
	\item
The contributions to the longitudinal $\tilde\chi^0_i$ polarization
are
\end{itemize}
\begin{eqnarray}
\Sigma_P^3(ZZ) & = &
	\frac{2 g^4}{\cos^4\theta_W} |\Delta^s(Z)|^2  
	(L_{e}^2 c_{L} - R_{e}^2 c_{R} ) E_b^2 \cos\theta
\nonumber\\ & & \times
	\Big[ |O^{''R}_{ij}|^2 (E_{\chi^0_i} E_{\chi^0_j}+q^2)
	-[(Re O^{''R}_{ij})^2-(Im O^{''R*}_{ij})^2] 
	m_{\chi^0_i}  m_{\chi^0_j} \Big], \label{s3_zz} \\
%----------------------------------------------------------
\Sigma_P^3(Z\tilde e_R) &=& 
	\frac{-g^4}{\cos^2\theta_W} R_{e} c_{R} E_b^2 
	\nonumber\\ & & \times
	\Big[ Re\Big\{ \Delta^s(Z) [f^{R}_{e i}f^{R*}_{e j} O^{''R*}_{ij} 
	\Delta^{u*}(\tilde{e}_R) -f^{R*}_{e i}f^{R}_{e j} O^{''R}_{ij} 
	\Delta^{t*}(\tilde{e}_R)] E_{\chi^0_j} q \Big\}
	\nonumber\\ & & %\mbox{\hspace*{-1.2cm}}
	+Re\Big\{ \Delta^s(Z) [f^{R}_{e i}f^{R*}_{e j} O^{''R*}_{ij}\Delta^{u*}(\tilde{e}_R)
	+f^{R*}_{e i}f^{R}_{e j} O^{''R}_{ij} \Delta^{t*}(\tilde{e}_R)]
	(E_{\chi^0_i} E_{\chi^0_j}+ q^2)\cos\theta \Big \}
\nonumber\\ & & %\mbox{\hspace*{-1.2cm}}
 -Re\Big\{ \Delta^s(Z) 
	[f^{R}_{e i}f^{R*}_{e j} O^{''R}_{ij} \Delta^{u*}(\tilde{e}_R)
	+f^{R*}_{e i}f^{R}_{e j} O^{''R*}_{ij} \Delta^{t*}(\tilde{e}_R)]
m_{\chi^0_i} m_{\chi^0_j} \cos\theta\Big \}
\nonumber\\ & & %\mbox{\hspace*{-1.2cm}}
+Re\Big\{ \Delta^s(Z) 
	[f^{R}_{e i}f^{R*}_{e j} O^{''R*}_{ij} \Delta^{u*}(\tilde{e}_R)
	-f^{R*}_{e i}f^{R}_{e j} O^{''R}_{ij}  \Delta^{t*}(\tilde{e}_R)]
	 E_{\chi^0_i} q \cos^2\theta \Big \}
 \Big ], \label{s3_zl} \\
%--------------------------------------------------------
\Sigma_P^3(\tilde{e}_R \tilde{e}_R) & = & 
	-\frac{g^4}{4} c_{R} E_b^2 \Big[ |f^R_{e i}|^2 |f^R_{e j}|^2 
	\nonumber \\  &&\times
	\Big\{ [|\Delta^u (\tilde{e}_R)|^2-|\Delta^t (\tilde{e}_R)|^2]
	E_{\chi^0_j} q+[|\Delta^u (\tilde{e}_R)|^2 -|\Delta^t (\tilde{e}_R)|^2]
	q E_{\chi^0_i}  \cos^2 \theta
	\nonumber\\ & &
	+[|\Delta^t (\tilde{e}_R)|^2+|\Delta^u (\tilde{e}_R)|^2]
	(E_{\chi^0_i} E_{\chi^0_j} +q^2) \cos\theta \Big\} 
	\nonumber\\ & &
	-2 Re\{(f^{R*}_{e i})^2 (f^{R}_{e j})^2
	\Delta^u (\tilde{e}_R)\Delta^{t*} (\tilde{e}_R)\}
	m_{\chi^0_i} m_{\chi^0_j} \cos\theta\Big]. \label{s3_ll}
\end{eqnarray}
To obtain the expressions for
$\Sigma_P^3(Z \tilde{e}_L)$ and
$\Sigma_P^3(\tilde{e}_L \tilde{e}_L)$
one has to apply the exchanges (\ref{exchange})
in (\ref{s3_zl}) and (\ref{s3_ll})
 and to change the overall
 sign of the right hand side of (\ref{s3_zl}) and (\ref{s3_ll}).

\section{Neutralino decay into sleptons
  \label{Neutralino decay into sleptons}}

For neutralino two-body decay into sleptons
\begin{eqnarray}\label{Appendix:decay_neutslepton}
	\tilde\chi^0_i(p_{\chi^0_i},\lambda_i) &\to& 
	\tilde{\ell} + \ell_1; \quad \ell =e,\mu,\tau, 
\end{eqnarray}
the neutralino decay matrix~(\ref{amplitude1_neut}) is given by
\begin{equation}\label{neut:rhoD}
\rho_{D_1}(\tilde \chi_i^0)_{\lambda_i' \lambda_i} = 
\delta_{\lambda_i' \lambda_i} D_1 +
\sigma^a_{\lambda_i' \lambda_i}\Sigma^a_{D_1},
\end{equation}
where we sum over a.
For the decay into right sleptons 
$\tilde\chi^0_i \to \tilde \ell_R^{\mp}\,\ell_1^{\pm}$, $\ell=e,\mu$,
the expansion coefficients are
\begin{eqnarray}
D_1 &=& \frac{g^2}{2} |f^{R}_{\ell i}|^2 (m_{\chi_i^0}^2 -m_{\tilde\ell}^2 ),\\
\Sigma^a_{D_1} &=& \pm g^2 |f^{R}_{\ell i}|^2 
m_{\chi_i^0} (s^a_{\chi_i^0} \cdot p_{\ell_1}).
	\end{eqnarray}
For the decay into the left sleptons
$\tilde\chi^0_i \to \tilde\ell_L^{\mp}\,\ell_1^{\pm}$, $\ell=e,\mu$,
the coefficients are
\begin{eqnarray}
D_1 &=& \frac{g^2}{2} |f^{L}_{\ell i}|^2 
		(m_{\chi_i^0}^2 - m_{\tilde\ell}^2 ),\\
\Sigma^a_{D_1} &=&  \mp g^2 |f^{L}_{\ell i}|^2 m_{\chi_i^0} 
		(s^a_{\chi_i^0} \cdot p_{\ell_1}).
   \end{eqnarray}
For the decay into the stau 
$\tilde\chi^0_i \to \tilde \tau_k^{\mp} \,\tau^{\pm}$, $k=1,2$, one
obtains
\begin{eqnarray}\label{plusterm}
D_1 &=& \frac{g^2}{2} (
			|a_{ki}^{\tilde \tau}|^2 +|b_{ki}^{\tilde \tau}|^2 )
				(m_{\chi_i^0}^2 - m_{\tilde{\tau}_k}^2 ),\\
\Sigma^a_{D_1} &=&  \mp g^2 (
			|a_{ki}^{\tilde \tau}|^2-|b_{ki}^{\tilde \tau}|^2 )
			m_{\chi_i^0}(s^a_{\chi_i^0} \cdot p_{\ell_1}).
\end{eqnarray}
The factor for the subsequent slepton decays 
$\tilde\ell_{R,L} \to \ell_2 \tilde \chi_1^0$, $\ell=e,\mu$, is given by
\begin{eqnarray}\label{sleptonD_2}
	D_2 &=& g^2 |f^{R,L}_{\ell 1}|^2 
	( m_{\tilde\ell}^2-m_{\chi_1^0}^2 ),
\end{eqnarray}
and that for stau decay $\tilde\tau_k\to \tau \tilde \chi_1^0$ by
\begin{eqnarray}
D_2 & = &g^2 (
		|a_{k1}^{\tilde\tau}|^2 +|b_{k1}^{\tilde \tau}|^2 )
			(m_{\tilde\tau_k}^2-m_{\chi_1^0}^2 ).
\end{eqnarray}

\section{Neutralino decay into staus
  \label{Neutralino decay into staus}}

For neutralino two-body decay into staus
\begin{eqnarray} \label{Appendix:decay_neutstau}
	\tilde\chi^0_i(p_{\chi^0_i}, \lambda_i) \to 
	\tilde\tau_m^{\pm}(p_{\tilde\tau_m}) +\tau^{\mp}(p_{\tau},
		\lambda_k); \quad m=1,2,
\end{eqnarray}
the decay matrix is 
   \begin{eqnarray} \label{stau:rhoD}
\rho_D(\tilde{\chi}^{0}_i)_{\lambda_i' \lambda_i}^{\lambda_k\lambda'_k} &=& 
           \delta_{\lambda_i' \lambda_i} D^{\lambda_k\lambda'_k} +
			  \sum_a \sigma^a_{\lambda_i'
				  \lambda_i}(\Sigma^a_D)^{\lambda_k\lambda'_k}.
    \end{eqnarray}
With the  spin basis vectors $s^b_{\tau}$
for the $\tau^{\mp}$, given in~(\ref{stau:polvec}), we can expand
\begin{eqnarray} 
D^{\lambda_k\lambda'_k} &=&  \delta_{\lambda_k\lambda'_k}D+
	\sigma^b_{\lambda_k\lambda'_k}D^b,\label{stau:D1}\\
(\Sigma_D^a)^{\lambda_k\lambda'_k} &=& 
	\delta_{\lambda_k\lambda'_k}\Sigma_D^a+
	\sigma^b_{\lambda_k\lambda'_k}\Sigma_D^{ab}. \label{stau:Sigma1}
\end{eqnarray}
The expansion coefficients are given by
\begin{eqnarray} 
D&=&g^2{\rm Re}({b^{\tilde \tau}_{mi}}^*a^{\tilde \tau}_{mi})
	m_{\tau} m_{\chi_i^0} 
	+\frac{g^2}{2}(|b^{\tilde \tau}_{mi}|^2+|a^{\tilde \tau}_{mi}|^2) 
	(p_{\tau}\cdot p_{\chi_i^0}), \label{stau:d1} \\
%	\end{equation}
%
%\begin{equation} \label{stau:db1}
D^b&=&\pm\frac{g^2}{2}m_{\tau}(|b^{\tilde \tau}_{mi}|^2-
	|a^{\tilde\tau}_{mi}|^2)(p_{\chi_i^0}\cdot s^b_{\tau}),
		 \label{stau:db1} \\
%\end{equation}
%
%\begin{equation} \label{stau:sa1}
\Sigma_D^a&=&\pm\frac{g^2}{2}m_{\chi_i^0}(|a^{\tilde \tau}_{mi}|^2-
	|b^{\tilde \tau}_{mi}|^2)(p_{\tau}\cdot s^a_{\chi_i^0}),
		\label{stau:sa1}\\
%\end{equation}
%\begin{eqnarray} \label{stau:sab1}
\Sigma_D^{ab}&=&g^2{\rm Re}({b^{\tilde \tau}_{mi}}^*a^{\tilde \tau}_{mi})
	(p_{\tau}\cdot s^a_{\chi_i^0})
	(p_{\chi_i^0}\cdot s^b_{\tau})
{}\nonumber\\[3mm]
{}&&-g^2(s^a_{\chi_i^0}\cdot s^b_{\tau})
\lbrack\frac{1}{2}(|b^{\tilde \tau}_{mi}|^2+
|a^{\tilde \tau}_{mi}|^2)m_{\tau} m_{\chi_i^0}+
{\rm Re}({b^{\tilde \tau}_{mi}}^*a^{\tilde \tau}_{mi})
(p_{\tau}\cdot p_{\chi_i^0})\rbrack
{}\nonumber\\[3mm]
{}&&\mp g^2{\rm Im}({b^{\tilde \tau}_{mi}}^*a^{\tilde \tau}_{mi})
\epsilon_{\mu\nu\rho\sigma}\,
p_{\tau}^{\mu} \,p_{\tilde\chi_i^0}^{\nu}\,s^{a,\,\rho}_{\chi_i^0}\,
s^{b,\,\sigma}_{\tau},\quad(\epsilon_{0123}=1).\label{stau:sab1}
\end{eqnarray}
%with $\epsilon_{0123}=1$.
%Only for $b=2$, the last term in Eq.~\ref{stau:sab1} 

\section{Neutralino decay into the $Z$ boson
  \label{Neutralino decay into the Z boson}}

For the neutralino two-body decay into the $Z$ boson
\begin{eqnarray} \label{Appendix:decay_neutZ}
\tilde\chi^0_i(p_{\chi^0_i}, \lambda_i) &\to& 
	\chi^0_n(p_{\chi_n^0}, \lambda_n) +Z(p_{Z}, \lambda_k); \quad n<i,
\end{eqnarray}
the decay matrix is given by
\begin{eqnarray}\label{Z:rhoD1A}
\rho_{D_1}(\tilde\chi^0_i)_{\lambda_i' \lambda_i}^{\lambda_k\lambda'_k} &=&
\sum_{\lambda_n}T_{D_1,\lambda_i}^{\lambda_n\lambda_k}
T_{D_1,\lambda_i'}^{\lambda_n\lambda_k'\ast},
\end{eqnarray}
with the helicity amplitude 
%$T_P^{\lambda_i \lambda_j}$
%for the on process are given in \cite{gudi1}.
%Those for the two-body decays, 
%Eq.~(\ref{decay_1}) and Eq.~(\ref{decay_2}), are
\begin{eqnarray}
	T_{D_1,\lambda_i}^{\lambda_n\lambda_k} &=& 
	\bar u(p_{\chi_n^0},\lambda_n)
	\gamma^{\mu}\frac{g}{\cos\theta_W}[O_{ni}^{''L}P_L + O_{ni}^{''R}P_R]
		u(p_{\chi_i^0}, \lambda_i)
	\varepsilon_{\mu}^{\lambda_k\ast}.
\end{eqnarray}
For the subsequent decay of the $Z$ boson
\begin{eqnarray} \label{Appendix:decay_Z}
	Z(p_{Z}, \lambda_k) &\to& f(p_{f}, \lambda_f) +\bar f(p_{\bar f}, \lambda_{\bar f});
	\quad f=\ell, q,
\end{eqnarray}
the decay matrix is
\begin{eqnarray}\label{Z:rhoD2A}
	\rho_{D_2}(Z)_{\lambda_k' \lambda_k}&=&
	\sum_{\lambda_f, \lambda_{\bar f}}
	T_{D_2,\lambda_k}^{\lambda_f \lambda_{\bar f} }
	T_{D_2,\lambda_k'}^{\lambda_f \lambda_{\bar f} \ast},
\end{eqnarray}
and the helicity amplitude
\begin{eqnarray}
	T_{D_2,\lambda_k}^{\lambda_f \lambda_{\bar f}} &=& 
	\bar u(p_f,\lambda_f)\gamma^{\mu}
	\frac{g}{\cos\theta_W}[L_f P_L + R_f P_R] v(p_{\bar f},\lambda_{\bar f})
	\varepsilon_{\mu}^{\lambda_k}.
\end{eqnarray}
The polarization vectors of the $Z$ boson
$\varepsilon_{\mu}^{\lambda_k},\lambda_k =0,\pm1$, are given 
in~(\ref{circularbasis}). With the set of neutralino 
spin vectors $s^{a}_{\chi_i^0}$, given in~(\ref{spinvec}), 
we obtain for the neutralino decay matrix
\begin{eqnarray} \label{Appendix:neutZrhoD1}
\rho_{D_1}(\tilde\chi^0_i)_{\lambda_i' \lambda_i}^{\lambda_k\lambda'_k} &=& 
(\delta_{\lambda_i' \lambda_i} D_1^{\mu\nu} 
+ \sigma^a_{\lambda_i'\lambda_i}  \Sigma^{a~\mu\nu}_{D_1})
		  \varepsilon_{\mu}^{\lambda_k\ast}\varepsilon_{\nu}^{\lambda'_k},
\end{eqnarray}
and for the $Z$ decay matrix
\begin{eqnarray}\label{Appendix:neutZrhoD2}
		  \rho_{D_2}(Z)_{\lambda'_k\lambda_k}&=& D_2^{\mu\nu}
\varepsilon_{\mu}^{\lambda_k}\varepsilon_{\nu}^{\lambda'_k\ast},
\end{eqnarray}
with
\begin{eqnarray}
	D_1^{\mu\nu} &=&
	\frac{2g^2}{\cos^2\theta_W}\Big\{[2~p^{\mu}_{\chi_i^0} p^{\nu}_{\chi_i^0}
		-(p^{\mu}_{\chi_i^0}p^{\nu}_Z +  p^{\nu}_{\chi_i^0} p^{\mu}_Z)
	- {\textstyle\frac{1}{2}}(m_{\chi_i^0}^2+m_{\chi_n^0}^2-m_Z^2)g^{\mu\nu} ]
	|O^{''L}_{ni}|^2 \nonumber\\
	&&-g^{\mu\nu}m_{\chi_i^0}m_{\chi_n^0}[(Re O^{''L}_{ni})^2 -(Im
			O^{''L}_{ni})^2]\Big\}, \\
\Sigma_{D_1}^{a~ \mu\nu} &=& 
	\frac{2ig^2}{\cos^2\theta_W}\Big\{-m_{\chi_i^0}\epsilon^{\mu\alpha\nu\beta} 
	s^a_{\chi_i^0,\alpha}(p_{\chi_i^0,\beta} -p_{Z,\beta} )|O^{''L}_{ni}|^2
	\nonumber\\&&
	+2m_{\chi_n^0}(s^{a,\mu}_{\chi_i^0}p^{\nu}_{\chi_i^0}-
		s^{a,\nu}_{\chi_i^0}p^{\mu}_{\chi_i^0})
	(Im O^{''L}_{ni})(Re O^{''L}_{ni})\nonumber\\
	&&-m_{\chi_n^0}\epsilon^{\mu\alpha\nu\beta}
	s^a_{\chi_i^0,\alpha}p_{\chi_i^0,\beta} 
	[(Re O^{''L}_{ni})^2 -(Im O^{''L}_{ni})^2]
		\Big\}; \quad (\epsilon_{0123}=1),
\end{eqnarray}
and
\begin{eqnarray}
	D_2^{\mu\nu} &=&\frac{2g^2}{\cos^2\theta_W}\Big\{ 
		(-2~p^{\mu}_{\bar f}p^{\nu}_{\bar f} 
		+p^{\mu}_Zp^{\nu}_{\bar f} +p^{\mu}_{\bar f}p^{\nu}_Z
		-{\textstyle\frac{1}{2}}m_Z^2 g^{\mu\nu})(L_f^2+R_f^2)
	\nonumber\\&& -i\epsilon^{\mu\alpha\nu\beta}p_{Z,\alpha}p_{\bar
	f,\beta}(L_f^2-R_f^2)\Big\}.
\end{eqnarray}
Due to the Majorana character of the neutralinos,
$D_1^{\mu\nu}$ is symmetric and $\Sigma^{a~\mu\nu}_{D_1}$
is antisymmetric  under interchange of $\mu$ and $\nu$.
In~(\ref{Appendix:neutZrhoD1}) and~(\ref{Appendix:neutZrhoD2}) we use the 
expansion~(\ref{expansion}) for the $Z$ polarization vectors
%We introduce also a complete orthonormal set of spin basis
%vectors $t^c_Z\;(c=1,2,3)$ for the $Z$ boson, 
%which fulfill the orthonormality relations 
%$t^c_{Z}\cdot t^d_{Z}=-\delta^{cd}$,
%$t^c_{Z}\cdot p_{Z}=0$ and the completeness relation 
%$\sum_{c}t^c_{Z\mu}t^c_{Z\nu}=-g_{\mu\nu}+p_{Z \mu}p_{Z \nu}/m_Z^2$.
\begin{eqnarray}\label{Z:expansion}
\varepsilon_{\mu}^{\lambda_k}\varepsilon_{\nu}^{\lambda'_k\ast}&=&
{\textstyle \frac{1}{3}}\delta^{\lambda_k'\lambda_k}I_{\mu\nu}
-\frac{i}{2m_Z}\epsilon_{\mu\nu\rho\sigma}
p_Z^{\rho}t_Z^{c,\sigma}(J^c)^{\lambda_k'\lambda_k}
-{\textstyle \frac{1}{2}}t_{Z,\mu}^ct_{Z,\nu}^d (J^{cd})^{\lambda_k'\lambda_k},
%\quad(\epsilon_{0123}=1),
\end{eqnarray}
summed over $c,d$. 
%
%Here, $J^c$ are the $3\times3$ spin 1 matrices with
%$[J^c,J^d]=i\epsilon_{cde}J^e$ and
%\begin{eqnarray}
%J^{cd}&=&J^cJ^d+J^dJ^c-{\textstyle \frac{4}{3}}\delta^{cd},
%\end{eqnarray}
%with $J^{11}+J^{22}+J^{33}=0$, are the components of a symmetric,
%traceless tensor, given in Appendix~\ref{Spin 1 matrices}, and 
%\begin{eqnarray}
%I_{\mu\nu}&=&-g_{\mu\nu}+\frac{p_{Z, \mu}p_{Z, \nu}}{m_Z^2}
%\end{eqnarray}
%guarantees the completeness relation of the polarization vectors
%\begin{eqnarray}\label{Z:completeness}
%\sum_{\lambda_k} \varepsilon^{\lambda_k\ast}_{\mu}
%\varepsilon^{\lambda_k}_{\nu}&=& -g_{\mu\nu}+\frac{p_{Z,\mu}p_{Z,\nu}}{m_Z^2}.
%\end{eqnarray}
%The second term in~(\ref{Z:expansion}) describes the vector
%polarization and the third term describes the tensor polarization
%of the $Z$ boson.
%
The decay matrices can be expanded in terms of the spin matrices
$J^c$ and $J^{cd}$, given in Appendix~\ref{Spin 1 matrices}. 
The first term  of the decay
matrix $\rho_{D_1}$~(\ref{Appendix:neutZrhoD1}), 
which is independent of the neutralino polarization, then gives
\begin{eqnarray}\label{Z:D1A}
D_1^{\mu\nu} 
\varepsilon_{\mu}^{\lambda_k\ast}\varepsilon_{\nu}^{\lambda'_k}&=&
  D_1 \delta^{\lambda_k\lambda_k'} 
  + \,^cD_1(J^c)^{\lambda_k\lambda_k'}
  + \,^{cd}D_1(J^{cd})^{\lambda_k\lambda_k'},
\end{eqnarray}
%summed over $c,d$, 
with
\begin{eqnarray}\label{Z:D1}
D_1&=&\frac{g^2}{\cos^2\theta_W}\Big\{\Big[m_{\chi_n^0}^2-{\textstyle\frac{1}{3}}m_{\chi_i^0}^2-m_Z^2+\frac{4}{3}
		\frac{(p_{\chi_i^0}\cdot p_Z)^2}{m_Z^2}\Big]|O^{''L}_{ni}|^2\nonumber\\
&&+2m_{\chi_i^0}m_{\chi_n^0}[(Re O^{''L}_{ni})^2 -(Im
		O^{''L}_{ni})^2]\Big\},\\
^{cd}D_1&=&-\frac{g^2}{\cos^2\theta_W}\Big\{
	\left[2(t^c_Z\cdot p_{\chi_i^0})(t^d_Z\cdot p_{\chi_i^0})+
		{\textstyle \frac{1}{2}}(m_{\chi_i^0}^2+m_{\chi_n^0}^2-m_Z^2)
		\delta^{cd}\right]|O^{''L}_{ni}|^2
	\nonumber\\
&&+\delta^{cd}m_{\chi_i^0}m_{\chi_n^0}[(Re O^{''L}_{ni})^2 -(Im
		O^{''L}_{ni})^2]\Big\},\label{Z:cdD1}
\end{eqnarray}
and $^{c}D_1=0$ due to the Majorana character of the neutralinos.
As a consequence of the completeness relation~(\ref{completeness}), 
the diagonal coefficients are linearly dependent
\begin{eqnarray}
	^{11}D_1+\,^{22}D_1+\,^{33}D_1&=&-{\textstyle \frac{3}{2}}D_1.
\end{eqnarray}
For large three momentum ${\bf p}_{\chi_i^0}$, the $Z$ boson will
mainly be emitted into the forward direction with respect to
${\bf p}_{\chi_i^0}$, i.e. $\hat {\bf p}_{\chi_i^0}\approx\hat {\bf p}_{Z}$, 
with $\hat {\bf p}={\bf p}/|{\bf p}|$, so that 
$(t^{1,2}_Z\cdot p_{\chi_i^0}) \approx 0$ in~(\ref{Z:cdD1}). 
Therefore, for high energies
$^{11}D_1\approx \,^{22}D_1$, and the contributions of the
non-diagonal coefficients $^{cd}D_1 (c \ne d)$ will be small.

For the second term of $\rho_{D_1}$~(\ref{Appendix:neutZrhoD1}), 
which depends on the polarization of the decaying neutralino,
we obtain
\begin{eqnarray}\label{Z:SigmaD1A}
\Sigma^{a~\mu\nu}_{D_1}
\varepsilon_{\mu}^{\lambda_k\ast}\varepsilon_{\nu}^{\lambda'_k}&=&
  \Sigma^{a}_{D_1}\delta^{\lambda_k\lambda_k'}
  +\,^c\Sigma^{a}_{D_1}(J^c)^{\lambda_k\lambda_k'}
  + \,^{cd}\Sigma^{a}_{D_1}(J^{cd})^{\lambda_k\lambda_k'},
\end{eqnarray}
%summed over $c$, $d$, 
with
 \begin{eqnarray}
^c\Sigma^{a}_{D_1}&=&\frac{2g^2}{m_Z \cos^2 \theta_W}\Big\{
	\Big[|O^{''L}_{ni}|^2m_{\chi_i^0}+[(Re O^{''L}_{ni})^2 -(Im
			O^{''L}_{ni})^2 ]m_{\chi_n^0}\Big]\nonumber\\ &&\times
	\left[(s^a_{\chi_i^0}\cdot p_Z)(t^c_Z\cdot p_{\chi_i^0})-
		(s^a_{\chi_i^0}\cdot t^c_Z)(p_Z\cdot p_{\chi_i^0})\right]
	+|O^{''L}_{ni}|^2m_{\chi_i^0}m_Z^2(s^a_{\chi_i^0}\cdot t^c_Z)\nonumber\\ &&
	-2(Im O^{''L}_{ni})(Re O^{''L}_{ni})m_{\chi_n^0}
	\epsilon_{\mu\nu\rho\sigma}s^{a,\mu}_{\chi_i^0}p_{\chi_i^0}^{\nu}
	p_Z^{\rho}t^{c,\sigma}_Z\Big\} \label{Z:csigmaaD1},
\end{eqnarray}
and $\Sigma^{a}_{D_1}=\,^{cd}\Sigma^{a}_{D_1}=0$ due to the Majorana
character of the neutralinos. Inserting~(\ref{Z:D1A}) and~(\ref{Z:SigmaD1A})
into~(\ref{Appendix:neutZrhoD1}), we obtain %for the neutralino decay matrix:
the expansion of the neutralino decay matrix
\begin{eqnarray}
\rho_{D_1}(\tilde\chi^0_i)_{\lambda_i' \lambda_i}^{\lambda_k\lambda'_k} &=& 
    \delta_{\lambda_i' \lambda_i} D_1~\delta^{\lambda_k\lambda_k'}+ 
%\nonumber \\  &&
  	\sigma^a_{\lambda_i' \lambda_i}~^c\Sigma^{a}_{D_1}~
	(J^c)^{\lambda_k\lambda_k'}+ 
%\nonumber \\  &&
\delta_{\lambda_i' \lambda_i} ~^{cd}D_1~
		(J^{cd})^{\lambda_k\lambda_k'},\label{Z:rhoD1expanded}
\end{eqnarray}
%summed over $c$, $d$, which is the expansion in 
into the scalar (first term),
vector (second term) and tensor part (third term).

A similar expansion for the $Z$ 
decay matrix~(\ref{Appendix:neutZrhoD2}) results in
\begin{eqnarray}\label{Z:rhoD2expanded}
	 \rho_{D_2}(Z)_{\lambda'_k\lambda_k}&=&   
D_2~ \delta^{\lambda_k'\lambda_k} 
  + \,^cD_2~(J^c)^{\lambda_k'\lambda_k}
  + \,^{cd}D_2~(J^{cd})^{\lambda_k'\lambda_k},
\end{eqnarray}
%where we sum over $c$, $d$, 
with 
\begin{eqnarray}
D_2&=&\frac{2g^2}{3\cos^2\theta_W}(R_f^2+L_f^2)m_Z^2,\label{Z:D2}\\
{c}D_2&=&\frac{2g^2}{\cos^2\theta_W}
	(R_f^2-L_f^2)m_Z(t^c_Z\cdot p_{\bar f}),\label{Z:cD2}\\
^{cd}D_2&=&\frac{g^2}{\cos^2\theta_W}(R_f^2+L_f^2)
	\left[2(t^c_Z\cdot p_{\bar f})(t^d_Z\cdot p_{\bar f})-
	{\textstyle \frac{1}{2}}m_Z^2\delta^{cd}\right].\label{Z:cdD2}
\end{eqnarray}
As a consequence of the completeness 
relation~(\ref{completeness}), the diagonal 
coefficients are linearly dependent
\begin{eqnarray}
	^{11}D_2+\,^{22}D_2+\,^{33}D_2&=&-{\textstyle \frac{3}{2}}D_2.
\end{eqnarray}
For large three-momentum ${\bf p}_{Z}$, the fermion $\bar f$ will
mainly be emitted into the forward direction with respect to
${\bf p}_{Z}$, i.e. $\hat {\bf p}_{Z}\approx\hat {\bf p}_{\bar f}$, so that 
$(t^{1,2}_Z\cdot p_{\bar f}) \approx 0$ in~(\ref{Z:cdD2}). 
Therefore, for high energies
$^{11}D_2\approx \,^{22}D_2$, and the contributions for the
non-diagonal coefficients $^{cd}D_2 (c \ne d)$ will be small.

	\chapter{Spin-density matrices for chargino production and decay 
	\label{Chargino production and decay matrices}}

We give the analytic formulae for the 
%differential cross section
squared amplitudes for chargino production 
$e^+~e^-\to\tilde\chi^+_i~\tilde\chi^-_j$,
with longitudinally polarized beams
and for different subsequent two-body decay chains 
of one chargino. 
%In order to calculate the squared amplitudes 
%for such processes of chargino production and decay,
%~(\ref{Appendix:charproduction}), 
%$e^+\,e^- \to\tilde\chi^0_i \, \tilde\chi^0_j$ 
%followed by a two-body decay chain of the  
%chargino~$\tilde\chi^+_i$, 
We use the spin density matrix formalism 
as in \cite{gudichargino,spinhaber,gudidiss}.
The amplitude squared can then be written
\begin{eqnarray}    \label{char:amplitude}
|T|^2 &=& |\Delta(\tilde\chi^+_i)|^2~
	\sum_{\lambda_i\lambda_i'}~  
	\rho_P(\tilde\chi^+_i)^{\lambda_i \lambda_i'}~
	\rho_D(\tilde\chi^+_i)_{\lambda_i'\lambda_i},
\end{eqnarray}
with $\rho_P(\tilde\chi^+_i)$  the spin density production matrix 
of chargino $ \tilde\chi^+_i$, the propagator 
$ \Delta(\tilde\chi^+_i)=i/[s_{\chi_i^+}-m_{\chi_i^+}^2
	+im_{\chi_i^+}\Gamma_{\chi_i^+}]$ and
the chargino decay matrix $\rho_D(\tilde\chi^+_i)$.

%We give the analytical formulae for
%$P,\Sigma_P^1,\Sigma_P^2,\Sigma_P^3$ 
%of the chargino production matrix
%$\rho_P(\tilde\chi^+_i)^{\lambda_i \lambda_i'} =
%  2(\delta_{\lambda_i \lambda_i'} P + 
%       \sigma^{a}_{\lambda_i \lambda_i'}\Sigma_P^a)$
%in the laboratory system. 
%Lorentz invariant expressions for these functions 
%can be found in \cite{gudichargino}.

\section{Chargino production 
     \label{Chargino production}}

\begin{figure}[h]
\begin{minipage}{3cm}
\begin{center}
{\setlength{\unitlength}{1cm}
\begin{picture}(2.5,2.5)
\put(0,-1.1){\includegraphics{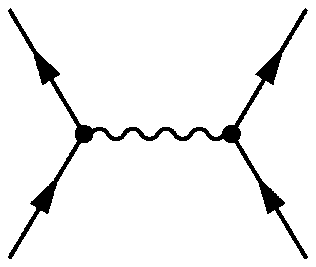}}
\put(0,-1.8){$e^-$}
\put(3.1,-1.8){$\tilde\chi^-_j$}
\put(0,1.7){$e^{+}$}
\put(3.1,1.7){$\tilde\chi^+_i$}
\put(1.7,.4){$\gamma$}
\end{picture}}
\end{center}
\end{minipage}
\hspace{3cm}
\vspace{.8cm}
\begin{minipage}{3cm}
\begin{center}
{\setlength{\unitlength}{1cm}
\begin{picture}(2.5,2.5)
\put(-1.2,-1.1){\includegraphics{./charprog.ps}}
\put(-1,-1.8){$e^-$}
\put(1.8,-1.8){$\tilde\chi^-_j$}
\put(-1,1.7){$e^{+}$}
\put(1.8,1.7){$\tilde\chi^+_i$}
\put(.4,.4){$Z^0$}
\end{picture}}
\end{center}
\end{minipage}
\hspace{2.5cm}
\vspace{.8cm}
\begin{minipage}{2.5cm}
\begin{center}
{\setlength{\unitlength}{1cm}
\begin{picture}(2.5,2)
\put(-1.2,-1.5){\includegraphics{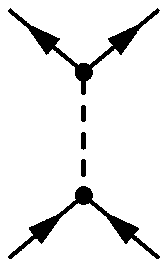}}
\put(-1.5,-2){$e^-$}
\put(-1.5,1.5){$e^+$}
\put(.5,-2){$\tilde\chi^-_j$}
\put(.5,1.5){$\tilde\chi^+_i$}
\put(-.9,-.3){$\tilde\nu$}
\end{picture}}
\end{center}
\end{minipage}
\par\vspace{1.0cm}
\caption{Feynman diagrams for chargino production
	\label{Feynman diagramms for chargino production}}
\end{figure}

For the production of charginos 
\begin{eqnarray} \label{Appendix:charproduction}
	e^++e^-&\to&\tilde\chi^+_i(p_{\chi_i^+}, \lambda_i)+
	            \tilde\chi^-_j(p_{\chi_j^-}, \lambda_j), 
\end{eqnarray}
with momentum $p$ and helicity $\lambda$, the unnormalized
spin-density matrix of chargino $\tilde\chi^+_i$  is defined as
\begin{eqnarray} \label{char:rhoPdef}
	\rho_P(\tilde\chi^+_i)^{\lambda_i \lambda_i'}&=&\sum_{\lambda_j}
	T_P^{\lambda_i \lambda_j}T_P^{\lambda_i' \lambda_j \ast}.
\end{eqnarray}
The helicity amplitudes are \cite{gudichargino,gudidiss}: 
\begin{eqnarray}
T_P^{\lambda_i \lambda_j}(\gamma)&=&- e^2 \Delta(\gamma)\delta_{ij} 
\bar{v}(p_{e^+}) \gamma^{\mu} u(p_{e^-}) \bar{u}(p_{\chi^+_i}, \lambda_i)
	\gamma_{\mu} v(p_{\chi^-_j}, \lambda_j),\label{char:T1}\\
T_P^{\lambda_i \lambda_j}(Z)&=&- \frac{g^2}{\cos^2\theta_W}\Delta(Z)
		\bar{v}(p_{e^+}) \gamma^{\mu} (L_{e} P_L+R_{e} P_R) u(p_{e^-}) 
		\nonumber\\ & & \times
		\bar{u}(p_{\chi^+_i}, \lambda_i) \gamma_{\mu} 
		(O^{'L}_{ij} P_L +O^{'R}_{ij} P_R)
		v(p_{\chi^-_j}, \lambda_j), \label {char:T2}\\
T_P^{\lambda_i \lambda_j}(\tilde{\nu})&=&
		-g^2 V_{i1} V_{j1}^{*}\Delta(\tilde{\nu})
		\bar{v}(p_{e^+}) P_R v(p_{\chi^+_i}, \lambda_i) 
		\bar{u}(p_{\chi^-_j}, \lambda_j) 
		P_L u(p_{e^-}),\label{char:T3}
	\end{eqnarray}
with the propagators
\begin{equation}\label{char:propagators}
        \Delta(\gamma) = \frac{i}{p_{\gamma}^2 },\quad
		      \Delta(Z)  = \frac{i}{p_Z^2-m^2_Z+im_Z\Gamma_Z},\quad
	\Delta(\tilde \nu)  = \frac{i}{p_{\tilde \nu}^2-
			  m^2_{\tilde \nu}}. 
\end{equation}
%where $p,m$ and $\Gamma$ denote the momentum, mass and width of the
%particle, respectively. 
The Feynman diagrams are shown in 
Fig.~\ref{Feynman diagramms for chargino production}.

For the polarization of the decaying chargino $\tilde\chi^+_i$
with momentum $p_{\chi_i^+}$ we have introduced 
three space like spin vectors $s^a_{\chi_i^+}$~(\ref{spinvec}).
%which together with $p_{\chi_i^+}^{\mu}/m_{\chi_i^+}$
%form an orthonormal set: 
%$s^a_{\chi_i^+}\cdot s^b_{\chi_i^+}=-\delta^{ab}$, 
%$s^a_{\chi_i^+}\cdot p_{\chi_i^+}=0$.
Then the chargino production matrix~(\ref{char:amplitude})
can be expanded in terms of the Pauli matrices,
see Appendix~\ref{Bouchiat-Michel formulae for spin 1/2 particles}:
\begin{eqnarray} \label{char:rhoP}
  \rho_P(\tilde\chi^+_i)^{\lambda_i \lambda_i'} &=&
      2\big(\delta_{\lambda_i \lambda_i'} P + 
        \sigma^a_{\lambda_i \lambda_i'}
		\Sigma_P^a\big),   
\end{eqnarray}
where we sum over $a$. The factor 2 in~(\ref{char:rhoP}) 
is due to the summation
of the helicities of the second chargino~$\tilde\chi^-_j$,
whose decay will not be considered. 
With our choice of the spin vectors,
$\Sigma^3_P/P$ is the longitudinal polarization of 
chargino~$\tilde\chi^+_i$,
$\Sigma^1_P/P$ is the transverse polarization in the 
production plane and $\Sigma^2_P/P$ is the polarization
perpendicular to the production plane. 
Only if there is a non-vanishing CP phase $\varphi_{\mu}$ 
in the chargino sector, and
only if two different charginos are produced,
$e^+~e^-\to\tilde\chi^{\pm}_1~\tilde\chi^{\mp}_2$,
the polarization $\Sigma^2_P/P$ perpendicular to the 
production plane is non-zero. Thus it is a probe for
CP violation in the production of an unequal pair of charginos.
Note that $\Sigma^2_P$ also gets contributions from
the finite $Z$ width, which however do not signal CP violation.
%which is however a higher order effect and will be small. 

We give the analytical 
formulae for $P$ and $\Sigma_P^1,\Sigma_P^2,\Sigma_P^3$ 
in the laboratory system in the following sections. 
Lorentz invariant expressions for these functions 
can be found in \cite{gudichargino,gudidiss}.

\subsection{Chargino polarization independent quantities 
     \label{Chargino polarization independent quantities}}
  
The coefficient $P$ is independent of the chargino polarization. 
It can be composed into contributions from the 
different production channels
\begin{equation}
	P =  P(\gamma \gamma)
		+ P(\gamma Z) 
		+ P(\gamma \tilde \nu)
		+ P(Z Z)
		+ P(Z\tilde \nu)
		+ P(\tilde \nu\tilde \nu)
\end{equation}
which read
\begin{eqnarray}
P(\gamma \gamma)&=&\delta_{ij}2e^4 |\Delta (\gamma)|^2
%	(1-P_{e^-}P_{e^+} )
	(c_{L} + c_{R})
	E_b^2(E_{\chi_i^+} E_{\chi_j^-}+m_{\chi_i^+} m_{\chi_j^-}+
	q^2\cos^2\theta),\\
P(\gamma Z)&=&\delta_{ij}2\frac{e^2g^2}{\cos^2 \theta_W}E_b^2
	Re\Big\{  \Delta (\gamma)\Delta (Z)^{\ast} \Big[   
	%	[L_{e}(1-P_{e^-})(1+P_{e^+}) -R_{e}(1+P_{e^-})(1-P_{e^+})]
	(L_e c_{L} - R_e c_{R})
	(O^{'R\ast}_{ij} - O^{'L\ast}_{ij})
	2E_bq\cos\theta
		\nonumber\\& & 
%	[L_{e}(1-P_{e^-})(1+P_{e^+}) +R_{e}(1+P_{e^-})(1-P_{e^+})]
	+(L_ec_{L} + R_e c_{R})
	(O^{'L\ast}_{ij}+O^{'R\ast}_{ij})
	(E_{\chi_i^+} E_{\chi_j^-}+m_{\chi_i^+} m_{\chi_j^-}+q^2\cos^2\theta)
	\Big]\Big\},\\
P(\gamma \tilde \nu)&=&\delta_{ij}e^2g^2E_b^2
%	(1-P_{e^-})(1+P_{e^+})
	c_{L}
	Re\Big\{ V^{\ast}_{i1}V_{j1}\Delta (\gamma)\Delta (\tilde \nu)^{\ast}
	\Big\}  \nonumber\\& & \times
	(E_{\chi_i^+} E_{\chi_j^-}+m_{\chi_i^+} m_{\chi_j^-}-2E_bq\cos\theta
	+q^2\cos^2\theta),\\
P(Z Z)&=&\frac{g^4}{\cos^4\theta_W}|\Delta (Z)|^2E_b^2\Big[
%	+[L_{e}^2(1-P_{e^-})(1+P_{e^+}) -R_{e}^2(1+P_{e^-})(1-P_{e^+})]
	(L_e^2 c_{L} - R_e^2  c_{R})
	(|O^{'R}_{ij}|^2-|O^{'L}_{ij}|^2)2E_bq\cos\theta
	\nonumber\\& &
%	[L_{e}^2(1-P_{e^-})(1+P_{e^+}) +R_{e}^2(1+P_{e^-})(1-P_{e^+})]
	+(L_e^2c_{L} + R_e^2c_{R})
	(|O^{'L}_{ij}|^2+|O^{'R}_{ij}|^2)
	(E_{\chi_i^+} E_{\chi_j^-}+ q^2\cos^2\theta)
	\nonumber\\& &
%	[L_{e}^2(1-P_{e^-})(1+P_{e^+}) +R_{e}^2(1+P_{e^-})(1-P_{e^+})]
	+(L_e^2 c_{L} + R_e^2 c_{R})
	2Re\{O^{'L}_{ij}O^{'R\ast}_{ij}\}m_{\chi_i^+} m_{\chi_j^-}
	\Big], \\
P(Z\tilde \nu)&=&\frac{g^4}{\cos^2\theta_W}
	L_e c_{L}E_b^2Re\Big\{
	V^{\ast}_{i1}V_{j1}\Delta (Z)\Delta (\tilde \nu)^{\ast} %\times
	\nonumber\\& & \times
	[O^{'L}_{ij}(E_{\chi_i^+} E_{\chi_j^-}-2E_bq\cos\theta+q^2\cos^2\theta) 
		+O^{'R}_{ij}m_{\chi_i^+} m_{\chi_j^-}]
	\Big\},\\
P(\tilde \nu\tilde \nu)&=& \frac{g^4}{4}c_{L}
|V_{i1}|^2|V_{j1}|^2 |\Delta (\tilde \nu)|^2
	E_b^2(E_{\chi_i^+} E_{\chi_j^-}-2E_bq\cos\theta+q^2\cos^2\theta ).
\end{eqnarray}
The longitudinal beam polarizations are included in the weighting factors  
\begin{equation}
c_L =(1-P_{e^-})(1+P_{e^+}), \quad c_R= (1+P_{e^-})(1-P_{e^+}).
\end{equation}
Sneutrino exchange is enhanced for $P_{e^-}<0$ and $P_{e^+}>0$.

\subsection{Chargino polarization 
     \label{Chargino polarization}}
 
The coefficients $\Sigma^a_P$, 
which describe the polarization of the chargino $\tilde{\chi}^+_i$, 
decompose into
   \begin{equation}
     \Sigma_P^a =
     \Sigma_P^a(\gamma \gamma)
   + \Sigma_P^a(\gamma Z)
   + \Sigma_P^a(\gamma \tilde \nu)
   + \Sigma_P^a(Z Z)
	+ \Sigma_P^a(Z\tilde \nu)
	+ \Sigma_P^a(\tilde \nu\tilde \nu).\label{charg:sigmaP}
\end{equation}
\begin{itemize}
\item
The contributions to the transverse 
%$\tilde{\chi}^+_i$ 
polarization in the production plane are
\end{itemize}
\begin{eqnarray}
\Sigma_P^1(\gamma \gamma)&=&\delta_{ij}2e^4 |\Delta (\gamma)|^2
	(c_{R} - c_{L})
	E_b^2\sin\theta(m_{\chi_i^+}E_{\chi_j^-}+m_{\chi_j^-}E_{\chi_i^+}),\\
\Sigma_P^1(\gamma Z)&=&\delta_{ij}2\frac{e^2g^2}{\cos^2 \theta_W}E_b^2
\sin\theta Re\Big\{  \Delta (\gamma)\Delta (Z)^{\ast} 
	\nonumber \\ && \times
	\Big[   
	-(L_e c_{L} + R_e c_{R})
	(O^{'R\ast}_{ij} - O^{'L\ast}_{ij})
	m_{\chi_i^+}q\cos\theta
		\nonumber\\& & 
%	[L_{e}(1-P_{e^-})(1+P_{e^+}) +R_{e}(1+P_{e^-})(1-P_{e^+})]
	+(R_e c_{R} - L_e c_{L})
	(O^{'L\ast}_{ij}+O^{'R\ast}_{ij})
	(m_{\chi_i^+}E_{\chi_j^-}+m_{\chi_j^-}E_{\chi_i^+})
	\Big]\Big\},\\
\Sigma_P^1(\gamma \tilde \nu)&=&-\delta_{ij} e^2 g^2 
	c_{L}E_b^2\sin\theta
%	\nonumber\\& & \times
	Re\Big\{ V^{\ast}_{i1}V_{j1}\Delta (\gamma)\Delta (\tilde \nu)^{\ast}
	\Big\} \nonumber\\& & \times
	[m_{\chi_i^+}(E_{\chi_j^-}-q\cos\theta)+m_{\chi_j^-}E_{\chi_i^+}],\\
\Sigma_P^1(Z Z)&=&\frac{g^4}{\cos^4\theta_W}|\Delta (Z)|^2E_b^2 \sin\theta\Big[
	(L_e^2 c_{L} + R_e^2 c_{R})
	(|O^{'L}_{ij}|^2-|O^{'R}_{ij}|^2)m_{\chi_i^+} q\cos\theta
	\nonumber\\& &
	+(R_e^2 c_{R} - L_e^2 c_{L})
	2Re\Big\{O^{'L}_{ij}O^{'R\ast}_{ij}\Big\}m_{\chi_j^-}E_{\chi_i^+}
	\nonumber\\& &
	+(R_e^2 c_{R} - L_e^2 c_{L})
	(|O^{'R}_{ij}|^2+|O^{'L}_{ij}|^2)m_{\chi_i^+}E_{\chi_j^-}
	\Big], \\
\Sigma_P^1(Z\tilde \nu)&=&-\frac{g^4}{\cos^2\theta_W}
	L_e c_{L} E_b^2\sin\theta Re\Big\{
	V^{\ast}_{i1}V_{j1}\Delta (Z)\Delta (\tilde \nu)^{\ast}%\times
	\nonumber\\& & \times
	[O^{'L}_{ij}m_{\chi_i^+}(E_{\chi_j^-}-q\cos\theta) 
		+O^{'R}_{ij}m_{\chi_j^-} E_{\chi_i^+}]
	\Big\},\\
\Sigma_P^1(\tilde \nu\tilde \nu)&=& -\frac{g^4}{4}c_{L}
	|V_{i1}|^2|V_{j1}|^2 |\Delta (\tilde \nu)|^2
	E_b^2\sin\theta m_{\chi_i^+} (E_{\chi_j^-}-q\cos\theta).
\end{eqnarray}
\begin{itemize}
	\item
The contributions to the transverse $\tilde{\chi}^+_i$ polarization
perpendicular to the production plane are
\end{itemize}
\begin{eqnarray}
\Sigma_P^2(\gamma \gamma)&=&\Sigma_P^2(\tilde \nu\tilde \nu)\;=\;0,\\
\Sigma_P^2(\gamma Z)&=&\delta_{ij}2\frac{e^2g^2}{\cos^2 \theta_W}
	(R_e c_{R} - L_e c_{L})
	Im\Big\{  \Delta (\gamma)\Delta (Z)^{\ast}    
		(O^{'R\ast}_{ij} - O^{'L\ast}_{ij})\Big\} 
	\nonumber\\& &
	\times E_b^2m_{\chi_j^-}q\sin\theta,\\
	\Sigma_P^2(\gamma \tilde \nu)&=&\delta_{ij} e^2 g^2 c_{L}
	Im\Big\{ V^{\ast}_{i1}V_{j1}\Delta (\gamma)\Delta (\tilde \nu)^{\ast}
	\Big\}E_b^2m_{\chi_j^-}q\sin\theta, \\
\Sigma_P^2(Z Z)&=&2\frac{g^4}{\cos^4\theta_W}|\Delta (Z)|^2
	(R_e^2 c_{R} - L_e^2 c_{L})
	Im\Big\{O^{'L}_{ij}O^{'R\ast}_{ij}\Big\}
	E_b^2m_{\chi_j^-}q\sin\theta, \label{char:ZZ}\\
\Sigma_P^2(Z\tilde \nu)&=&\frac{g^4}{\cos^2\theta_W}L_e c_{L} 
	Im\Big\{V^{\ast}_{i1}V_{j1}O^{'R}_{ij}
	\Delta (Z)\Delta (\tilde \nu)^{\ast}\Big\}
	E_b^2m_{\chi_j^-}q\sin\theta.\label{char:Zsnu}
\end{eqnarray}
\begin{itemize}
	\item
The contributions to the longitudinal $\tilde{\chi}^+_i$ polarization are
\end{itemize}
\begin{eqnarray}
\Sigma_P^3(\gamma \gamma)&=&\delta_{ij}2e^4 |\Delta (\gamma)|^2
	(c_{L} - c_{R})
	E_b^2\cos\theta (q^2+  E_{\chi_i^+} E_{\chi_j^-}+m_{\chi_i^+} m_{\chi_j^-}),\\
\Sigma_P^3(\gamma Z)&=&\delta_{ij}2\frac{e^2g^2}{\cos^2 \theta_W}E_b^2
	Re\Big\{  \Delta (\gamma)\Delta (Z)^{\ast} 
	\nonumber\\& &\times \Big[   
	(L_e c_{L} - R_e c_{R})
	(O^{'R\ast}_{ij} + O^{'L\ast}_{ij})
	(q^2+  E_{\chi_i^+} E_{\chi_j^-}+m_{\chi_i^+} m_{\chi_j^-})\cos\theta
		\nonumber\\& & 
	+(L_e c_{L} + R_e c_{R})
	(O^{'R\ast}_{ij}-O^{'L\ast}_{ij})
	q(E_{\chi_j^-}+E_{\chi_i^+}\cos^2\theta)
	\Big]\Big\},\\
\Sigma_P^3(\gamma \tilde \nu)&=&-\delta_{ij}e^2g^2 c_{L}E_b^2
	Re\Big\{ V^{\ast}_{i1}V_{j1}\Delta (\gamma)\Delta (\tilde \nu)^{\ast}
	\Big\} \nonumber\\& & \times
	[qE_{\chi_j^-} - (q^2+E_{\chi_i^+} E_{\chi_j^-})\cos\theta
	+qE_{\chi_i^+}\cos^2\theta- m_{\chi_i^+}  m_{\chi_j^-}\cos\theta],\\
\Sigma_P^3(Z Z)&=&\frac{g^4}{\cos^4\theta_W}|\Delta (Z)|^2E_b^2\Big[
	(L_e^2 c_{L} + R_e^2 c_{R})
	(|O^{'R}_{ij}|^2-|O^{'L}_{ij}|^2)q(E_{\chi_j^-}+E_{\chi_i^+}\cos^2\theta)
	\nonumber\\& &
	+(L_e^2 c_{L} - R_e^2 c_{R})
	2Re\Big\{O^{'L}_{ij}O^{'R\ast}_{ij}\Big\}
	m_{\chi_i^+} m_{\chi_j^-}\cos\theta
	\nonumber\\& &
	+(L_e^2 c_{L} - R_e^2c_{R})
	(|O^{'L}_{ij}|^2+|O^{'R}_{ij}|^2)
	(q^2+E_{\chi_i^+} E_{\chi_j^-})\cos\theta\Big], \\
\Sigma_P^3(Z\tilde \nu)&=&\frac{g^4}{\cos^2\theta_W} L_e c_{L} E_b^2Re\Big\{
	V^{\ast}_{i1}V_{j1}\Delta (Z)\Delta (\tilde \nu)^{\ast}
	[O^{'R}_{ij}m_{\chi_i^+} m_{\chi_j^-}\cos\theta
	\nonumber\\& & 
	-O^{'L}_{ij}(q E_{\chi_j^-}-(q^2+E_{\chi_i^+} E_{\chi_j^-})\cos\theta
	+qE_{\chi_i^+}\cos^2\theta)] 
	\Big\},\\
\Sigma_P^3(\tilde \nu\tilde \nu)&=& -\frac{g^4}{4}c_{L}
	|V_{i1}|^2|V_{j1}|^2 |\Delta (\tilde \nu)|^2 E_b^2%\times
	\nonumber\\& & \times
	[q E_{\chi_j^-}-(q^2+E_{\chi_i^+} E_{\chi_j^-})\cos\theta
	+qE_{\chi_i^+}\cos^2\theta].
\end{eqnarray}

\section{Chargino decay into sneutrinos
  \label{Chargino decay into sneutrinos}}

For chargino two-body decay into sneutrinos
\begin{eqnarray}\label{Appendix:decay_charsneut}
	\tilde\chi^+_i (p_{\chi^+_i},\lambda_i)&\to& 
	\ell^+ + \tilde\nu_{\ell}; \quad \ell =e,\mu,\tau, 
\end{eqnarray}
the chargino decay matrix is given by
\begin{eqnarray}\label{sneut:rhoD}
	\rho_D(\tilde\chi^+_i)_{\lambda_i' \lambda_i} &=&
  \delta_{\lambda_i' \lambda_i} D + 
       \sigma^{a}_{\lambda_i' \lambda_i}
		\Sigma_D^a.   
\end{eqnarray}
%where we sum over a.
For the chargino decay into an electron or muon sneutrino 
the coefficients are
\begin{eqnarray}\label{sneut:D_1A}
D &= & \frac{g^2}{2} |V_{i1}|^2 
	      (m_{\chi_i^+}^2 -m_{\tilde\nu_{\ell}}^2 ),\\
\Sigma^a_{D} &=&\;  \,^{\;\,-}_{(+)} g^2 |V_{i1}|^2 
	m_{\chi_i^+} (s^a_{\chi^+_i} \cdot p_{\ell});
	\quad {\rm for} \;\ell=e,\mu, 
\end{eqnarray}
where the sign in parenthesis holds for the conjugated process 
$\tilde\chi^-_i \to \ell^-\bar{\tilde\nu}_{\ell}$.
For the decay into the tau sneutrino the coefficients are given by
\begin{eqnarray}\label{sneut:D_1B}
D  &=&  \frac{g^2}{2} (|V_{i1}|^2 +Y_{\tau}^2|U_{i2}|^2)
	(m_{\chi_i^+}^2 -m_{\tilde\nu_{\tau}}^2 ),\\ 
\Sigma^a_{D} &=&  \,^{\;\,-}_{(+)} g^2 (|V_{i1}|^2 -Y_{\tau}^2|U_{i2}|^2)
	m_{\chi_i^+} (s^a_{\chi^+_i} \cdot
		p_{\tau}),\label{sneut:SD_1B}
\end{eqnarray}
where $Y_{\tau}= m_{\tau}/(\sqrt{2}m_W\cos\beta)$ is the
$\tau$ Yukawa coupling, and
the sign in parenthesis holds for the conjugated process 
$\tilde\chi^-_i \to \tau^- \bar{\tilde\nu}_{\tau}$.

\section{Chargino decay into the $W$ boson
  \label{Chargino decay into the W boson}}

For the chargino two-body decay into the $W$ boson
\begin{eqnarray} \label{W:decay_1}
	\tilde\chi^+_i(p_{\chi^+_i}, \lambda_i) &\to&
	\tilde \chi^0_n(p_{\chi^0_n}, \lambda_n) 
	+W^+(p_W, \lambda_k), 
\end{eqnarray}
the decay matrix is given by
\begin{eqnarray}\label{W:rhoD1}
\rho_{D_1}(\tilde\chi^+_i)_{\lambda_i' \lambda_i}^{\lambda_k\lambda'_k} &=&
\sum_{\lambda_n}T_{D_1,\lambda_i}^{\lambda_n\lambda_k}
T_{D_1,\lambda_i'}^{\lambda_n\lambda_k'\ast},
\end{eqnarray}
with helicity amplitude 
\begin{eqnarray}
	T_{D_1,\lambda_i}^{\lambda_n\lambda_k} &=& 
	ig\bar u(p_{\chi^0_n},\lambda_n)
	\gamma^{\mu}[O_{ni}^{L}P_L + O_{ni}^{R}P_R]
		u(p_{\chi^+_i}, \lambda_i)
	\varepsilon_{\mu}^{\lambda_k\ast}.
\end{eqnarray}
For the subsequent decay of the $W$ boson
 \begin{eqnarray}\label{W:decay_2}
	 W^+(p_W, \lambda_k) &\to& f^{'}(p_{f^{'}}, \lambda_{f^{'}}) +
	 \bar f(p_{\bar f}, \lambda_{\bar f}),
\end{eqnarray}
the decay matrix is
\begin{eqnarray}\label{W:rhoD2}
	\rho_{D_2}(W^+)_{\lambda_k' \lambda_k}&=&
	\sum_{\lambda_{f^{'}}, \lambda_{\bar f}}
	T_{D_2,\lambda_k}^{\lambda_{f^{'}} \lambda_{\bar f} }
	T_{D_2,\lambda_k'}^{\lambda_{f^{'}} \lambda_{\bar f} \ast}
\end{eqnarray}
and the helicity amplitude
\begin{eqnarray}
	T_{D_2,\lambda_k}^{\lambda_{f^{'}} \lambda_{\bar f}} &=& 
	i\frac{g}{\sqrt{2}}\bar u(p_{f^{'}},\lambda_{f^{'}})
	\gamma^{\mu} P_L v(p_{\bar f},\lambda_{\bar f})
	\varepsilon_{\mu}^{\lambda_k}.
\end{eqnarray}
The $W$ polarization vectors 
$\varepsilon_{\mu}^{\lambda_k},\lambda_k =0,\pm1$, are defined 
in~(\ref{circularbasis}). With the set of chargino spin 
vectors $s^{a}_{\chi_i^+}$, given  in~(\ref{spinvec}),
we obtain for the chargino decay matrix
\begin{eqnarray} \label{W:rhoD1A}
\rho_{D_1}(\tilde\chi^+_i)_{\lambda_i' \lambda_i}^{\lambda_k\lambda'_k} &=& 
(\delta_{\lambda_i' \lambda_i} D_1^{\mu\nu} 
+ \sigma^a_{\lambda_i'\lambda_i}  \Sigma^{a~\mu\nu}_{D_1})
		  \varepsilon_{\mu}^{\lambda_k\ast}\varepsilon_{\nu}^{\lambda'_k}
\end{eqnarray}
and for the $W$ boson decay matrix
\begin{eqnarray}\label{W:rhoD2A}
		  \rho_{D_2}(W^+)_{\lambda'_k\lambda_k}&=& D_2^{\mu\nu}
\varepsilon_{\mu}^{\lambda_k}\varepsilon_{\nu}^{\lambda'_k\ast}.
\end{eqnarray}
The expansion coefficients are
\begin{eqnarray}
	D_1^{\mu\nu} &=& g^2(|O^{R}_{ni}|^2+|O^{L}_{ni}|^2)
	[2 p^{\mu}_{\chi^+_i} p^{\nu}_{\chi^+_i}
	 -(p^{\mu}_{\chi^+_i} p^{\nu}_W + p^{\nu}_{\chi^+_i} p^{\mu}_W)
	 -{\textstyle\frac{1}{2}}(m_{\chi^+_i}^2+m_{\chi^0_n}^2-m_W^2)
	 g^{\mu\nu} ] \nonumber\\
 && +2g^2 Re(O^{R\ast}_{ni}O^{L}_{ni})
	m_{\chi^+_i}m_{\chi^0_n}g^{\mu\nu} 
	\,^{\;\,+}_{(-)}  
	ig^2(|O^{R}_{ni}|^2-|O^{L}_{ni}|^2)\epsilon^{\mu\alpha\nu\beta}~
          p_{\chi^+_i,\alpha}~p_{W,\beta} , \\
 \Sigma_{D_1}^{a~\mu\nu}&=&
			 \,^{\;\,+}_{(-)}
			 g^2(|O^{R}_{ni}|^2-|O^{L}_{ni}|^2)m_{\chi^+_i}
			 [s^{a,\mu}_{\chi^+_i}(p^{\nu}_{\chi^+_i} -p^{\nu}_{W})
			+s^{a,\nu}_{\chi^+_i}(p^{\mu}_{\chi^+_i}-p^{\mu}_{W})
			+(s^{a}_{\chi^+_i}\cdot p_{W})g^{\mu\nu}]\nonumber\\
	&&-ig^2(|O^{R}_{ni}|^2+|O^{L}_{ni}|^2)m_{\chi^+_i}
		\epsilon^{\mu\alpha\nu\beta}s^a_{\chi^+_i,~\alpha}
		(p_{\chi^+_i,~\beta}-p_{W,\beta})\nonumber\\
	&&+2ig^2 Re(O^{R\ast}_{ni}O^{L}_{ni})m_{\chi^0_n}
		\epsilon^{\mu\alpha\nu\beta}s^a_{\chi^+_i,\alpha}~
		p_{\chi^+_i,\beta}\nonumber\\
&&-2ig^2 Im(O^{R\ast}_{ni}O^{L}_{ni})m_{\chi^0_n}
			(s^{a,\mu}_{\chi^+_i}p^{\nu}_{\chi^+_i}-
			s^{a,\nu}_{\chi^+_i}p^{\mu}_{\chi^+_i}	)
	; \quad (\epsilon_{0123}=1),
\end{eqnarray}
and
\begin{eqnarray}
	D_2^{\mu\nu} &=&g^2(-2p^{\mu}_{\bar f}p^{\nu}_{\bar f} 
		+p^{\mu}_Wp^{\nu}_{\bar f} +p^{\mu}_{\bar f}p^{\nu}_W
		-{\textstyle\frac{1}{2}}m_W^2 g^{\mu\nu})
\,^{\;\,-}_{(+)}ig^2\epsilon^{\mu\alpha\nu\beta}p_{W,\alpha}~p_{\bar f,\beta},
\end{eqnarray}
where here, and in the following, 
the signs in parenthesis hold for the charge conjugated processes, 
$\tilde\chi^-_i\to W^-\tilde\chi^0_n$ and
$W^- \to \bar f^{'} f$, respectively.
In~(\ref{W:rhoD1A}) and~(\ref{W:rhoD2A}) we use the 
expansion~(\ref{expansion}) for the $W$ polarization vectors
\begin{eqnarray}\label{W:expansion}
\varepsilon_{\mu}^{\lambda_k}\varepsilon_{\nu}^{\lambda'_k\ast}&=&
{\textstyle \frac{1}{3}}\delta^{\lambda_k'\lambda_k}I_{\mu\nu}
-\frac{i}{2m_W}\epsilon_{\mu\nu\rho\sigma}
p_W^{\rho}t_W^{c,\sigma}(J^c)^{\lambda_k'\lambda_k}
-{\textstyle \frac{1}{2}}t_{W,\mu}^c t_{W,\nu}^d
(J^{cd})^{\lambda_k'\lambda_k}.
%\quad(\epsilon_{0123}=1),
\end{eqnarray}
%summed over $c,d$. 
The decay matrices can be expanded in terms 
of the spin matrices $J^c$ and $J^{cd}$, given in 
Appendix~\ref{Spin 1 matrices}. The first term  of the decay
matrix $\rho_{D_1}$ (\ref{W:rhoD1A}), 
which is independent of the chargino polarization, then is
\begin{eqnarray}\label{W:D1A}
D_1^{\mu\nu} \varepsilon_{\mu}^{\lambda_k\ast}\varepsilon_{\nu}^{\lambda'_k}&=&
  D_1 \delta^{\lambda_k\lambda_k'} 
  + \,^cD_1(J^c)^{\lambda_k\lambda_k'}
  + \,^{cd}D_1(J^{cd})^{\lambda_k\lambda_k'},
\end{eqnarray}
%summed over $c,d$, 
with
\begin{eqnarray}\label{W:D1}
	D_1&=&{\textstyle \frac{1}{6}}g^2(|O^{R}_{ni}|^2+|O^{L}_{ni}|^2)
	\Big[m_{\chi^+_i}^2+m_{\chi^0_n}^2-2 m_W^2
		+\frac{(m_{\chi^+_i}^2-m_{\chi^0_n}^2)^2}{m_W^2}\Big]
	\nonumber\\
&&-2g^2Re( O^{R\ast}_{ni}O^{L}_{ni})m_{\chi^+_i}m_{\chi^0_n},\\
^{c}D_1&=&\,^{\;\,+}_{(-)}g^2(|O^{R}_{ni}|^2-|O^{L}_{ni}|^2)m_W(t^c_W\cdot
         	p_{\chi^+_i}),\label{W:cD1}\\
^{cd}D_1&=&-g^2 (|O^{R}_{ni}|^2+|O^{L}_{ni}|^2)
			\left[(t^c_W\cdot p_{\chi^+_i})(t^d_W\cdot p_{\chi^+_i})+
	{\textstyle \frac{1}{4}}(m_{\chi^+_i}^2+m_{\chi^0_n}^2-m_W^2)
		\delta^{cd}\right]\nonumber \\
&& + g^2Re( O^{R\ast}_{ni}O^{L}_{ni})m_{\chi^+_i}m_{\chi^0_n}
		\delta^{cd}.\label{W:cdD1}
\end{eqnarray}
%where the sign in parenthesis holds for the conjugated process 
%$\tilde\chi^-_i\to W^-\tilde\chi^0_n$.
As a consequence of the completeness relation~(\ref{completeness}), 
the diagonal coefficients are linearly dependent
\begin{eqnarray}
	^{11}D_1+\,^{22}D_1+\,^{33}D_1&=&-{\textstyle \frac{3}{2}}D_1.
\end{eqnarray}
For large chargino momentum ${\mathbf p}_{\chi^+_i}$, the $W$ boson will
mainly be emitted into the forward direction with respect to
${\mathbf p}_{\chi^+_i}$, i.e. 
$\hat {\mathbf p}_{\chi^+_i}\approx\hat {\mathbf p}_{W}$, 
with $\hat {\mathbf p}={\mathbf p}/|{\mathbf p}|$.
Therefore, for high energies we have 
$(t^{1,2}_W\cdot p_{\chi^+_i}) \approx 0$ in~(\ref{W:cdD1}),
and in $^{11}D_1\approx \,^{22}D_1$.
%and the contributions for the
%non-diagonal coefficients $^{cd}D_1 (c \slashed{=}d)$
%will be small.

For the second term of $\rho_{D_1}$~(\ref{W:rhoD1A}), 
which depends on the polarization of the decaying chargino,
we obtain
\begin{eqnarray}
\Sigma^{a~\mu\nu}_{D_1}\label{W:SigmaD1A}
\varepsilon_{\mu}^{\lambda_k\ast}\varepsilon_{\nu}^{\lambda'_k}&=&
  \Sigma^{a}_{D_1}\delta^{\lambda_k\lambda_k'}
  +\,^c\Sigma^{a}_{D_1}(J^c)^{\lambda_k\lambda_k'}
  + \,^{cd}\Sigma^{a}_{D_1}(J^{cd})^{\lambda_k\lambda_k'},
\end{eqnarray}
%summed over $c$, $d$, 
with
\begin{eqnarray}
\Sigma^{a}_{D_1}&=&\,^{\;\,+}_{(-)}
	{\textstyle \frac{2}{3}}g^2 (|O^{R}_{ni}|^2-|O^{L}_{ni}|^2)
		m_{\chi^+_i}(s^a_{\chi^+_i}\cdot p_W)[
			\frac{m_{\chi^+_i}^2-m_{\chi^0_n}^2}{2 m_W^2}-1],
		\label{W:SigmaaD1}\\
^c\Sigma^{a}_{D_1}&=& \frac{g^2}{m_W}
		\left[(|O^{R}_{ni}|^2+|O^{L}_{ni}|^2)m_{\chi^+_i}
				-2Re(O^{R\ast}_{ni}O^{L}_{ni})m_{\chi^0_n}
		\right]
		\times\nonumber\\&&
		\left[(t^c_W\cdot p_{\chi^+_i})(s^a_{\chi^+_i}\cdot p_W)
			+{\textstyle \frac{1}{2}}(t^c_W\cdot s^a_{\chi^+_i})
			(m^2_{\chi^0_n}-m^2_{\chi^+_i}+m_W^2)
		\right]\nonumber\\
		&&+\frac{2g^2}{m_W}Im(O^{R\ast}_{ni}O^{L}_{ni})
		m_{\chi^0_n}\epsilon_{\mu\nu\rho\sigma}~
		s^{a,\mu}_{\chi^+_i}~p^{\nu}_{\chi^+_i}~
		p^{\rho}_W~ t^{c,\sigma}_W, \label{W:csigmaaD1}\\
^{cd}\Sigma^{a}_{D_1}&=&\,^{\;\,+}_{(-)}{\textstyle \frac{1}{2}}g^2
		(|O^{R}_{ni}|^2-|O^{L}_{ni}|^2)m_{\chi^+_i} \times \nonumber\\
&&\left[(s^a_{\chi^+_i}\cdot p_W)\delta^{cd}
			-(t^c_W\cdot p_{\chi^+_i})(t^d_W\cdot s^a_{\chi^+_i})
			-(t^d_W\cdot p_{\chi^+_i})(t^c_W\cdot s^a_{\chi^+_i})
			\right].
\end{eqnarray}
%where the signs in parenthesis hold for the conjugated process 
%$\tilde\chi^-_i\to W^-\tilde\chi^0_n $. 
Inserting~(\ref{W:D1A}) and~(\ref{W:SigmaD1A})
into~(\ref{W:rhoD1A}), we obtain the expansion of
the chargino decay matrix in the scalar (first term),
vector (second term) and tensor part (third term):
\begin{eqnarray}
\rho_{D_1}(\tilde\chi^+_i)_{\lambda_i' \lambda_i}^{\lambda_k\lambda'_k} &=& 
(    \delta_{\lambda_i' \lambda_i}~~ D_1+
	\sigma^a_{\lambda_i' \lambda_i}~~\Sigma^{a}_{D_1}
	)~\delta^{\lambda_k\lambda_k'}+ 
\nonumber \\  &&
(   \delta_{\lambda_i' \lambda_i} ~~^{c}D_1+
	\sigma^a_{\lambda_i' \lambda_i}~~^c\Sigma^{a}_{D_1}
	)~(J^c)^{\lambda_k\lambda_k'}+ \nonumber \\  &&
(\delta_{\lambda_i' \lambda_i} ~^{cd}D_1+
	\sigma^a_{\lambda_i' \lambda_i}~^{cd}\Sigma^{a}_{D_1}
	)~(J^{cd})^{\lambda_k\lambda_k'}.\label{W:rhoD1expanded}
\end{eqnarray}
%summed over $c$, $d$,
% which is the expansion in the scalar (first term),
%vector (second term) and tensor part (third term).

A similar expansion for 
the $W$ decay matrix~(\ref{W:rhoD2A}), results in
\begin{eqnarray}\label{W:rhoD2expanded}
	 \rho_{D_2}(W^+)_{\lambda'_k\lambda_k}&=&   
D_2 ~\delta^{\lambda_k'\lambda_k} 
  + \,^cD_2~(J^c)^{\lambda_k'\lambda_k}
  + \,^{cd}D_2~(J^{cd})^{\lambda_k'\lambda_k},
\end{eqnarray}
%where we sum over $c$, $d$, 
with 
\begin{eqnarray}
D_2&=&{\textstyle\frac{1}{3}}g^2 m_W^2,\label{W:D2}\\
^{c}D_2&=&\,^{\;\,-}_{(+)}g^2 m_W(t^c_W\cdot p_{\bar f}),\label{W:cD2}\\
^{cd}D_2&=&g^2 \left[(t^c_W\cdot p_{\bar f})(t^d_W\cdot p_{\bar f})-
	{\textstyle \frac{1}{4}}m_W^2\delta^{cd}\right].\label{W:cdD2}
\end{eqnarray}
%where the sign in parenthesis holds for the conjugated process 
%$W^- \to \bar f^{'} f$. 
%As a consequence of the completeness relation (\ref{completeness}), 
The diagonal coefficients are linearly dependent
\begin{eqnarray}
	^{11}D_2+\,^{22}D_2+\,^{33}D_2&=&-{\textstyle \frac{3}{2}}D_2.
\end{eqnarray}
%For large three-momentum $p_{W}$, the fermion $\bar f$ will
%mainly be emitted into the forward direction with respect to
%$p_{W}$, i.e. $\hat p_{W}\approx\hat p_{\bar f}$, so that 
%$(t^{1,2}_W\cdot p_{\bar f}) \approx 0$ in 
%Eq.~(\ref{cdD2}). Therefore, for high energies
%$^{11}D_2\approx \,^{22}D_2$, and the contributions for the
%non-diagonal coefficients $^{cd}D_2 (c \slashed{=}d)$ will be small.

	\chapter{Neutralino and Chargino two-body decay widths
	\label{Chargino and neutralino decay widths}}

%\section{Two-body decay  widths 
%     \label{Two-body decay  widths}}

For the two-body decay of a massive particle in its
rest frame
\begin{eqnarray} \label{sampledecay}
a\to b+c
\end{eqnarray}  
the decay width of particle $a$ is 
\begin{eqnarray} 
	\Gamma(a\to b~c) = \frac{|{\bf p}_b|}{32~ \pi^2~ m_a^2}
				\int |T|^2 d\Omega =
				\frac{\sqrt{\lambda(m_a^2,m_b^2,m_c^2)}}{16 ~\pi~
					m_a^3}|T|^2,
\end{eqnarray}  
with $\lambda(x,y,z) = x^2+y^2+z^2-2(xy+xz+yz)$
and $|T|^2$ the amplitude squared for decay~(\ref{sampledecay}),
where we average over the spins of particle $a$ and
sum over the spins of particles $b$,$c$.

\section{Neutralino decay widths 
     \label{Neutralino decay widths}}

We give the tree-level formulae for the neutralino 
two-body decay widths $\Gamma_{\chi_i^0}$ for the decays
\begin{eqnarray}
	\tilde\chi^0_i &\to& \tilde e_{R,L} e,~ 
	\tilde \mu_{R,L}\mu,~
	\tilde\tau_{m}\tau,~
	\tilde\nu_{\ell} \bar\nu_{\ell},~
	\tilde\chi^0_n Z,~
	\tilde\chi^{\mp}_m W^{\pm},~
	\tilde\chi^0_n H_1^0;~
	\ell=e,\mu,\tau; ~ m=1,2.
%	~ n<i
\end{eqnarray}

  \begin{itemize}

\item{Neutralino decay into right selectrons or smuons:
		$\tilde\chi^0_i \to\tilde \ell_{R}^+ +\ell^-;~\ell=e,\mu$
\begin{eqnarray}
|T|^2(\tilde\chi^0_i \to\tilde \ell_{R}^+ \ell^-)&=&
\frac{g^2}{2}|f^R_{\ell i}|^2(m_{\chi^0_i}^2-m_{\tilde\ell_R}^2),\\[3mm]
\Gamma(\tilde\chi^0_i \to\tilde \ell_{R}^+ \ell^-)&=&
\frac{(m_{\chi^0_i}^2-m_{\tilde\ell_R}^2)^2}{32~\pi~
	m_{\chi^0_i}^3}g^2|f^R_{\ell i}|^2.
\end{eqnarray}
}

\item{Neutralino decay into left selectrons or smuons:
		$\tilde\chi^0_i \to\tilde \ell_{L}^+ +\ell^-;~\ell=e,\mu$
\begin{eqnarray}
|T|^2(\tilde\chi^0_i \to\tilde \ell_{L}^+ \ell^-)&=&
\frac{g^2}{2}|f^L_{\ell i}|^2(m_{\chi^0_i}^2-m_{\tilde\ell_L}^2),\\[3mm]
\Gamma(\tilde\chi^0_i \to\tilde \ell_{L}^+ \ell^-)&=&
\frac{(m_{\chi^0_i}^2-m_{\tilde\ell_L}^2)^2}{32~\pi~
	m_{\chi^0_i}^3}g^2|f^L_{\ell i}|^2.
\end{eqnarray}
}

\item{Neutralino decay into staus: 
	$\tilde\chi^0_i \to\tilde  \tau_m^+ +\tau^-;~m=1,2$
\begin{eqnarray}
|T|^2(\tilde\chi^0_i \to\tilde  \tau_m^+ \tau^-)&=&
	\frac{g^2}{2}(|a^{\tilde\tau}_{m i}|^2+|b^{\tilde\tau}_{m i}|^2 )
	(m_{\chi^0_i}^2-m_{\tilde\tau_m}^2),\\[3mm]
\Gamma(\tilde\chi^0_i \to\tilde  \tau_m^+ \tau^-)&=&
\frac{(m_{\chi^0_i}^2-m_{\tilde\tau_m}^2)^2}{32~\pi~
	m_{\chi^0_i}^3}g^2(|a^{\tilde\tau}_{m i}|^2+|b^{\tilde\tau}_{m i}|^2 ).
\end{eqnarray}
}

\item{Neutralino decay into sneutrinos: 
	$\tilde\chi^0_i \to\tilde\nu_{\ell}+\bar\nu_{\ell};~\ell=e,\mu,\tau$
\begin{eqnarray}
|T|^2(\tilde\chi^0_i \to\tilde \nu_{\ell}\bar\nu_{\ell} )&=&
	\frac{g^2}{2}|f^{L}_{\nu i}|^2
	(m_{\chi^0_i}^2-m_{\tilde\nu_{\ell}}^2),\\[3mm]
\Gamma(\tilde\chi^0_i \to\tilde \nu_{\ell}\bar\nu_{\ell})&=&
\frac{(m_{\chi^0_i}^2-m_{\tilde\nu_{\ell}}^2)^2}{32~\pi~
	m_{\chi^0_i}^3}g^2|f^{L}_{\nu i}|^2.
\end{eqnarray}
}

\item{Neutralino decay into $Z$ boson: $\tilde\chi^0_i \to Z+\tilde\chi^0_n$
	\begin{eqnarray}
		\lefteqn{
			|T|^2(\tilde\chi^0_i \to Z\tilde\chi^0_n) =
	\frac{g^2}{\cos^2\theta_W}\bigg\{
		6~m_{\chi^0_i}m_{\chi^0_n}
		[(Re O^{''L}_{ni})^2 - (Im O^{''L}_{ni})^2]~+}
	\nonumber\\[3mm]&&
	+~|O^{''L}_{ni}|^2\Big[m_{\chi^0_i}^2 + m_{\chi^0_n}^2 
		-2m_{Z}^2+\frac{(m_{\chi^0_i}^2 - m_{\chi^0_n}^2)^2}{m_{Z}^2}\Big]
	\bigg\} ,\\[3mm]
&&\Gamma(\tilde\chi^0_i \to Z \tilde\chi^0_n) = 
\frac{\sqrt{\lambda(m_{\chi^0_i}^2,m_{\chi^0_n}^2,m_{Z}^2)}}{16 ~\pi~
					m_{\chi^0_i}^3}|T|^2(\tilde\chi^0_i \to Z\tilde\chi^0_n).
\end{eqnarray}
	}
\item{Neutralino decay into $W$ boson: $\tilde\chi^0_i \to W^++\tilde\chi^-_j$
	\begin{eqnarray}
		\lefteqn{
			|T|^2(\tilde\chi^0_i \to W^+\tilde\chi^-_j) =
			-6~g^2m_{\chi^0_i}m_{\chi^-_j} Re(O^{R \ast}_{ij} O^{L}_{ij})~+}
	\nonumber\\[3mm]&&
		+~\frac{g^2}{2} ( |O^{R}_{ij}|^2+|O^{L}_{ij}|^2)
		 \Big[m_{\chi^0_i}^2 + m_{\chi^-_j}^2
		-2m_{W}^2+\frac{(m_{\chi^0_i}^2 - m_{\chi^-_j}^2)^2}{m_{W}^2}\Big]
	 ,\\[3mm]
&&\Gamma(\tilde\chi^0_i \to W^+\tilde\chi^-_j) = 
\frac{\sqrt{\lambda(m_{\chi^0_i}^2,m_{\chi^-_j}^2,m_{W}^2)}}{16 ~\pi~
					m_{\chi^0_i}^3}|T|^2(\tilde\chi^0_i \to W^+\tilde\chi^-_j).
\end{eqnarray}
	}
	\item{Neutralino decay into Higgs boson: 
	$\tilde\chi^0_i\to H_1^0+\tilde\chi^0_n$
	\begin{eqnarray}
	|T|^2(\tilde\chi^0_i\to H_1^0\tilde\chi^0_n)&=&
	2g^2m_{\chi^0_i}m_{\chi^0_n}[ Re( H^{L}_{ni})Re( H^{R}_{ni}) 
			+Im( H^{L}_{ni})Im( H^{R}_{ni})]~+
	\nonumber\\[3mm]&&
	\frac{g^2}{2}(m_{\chi^0_i}^2+m_{\chi^0_n}^2-m_{H_1^0}^2)
		( |H^{L}_{ni}|^2 + |H^{R}_{ni}|^2),\\[3mm]
	\Gamma(\tilde\chi^0_i\to H_1^0\tilde\chi^0_n) &=& 
\frac{\sqrt{\lambda(m_{\chi^0_i}^2,m_{\chi^0_n}^2,m_{H_1^0}^2)}}{16
	~\pi~m_{\chi^0_i}^3}|T|^2(\tilde\chi^0_i\to H_1^0\tilde\chi^0_n),
\end{eqnarray}
with $H^{L}_{ij}=Q^{''\ast}_{ij}\cos\alpha-S^{''\ast}_{ij}\sin\alpha$,
$H^{R}_{ij}=H^{ L\ast}_{ij}$ and
\begin{eqnarray}
	Q^{''}_{ij}&=&\frac{1}{2\cos\theta_W}\left[
		 ( N_{i3}\cos\beta +N_{i4}\sin\beta )N_{j2}
		+( N_{j3}\cos\beta +N_{j4}\sin\beta )N_{i2}\right],\\
	S^{''}_{ij}&=&\frac{1}{2\cos\theta_W}\left[
		 ( N_{i4}\cos\beta -N_{i3}\sin\beta )N_{j2}
		+( N_{j4}\cos\beta -N_{j3}\sin\beta )N_{i2}\right].
\end{eqnarray}
%and $\alpha$ Higgs mixing angle.
}
The Higgs mixing angle $\alpha$ for small $\tan \beta $ can be 
obtained approximately by diagonalization of the neutral Higgs 
mass matrix
\begin{equation}
M^H =
\left(\begin{array}{cc}
		m_Z^2\cos^2\beta +m_A^2\sin ^2\beta  &
		\qquad- (m_Z^2 +m_A^2)\cos\beta\sin \beta \\[5mm]
				- (m_Z^2 +m_A^2)\cos\beta\sin \beta& 
		\qquad~m_A^2\cos^2\beta +m_Z^2\sin ^2\beta  +\delta_t
		\end{array}\right),
\end{equation}
which includes the largest term (top-loop) of the one-loop
radiative corrections
\begin{eqnarray}
	\delta_t &=& \frac{3 g^2 m_t^4}{16~\pi^2m_W^2 \sin^2\beta}
	\log \left(\frac{m_{\tilde t_1}^2 m_{\tilde t_2}^2 }{m_t^4}\right).
\end{eqnarray}
We obtain for the Higgs masses
\begin{eqnarray}
	(m_{H_1^0})^2&=&
	\frac{1}{2}\left[
		M^H_{11}+M^H_{22}
		-\sqrt{(M^H_{11}-M^H_{22})^2 +4~(M^H_{12})^2}~\right],
	\label{Higgsm1}\\
(m_{H_2^0})^2&=&
	\frac{1}{2}\left[
		M^H_{11}+M^H_{22}
		+\sqrt{(M^H_{11}-M^H_{22})^2 +4~(M^H_{12})^2}~
	\right].\label{Higgsm2}
\end{eqnarray}
For the mixing angle we obtain
\begin{eqnarray}\label{Higgsmixcos}
	\cos\alpha&=&
	\frac{-M^H_{12}}{\sqrt{(M^H_{12})^2
			+[M^H_{11}-(m_{H_1^0})^2]^2~}~},\\
\sin\alpha&=&
	\frac{M^H_{11}-(m_{H_1^0})^2}{\sqrt{(M^H_{12})^2
			+[M^H_{11}-(m_{H_1^0})^2]^2~}~}.\label{Higgsmixsin}
\end{eqnarray}
If we choose a large Higgs mass parameter, e.g.  $m_{A}=1$~TeV,
we have very approximately 
$m_{H_1^0}\approx 115-130$~GeV and 
$m_{H_2^0}\approx M_A$, which follows from (\ref{Higgsm1}) 
and (\ref{Higgsm2}).
In addition,  explicit CP violation is not relevant
for the lightest Higgs state \cite{ref3}.

\end{itemize}

\section{Chargino decay widths 
     \label{Chargino decay widths}}

We give the tree-level formulae for the chargino 
two-body decay widths $\Gamma_{\chi_i^+}$ for the decays
\begin{eqnarray}
\tilde\chi^+_i &\to& 
	W^+\tilde\chi^0_n,~
	\tilde e_{L}^+\nu_{e},~
	\tilde\mu_{L}^+\nu_{\mu},~
	\tilde\tau_{1,2}^+\nu_{\tau},~
	e^+\tilde\nu_{e},~
	\mu^+\tilde\nu_{\mu},~
	\tau^+\tilde\nu_{\tau}.
\end{eqnarray}
For the heavy chargino $\tilde\chi_2^+$ also the decays
into the lightest neutral Higgs boson $H_1^0$ and the $Z$ boson
are possible
\begin{eqnarray}
	\tilde\chi^+_2 &\to&
	\tilde\chi^+_1 Z,~
	\tilde\chi^+_1 H_1^0.
\end{eqnarray}
%are possible, where $H_1^0$ denotes the lightest neutral Higgs boson.
\begin{itemize}

\item{Chargino decay into $W$ boson: $\tilde\chi^+_i \to W^+ +\tilde\chi^0_n$
	\begin{eqnarray}
	|T|^2(\tilde\chi^+_i \to W^+ \tilde\chi^0_n) &=&
	 \frac{g^2}{2}(|O^{R}_{ni}|^2+|O^{L}_{ni}|^2)
	\Big[m_{\chi^+_i}^2+m_{\chi^0_n}^2-2 m_W^2
		+\frac{(m_{\chi^+_i}^2-m_{\chi^0_n}^2)^2}{m_W^2}\Big]
	\nonumber\\[3mm]&&
	-6g^2m_{\chi^+_i}m_{\chi^0_n}Re( O^{R\ast}_{ni}O^{L}_{ni}),\\[3mm]
\Gamma(\tilde\chi^+_i \to W^+ \tilde\chi^0_n) &=& 
\frac{\sqrt{\lambda(m_{\chi^+_i}^2,m_{\chi^0_n}^2,m_{W}^2)}}{16~ \pi~
					m_{\chi^+_i}^3}|T|^2(\tilde\chi^+_i \to W^+ \tilde\chi^0_n).
\end{eqnarray}
	}

\item{Chargino decay into selectrons or smuons:
		$\tilde\chi^+_i \to\tilde \ell_{L}^+ +\nu_{\ell};~\ell=e,\mu$
\begin{eqnarray}
|T|^2(\tilde\chi^+_i \to\tilde \ell_{L}^+ \nu_{\ell})&=&
	\frac{g^2}{2}|U_{i1}|^2(m_{\chi^+_i}^2-m_{\tilde\ell}^2),\\[3mm]
\Gamma(\tilde\chi^+_i \to\tilde \ell_{L}^+ \nu_{\ell})&=&
\frac{(m_{\chi^+_i}^2-m_{\tilde\ell}^2)^2}{32~\pi~
	m_{\chi^+_i}^3}g^2|U_{i1}|^2.
\end{eqnarray}
}

\item{Chargino decay into staus:
		$\tilde\chi^+_i \to \tilde \tau_{m}^+ +\nu_{\tau};~m=1,2$
\begin{eqnarray}
|T|^2(\tilde\chi^+_i \to \tilde \tau_{m}^+ \nu_{\tau})&=&
\frac{g^2}{2}|\ell_{mi}^{\tilde \tau}|^2(m_{\chi^+_i}^2-m_{\tilde\tau_m}^2),\\[3mm]
\Gamma(\tilde\chi^+_i \to\tilde \tau_{m}^+ \nu_{\tau})&=&
\frac{(m_{\chi^+_i}^2-m_{\tilde\tau_m}^2)^2}{32~\pi~
	m_{\chi^+_i}^3}g^2|\ell_{mi}^{\tilde \tau} |^2,
\end{eqnarray}
and $\ell_{mi}^{\tilde \tau}$ defined in~(\ref{eq:coupl1}). 
}

\item{Chargino decay into electron or muon sneutrinos:
		$\tilde\chi^+_i \to \ell^+ +\tilde\nu_{\ell};~\ell=e,\mu$
\begin{eqnarray}
|T|^2(\tilde\chi^+_i \to \ell^+ \tilde\nu_{\ell})&=&
\frac{g^2}{2}|V_{i1}|^2(m_{\chi^+_i}^2-m_{\tilde\nu}^2),\\[3mm]
\Gamma(\tilde\chi^+_i \to \ell^+ \tilde\nu_{\ell})&=&
\frac{(m_{\chi^+_i}^2-m_{\tilde\nu}^2)^2}{32~\pi~
	m_{\chi^+_i}^3}g^2|V_{i1} |^2.
\end{eqnarray}
%and $\ell_{mi}^{\tilde \tau}$ defined in~(\ref{eq:coupl1}). 
}
\item{Chargino decay into tau sneutrino:
		$\tilde\chi^+_i \to \tau^+ +\tilde\nu_{\tau}$
\begin{eqnarray}
|T|^2(\tilde\chi^+_i \to  \tau^+ \tilde\nu_{\tau})&=&
\frac{g^2}{2}(|V_{i1}|^2 +Y_{\tau}^2|U_{i2}|^2)(m_{\chi^+_i}^2-m_{\tilde\nu}^2),\\[3mm]
\Gamma(\tilde\chi^+_i \to  \tau^+ \tilde\nu_{\tau})&=&
\frac{(m_{\chi^+_i}^2-m_{\tilde\nu}^2)^2}{32~\pi~
	m_{\chi^+_i}^3}g^2(|V_{i1}|^2 +Y_{\tau}^2|U_{i2}|^2),
\end{eqnarray}
and $Y_{\tau}$ defined in~(\ref{eq:coupl4}). 
}
\item{Chargino decay into $Z$ boson: $\tilde\chi^+_2 \to Z+\tilde\chi^+_1$
	\begin{eqnarray}
		\lefteqn{
			|T|^2(\tilde\chi^+_2 \to Z \tilde\chi^+_1) =
	\frac{g^2}{\cos^2\theta_W}\bigg\{
		-6~m_{\chi^+_1}m_{\chi^+_2}Re( O^{R\ast}_{12}O^{L}_{12})~+}
	\nonumber\\[3mm]&&
	+~\frac{1}{2}(|O^{R}_{12}|^2+|O^{L}_{12}|^2)
	\Big[m_{\chi^+_2}^2+m_{\chi^+_1}^2-2 m_Z^2
		+\frac{(m_{\chi^+_2}^2-m_{\chi^+_1}^2)^2}{m_Z^2}\Big]\bigg\} ,\\[3mm]
&&\Gamma(\tilde\chi^+_2 \to Z \tilde\chi^+_1) = 
\frac{\sqrt{\lambda(m_{\chi^+_2}^2,m_{\chi^+_1}^2,m_{Z}^2)}}{16 ~\pi~
					m_{\chi^+_2}^3}|T|^2(\tilde\chi^+_2 \to Z \tilde\chi^+_1).
\end{eqnarray}
	}
\item{Chargino decay into Higgs boson: $\tilde\chi^+_2\to H_1^0+\tilde\chi^+_1$
	\begin{eqnarray}
	|T|^2(\tilde\chi^+_2\to H_1^0\tilde\chi^+_1)&=&
	2g^2m_{\chi^+_1}m_{\chi^+_2}[ Re( F^{L}_{12})Re( F^{R}_{12}) 
			+Im( F^{L}_{12})Im( F^{R}_{12})]~+
	\nonumber\\[3mm]&&
	\frac{g^2}{2}(m_{\chi^+_1}^2+m_{\chi^+_2}^2-m_{H_1^0}^2)
		( |F^{L}_{12}|^2 + |F^{R}_{12}|^2),\\[3mm]
	\Gamma(\tilde\chi^+_2 \to H_1^0\tilde\chi^+_1) &=& 
\frac{\sqrt{\lambda(m_{\chi^+_2}^2,m_{\chi^+_1}^2,m_{H_1^0}^2)}}{16
	~\pi~
					m_{\chi^+_2}^3}|T|^2(\tilde\chi^+_2 \to H_1^0\tilde\chi^+_1),
\end{eqnarray}
with $F^{L}_{ij}=\frac{1}{\sqrt 2}(U_{i2}^{\ast}V_{j1}^{\ast}\sin\alpha 
	-U_{i1}^{\ast}V_{j2}^{\ast}\cos\alpha)$ and
$F^{R}_{ij}=F^{L\ast}_{ji}$.
%and $\alpha$ Higgs mixing angle.
}
\end{itemize}
The Higgs mixing angle $\alpha$ is given in~(\ref{Higgsmixcos})
and~(\ref{Higgsmixsin}).

	\chapter{Spin formalism for fermions and bosons
	\label{Spin formalism for fermions and bosons}}

\section{Bouchiat-Michel formulae for spin $\frac{1}{2}$ particles
	\label{Bouchiat-Michel formulae for spin 1/2 particles}}

For the calculation of cross sections we expand the 
spin density matrices in terms of the Pauli
matrices, see e.g.~(\ref{neut:amplitude}), (\ref{neut:rhoP}) 
for neutralinos. This expansion is straight forward
if for the neutralinos or charginos
a set of spin-basis vectors $s^{a,\mu}$ 
has been introduced, see~(\ref{spinvec}).
Together with $\hat p^{\mu}= p^{\mu}/m$ they form an orthonormal set  
\begin{eqnarray}
\hat p\cdot s^a &=& 0,\\
s^a\cdot s^b &=& -\delta^{ab},\\
s^{a}_{\mu} s^{a}_{\nu} &=& -g_{\mu\nu} +\frac{p_{\mu}p_{\nu}}{m^2},
								\quad ({\rm sum~ over}~a).
\end{eqnarray}
The helicity spinors are normalized by
\begin{eqnarray}
	\bar u(p,\lambda)~u(p,\lambda')&=& 2~m~\delta_{\lambda\lambda'},\\
	\bar v(p,\lambda)~v(p,\lambda')&=& -2~m~\delta_{\lambda\lambda'}.
\end{eqnarray}
The Bouchiat-Michel formulae
for massive spin $1/2$ particles are then \cite{Bouchiat-Michel}
\begin{eqnarray}
u(p,\lambda')~\bar u(p,\lambda)&=&\frac{1}{2}~[\delta_{\lambda\lambda'}+
		\gamma_5\slashed s^a \sigma^a_{\lambda\lambda'}](\slashed p+m),\\
v(p,\lambda')~\bar v(p,\lambda)&=&\frac{1}{2}~[\delta_{\lambda'\lambda}+
		\gamma_5\slashed s^a \sigma^a_{\lambda'\lambda}](\slashed p-m),
\quad ({\rm sum~ over}~a).
\end{eqnarray}
%with sum over $a$.
%In calculating products of density matrices, these relations
%are helpful
%\begin{eqnarray}
%	{\rm Tr}\{ \delta ~\sigma^a\}=0, \quad 
%	{\rm Tr}\{ \sigma^a ~\sigma^b\}=2 \delta^{ab}.
%\end{eqnarray}

\section{Spin formulae for spin $1$ particles
	\label{Spin 1 matrices}}

The Bouchiat-Michel formulae for spin $1/2$ particles
can be generalized for higher spins \cite{choiBM}.
In order to describe the polarization states of a spin 1 boson,
we have introduced a set of spin vectors
$t^a_{\mu}$, see~(\ref{defoft}).
Note that they are are not helicity eigenstates like 
the polarization vectors $\varepsilon_{\mu}^{\lambda_k}$,
defined in~(\ref{circularbasis}).
The spin vectors $t^a_{\mu}$ and 
%the unit momentum four-vector of the boson 
$\hat k^{\mu}= k^{\mu}/m$ form an orthonormal set:
\begin{eqnarray}
\hat k\cdot t^a &=& 0,\\
t^a\cdot t^b &=& -\delta^{ab},\\
t^{a}_{\mu} ~t^{a}_{\nu} &=& -g_{\mu\nu} +\frac{k_{\mu}k_{\nu}}{m^2},
							\quad ({\rm sum~ over}~a).
\end{eqnarray}
%The analog to the Pauli matrices $\sigma^a$ are the 
The $3\times3$ spin $1$ matrices $J^c$   
%\begin{eqnarray}
obey $[J^c,J^d]=i\epsilon_{cde}J^e$ and are given below. 
We can define six further matrices 
\begin{eqnarray}
	J^{cd}&=&J^cJ^d+J^dJ^c-{\textstyle \frac{4}{3}}\delta^{cd},
\end{eqnarray}
with $J^{11}+J^{22}+J^{33}=0$. They are the components of a 
symmetric, traceless tensor. % and are also given below.
We now can expand  \cite{choiBM}
\begin{eqnarray}\label{expansion}
\varepsilon_{\mu}^{\lambda_k}\varepsilon_{\nu}^{\lambda'_k\ast}&=&
{\textstyle \frac{1}{3}}\delta^{\lambda_k'\lambda_k}I_{\mu\nu}
-\frac{i}{2m}\epsilon_{\mu\nu\rho\sigma}
p_Z^{\rho}t^{c,\sigma}(J^c)^{\lambda_k'\lambda_k}
-{\textstyle \frac{1}{2}}t_{\mu}^ct_{\nu}^d (J^{cd})^{\lambda_k'\lambda_k},
%\quad(\epsilon_{0123}=1),
\end{eqnarray}
summed over $c,d$. The tensor 
\begin{eqnarray}
I_{\mu\nu}&=&-g_{\mu\nu}+\frac{k_{ \mu}k_{ \nu}}{m^2}
\end{eqnarray}
guarantees the completeness relation of the polarization vectors
\begin{eqnarray}\label{completeness}
\sum_{\lambda_k} \varepsilon^{\lambda_k\ast}_{\mu}
\varepsilon^{\lambda_k}_{\nu}&=&
-g_{\mu\nu}+\frac{k_{\mu}k_{\nu}}{m^2}.
\end{eqnarray}
The second term of~(\ref{expansion}) describes the vector
polarization and the third term describes the tensor polarization
of the  boson.

In the linear basis~(\ref{defoft}) the spin-$1$ matrices are defined as
$(J_L^c)^{jk}=-i\epsilon_{cjk}$:
%(linear basis) and
%$J^{cd}_L=J^c_LJ^d_L+J^d_LJ^c_L-{\textstyle \frac{4}{3}}\delta^{cd}$:
 \begin{eqnarray}
  J^1_L=
  \left(
        \begin{array}{rrr}
         0&0&0\\0&0&-i\\0&i&0
        \end{array}
	  \right),
&
  J^2_L=
  \left(
        \begin{array}{rrr}
         0&0&i\\0&0&0\\-i&0&0
        \end{array}
	  \right),
&
  J^3_L=
  \left(
        \begin{array}{rrr}
         0&-i&0\\i&0&0\\0&0&0
        \end{array}
	\right), \label{Jlinear1}\\
%-------------------
	J^{11}_L=
  \left(
        \begin{array}{rrr}
         -\frac{4}{3}&0&0\\0&\frac{2}{3}&0\\0&0&\frac{2}{3}
        \end{array}
	  \right),
&
	J^{22}_L=
  \left(
        \begin{array}{rrr}
         \frac{2}{3}&0&0\\0&-\frac{4}{3}&0\\0&0&\frac{2}{3}
        \end{array}
	  \right),
&
	J^{33}_L=
  \left(
        \begin{array}{rrr}
         \frac{2}{3}&0&0\\0&\frac{2}{3}&0\\0&0&-\frac{4}{3}
        \end{array}
  \right),\label{Jlinear2}\\
%-------------------
J^{12}_L=
%J^{12}=
  \left(
        \begin{array}{rrr}
         0&-1&0\\-1&0&0\\0&0&0
        \end{array}
	  \right),
&
J^{23}_L=
%J^{32}=
  \left(
        \begin{array}{rrr}
         0&0&0\\0&0&-1\\0&-1&0
        \end{array}
	  \right),
&
J^{13}_L=
%J^{31}=
  \left(
        \begin{array}{rrr}
         0&0&-1\\0&0&0\\-1&0&0
        \end{array}
  \right).\label{Jlinear3}
\end{eqnarray}
The matrices $J^c$ and $J^{cd}$ in the circular basis, 
see~(\ref{circularbasis}), are obtained 
by the unitary transformations
$J^{c} = A^{\dagger}\cdot J^c_L \cdot A $ and
$J^{cd} = A^{\dagger} \cdot J^{cd}_L \cdot A $, respectively, with 
\begin{eqnarray}\label{circtrafo}
A&=&\left(
        \begin{array}{rrr}
			  \frac{1}{\sqrt{2}}&0&-\frac{1}{\sqrt{2}}\\
		  -\frac{i}{\sqrt{2}}&0&-\frac{i}{\sqrt{2}}\\0&1&0
        \end{array}
  \right);
  \quad  A^{\dagger}= A^{-1},
\end{eqnarray}
%-------------------
 \begin{eqnarray}
  J^1=
  \left(
        \begin{array}{rrr}
			  0&\frac{1}{\sqrt{2}}&0\\
			  \frac{1}{\sqrt{2}}&0&\frac{1}{\sqrt{2}}\\
			  0&\frac{1}{\sqrt{2}}&0
        \end{array}
	  \right),
&
  J^2=
  \left(
        \begin{array}{rrr}
			  0&\frac{i}{\sqrt{2}}&0\\
			  -\frac{i}{\sqrt{2}}&0&\frac{i}{\sqrt{2}}
			  \\0&-\frac{i}{\sqrt{2}}&0
        \end{array}
	  \right),
&
  J^3=
  \left(
        \begin{array}{rrr}
         -1&0&0\\0&0&0\\0&0&1
        \end{array}
	  \right), \label{Jcircular1}\\
%-------------------
J^{11}=
  \left(
        \begin{array}{rrr}
			  -\frac{1}{3}&0&1\\
			  0&\frac{2}{3}&0\\
			  1&0&-\frac{1}{3}
        \end{array}
	  \right),
&
J^{22}=
\left(
	\begin{array}{rrr}
	  -\frac{1}{3}&0&-1\\
	  0&\frac{2}{3}&0\\
	  -1&0&-\frac{1}{3}
	  \end{array}
	  \right),
	  &
J^{33}=  
\left(
	\begin{array}{rrr}
	  \frac{2}{3}&0&0\\
	  0&-\frac{4}{3}&0\\
	  0&0&\frac{2}{3}
	  \end{array}
  \right),\label{Jcircular2}\\
%----------------
J^{12}=
  \left(
        \begin{array}{rrr}
         0&0&i\\0&0&0\\-i&0&0
        \end{array}
	  \right),
&
J^{23}=
  \left(
        \begin{array}{rrr}
			  0&-\frac{i}{\sqrt{2}}&0\\
			  \frac{i}{\sqrt{2}}&0&\frac{i}{\sqrt{2}}
			  \\0&-\frac{i}{\sqrt{2}}&0
        \end{array}
	  \right),
&
J^{13}=
  \left(
        \begin{array}{rrr}
			  0&-\frac{1}{\sqrt{2}}&0\\
			  -\frac{1}{\sqrt{2}}&0&\frac{1}{\sqrt{2}}\\
			  0&\frac{1}{\sqrt{2}}&0
        \end{array}
	  \right). \label{Jcircular3}
\end{eqnarray}

In calculating products of density matrices, the following relations
are helpful
\begin{eqnarray}
&&	{\rm Tr}\{ J^a\}=0, \quad 
	{\rm Tr}\{ J^{ab}\}=0, \quad 
	{\rm Tr}\{ J^a J^{bc} \}=0, \\ 
&&{\rm Tr}\{ J^a J^b\}=2 \delta^{ab}, \quad 
	{\rm Tr}\{ J^a J^b J^c \} = i\epsilon^{abc}, \quad 
	{\rm Tr}\{ J^a J^b J^c J^d \} = 
	\delta^{ab}\delta^{cd}+\delta^{ad}\delta^{bc}, \\
&&	{\rm Tr}\{ J^{ab}  J^{cd}\}=
	{\textstyle -\frac{4}{3}}\delta^{ab}\delta^{cd}
	+2 \delta^{ad}\delta^{bc} +2 \delta^{ac}\delta^{bd}.
\end{eqnarray}

	\chapter{Definitions and conventions}

We use  natural units $c=1$, $h/2 \pi=1$.

The metric tensor 
\begin{equation}
\begin{array}{ccccc}
(g_{\mu\nu}) & = & (g^{\mu\nu}) & := &
\left( \begin{array}{cccc}
1 & 0 & 0 & 0\\
0 & -1 & 0 & 0\\
0 & 0 & -1 & 0\\
0 & 0 & 0 & -1
\end{array}\right)
\end{array}
\end{equation}
defines scalar products
\begin{equation}
	(a\cdot b):=g_{\mu\nu}a^{\mu} b^{\nu}=a^{\mu} b_{\mu}=
	a^0b^0-{\bf a}{\bf b}
\end{equation}
between covariant and contravariant four-vectors
\begin{equation}
	a^{\mu}:= (a^0,a^1,a^2,a^3)= (a^0,{\bf a}), \quad
	a_{\mu}:= g_{\mu\nu}a^{\nu}=(a_0,a_1,a_2,a_3)= (a_0,-{\bf a}).
\end{equation}
The total antisymmetric $\epsilon$-tensor is defined as
\begin{equation}
\epsilon_{\mu \nu \rho \sigma} = -\epsilon^{\mu \nu \rho \sigma} := 
\left\{ 
\begin{array}{rcl}
+1 & , & \hspace*{.5cm}\mbox{if $\mu \nu \rho \sigma$ is an even
											permutation of } 0123,\\
-1 & , & \hspace*{.5cm}\mbox{if $\mu \nu \rho \sigma$ is an odd
											permutation},\\
0 & , & \hspace*{.5cm}\mbox{if any two indices are the same}.
\end{array}\right. 
\end{equation}
The analog definition in three dimensions, the 
\emph{Levi-Cevita-Tensor} is
\begin{equation}
\epsilon_{ijk} = \epsilon^{ijk} :=\left\{ 
\begin{array}{rcl}
+1 & , & \hspace*{.5cm}\mbox{if $ijk$ is an even
											permutation of } 123,\\
-1 & , & \hspace*{.5cm}\mbox{if $ijk$ is an odd
											permutation},\\
0 & , & \hspace*{.5cm}\mbox{if any two indices are the same}.
\end{array}\right. 
\end{equation}
Useful relations of the $\epsilon$-tensor
\begin{equation}
-\epsilon^{\alpha \beta \mu \nu}\epsilon_{\alpha \beta \rho \sigma}
= 2(\delta^{\mu}_{\rho} \delta^{\nu}_{\sigma} 
	-\delta^{\mu}_{\sigma} \delta^{\nu}_{\rho}),\quad
-\epsilon^{\alpha \beta \mu \nu}\epsilon_{\alpha \beta \mu \rho}
= 6~ \delta^{\nu}_{\rho},\quad
-\epsilon^{\alpha \beta \mu \nu}\epsilon_{\alpha \beta \mu \nu}
=24,
\end{equation}
with $\delta^{\mu}_{\rho} = g^{\mu \nu}g_{\nu\rho}$.

The Pauli matrices are
\begin{equation}
\begin{array}{cccccccccccc}
\sigma_{1} & = & 
\left(\begin{array}{cc}
0 & 1\\
1 & 0 
\end{array}\right) & ,\qquad & \sigma_{2} & = &
\left(\begin{array}{cc}
0 & -i\\
i & 0
\end{array}\right) & ,\qquad & \sigma_{3} & = & 
\left(\begin{array}{cc}
1 & 0\\
0 & -1
\end{array}\right) &.
\end{array}
\end{equation}
The Dirac matrices obey the commutation relations
\begin{equation}
\{ \gamma^{\mu},\gamma^{\nu}\}= \gamma^{\mu}\gamma^{\nu}+
\gamma^{\nu}\gamma^{\mu}=2g^{\mu\nu}.
\end{equation}
In the Dirac representation they read
\begin{equation}
\begin{array}{ccccccc}
\gamma^0 & = & 
\left(\begin{array}{cc}
I & 0\\
0 & -I
\end{array}\right),
\qquad  & \gamma^j & = &
\left(\begin{array}{cc}
0 & \sigma^j\\
-\sigma^j & 0 
\end{array}\right),\qquad  & j=1,2,3,
\end{array}
\end{equation}
and $\gamma_5:= -i\gamma_0\gamma_1\gamma_2\gamma_3 = 
i\gamma^0\gamma^1\gamma^2\gamma^3= \gamma^5$.

Trace theorems:
%$P_{L,R}:=\frac{1}{2}(1\mp \gamma_5)$:
%and $\slashed a$ 
\begin{equation}
\begin{array}{ccc}
	P_{L,R}:=\frac{1}{2}(1\mp \gamma_5),&\slashed
	a:=\gamma^{\mu}a_{\mu},& {\rm Tr}\{P_{L,R} \} =2, \\ [3mm]
	{\rm Tr}\{\slashed a ~P_{L,R} \}=0, &
	{\rm Tr}\{\slashed a \slashed b~P_{L,R} \}=2~(a \cdot b),&
	{\rm Tr}\{\slashed a \slashed b \slashed c ~P_{L,R} \}= 0,\\[3mm]
	\multicolumn{3}{c}{
		{\rm Tr}\{\slashed a \slashed b \slashed c \slashed d ~P_{L,R} \}=
		2~[ (a \cdot b) (c \cdot d) -(a \cdot c) (b \cdot d)
			+(a \cdot d) (b \cdot c)]\mp 2 i [a,b,c,d], 
	}
	\end{array}
\end{equation}
and $[a,b,c,d]:=\epsilon_{\mu \nu\rho\sigma}
a^{\mu} b^{\nu} c^{\rho} d^{\sigma}$.

For numerical calculations we have used the values
%Standard Model parameters: 
%given in table~\ref{Parameters for numerical calculations}.
\begin{eqnarray}
	\begin{array}{ccll}
		\alpha     &=& 1/128 & {\rm fine-structure~constant~at~500~GeV}\\
  \sin^2\theta_W &=& 0.2315& {\rm weak~mixing~angle}\\
					m_W&=& 80.41 ~{\rm GeV}& W ~{\rm boson~mass}\\
			\Gamma_W &=& 2.12 ~{\rm GeV} & W ~{\rm boson~width}\\
					m_Z&=& 91.187~ {\rm GeV} & Z~{\rm boson~mass}\\
			\Gamma_Z &=& 2.49 ~{\rm GeV}& Z ~{\rm boson~width}\\
\end{array}
\end{eqnarray}
%\begin{table}[h]
%	 \caption{Parameters for numerical calculations
%		\label{Parameters for numerical calculations}}
%\begin{tabular}{cccc}
%\hline \\
%$\alpha$&$=$&$1/128$& fine-structure constant at 500 GeV\\
%$\sin^2\theta_W$&=&$0.2315$& weak mixing angle\\
%$m_W$&$=$&$80.41$ GeV & $W$ boson mass\\
%$\Gamma_W$ &$=$& $2.12$ GeV& $W$ boson width\\
%$m_Z$&$=$& $91.187$ GeV & $Z$ boson mass\\
%$\Gamma_Z$ &$=$& $2.49$ GeV& $Z$ boson width\\
%\hline
%\end{tabular}
%\end{table}

\end{appendix}
\addcontentsline{toc}{chapter}{\numberline{}Bibliography}

\end{document}